%% file: 00-main-arxiv.tex
\documentclass[acmtog]{acmart}

\setcitestyle{square}

\input{99-macros}

\setcopyright{acmlicensed}



\title{Polyhedral Discretizations for Elliptic PDEs}

\author{Junyu Liu}
\affiliation{%
  \institution{Brown University}
  \country{USA}
}

\author{Daniele Panozzo}
\affiliation{%
  \institution{New York University}
  \country{USA}
}

\author{Mario Botsch}
\affiliation{%
  \institution{TU Dortmund University}
  \country{Germany}
}

\author{Teseo Schneider}
\affiliation{%
  \institution{University of Victoria}
  \country{Canada}
}

\usepackage[utf8]{inputenc}
\usepackage{amsmath}
\usepackage{amsfonts}
\usepackage{enumitem}

\usepackage{algorithm}

\usepackage{mathrsfs}
\usepackage{subcaption}

\usepackage{siunitx}

\usepackage{graphicx}

\usepackage{xcolor}

\usepackage{float}

\begin{document}

\input{01-abstract}

\maketitle

\input{01-teaser}

\input{01-introduction}
\input{02-related}
\input{04-benchmark}
\input{05-experiments}
\input{051-complex}
\input{06-concluding}

\bibliographystyle{ACM-Reference-Format}
\bibliography{00-main-arxiv}

\appendix
\input{21-related-addiitonal}
\input{42-problems}

\input{41-tables}

\end{document}

%% file: 99-macros.tex
\usepackage{amsmath}
\usepackage{amsfonts}
\usepackage{enumitem}

\usepackage{algorithm}

\usepackage{natbib}
\usepackage{mathrsfs}
\usepackage{subcaption}

\usepackage{siunitx}

\usepackage{graphicx}

\usepackage{hyperref}

\usepackage{xcolor}

\DeclareMathOperator{\tr}{tr}
\DeclareMathOperator{\Div}{div}

%% file: 01-abstract.tex
\begin{abstract}

We study the use of polyhedral discretizations for the solution of heat diffusion and elastodynamic problems in computer graphics. Polyhedral meshes are more natural for certain applications than pure triangular or quadrilateral meshes, which thus received significant interest as an alternative representation.
We consider finite element methods using barycentric coordinates as basis functions and the modern virtual finite element approach. We evaluate them on a suite of classical graphics problems to understand their benefits and limitations compared to standard techniques on simplicial discretizations. 
Our analysis provides recommendations and a benchmark for developing polyhedral meshing techniques and corresponding analysis techniques.

\end{abstract}

%% file: 01-teaser.tex
\begin{figure}
\centering\footnotesize
\parbox{.74\linewidth} {\centering
Voronoi mesh\\
\includegraphics[width=.24\linewidth]{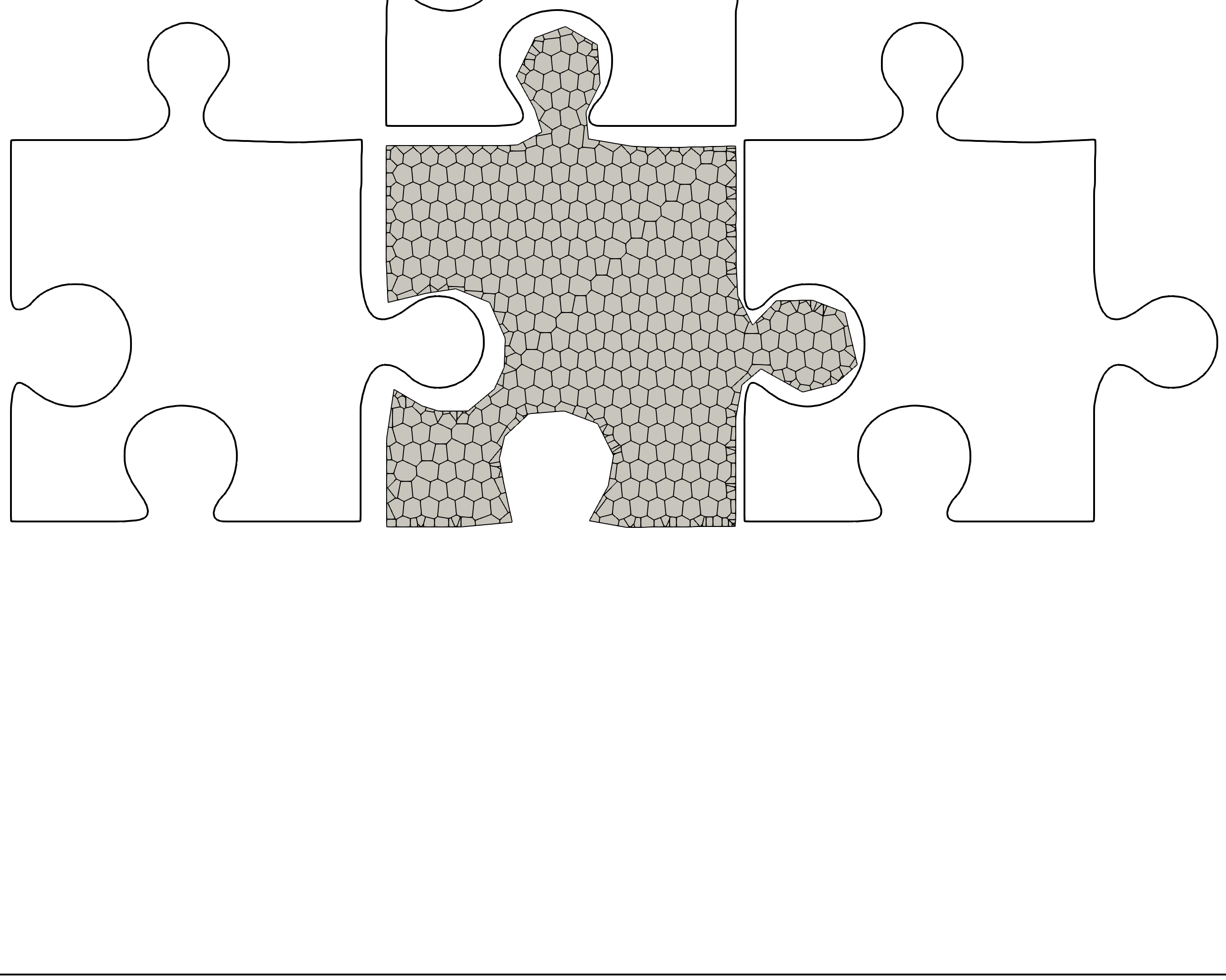}\hfill
\includegraphics[width=.24\linewidth]{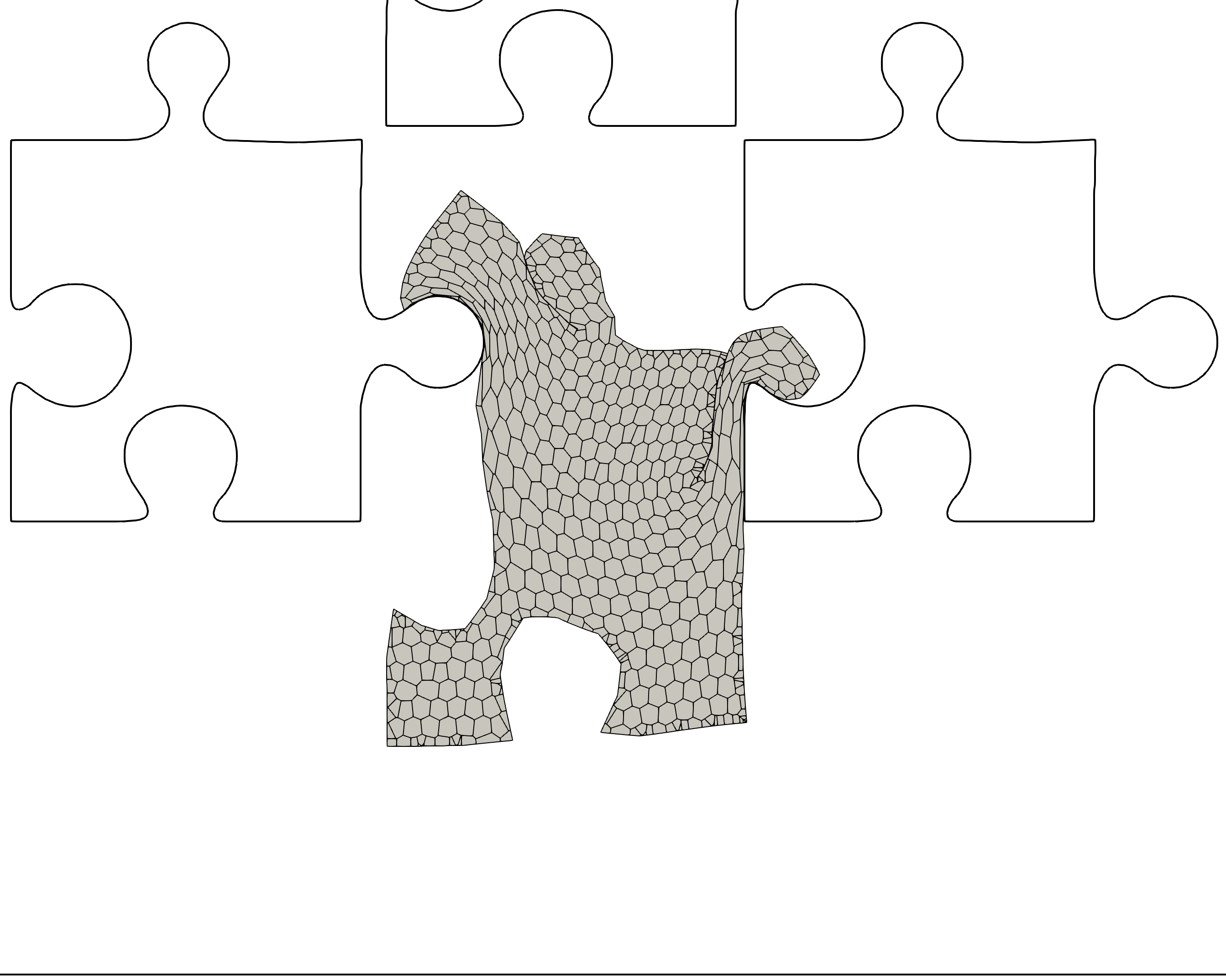}\hfill
\includegraphics[width=.24\linewidth]{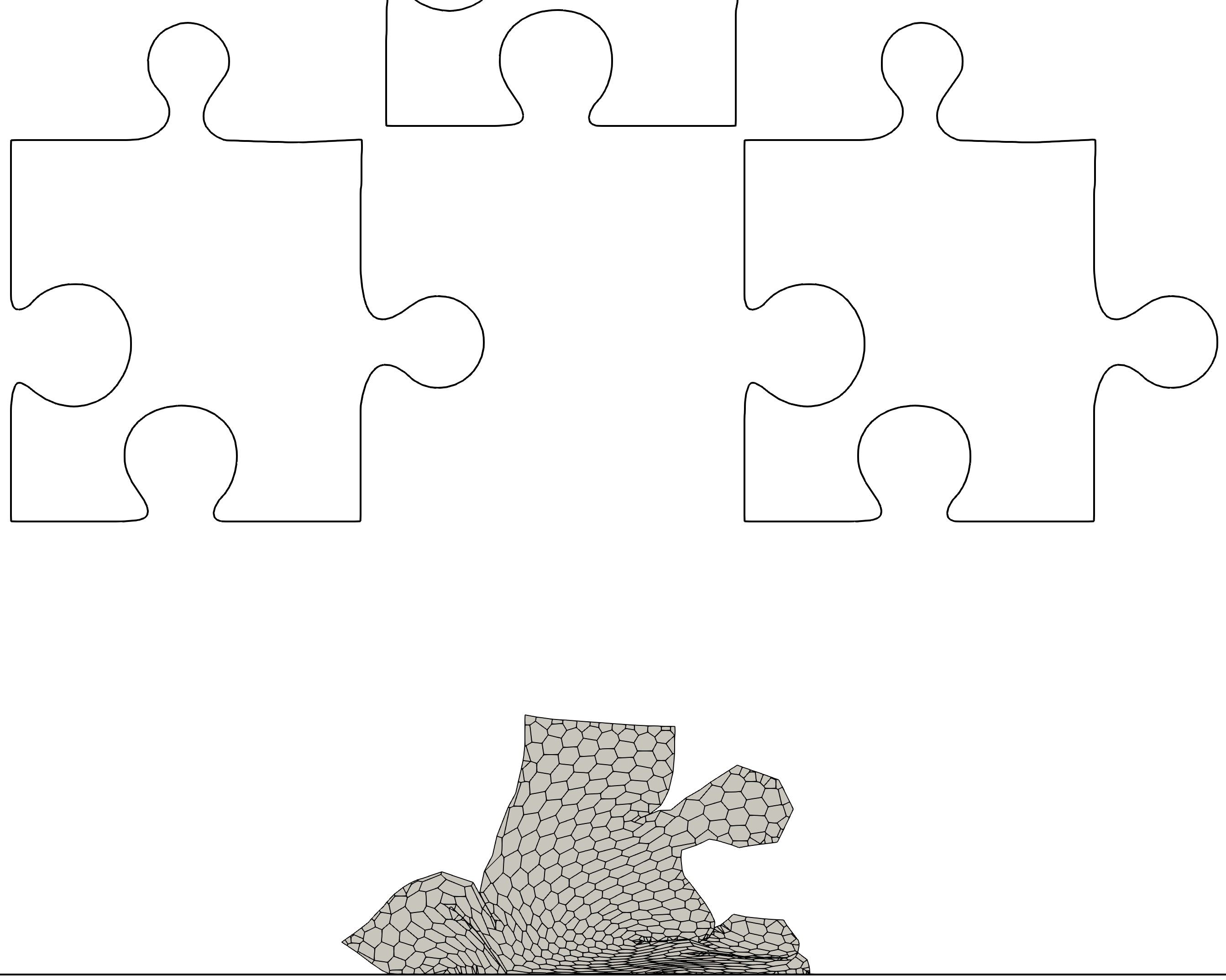}\hfill
\includegraphics[width=.24\linewidth]{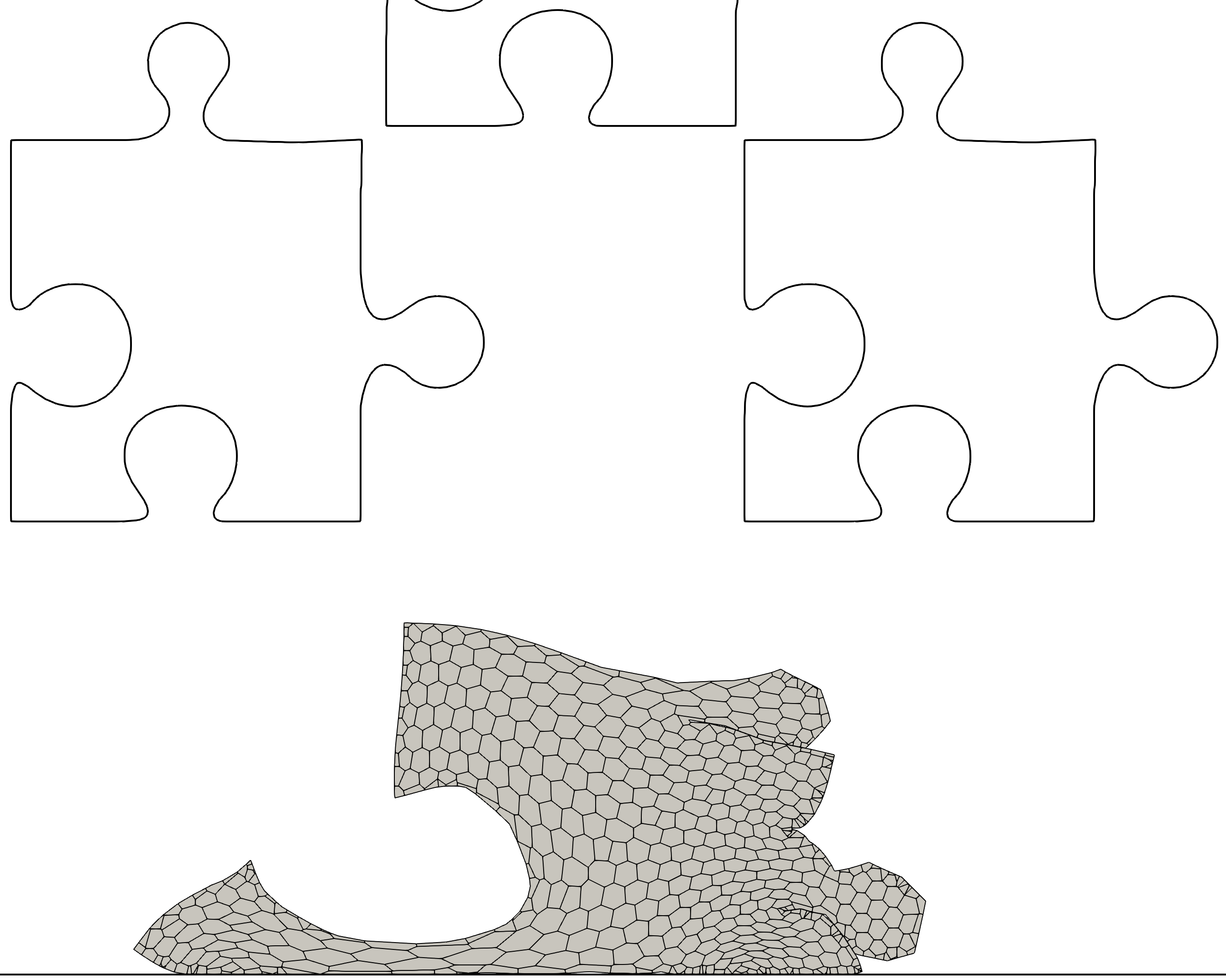}\\[1em]
Tiled mesh\\
\includegraphics[width=.24\linewidth]{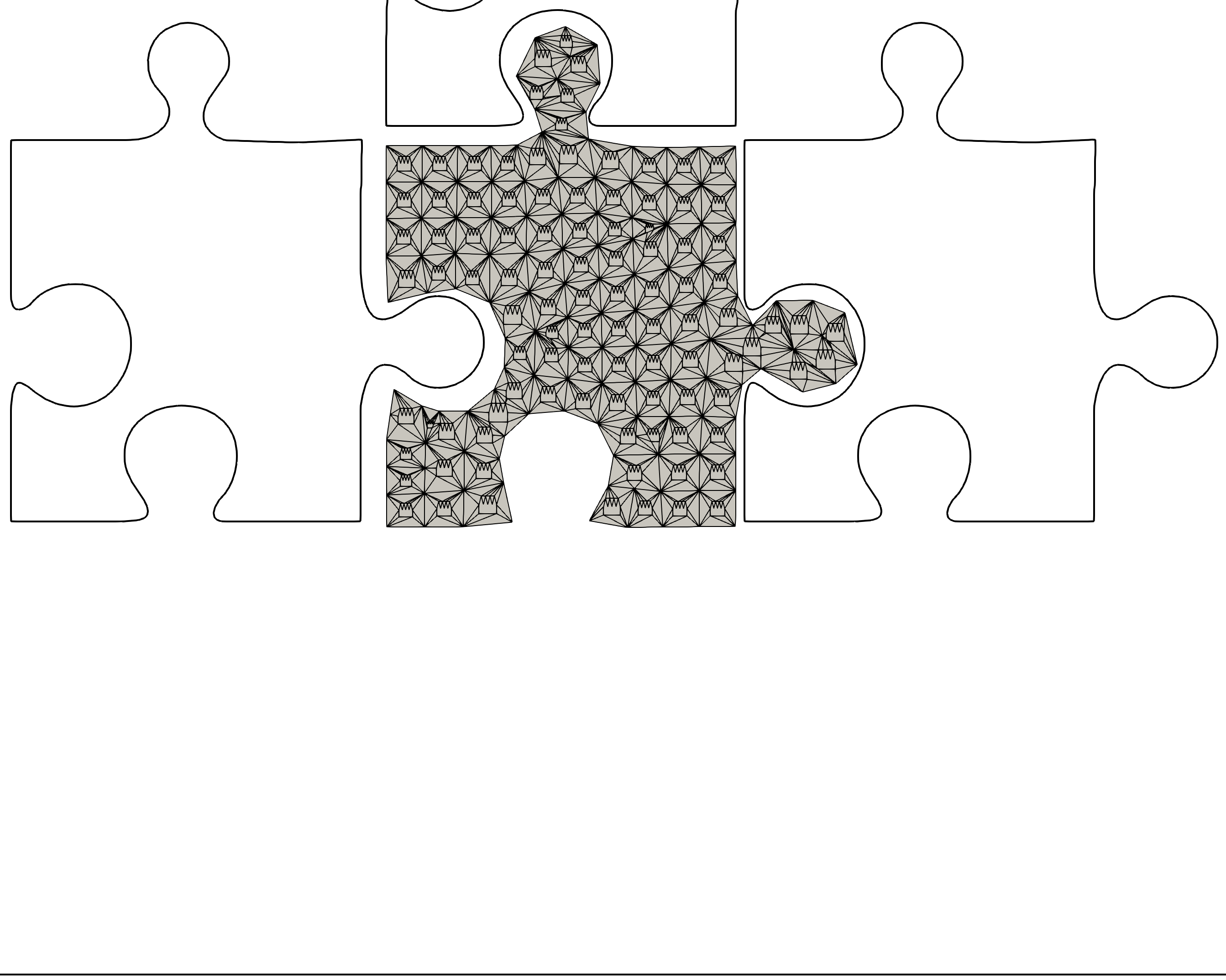}\hfill
\includegraphics[width=.24\linewidth]{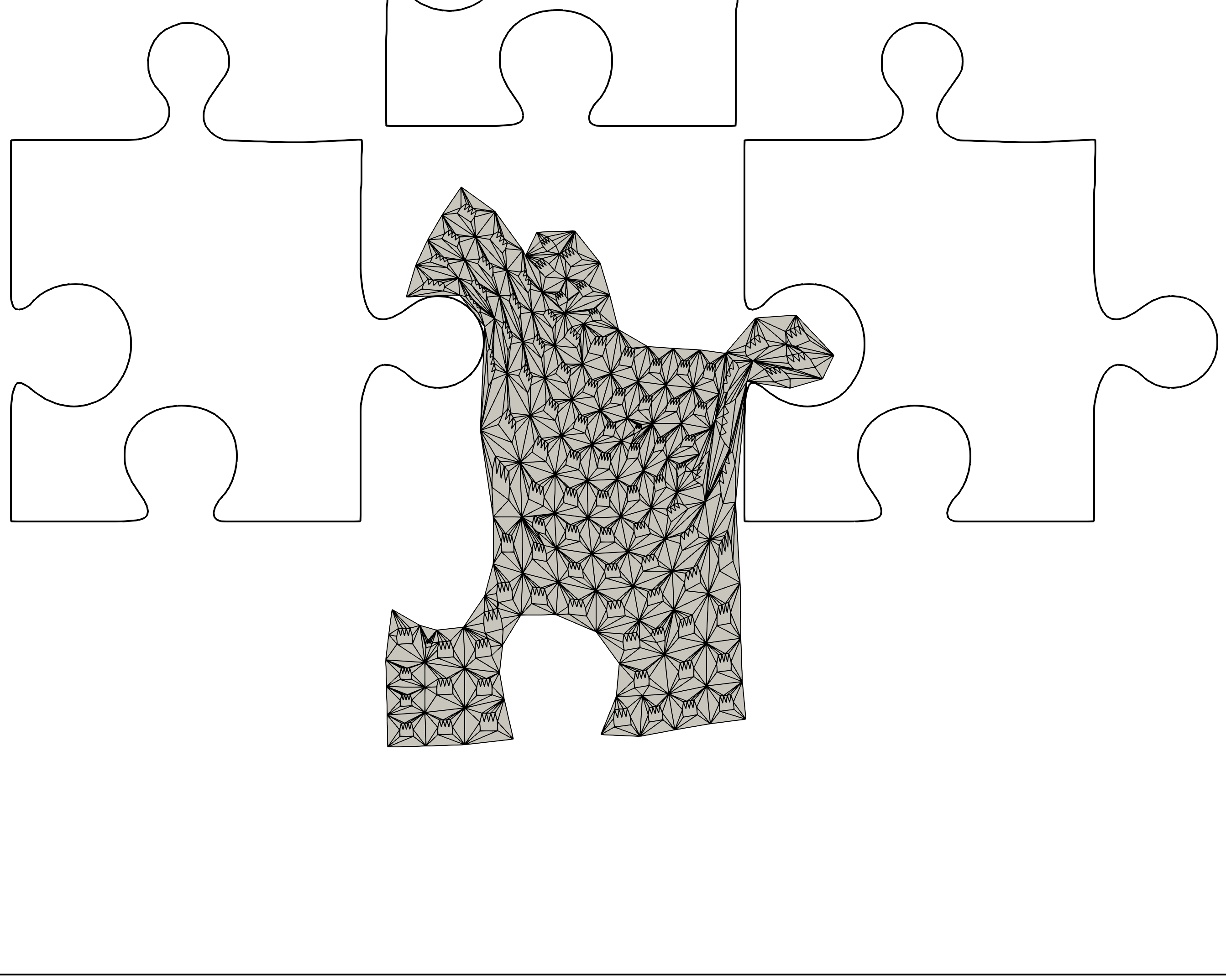}\hfill
\includegraphics[width=.24\linewidth]{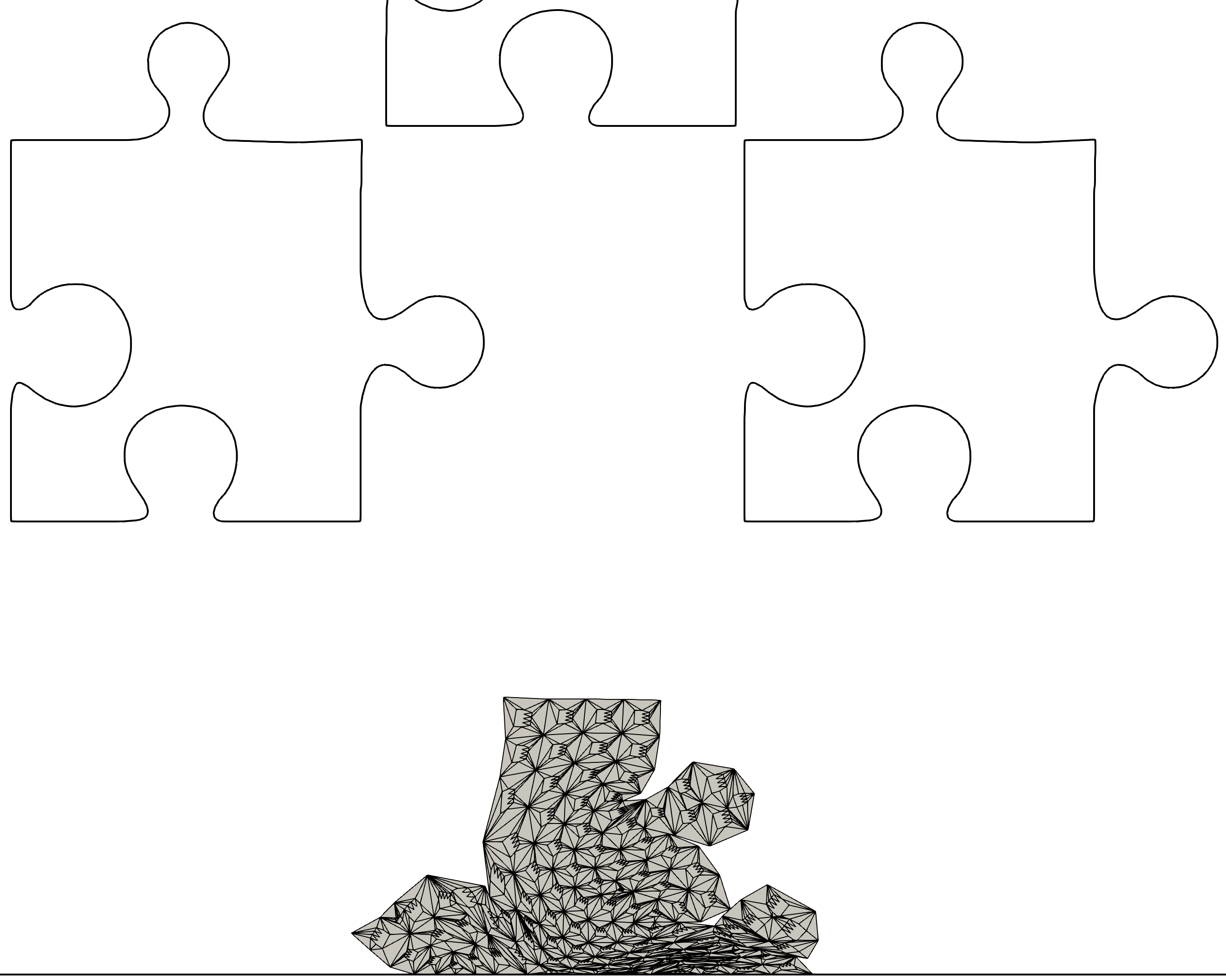}\hfill
\includegraphics[width=.24\linewidth]{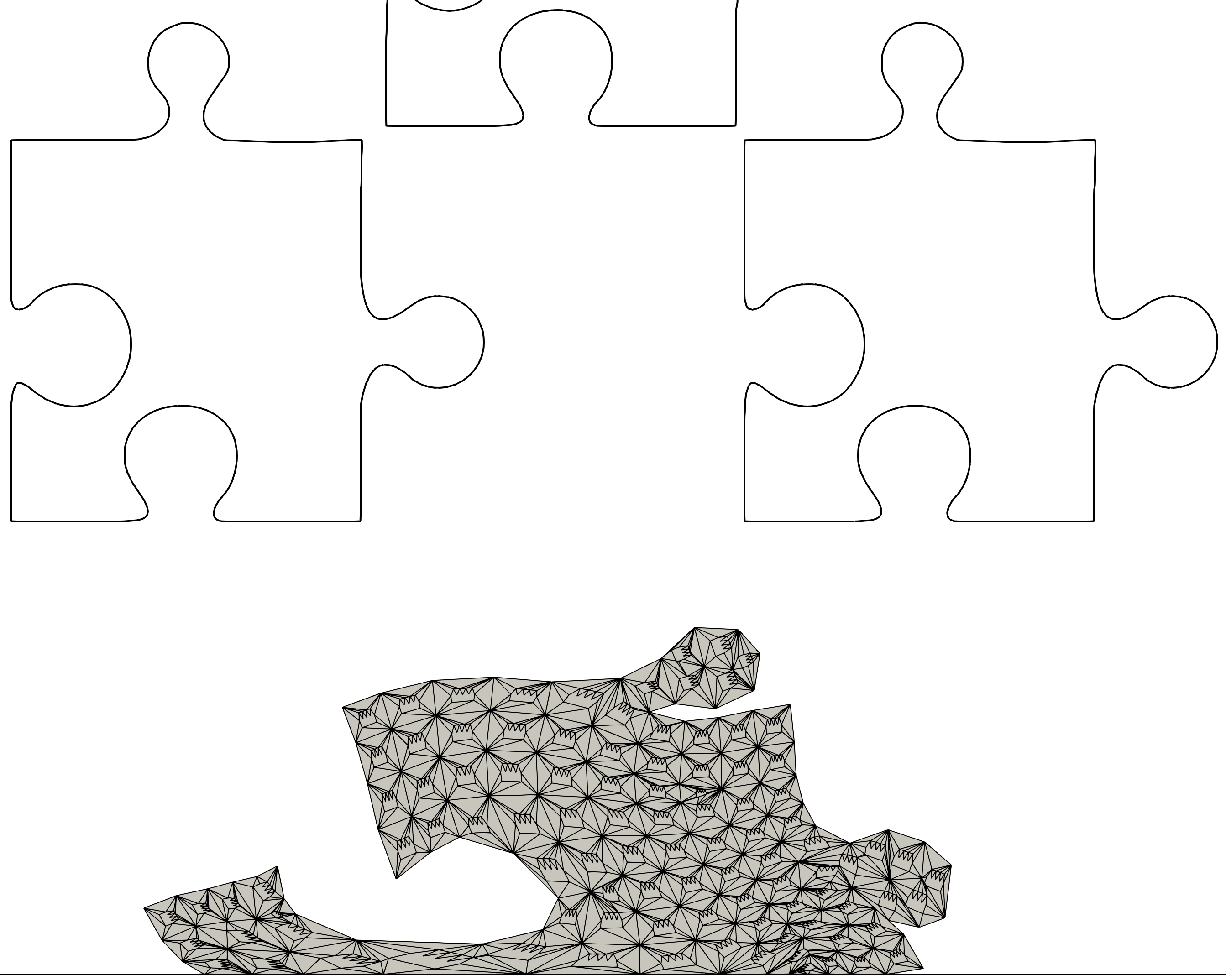}\\[1em]
Triangle mesh\\
\includegraphics[width=.24\linewidth]{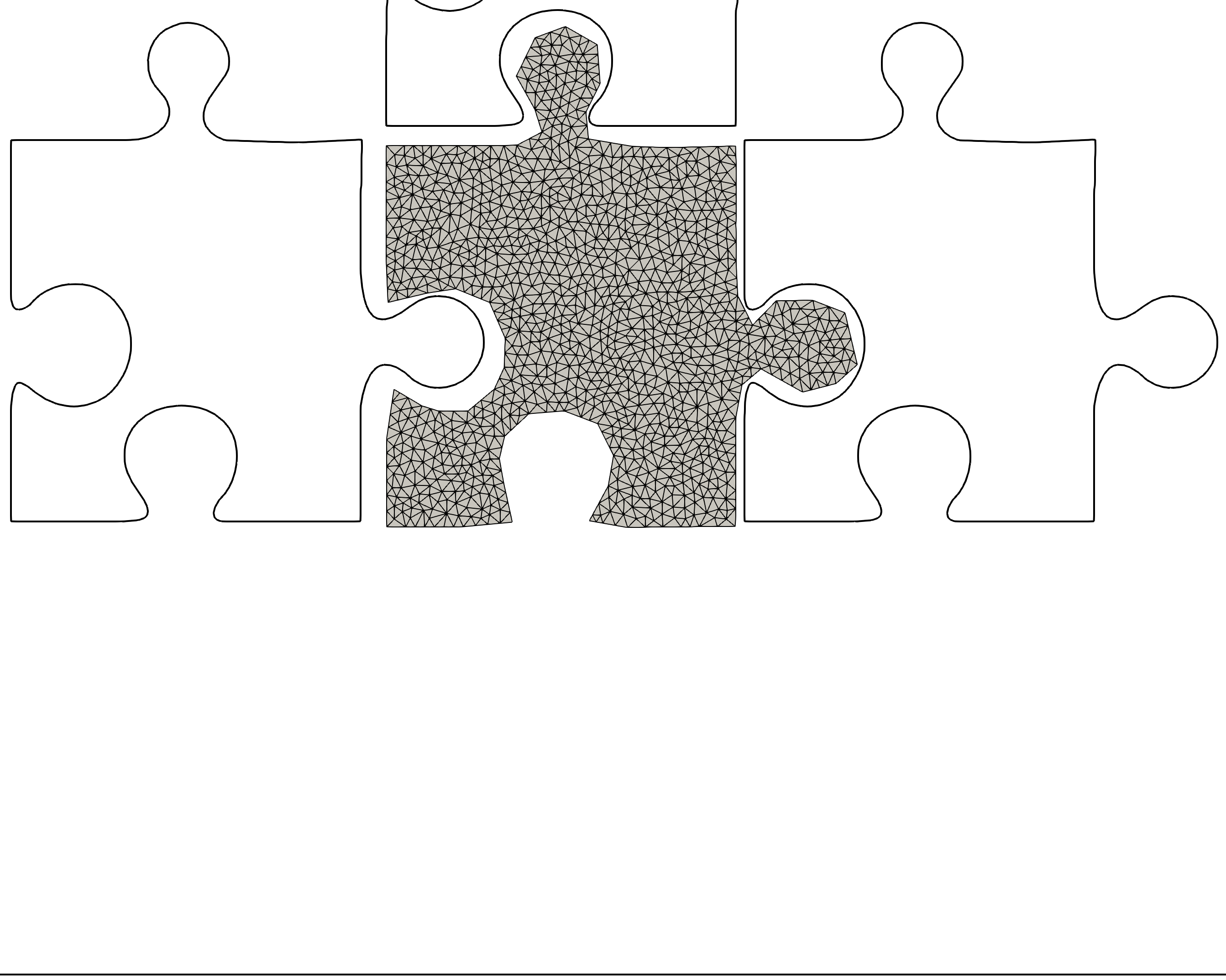}\hfill
\includegraphics[width=.24\linewidth]{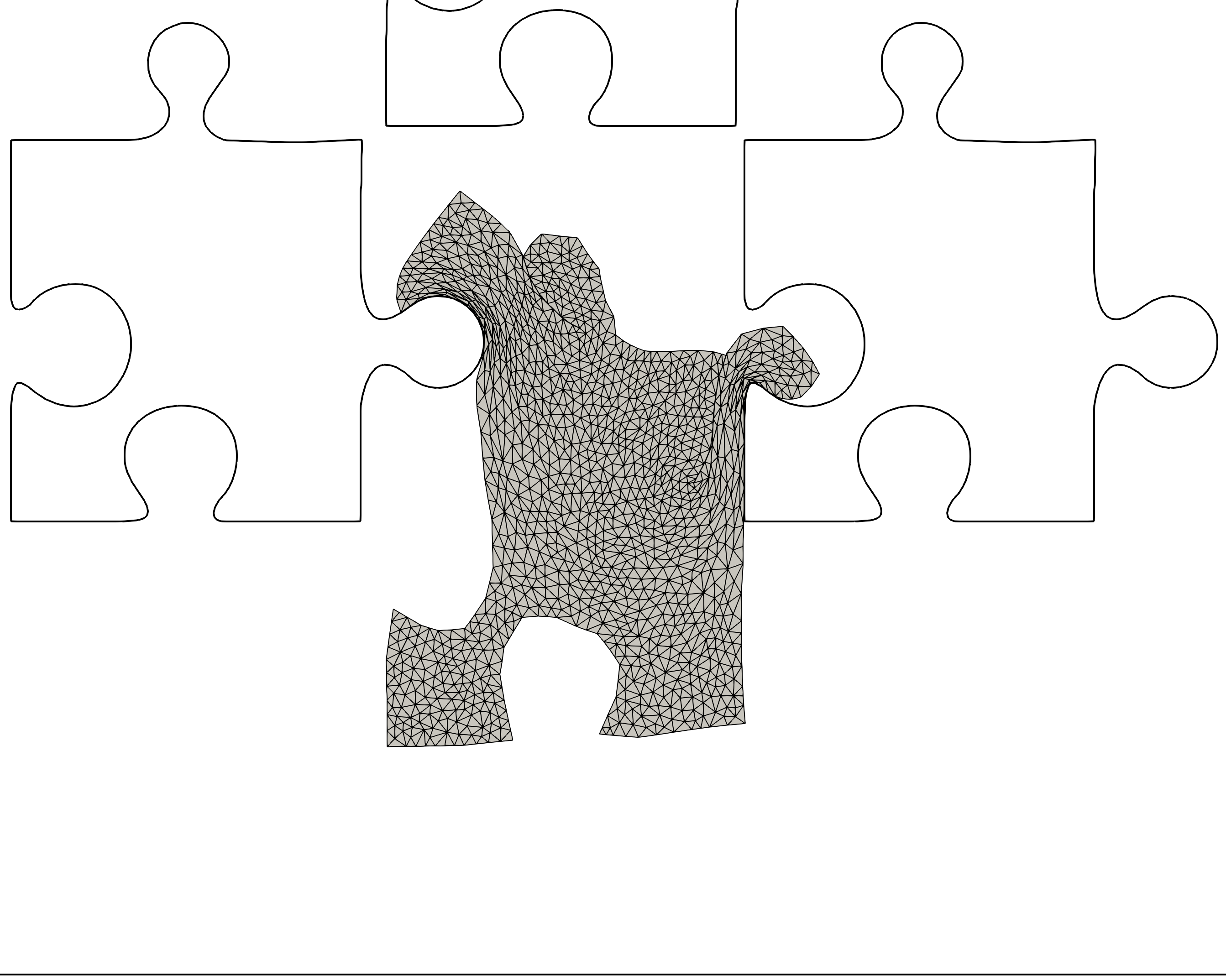}\hfill
\includegraphics[width=.24\linewidth]{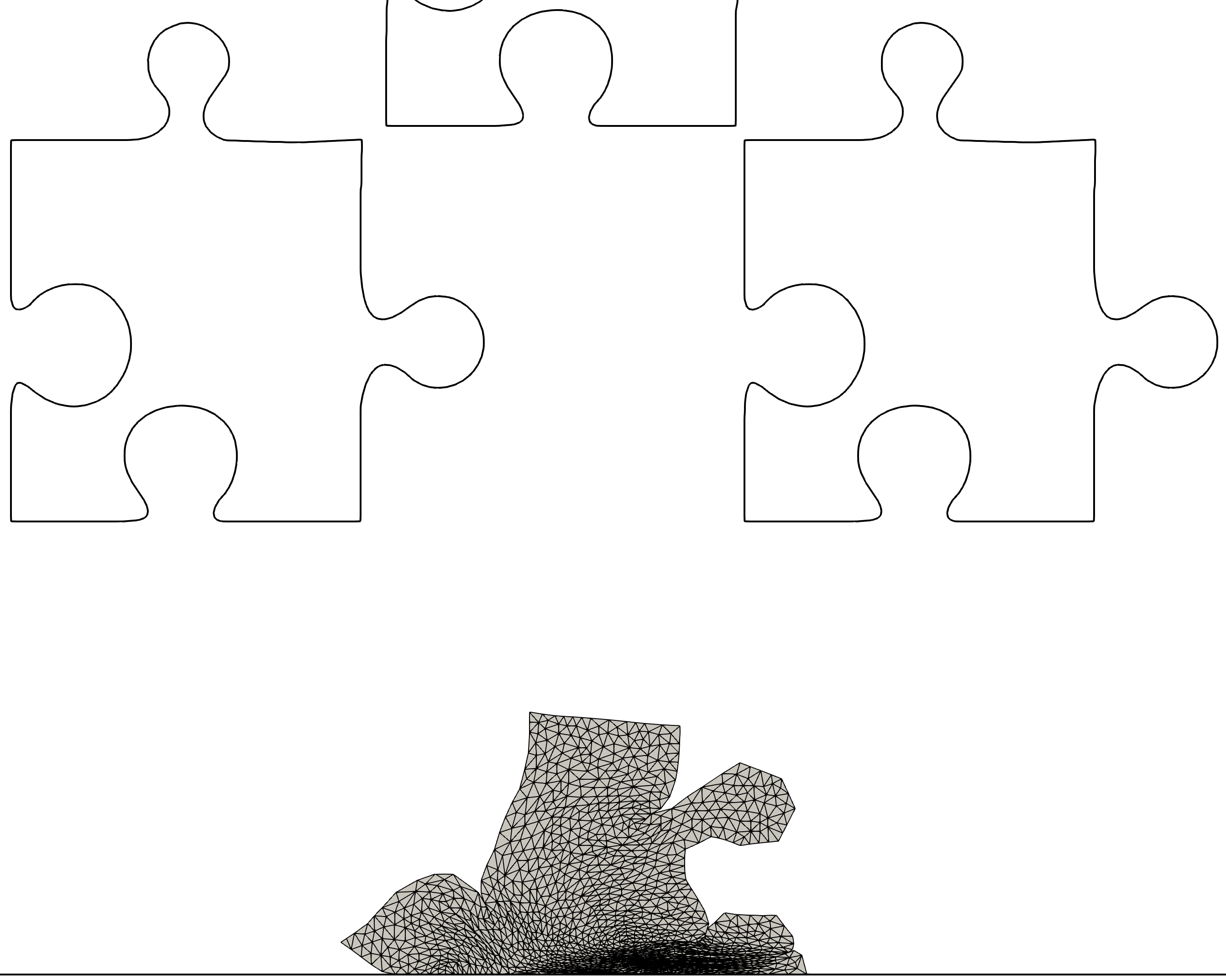}\hfill
\includegraphics[width=.24\linewidth]{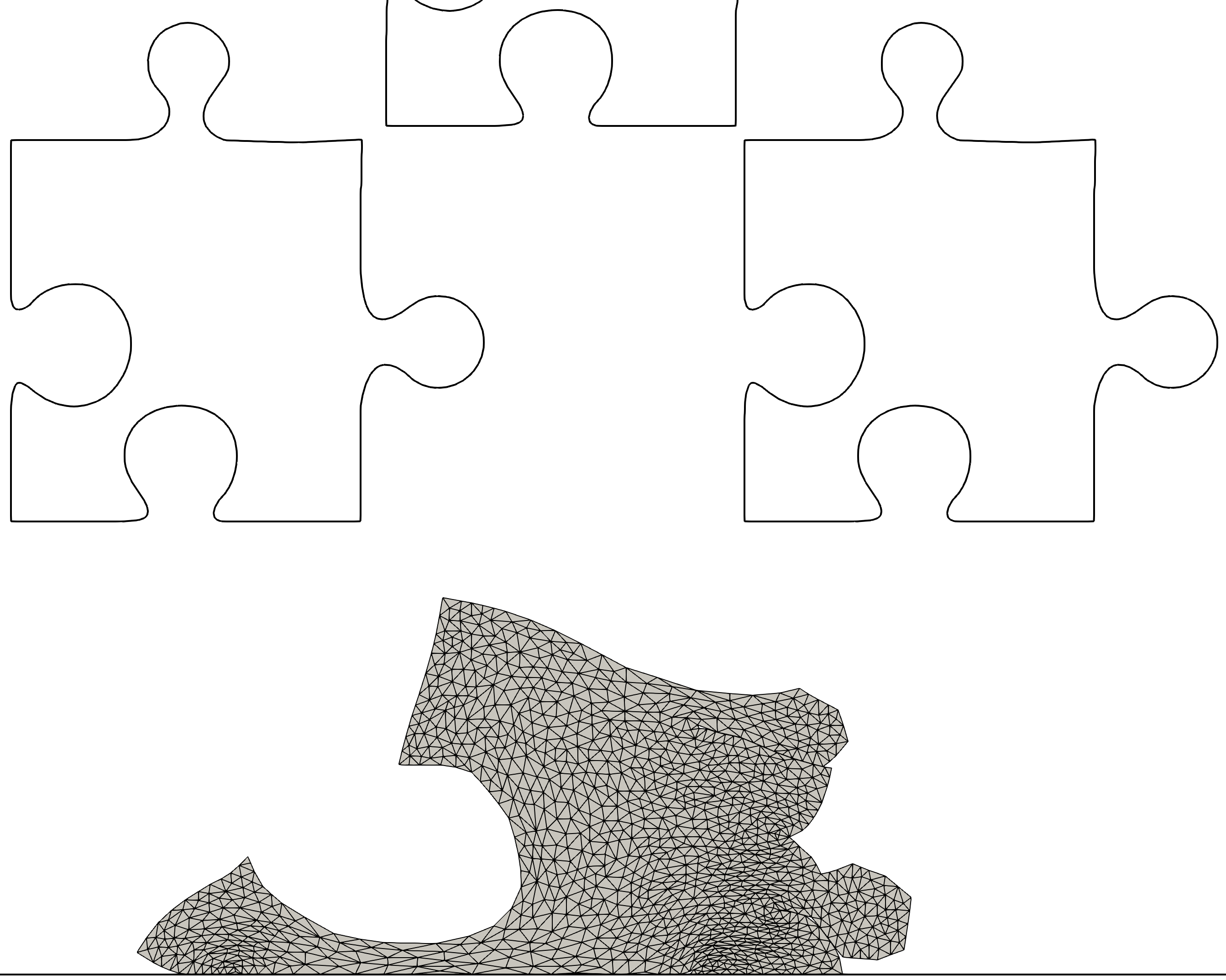}\\
\parbox{.24\linewidth}{\centering $t=0$\si{\second}}\hfill
\parbox{.24\linewidth}{\centering $t=0.5$\si{\second}}\hfill
\parbox{.24\linewidth}{\centering $t=1$\si{\second}}\hfill
\parbox{.24\linewidth}{\centering $t=1.5$\si{\second}}
}\hfill
\parbox{.23\linewidth} {\centering
Newton iterations
\includegraphics[width=\linewidth]{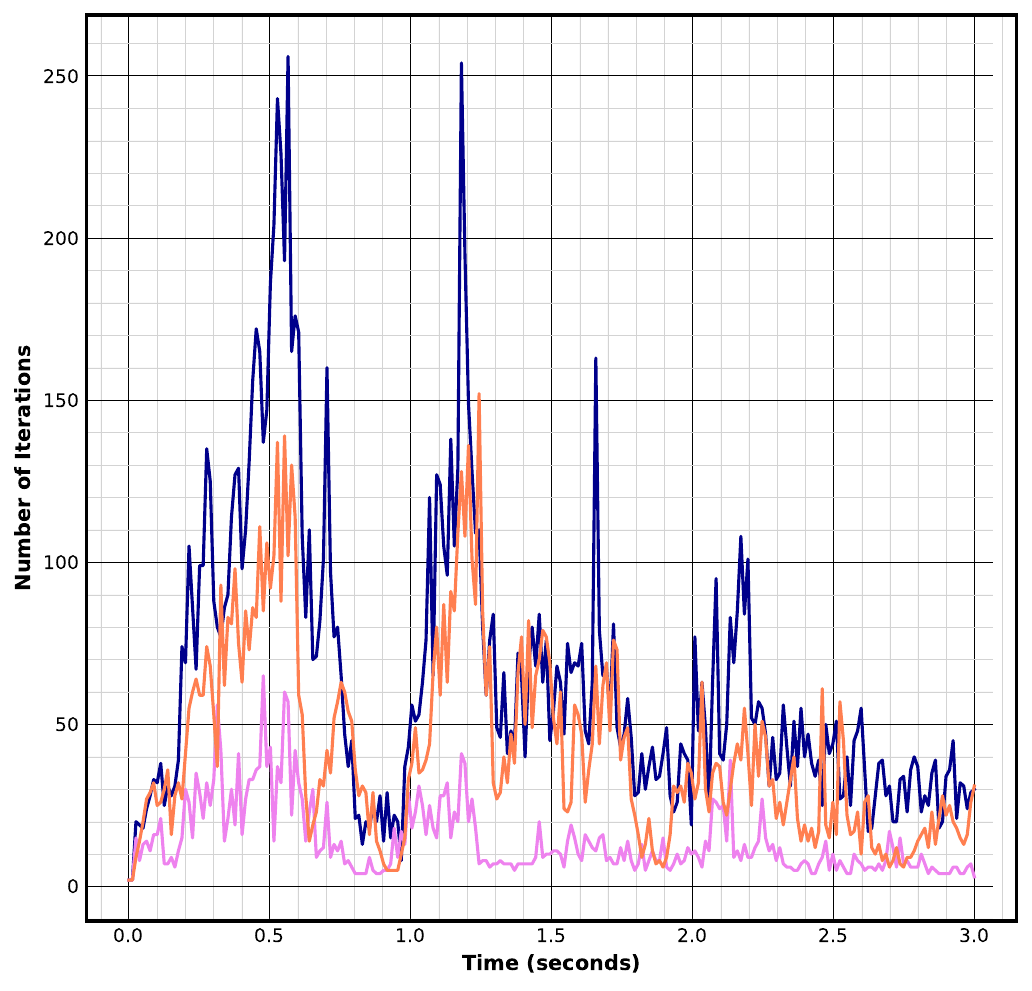}
Solve time
\includegraphics[width=\linewidth]{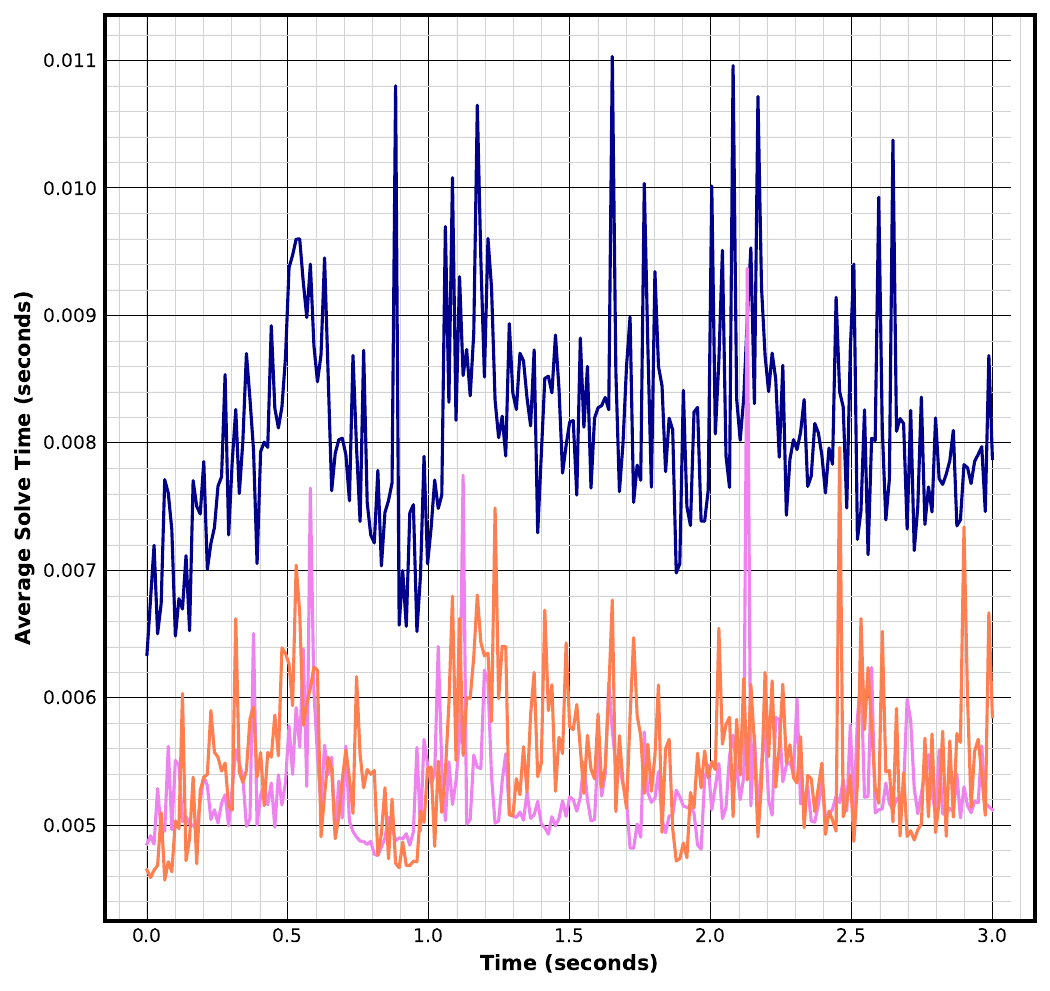}
}
\caption{Several frames of a soft stuck puzzle piece falling under gravity using a Voronoi (top), tiled (middle), and triangular (bottom) mesh with a similar number of vertices. The three simulations require approximately the same time to solve the linear system, while the tiled simulation has fewer Newton iterations.}
\label{fig:teaser}
\end{figure}

%% file: 01-introduction.tex
\section{Introduction}

The solution of partial differential equations (PDEs) is a core building block in many branches of science and engineering to simulate physical systems. A plethora of methods has been developed to discretize both the spatial and temporal domains. 
From a practical side, there are many issues to address while developing a PDE solver, including how to discretize the domain, how to model the physical system, and how to compute solutions using non-exact computation to make the solution tractable.

The most popular approach, due to its generality and efficiency for a wide range of PDEs, is the finite element method, where the spatial domain is partitioned into a collection of cells (usually triangles or quadrilaterals for 2D domains). Allowing some polygon/polyhedra in the partition simplifies the meshing process. For instance, generating pure quadrilateral or hexahedral meshes is challenging, while allowing a few polygonal elements simplifies the task. Similarly, adding geometrical cuts (e.g., coming from mesh booleans) naturally produce polygonal cells, which are, in general, difficult to tesselate with high-quality elements, especially in 3D.
FEM using barycentric coordinates (BFEM) and the more modern virtual element method (VEM) has been introduced to allow for more flexibility in domain discretizations, enabling the use of general polygonal (polyhedral in 3D) elements while sharing many of the other benefits of simplicial discretizations.

In this work, we provide a benchmark of problems to evaluate BFEM, VEM, and simplicial FEM in solving elliptic PDEs on geometrically complex geometries. We decide to focus exclusively on the method's efficiency, measured as the computational effort to obtain a solution with a given accuracy, as all methods converge to the same result when given a sufficiently dense and high-quality discretization. Since a typical analysis pipeline includes many stages, and especially for polygonal elements, many of them are not yet well established, we ignore the time required to create a discretization and to assemble the stiffness matrix and right-hand side --- we focus exclusively on the time required to solve the resulting linear system, which can be performed with the same heavily optimized solver for all methods. Our study aims at identifying the pros and cons of the different methods in an ideal setting where solving is the time bottleneck, which is usually the case as it is the only step whose complexity does not scale linearly with the number of DOFs.

The goal of our study is to answer, experimentally and for our specific set of elliptic PDEs, several questions; in particular:
\begin{itemize}
\item \textbf{Mesh} How do the choice of discretization methods (simplicial and polyhedral) and mesh quality affect the efficiency?
\item \textbf{Simulation} How do the simulation methods (FEM, BEFM, and VEM), basis order, and method-specific techniques affect the efficiency?
\item \textbf{Generalization} How do other common numerical analysis decisions (e.g., solver choices, PDE types) affect the generalizations of our findings?
\item \textbf{Future} What challenges need to be overcome in polyhedral meshing and solvers, and what would the benefit be?
\end{itemize}

We performed extensive experiments in two and three dimensions, for which we release the reference source-code implementation and data for replicability, and, consistently with previous studies, we anecdotally observe similar behavior in 3D, suggesting that our conclusions likely generalize to the third dimension. 

We believe our study will guide research efforts in polygonal or polyhedral meshing and its use for VE methods while providing an automated and replicable way of comparing its efficiency with a reference FE implementation.

%% file: 02-related.tex
\section{Related Work}\label{sec:related}

\paragraph*{Basis Functions}
Finite Element Methods (FEMs) have been one of the most popular numerical methods for solving PDEs
\cite{FEM-physics,FEM-mechanics}. Classical FEMs \cite{FEM} discretize one field's weak formulation, by means of shape functions inside mesh elements. Newer formulations of FEM, named Mixed Finite Element Methods (MFEM) \cite{MFEM}, discretize more than one field, such as pressure, stress, or strain formulations \cite{FEM-vs-VEM-elasticity}. MFEM shows advantages in certain scenarios over classic FEM, including overcoming locking problems \cite{locking-problem} when modeling nearly incompressible materials.
A common feature for both FEM and MFEM is that they all explicitly specify shape functions over triangular/tetrahedral or quadrilateral/hexahedral meshes.

While robust triangular/tetrahedral meshing is possible~\cite{FEM-analysis, TetGen, CGAL, TetWild, fTetWild, Mandoline}; polygonal and polyhedral meshes could enable a more natural representation of complex domains \cite{Polygon-application-complex} and give designers more freedom to achieve the expected attributes of the models \cite{Polygon-application-art}. Additionally, if a triangular/tetrahedral mesh is distorted, there will be poor quality elements with very thin shapes, which requires re-meshing \cite{FEM-distortion-unrobust, Decouple-Accuracy-Quality, Finite-Element-Quality}. Finally, polygonal meshes naturally allow for non-conforming nodes (e.g., having hanging nodes), where traditional FEM will fail to provide accurate solutions \cite{FEM-vs-VEM-engineering}. 
Despite the potential advantage of polygonal/polyhedral, the robust generation of such meshes is not as well explored as for simplicial meshes; only a few polygonal meshing methods exist: 2D Voronoi tessellations (PolyMesher \cite{PolyMesher} and VoroCrust~\cite{VoroCrust}); displacing from a structured grid 
(used in VEM-related works)
\cite{Displacement-Polygon-Meshing-1,Displacement-Polygon-Meshing-2}. 
Similarly, large-scale datasets of such meshes are still in their infancy (with \citet{VEM-quality-indicator} containing tiled 2D polygonal meshes with different levels of regularity on square domains; \citet{Polyheral-quality-indicator} containing 3D tetrahedral, hexahedral, Voronoi, and displaced polygonal meshes on cube domains), in particular covering complex domains and different meshing methods.

The flexibility of polygonal meshes enables better shape deformation and a more realistic stretch factor \cite{FEM-vs-VEM-engineering,basic-FEM-vs_VEM}. Thus, several works have explored how to use polygonal and polyhedral meshes and proposed adapted FEM-based methods, naming as Polytope Element Methods (PEMs) \cite{BEM-FEM,Poly-Spline-FEM,Polygonal-FEM}. A key challenge in using polygonal and polyhedral meshes is the design of basis functions. 
Two major approaches have been proposed in current PEMs: (1) generalizing basis functions using barycentric coordinates; and (2) avoiding explicit specification of basis functions through projection to decouple the solution to an exact and approximation part. 
We include a detailed introduction of these two approaches in Appendix~\ref{app:two-approaches-PEM}.

\paragraph*{Different Order of Basis}
For relatively restricted discretizations (i.e., triangular/tetrahedral or quadrilateral/hexahedral meshes), higher-order FEMs have been shown to have better accuracy than linear FEMs for elliptic PDEs \cite{tet-vs-hex}. \citet{Quadratic-Basis} extend the quadratic shape functions to arbitrary polygonal/polyhedral discretizations and break the limit of PDE-dependent modifications. On  the VEM side, \citet{Higher-Order-VEM} investigated higher-order polynomial degrees for VEM. In comparison to higher-order FEM, linear VEM and higher-order VEM are seen to have optimal or at least similar performance on 2D meshes and problems \cite{Polyheral-quality-indicator}. Higher-order VEMs on 3D meshes have not been implemented into open-source software, and the currently most interesting choice for engineering applications is still linear VEM \cite{Polyheral-quality-indicator}.

\paragraph*{Large Scale Comparison Study}
Multiple works have conducted comparisons between different elements of FEM in specific physical problems, including structural problems \cite{tet-vs-hex-simple-structure}, elastoplastic experiments~\cite{tet-vs-hex-elastic}, nonlinear incompressible materials~\cite{tet-vs-hex-footwear}, and linear static problems~\cite{tet-vs-hex-linearstatic} (More details in Appendix~\ref{app:compare-fem-physical}).
However, all these studies are restricted to certain problems, certain discretization, and a small set of geometries, limiting the ability to generalize the conclusions to general problems and domains.

\citet{tet-vs-hex} performed the first large-scale comparison of the overall performance between tetrahedral and hexahedral meshes in FEM. This study covers several elliptic PDEs 
on various commonly used test problems
and large-scale benchmarks of manufactured solutions in complex real-world models. The experiments show consistent results of the well-known problematic linear triangular/tetrahedral meshes, and further conclude that tetrahedral meshes with quadratic elements have similar or better performance than semi-structured hexahedral meshes with Lagrangian elements, but are slightly inferior to regular lattices with spline elements. The study does not consider general polyhedral meshes.

Several works have compared the impact of polygonal/polyhedral meshes on the performance of VEM solutions. 
We include a detailed introduction in Appendix~\ref{app:poly-impact-vem}.

Current comparisons of FEM and VEM mainly focus on the comparison of inherent characteristics of VEM (e.g., the ability to handle hanging nodes and robustness to mesh distortion) \cite{VEM-50lines-code,VEM-stability} and implementation differences (e.g., the specification of implicit degrees of freedom and discussion of complex boundary conditions) \cite{VEM,FEM-vs-VEM-elasticity}, but the comparison of using general elements has not been thoroughly discussed. \citet{FEM-vs-VEM-engineering} focuses on linear electrostatics problems by comparing VEM to p-FEM in the context of higher-order interpolations. This study concludes that VEM can achieve a similar level of convergence at a faster rate when compared to p-FEM. \citet{FEM-vs-VEM-elasticity} also focuses on linear elasticity by comparing p-FEM with triangular meshes and mVEM with triangular, quad, and perturbed polygon meshes. This study concludes that mixed polynomial FEM and VEM behave similarly on distorted meshes, but stable discretizations are needed for polygonal meshes for optimal approximations. These works have not fully covered the wide range of choices of polygonal meshes 
and are restricted to certain problems and relatively simple domains. Finally, all previous work focuses on convergence rate, while for practical applications, the most relevant metric is solve-time versus error.

%% file: 04-benchmark.tex
\section{Dataset Description}\label{sec:dataset}

To carry out a benchmark study regarding various polyhedral discretization on elliptic PDEs, we need to construct a large-scale dataset containing various geometry domains, mesh discretization methods, and problems. We introduce our efforts in each aspect in detail (Section~\ref{subsec:dataset:domains}, \ref{subsec:dataset:problems}, and \ref{subsec:dataset:meshsing}) and then explain the combination of them in a large dataset used in our benchmark study in Section~\ref{subsec:dataset:benchmark}.

We use the Poisson and linear elasticity PDE to test the performance of different meshing and collect a number of standard test cases to simulate various physical phenomena and scenarios; we refer to Appendix~\ref{sec:pde-definition} for a formal definition of the bases we use.

\subsection{Domains}
\label{subsec:dataset:domains}

\begin{figure}
    \centering\footnotesize
    \includegraphics[width=.16\linewidth]{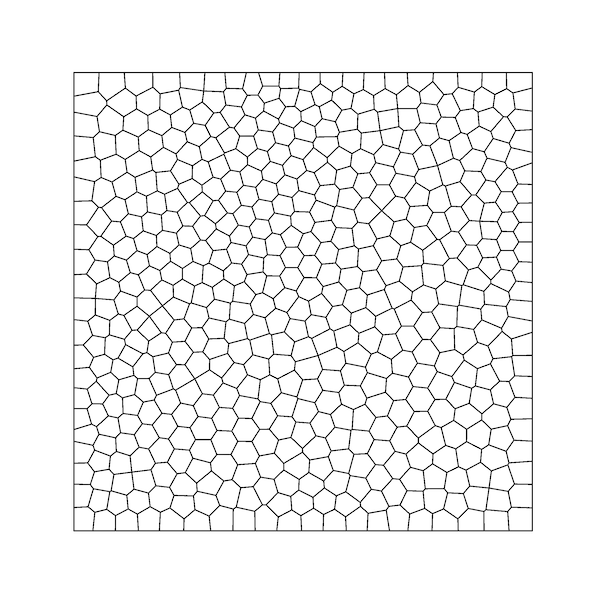}\hfill
    \includegraphics[width=.16\linewidth]{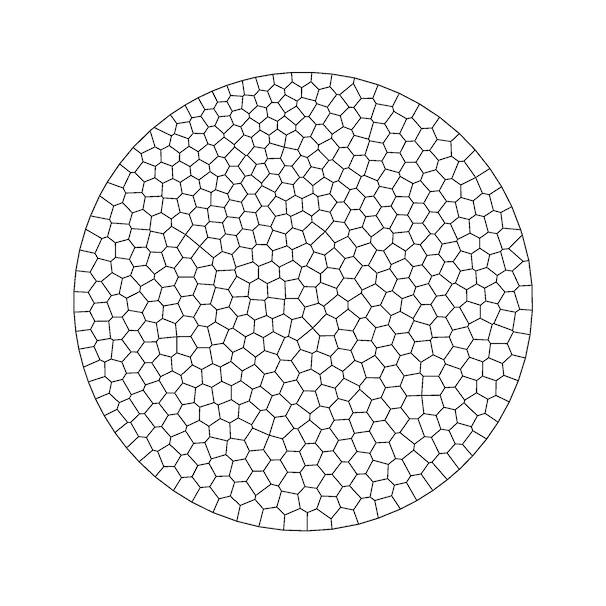}\hfill
    \includegraphics[width=.16\linewidth]{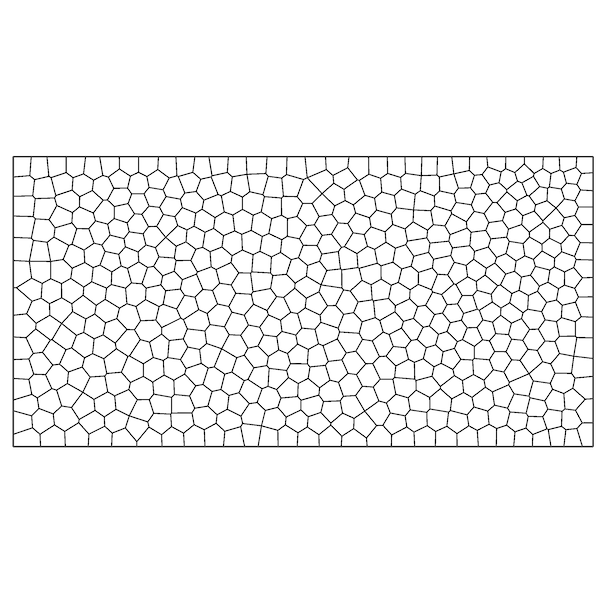}\hfill
    \includegraphics[width=.16\linewidth]{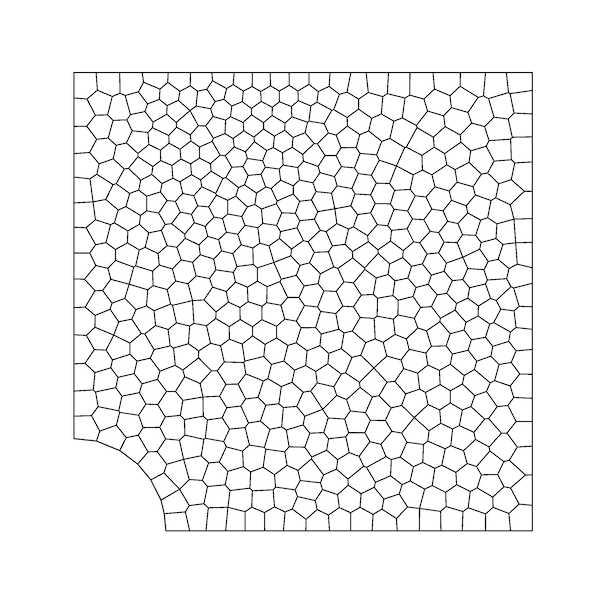}\hfill
    \includegraphics[width=.16\linewidth]{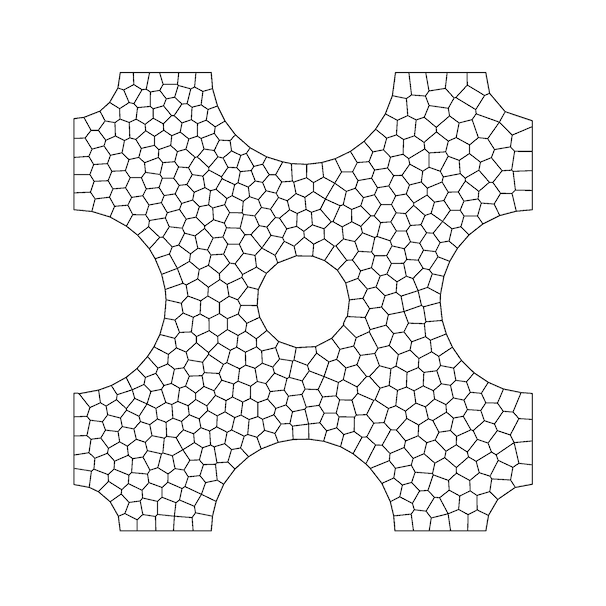}\hfill
    \includegraphics[width=.16\linewidth]{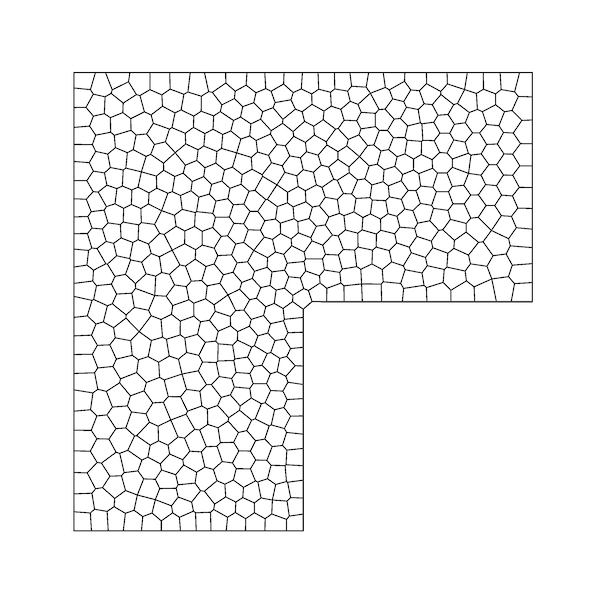}\par
    \parbox{.16\linewidth}{\centering Unit Square (US)}\hfill
    \parbox{.16\linewidth}{\centering Unit Disk (UD)}\hfill
    \parbox{.16\linewidth}{\centering Beam\\(BE)}\hfill
    \parbox{.16\linewidth}{\centering Plate with Hole (PH)}\hfill
    \parbox{.16\linewidth}{\centering Swiss Cheese (SC)}\hfill
    \parbox{.16\linewidth}{\centering L-Shape\\(LS)}\par
    \caption{Domains used in our case study.}
    \label{fig:domains}
\end{figure}

We test the problem on four different domains with different complexity (Figure~\ref{fig:domains}) to ensure that our results can be generalized to many real-world problems: a unit square (US), a unit disk (UD), a beam of size $8\times4$ (BE), a plate with a hole of size $2\times2$ with hole radius $0.4$ (PH), a unit ``Swiss cheese''-shape (SC), and an L-shape of size $2\times2$ (LS). We use a combination of the problem name followed by the domain name to specify a particular setup. For instance, LEB-SC stands for linear elasticity with biaxial force on the ``Swiss cheese'' domain.

\begin{figure}
    \centering\footnotesize
    \includegraphics[width=.16\linewidth]{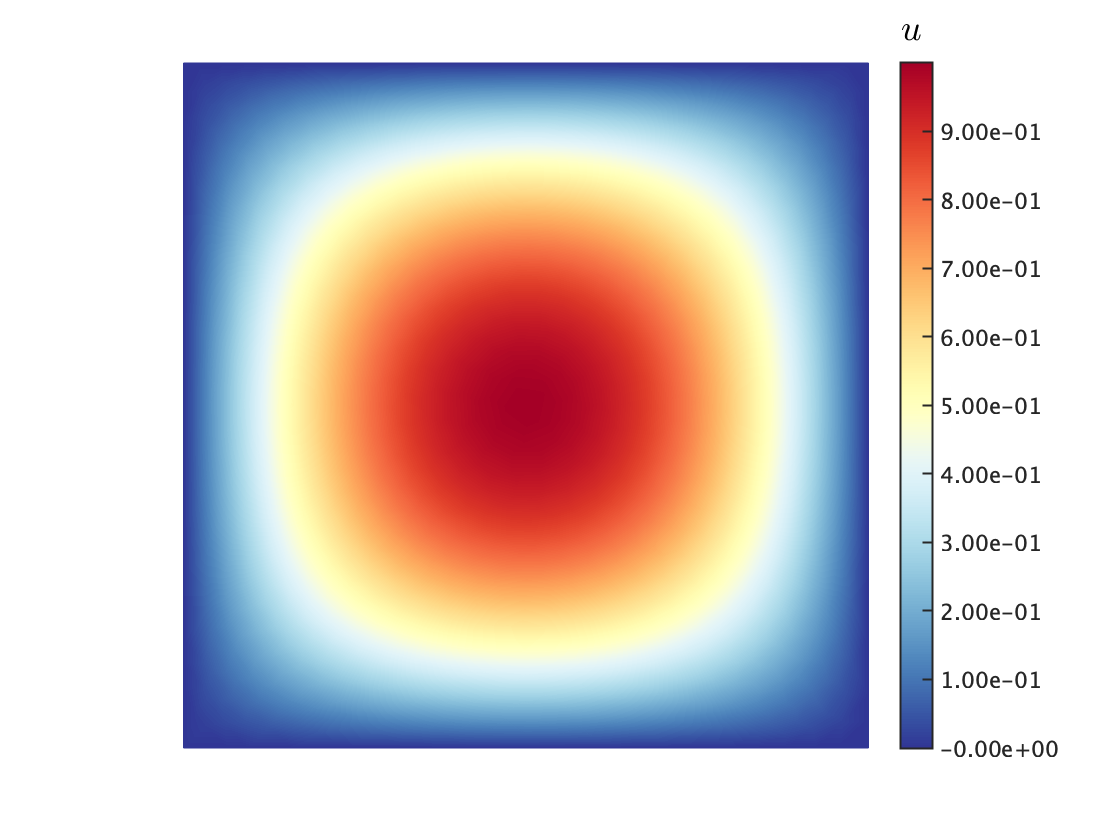}\hfill
    \includegraphics[width=.16\linewidth]{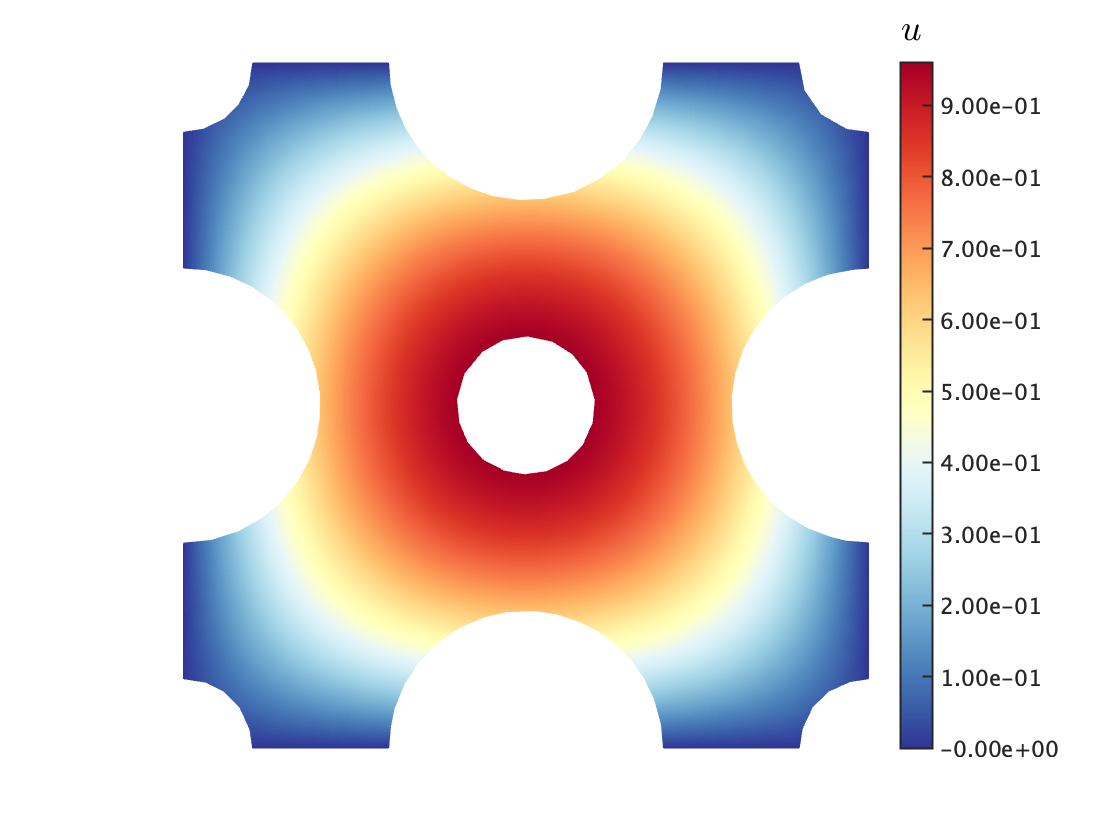}\hfill
    \includegraphics[width=.16\linewidth]{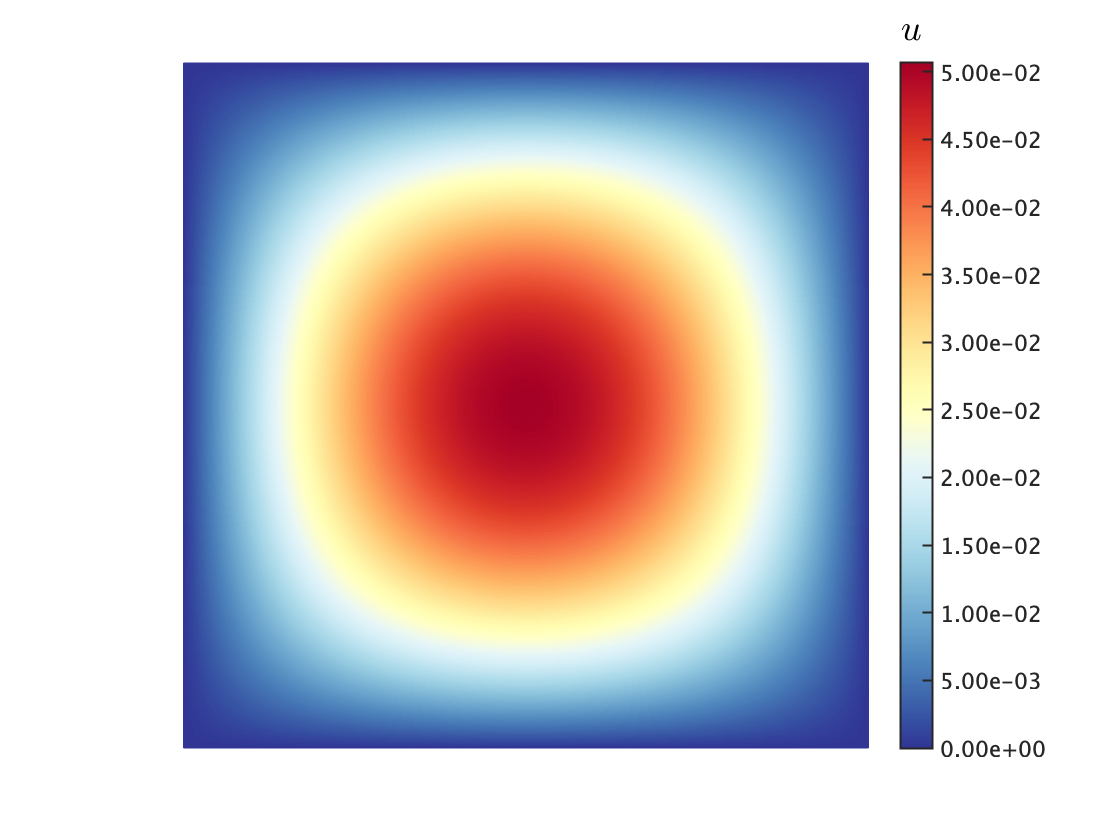}\hfill
    \includegraphics[width=.16\linewidth]{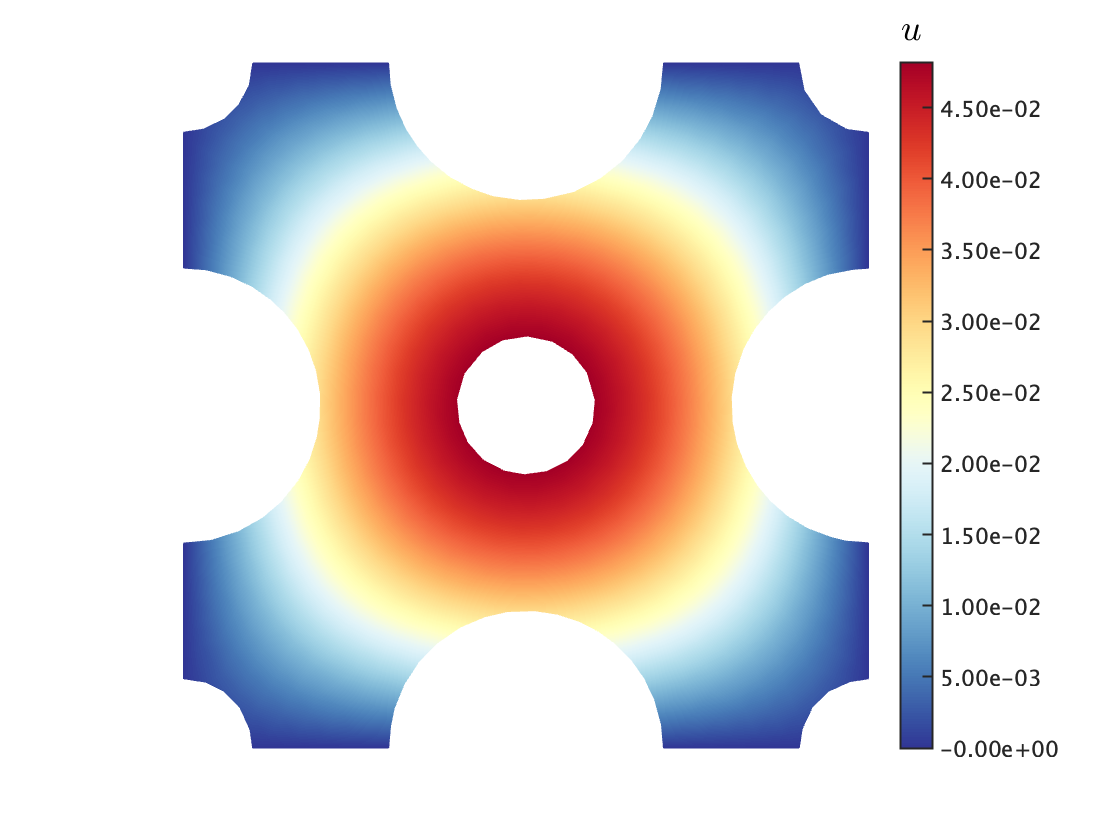}\hfill
    \includegraphics[width=.16\linewidth]{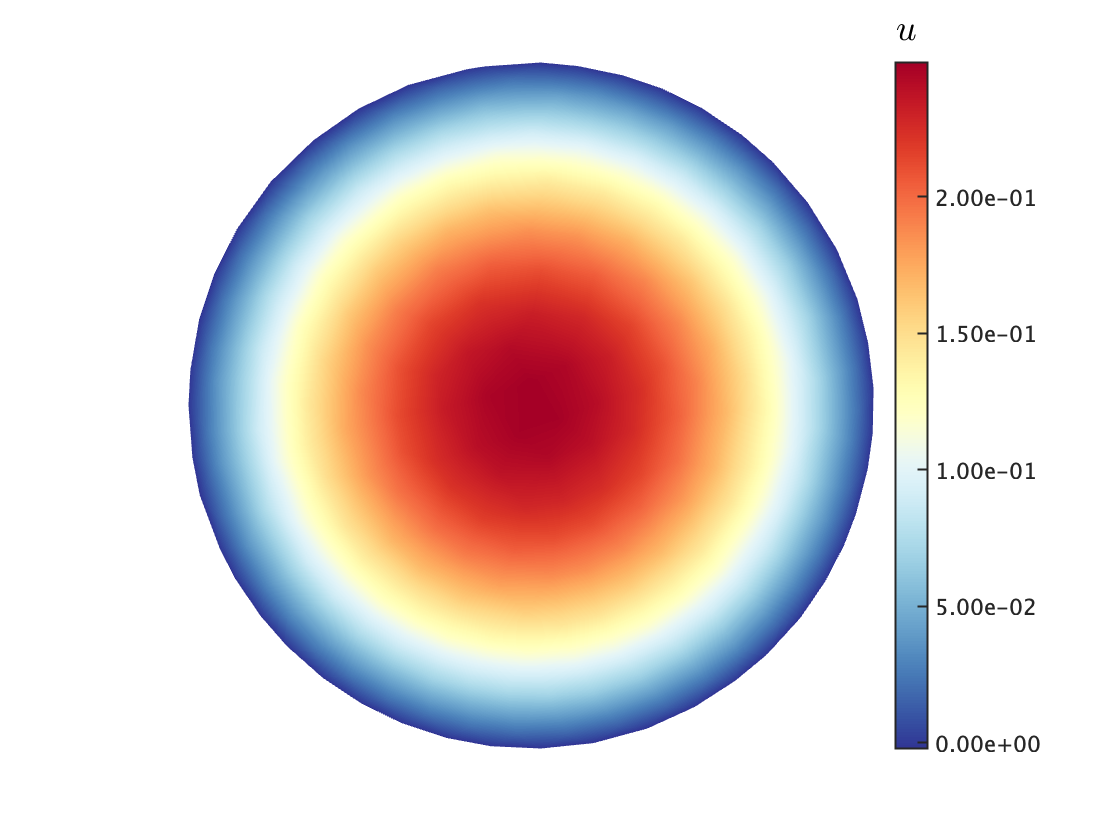}\hfill
    \includegraphics[width=.16\linewidth]{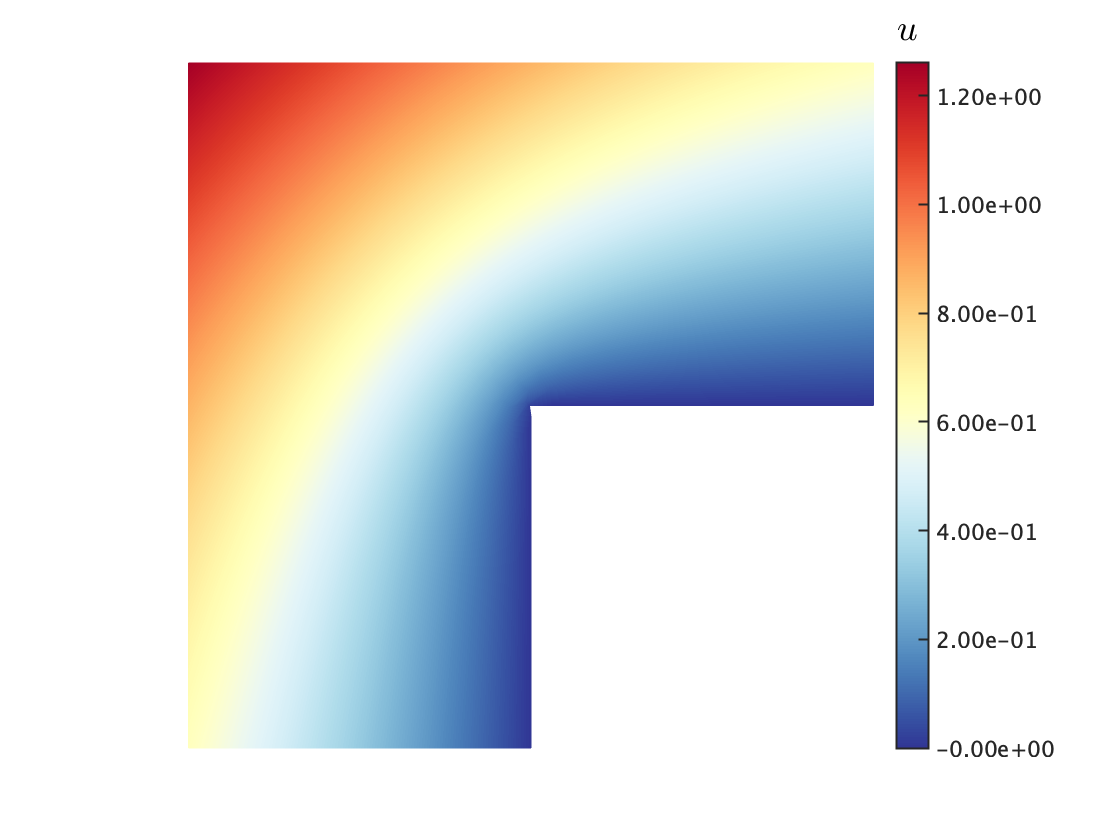}\par
    \parbox{.16\linewidth}{\centering PS\#1-US}\hfill
    \parbox{.16\linewidth}{\centering PS\#1-SC}\hfill
    \parbox{.16\linewidth}{\centering PS\#2-US}\hfill
    \parbox{.16\linewidth}{\centering PS\#2-SC}\hfill
    \parbox{.16\linewidth}{\centering PS\#3-UD}\hfill
    \parbox{.16\linewidth}{\centering PB-LS}\\[1em]
    \caption{Visualization of the solution for different heat sources (first five figures) and the biaxial load (last figure). }
    \label{fig:shape_poisson}
\end{figure}

\subsection{Problems}
\label{subsec:dataset:problems}
\paragraph*{Poisson Problem} For this problem we use manufactured solutions~\cite{SALARI:2000:CVB} (Figure~\ref{fig:shape_poisson}); we define three different heat sources (named PS\#1, PS\#2, and PS\#3)
{\scriptsize\[
    u = 0.25(1-x^2-1-y^2),~
    u = 16xy(1-x)(1-y),
 \text{ and }
    u = -\frac{1}{2\pi^2}\sin(\pi x)\sin(\pi y),
\]}
and a biaxial load (named PB) designed to work on an L-shaped domain
\[
    u = r^{\frac{2}{3}}\sin({2}/{3}\theta),
\]
where $r = \sqrt{x^2 + y^2}$ and $\theta = \mathrm{atan}({y}/{x})$. In 3D we use (named PS3D\#1, PS3D\#2, and PS3D\#3)
{\footnotesize\[
    u = \sin(2xy)\cos(z),~
    u = x^2 + y^2 + z^2, \text{ and }
    u = \sin(\pi x)\cos(\pi y)\cos(\pi z).
\]}

\begin{figure}
    \centering\footnotesize
    \parbox{.3\linewidth}{\centering
    \includegraphics[width=.48\linewidth]{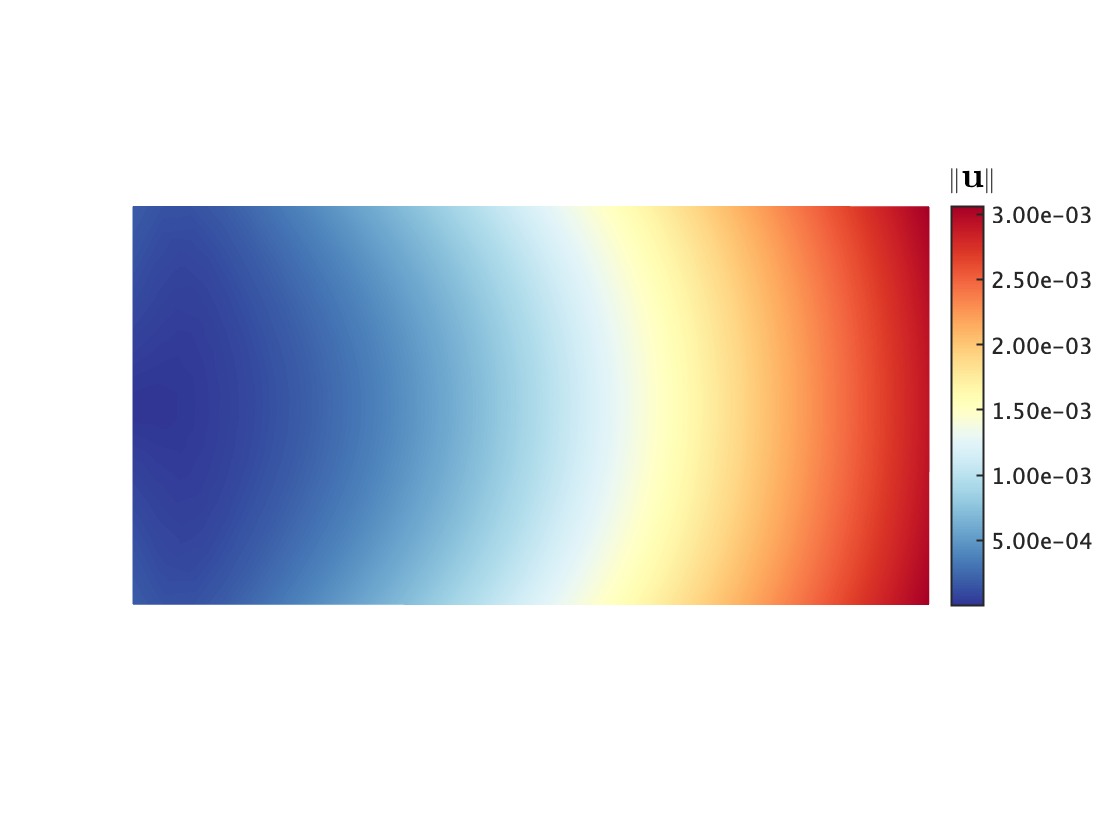} \hfill
    \includegraphics[width=.48\linewidth]{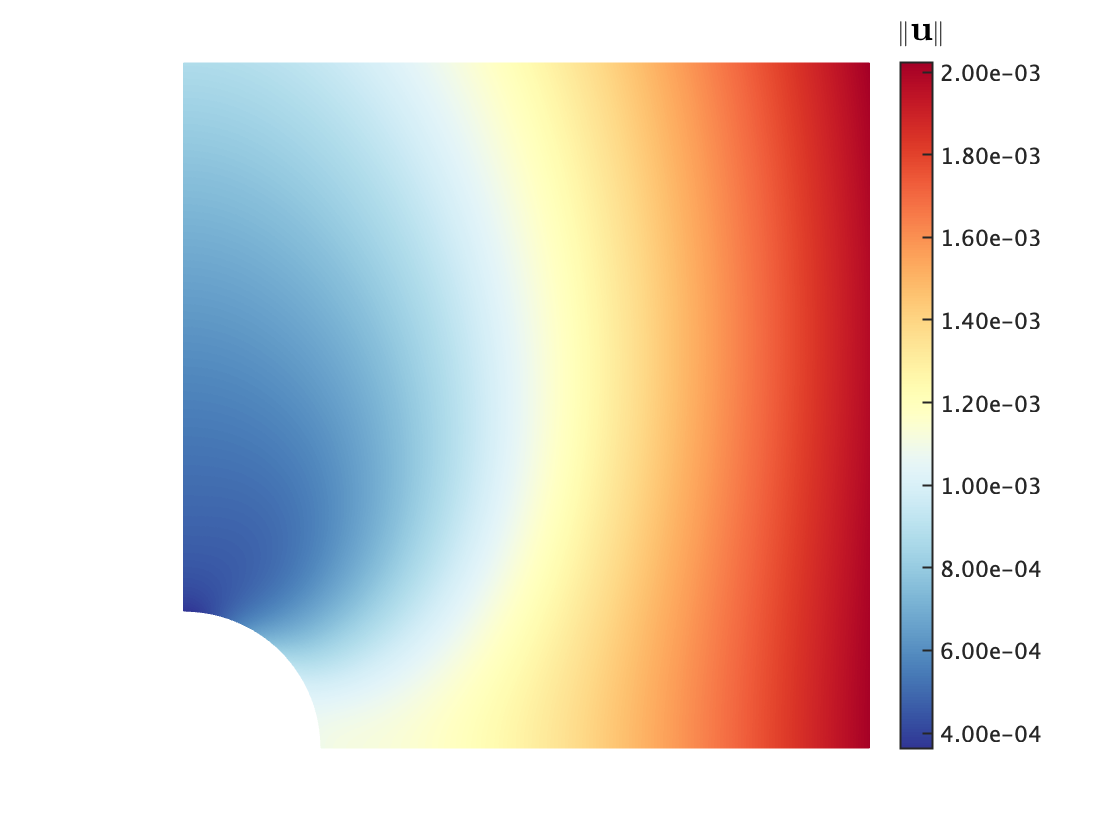}\par
    \parbox{.48\linewidth}{\centering  LEP-BE}\hfill
    \parbox{.48\linewidth}{\centering  LEB-PH}
    }
    \caption{Visualization of the solution for elastic problems.}
    \label{fig:elast}
\end{figure}

\paragraph*{Linear Elasticity}
A commonly used 2D linear elasticity test problem is applying a parabolic load (named LEP, Figure \ref{fig:elast}, left) on a cantilever beam, with $E=10^7$\si{\pascal} and $\nu=0.3$. The Neumann boundary conditions on the clamped edges are applied using the analytical solution provided by Timoshenko and Goodier
{\footnotesize\begin{equation*}
    \begin{aligned}
        &u_x(x,y) = -\frac{P_y}{6\bar{E} I} \big((6 \ell - 3x) x + (2+\bar{\nu}) y^2 - \frac{3 h^2}{2}(1+\bar{\nu})\big),\\
        &u_y(x, y) = \frac{P_y}{6\bar{E} I} \big(3\bar{\nu}y^2(\ell-x) + (3\ell-x)x^2\big),
    \end{aligned}
\end{equation*}}
where $P_y = -1000$, $\bar{E} = E/(1-\nu^2)$, $\bar{\nu} = \nu/(1-\nu)$, $\ell$ and $h$ respectively denotes the length and height of the beam, and $I=\ell h^3 /12$ indicates the second-area moment of the beam section. 
Our second 2D problem (named LEB, Figure \ref{fig:elast}, right) applies to a domain with a hole, where orthogonal in-plane stress is used on the left and bottom boundaries. The material parameters are the same as those used above, with $E = 10^7$\si{\pascal} and $\nu = 0.3$. The Neumann boundary conditions are applied on the left and bottom sides of the plate
{\footnotesize\[
    f_{xy}(x, y) = - \big((\frac{a^2}{r^2} 0.5(\sin(2\theta) + \sin(4\theta))) - \frac{3a^4}{2r^4}\sin(4\theta)\big),
\]}
where $a$ ($a=0.4$ for our domain) is the radius of the hole, $r=\sqrt{x^2+y^2}$, and $\theta=\mathrm{atan}({y}/{x})$. The analytical solution is
{\footnotesize\[
    u_x = \frac{1}{4\mu} \big((\lambda+1.0)r\frac{\cos(\theta)}{2} + \frac{a^2}{r} ((1+\lambda)\cos\theta + \cos(3\theta)) - \frac{a^4}{r^3}\cos(3\theta)\big),
\]}
where $\mu={E}/(1+2\nu)$, $\lambda=3-4\nu$.

\subsection{Meshing}\label{sec:meshing}
\label{subsec:dataset:meshsing}
\begin{figure}
     \centering\footnotesize
     Voronoi polygon tessellation of a square domain with 800 cells.\\
     \includegraphics[width=.24\linewidth]{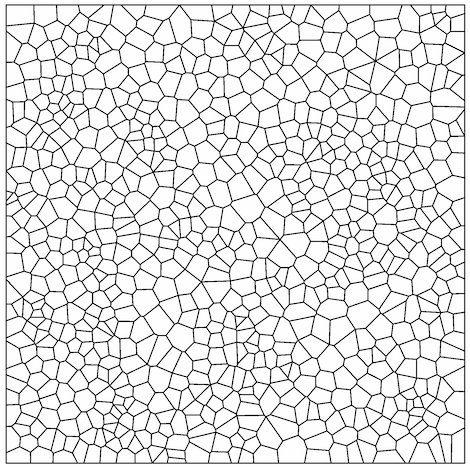}\hfill
     \includegraphics[width=.24\linewidth]{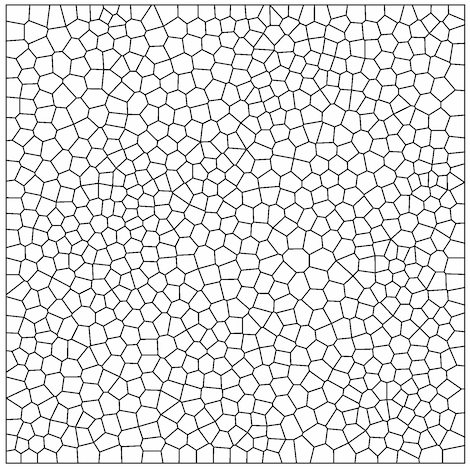}\hfill
     \includegraphics[width=.24\linewidth]{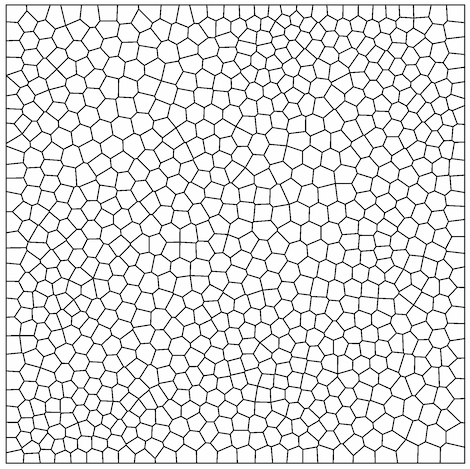}\hfill
     \includegraphics[width=.24\linewidth]{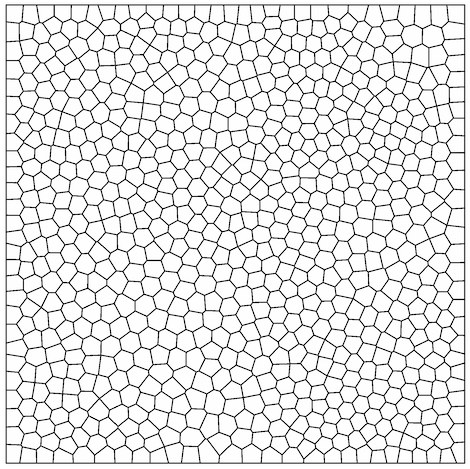}\par
     \parbox{.24\linewidth}{\centering 1 CVT (VP1)}\hfill
     \parbox{.24\linewidth}{\centering 5 CVT (VP2)}\hfill
     \parbox{.24\linewidth}{\centering 10 CVT (VP3)}\hfill
     \parbox{.24\linewidth}{\centering 20 CVT (VP4)}\\[1em]
     Displacement polygon tessellation of a square domain with 400 cells.\\
     \includegraphics[width=.19\linewidth]{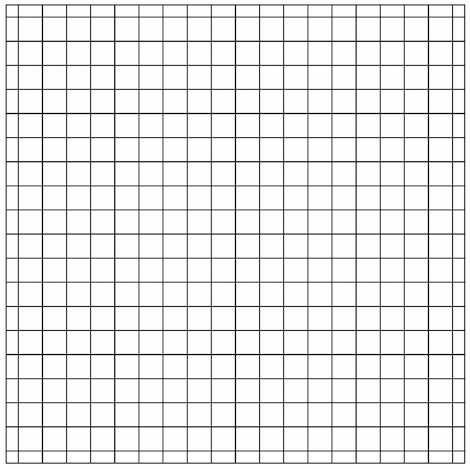}\hfill
     \includegraphics[width=.19\linewidth]{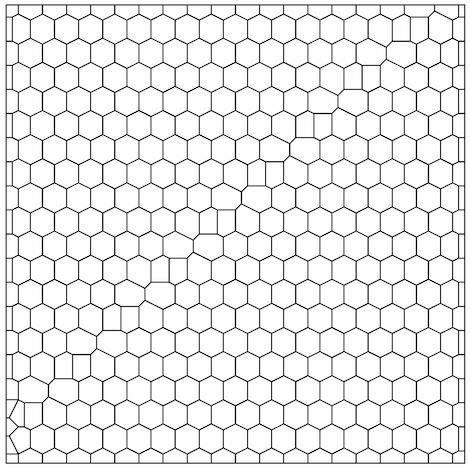}\hfill
     \includegraphics[width=.19\linewidth]{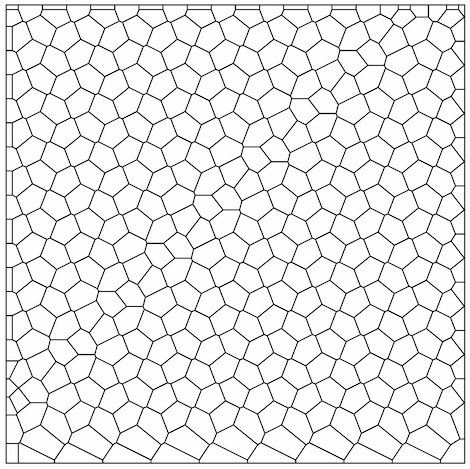}\hfill
     \includegraphics[width=.19\linewidth]{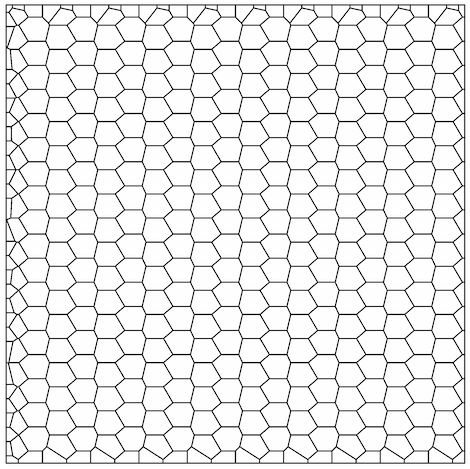}\hfill
     \includegraphics[width=.19\linewidth]{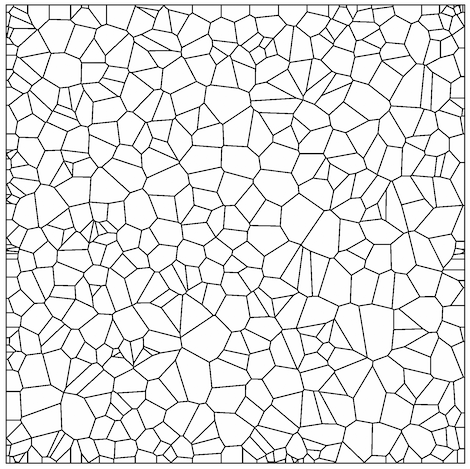}\par
     \parbox{.19\linewidth}{\centering Const (DP1)}\hfill
     \parbox{.19\linewidth}{\centering Uniform (DP2)}\hfill
     \parbox{.19\linewidth}{\centering Alter (DP3)}\hfill
     \parbox{.19\linewidth}{\centering Displace (DP4)}\hfill
     \parbox{.19\linewidth}{\centering Random (DP5)}\\[1em]
     Tiled polygon tessellation of a square domain from a $4\times4$ grid.\\
     \includegraphics[width=.19\linewidth]{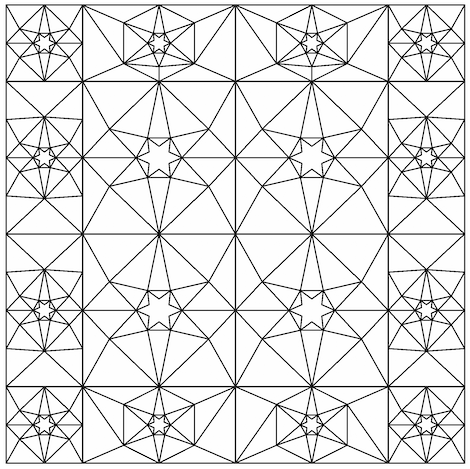}\hfill
     \includegraphics[width=.19\linewidth]{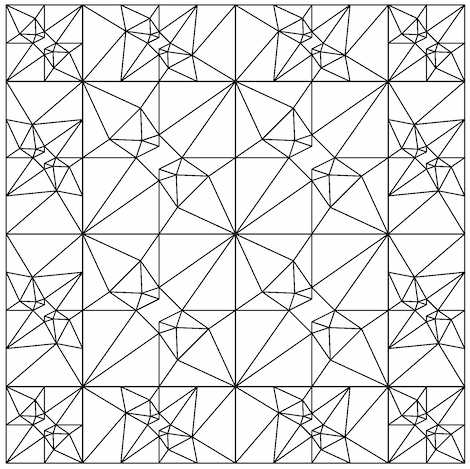}\hfill
     \includegraphics[width=.19\linewidth]{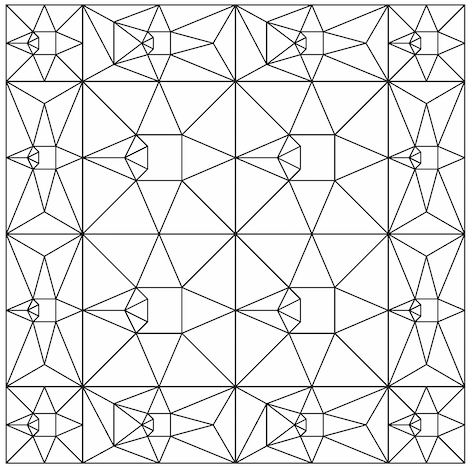}\hfill
     \includegraphics[width=.19\linewidth]{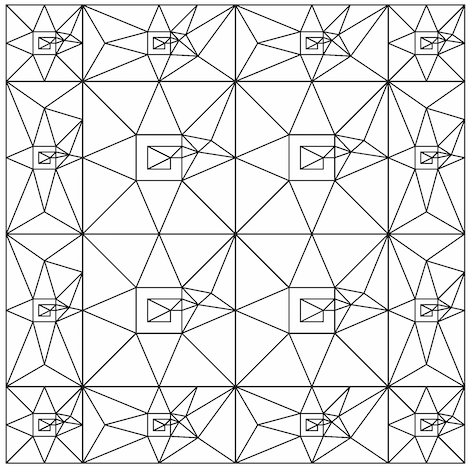}\hfill
     \includegraphics[width=.19\linewidth]{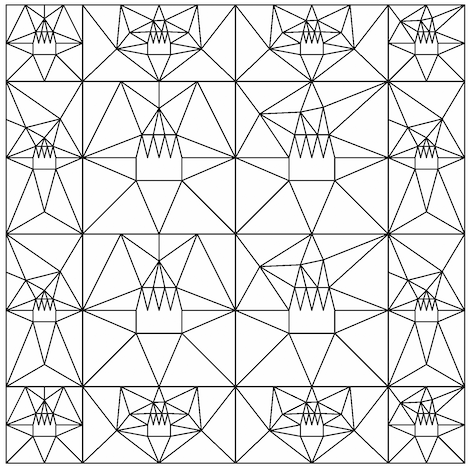}\par
     \includegraphics[width=.19\linewidth]{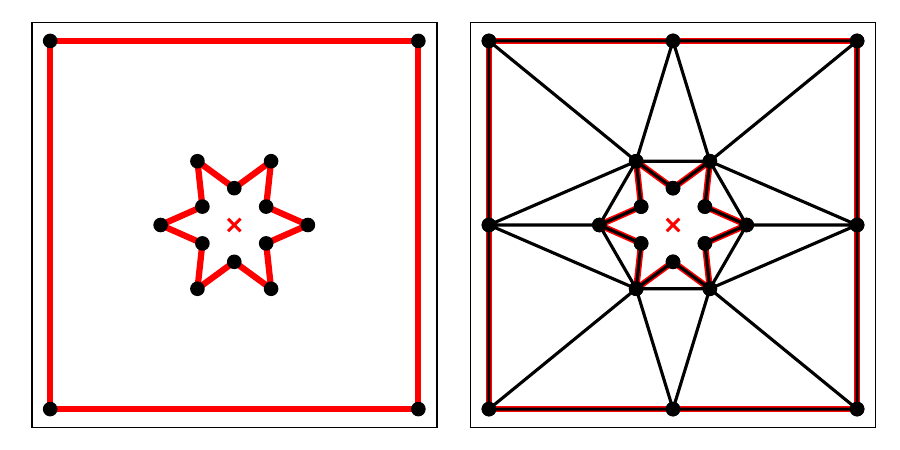}\hfill
     \includegraphics[width=.19\linewidth]{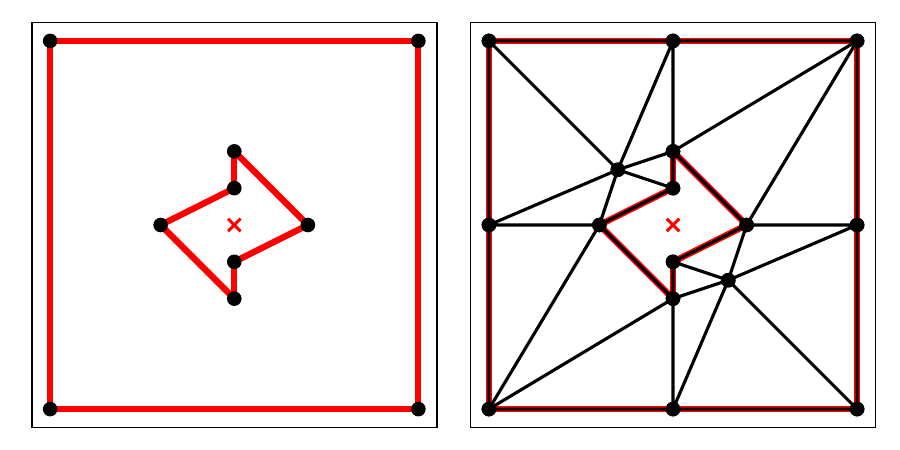}\hfill
     \includegraphics[width=.19\linewidth]{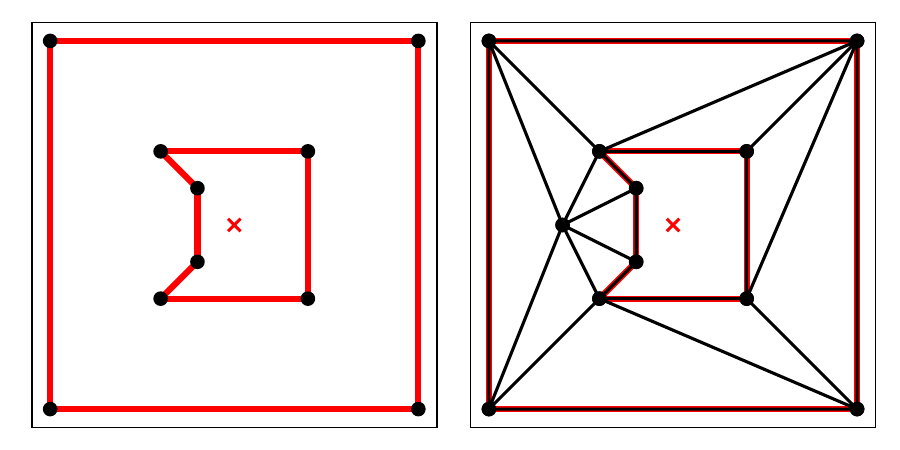}\hfill
     \includegraphics[width=.19\linewidth]{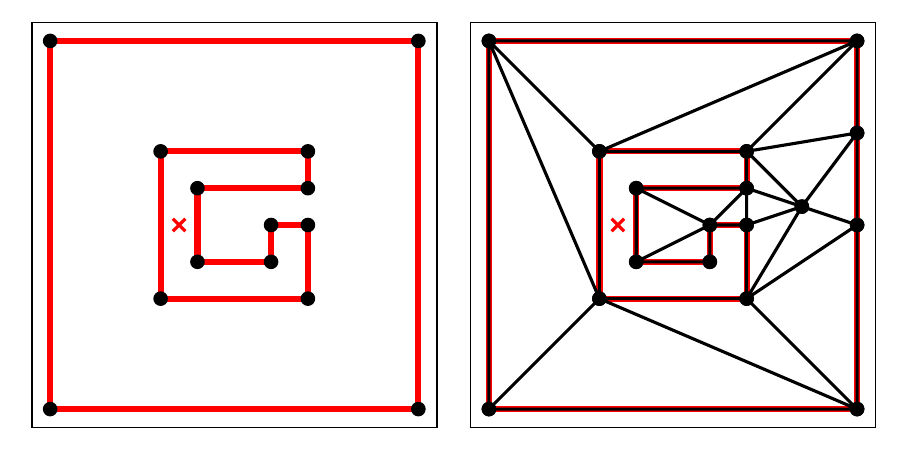}\hfill
     \includegraphics[width=.19\linewidth]{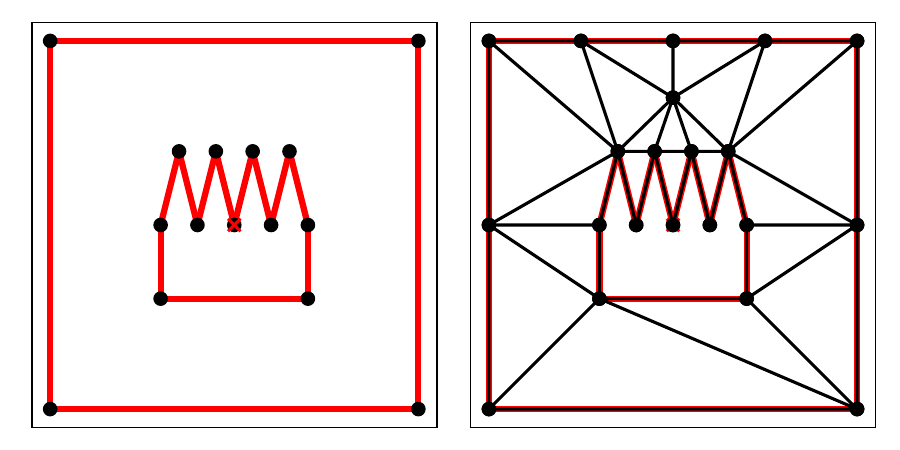}\par
     \parbox{.19\linewidth}{\centering Star (TP1)}\hfill
     \parbox{.19\linewidth}{\centering Z (TP2)}\hfill
     \parbox{.19\linewidth}{\centering U (TP3)}\hfill
     \parbox{.19\linewidth}{\centering Maze (TP4)}\hfill
     \parbox{.19\linewidth}{\centering Comb (TP5)}
     \caption{Example of all types of polygonal meshes in our dataset.}
     \label{fig:poly2d}
 \end{figure}

\begin{figure}
     \centering\footnotesize
     Triangular mesh of the dual of Voronoi polygon tessellation\\
     \includegraphics[width=.24\linewidth]{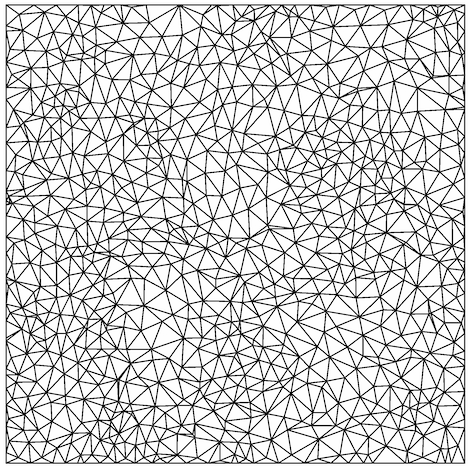}\hfill
     \includegraphics[width=.24\linewidth]{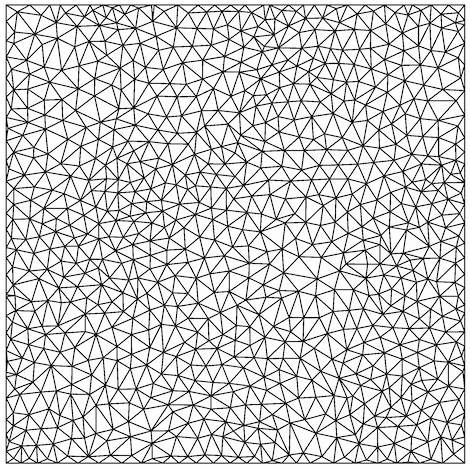}\hfill
     \includegraphics[width=.24\linewidth]{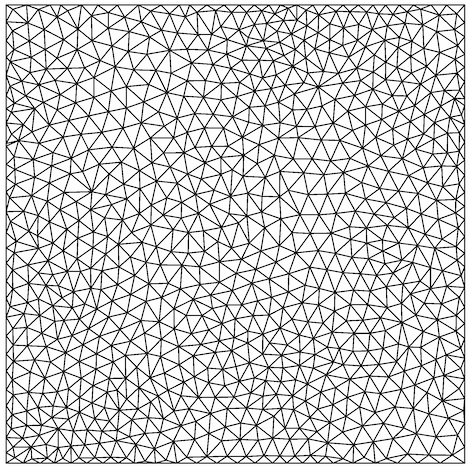}\hfill
     \includegraphics[width=.24\linewidth]{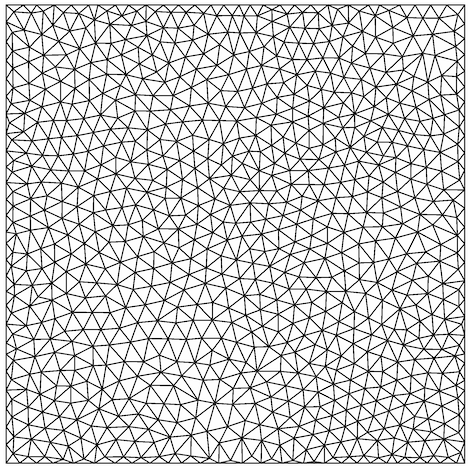}\par
     \parbox{.24\linewidth}{\centering 1 CVT (DT1)}\hfill
     \parbox{.24\linewidth}{\centering 5 CVT (DT2)}\hfill
     \parbox{.24\linewidth}{\centering 10 CVT (DT3)}\hfill
     \parbox{.24\linewidth}{\centering 20 CVT (DT4)}\\[1em]
     Triangulation inside polygonal elements.\\
     \includegraphics[width=.24\linewidth]{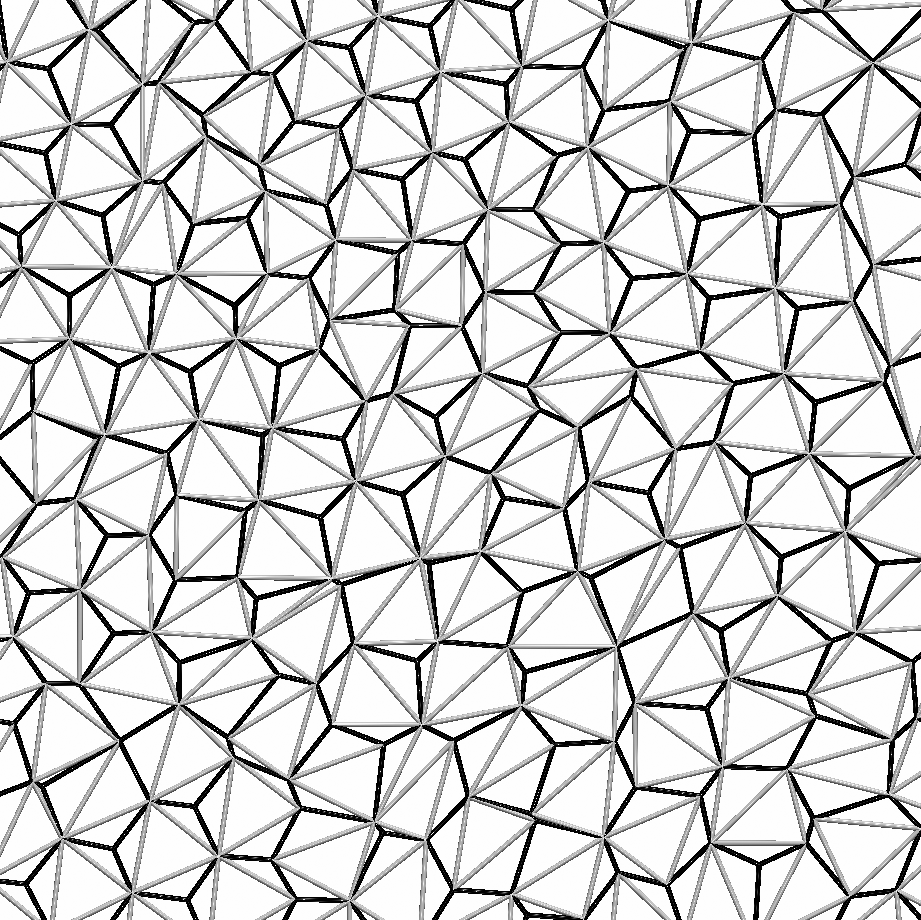}\hfill
     \includegraphics[width=.24\linewidth]{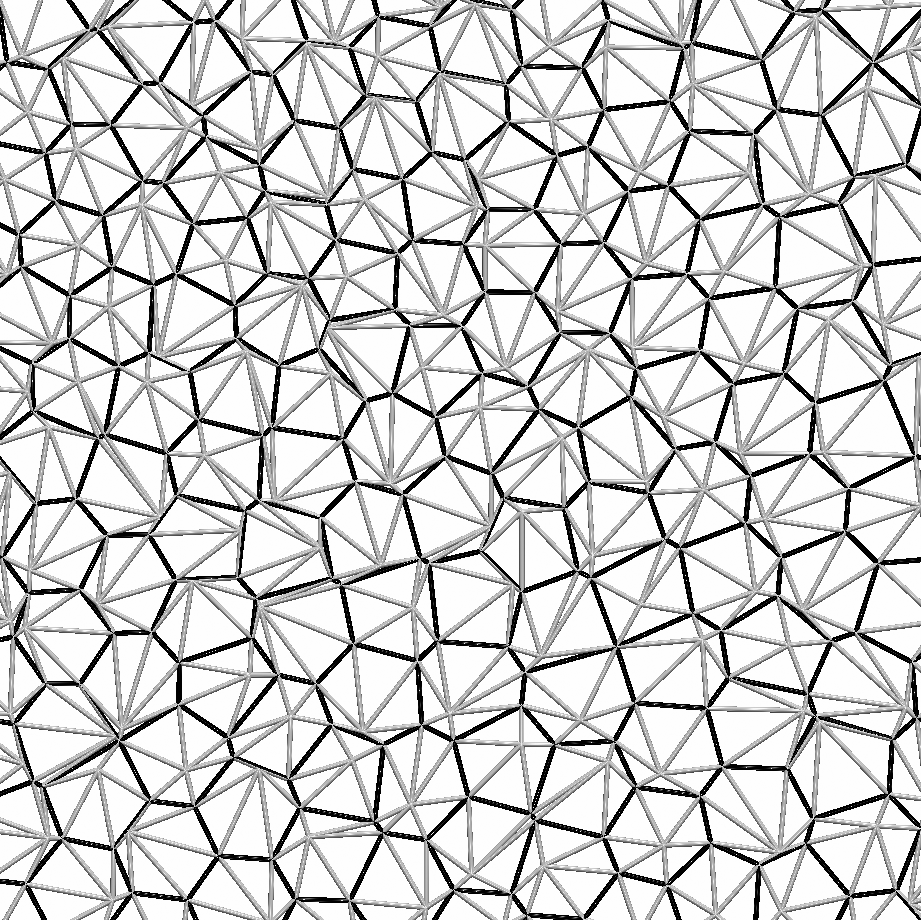}\hfill
     \includegraphics[width=.24\linewidth]{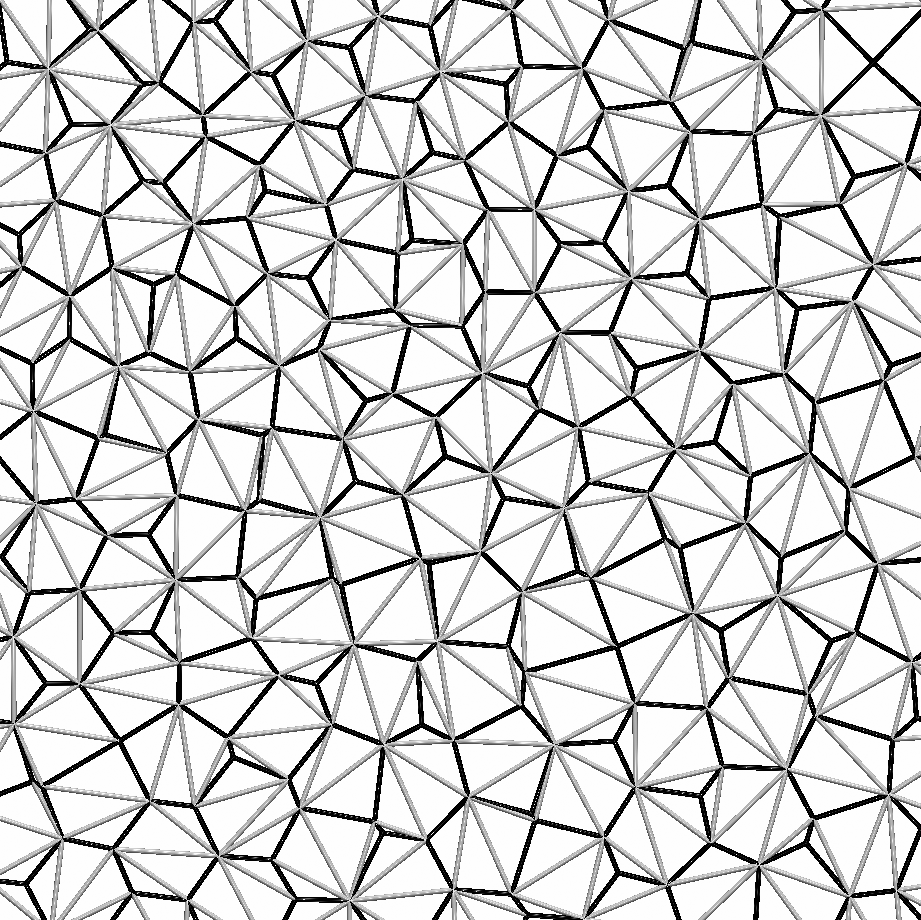}\hfill
     \includegraphics[width=.24\linewidth]{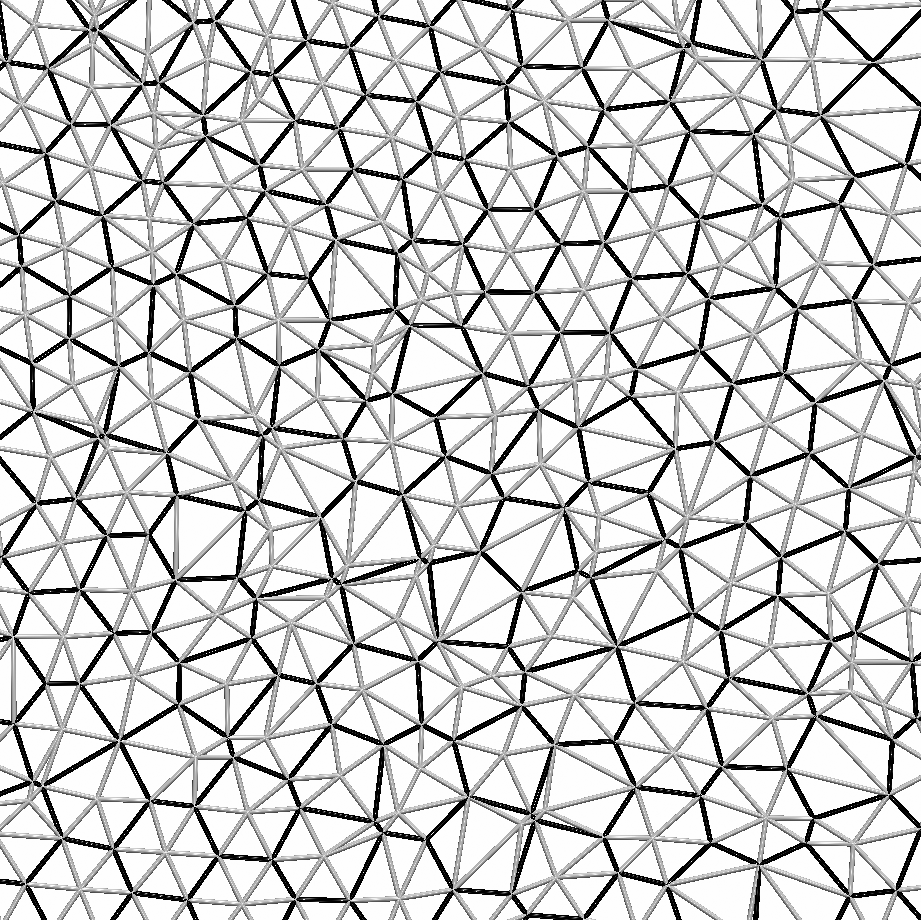}\par
     \parbox{.24\linewidth}{\centering Ear Clip (PT1)}\hfill
     \parbox{.24\linewidth}{\centering Insert (PT2)}\hfill
     \parbox{.24\linewidth}{\centering DT (PT3)}\hfill
     \parbox{.24\linewidth}{\centering CDT (PT4)}
     \caption{Example of triangular meshes in our dataset.}
     \label{fig:poly_tris}
 \end{figure}

\begin{figure}
     \centering\footnotesize
     \parbox{.48\linewidth}{\centering
     Voronoi polyhedral tessellation of a cube domain.\\
     \includegraphics[width=.32\linewidth]{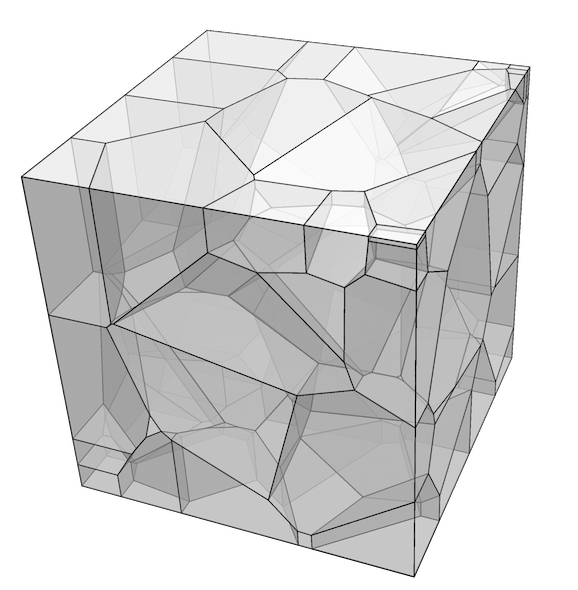}\hfill
     \includegraphics[width=.32\linewidth]{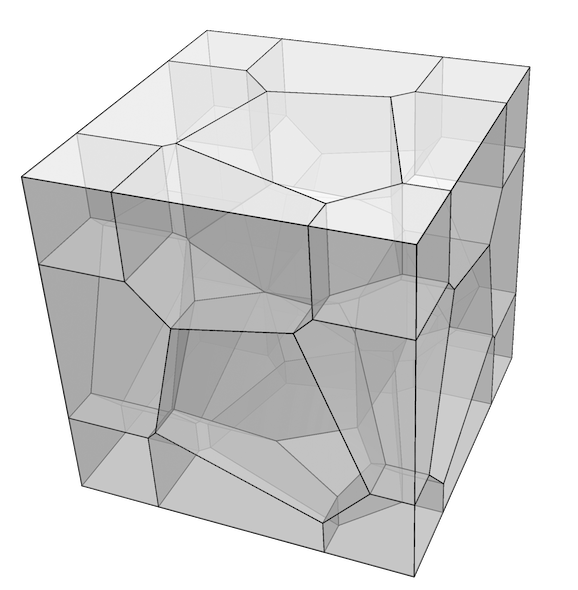}\hfill
     \includegraphics[width=.32\linewidth]{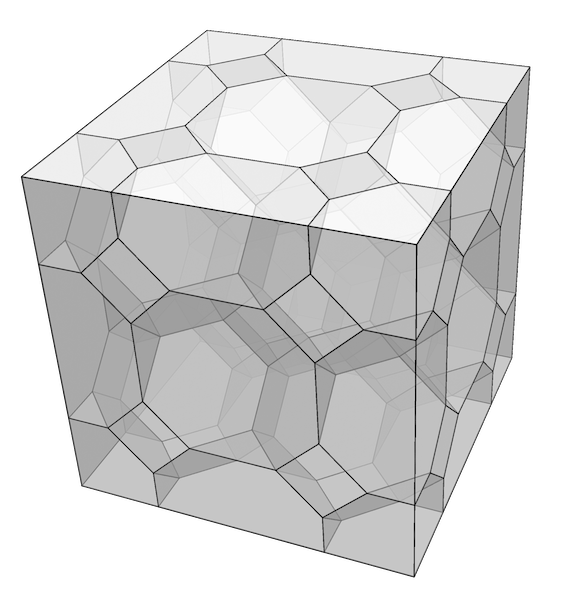}\par
     \parbox{.33\linewidth}{\centering Random (VP3D1)}\hfill
     \parbox{.33\linewidth}{\centering Poisson (VP3D2)}\hfill
     \parbox{.33\linewidth}{\centering BCL (VP3D3)}}\hfill
     \parbox{.48\linewidth}{\centering
     Displacement polyhedral meshes generated by displacing a grid.\\
     \includegraphics[width=.33\linewidth]{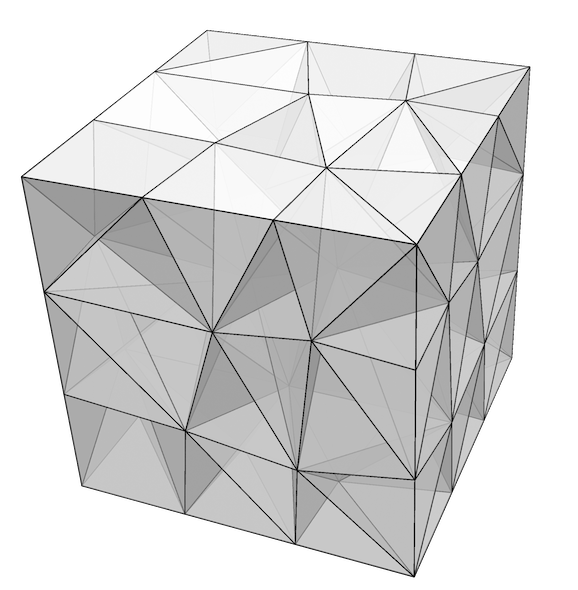}\hfill
     \includegraphics[width=.33\linewidth]{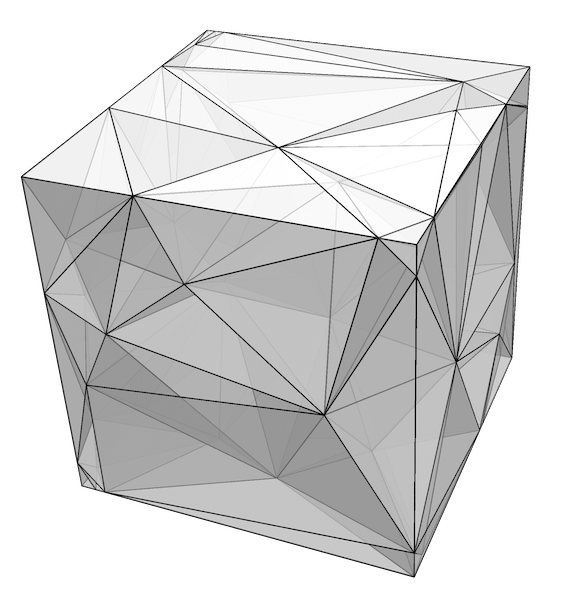}\hfill
     \includegraphics[width=.33\linewidth]{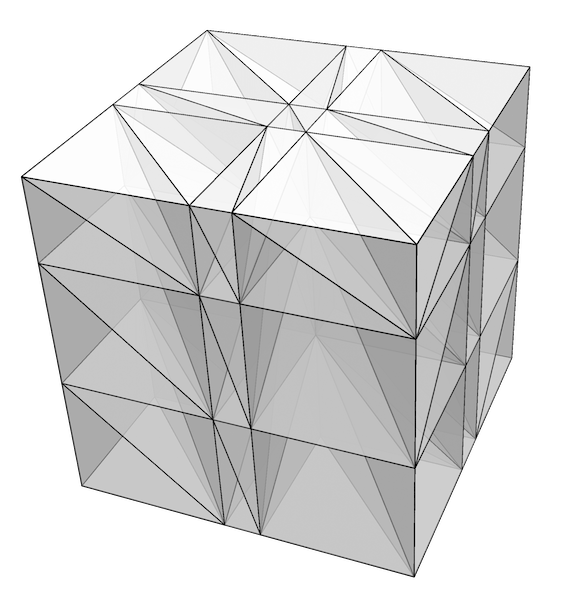}\par
     \parbox{.33\linewidth}{\centering Uniform (DP3D1)}\hfill
     \parbox{.33\linewidth}{\centering Anisotropic (DP3D2)}\hfill
     \parbox{.33\linewidth}{\centering Parallel (DP3D3)}}
     \caption{Example of all types of polyhedral meshes in our dataset.}
     \label{fig:poly3d}
 \end{figure}

\begin{figure}
     \centering\footnotesize
     \parbox{.48\linewidth}{\centering
     Tetrahedral mesh of the dual of Voronoi polyhedral tessellation.\\
     \includegraphics[width=.33\linewidth]{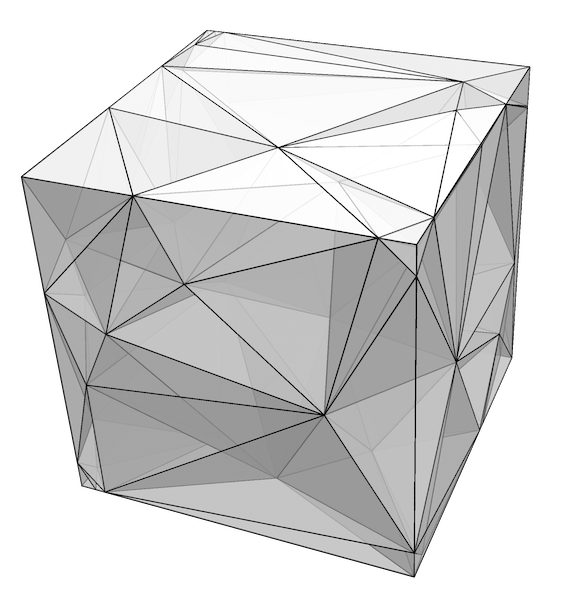}\hfill
     \includegraphics[width=.33\linewidth]{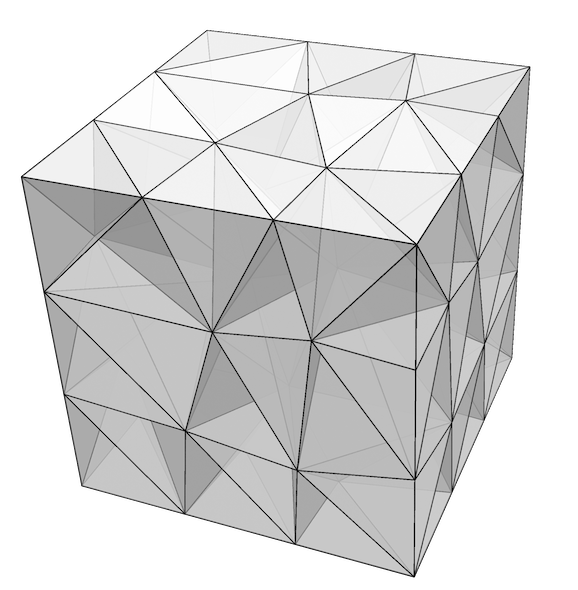}\hfill
     \includegraphics[width=.33\linewidth]{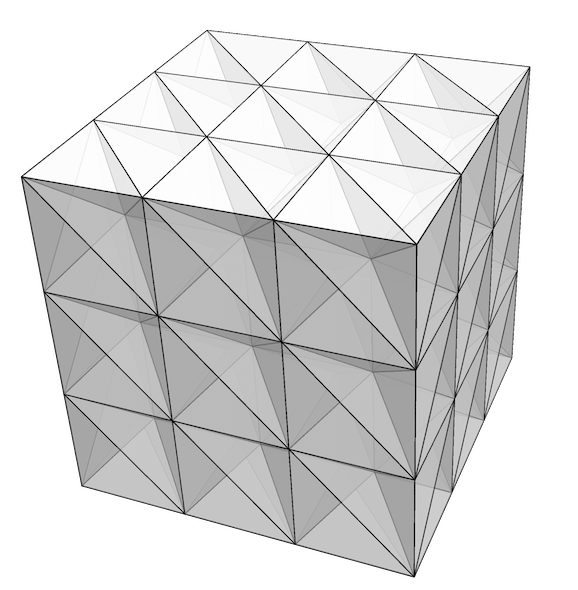}\par
     \parbox{.33\linewidth}{\centering Random (DT3D1)}\hfill
     \parbox{.33\linewidth}{\centering Poisson (DT3D2)}\hfill
     \parbox{.33\linewidth}{\centering BCL (DT3D3)}}\hfill
     \parbox{.48\linewidth}{\centering
     Tetrahedral mesh generated by tetrahedralizing inside polyhedral elements\\
     \includegraphics[width=.33\linewidth]{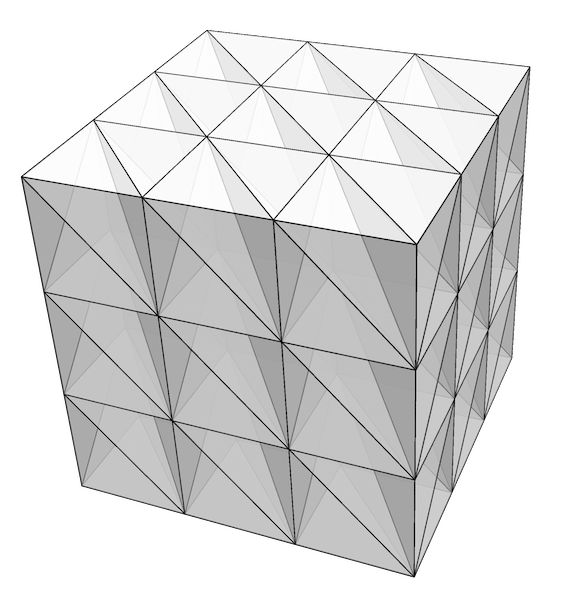}\hfill
     \includegraphics[width=.33\linewidth]{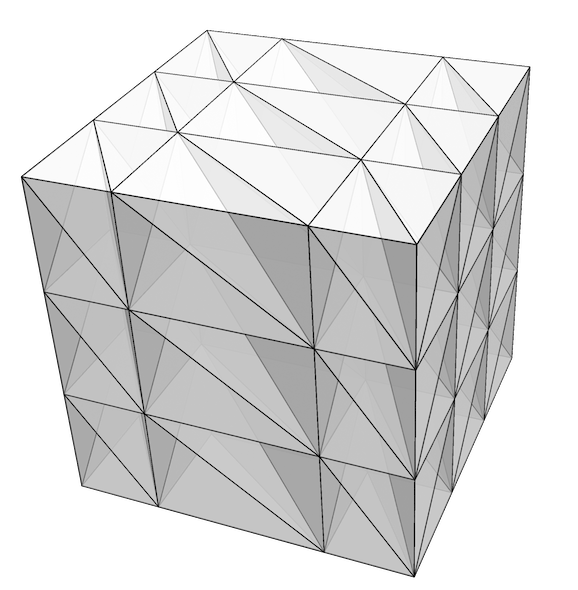}\hfill
     \includegraphics[width=.33\linewidth]{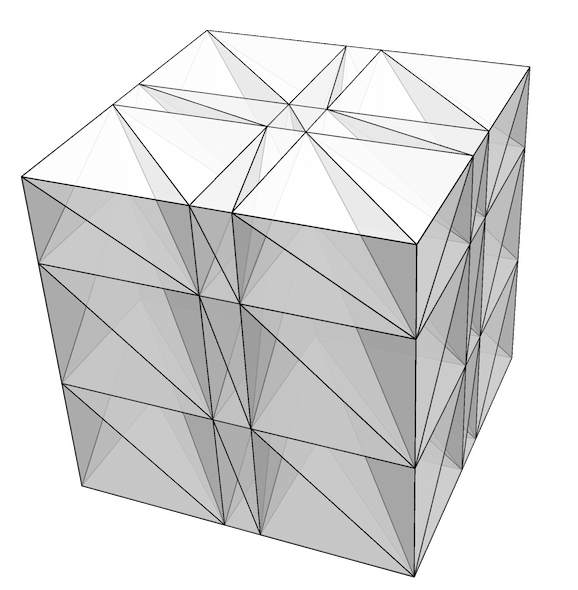}\par
     \parbox{.33\linewidth}{\centering Uniform (PT3D1)}\hfill
     \parbox{.33\linewidth}{\centering Anisotropic (PT3D2)}\hfill
     \parbox{.33\linewidth}{\centering Parallel (PT3D3)}}
     \caption{Example of tetrahedral meshes in our dataset.}
     \label{fig:poly_tets}
 \end{figure}

For every domain, we devise three strategies to generate polygonal meshes (Figure~\ref{fig:poly2d}): (1) Voronoi (VP) with four different CVT iterations; (2) Displacement (DP) with five different fields; and (3) Tiled (TP) with five tiles. We convert every polygonal mesh into simplices (Figure~\ref{fig:poly_tris}) by (1), for the four Voronoi meshes, using the dual operation (DT); and (2) by four different triangulations of the polygons (PT).
In 3D, we use similar strategies (Figure~\ref{fig:poly3d}): (1) Voronoi (VP3D) with three different samplings; and (2) Displacement (DP3D) with three different fields. We convert every polyhedral mesh into tetrahedra (Figure~\ref{fig:poly_tets}) by (1), for three Voronoi meshes, using the dual operation (DT3D); and (2) by three tetrahedralizations of the polyhedra (PT3D). As every meshing technique has a 2D and 3D version and several variants, we use the method's name followed by the variant number. For instance, VP3D2 is the second variant of the 3D Voronoi mesh (Voronoi mesh based on Poisson disk sampling). To generate a fair dataset, we design several methods to ``convert'' our polygonal meshes into simplicials. Every approach strives to maintain the same number of vertices.

\paragraph*{Voronoi}
In our experiments, we generate four Voronoi polygonal meshes using PolyMesher~\cite{PolyMesher} for each domain, with the CVT iteration number set as 1, 5, 10, and 20.
In 3D, Voro++ \cite{Voro++} provides three distributions of points: (1) Random sampling (similar to PolyMesher \cite{PolyMesher} with one iteration); (2) Poisson sampling with $1/n_P$ with $n_P$ the number of points; and (3). Body-Centered Lattice (BCL) sampling point distribution (similar to PolyMesher \cite{PolyMesher} with 20 iterations).

\begin{figure}
    \centering\footnotesize
    \parbox{.65\linewidth}{\centering
    \includegraphics[width=.33\linewidth]{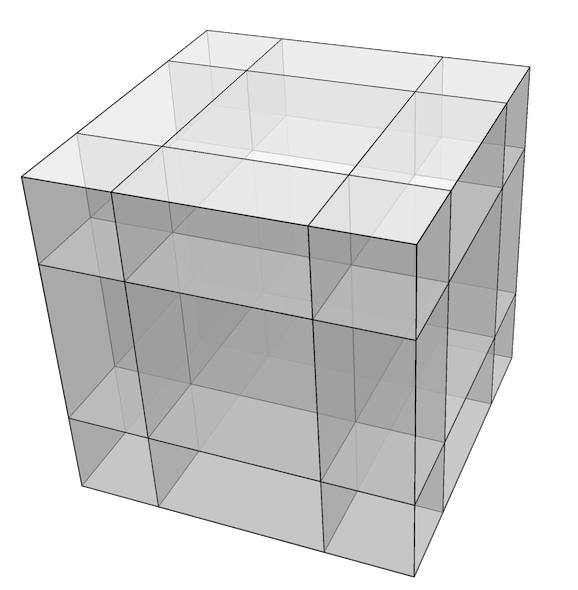}\hfill
    \includegraphics[width=.33\linewidth]{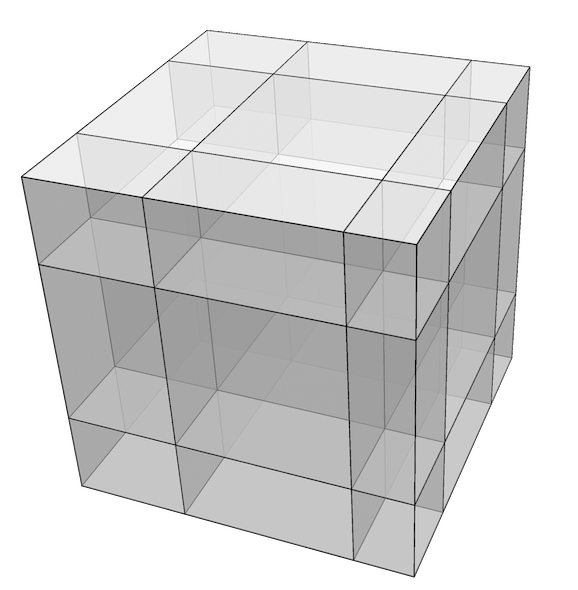}\hfill
    \includegraphics[width=.33\linewidth]{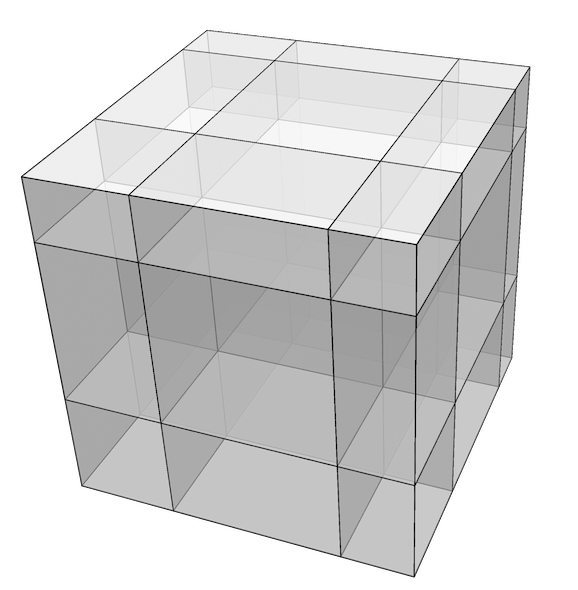}\par
    \parbox{.33\linewidth}{\centering Uniform (HP3D1)}\hfill
    \parbox{.33\linewidth}{\centering Anisotropic (HP3D2)}\hfill
    \parbox{.33\linewidth}{\centering Parallel (HP3D3)}
    }
    \caption{Displacement polyhedral meshes by displacing a grid.}
    \label{fig:polyhedra_displace}
\end{figure}

\paragraph*{Displacement}
Displacement tessellations of 2D domains use several displacement functions to change the distribution of vertices from regularly structured meshes~\cite{Displacement-Polygon-Meshing-1,Displacement-Polygon-Meshing-2}. In our experiments, we use Veamy \cite{Veamy}, a software that collects commonly used displacement functions, including five functions for the distribution, (1) constant distribution, (2) uniform distribution, (3) altered distribution, (4) displaced distribution, and (5) random noise distribution along both the $x$ and $y$ axis.
Displacement meshes can be naturally extended to 3D, and \cite{Polyheral-quality-indicator} provides three kinds of displaced grids with different point distribution (Figure \ref{fig:polyhedra_displace}): (1) Uniform; (2) Anisotropic (the distribution is fixed along two axes and incrementally enlarged along the other axis); and (3) Parallel (the points are first uniformly sampled and then randomly moved to another inner plane which is parallel to the original plane. For example, the points sharing the same $x$ coordinate are randomly changed with $x$ value by the same quantity, and the same operation iterates to the $y$ and $z$ coordinates). After we generate the initial displaced grid, we obtain the polyhedral meshes by tetrahedralizing every cell and randomly aggregate $20\%$ of the elements to generate possibly non-convex polyhedral elements.

\paragraph*{Tiled}
Tiled polygonal meshing generates challenging meshes since most Voronoi and Displacement meshing produces mostly convex polygonal tessellations.
In our experiments, we use a similar approach as \citet{VEM-quality-benchmark}, in which we generate one concave-shaped polygon inside each element of structured quadrilateral meshes and then we perform a constrained Delaunay triangulation to fill the empty space. We use similar settings as in \citet{VEM-quality-benchmark} to generate five concave polygonal shapes (star-shaped, Z-shaped, U-shaped, maze-shaped, and comb-shaped) for each domain.


\paragraph*{Dual}
For Voronoi meshes, the easiest triangulation method consists of using the dual to obtain a Delaunay mesh. In 3D, we use \citet{Polyheral-quality-indicator} to obtain the dual tetrahedral meshes.

\paragraph*{Triangulation}
Another method consists of triangulating the inside of each polygonal cell. In our experiment, we use four triangulation methods: (1) Ear Clip \cite{mapbox-earclip} using Mapbox; (2) insert a random point in each polygon cell and connect it with the boundary; (3) Constraint Delaunay Triangulation using~\cite{Triangle} with no additional Stainer points; and (4) Conforming Delaunay Triangulation using~\cite{Triangle}. We note that methods 1 and 2 work only with Voronoi and displacement meshes as the cells are convex.
\paragraph*{Tetrahedralization}
\citet{Polyheral-quality-indicator} tetrahedralize each polyhedral cell by iteratively splitting a polyhedral element with a diagonal plane until the left element is a tetrahedron.

\subsection{Our Benchmark Study}
\label{subsec:dataset:benchmark}
We summaries the large-scale dataset for our benchmark study by the combinations of the domains, problems, meshing techniques, and meshing instances. We also provide tables in Appendix~\ref{append:sec:dataset-tables} to summarize combinations of our notations introduced above, since these combinations would be frequently referred to in our results analysis in Section~\ref{sec:results}.

We conduct experiments for all the combinations of problems and domains introduced above, resulting in 36 problem-domain pairs (Poisson problems shown in Table~\ref{tab:poisson-problem-domain} and linear elasticity problems  shown in Table~\ref{tab:elasticity-problem-domain} ).

For each problem-domain pair, we conduct experiments on a large collection of mesh tessellations, namely 14 types of polygon meshes and their corresponding triangle meshes with 9 triangulation methods. Table~\ref{tab:mesh:voro-tri} and Table~\ref{tab:mesh:dp-tri} show all possible in-element triangulation meshing options from polygon meshes. Table~\ref{tab:mesh:voro-dual} shows triangulation through dual of Voronoi meshes.

For each type of meshing, we generated 20-25 different resolutions. We perform our benchmark study on this large-scale dataset of combinations of domains, problems, meshing techniques, and mesh resolutions.

%% file: 05-experiments.tex
\section{Case Study Results}
\label{sec:results}

For all the problems, we collect several metrics, in particular the $L^2$ error, $H_1$ error, and the time required to solve the linear system.  All experiments are run on a cluster node with 2 Intel E5-2683v4 2.1GHz CPUs and 250GB memory, each with 16 cores and a max 125GB of reserved memory. 

\paragraph{Overview}
We showcase our comparisons and findings from the mesh discretizations perspective (Section~\ref{subsec:mesh}) and simulation techniques perspective (Section~\ref{subsec:simulation}), and discuss the generalization of our findings in Section~\ref{subsec:extensive}. In each perspective, we conduct detailed comparisons of various technique choices and showcase only the most representative plots. We refer to our additional material for all the data. For instance, we only report $L^2$ as $H_1$ show similar trends. To facilitate the readability of the plots, we aggregate all related data in a unique band. For instance, in Figure~\ref{fig:compare_PDEs}, we aggregate all results coming from different Delaunay meshes (DT) into a unique band bounded by the largest/smallest error.

\paragraph{Plots}
We plot most of our diagrams showing the convergence of solutions as we increase the mesh resolution, resulting in plots comparing the $L^2$ error with solving time. The convergence efficiency can be interpreted by selecting an $L^2$ error level and comparing the required solving time to achieve such error (slicing a horizontal line) or by selecting a solving time and comparing the $L^2$ error (slicing a vertical line). More intuitively, the lines/bands that appear more towards the upper-right of our plots (i.e., requiring more time to achieve a specific $L^2$ error) perform worse than the lines/bands more towards the lower-left corner. For each of our studies, instead of showing just one plot for a specific problem-domain pair using a specific solver, we show several to showcase that our findings generalize and are not restricted to a particular problem or domain.

\paragraph{Assembly time}
We omit the assembly time in our comparison since it depends on specific implementations (e.g., programming language, libraries, or specific optimizations, could have significant impacts) and scales linearly with respect to mesh size, while solving time does not. On the other hand, solving time is controllable since we export the matrices and use the \emph{same} solver and the \emph{same} hardware. 

\subsection{Mesh Discretizations}
\label{subsec:mesh}
We report our comparisons of the time efficiency of different mesh discretizations. We first compare the polygon meshes with their triangulated simplicial meshes (Section~\ref{subsec:mesh:tripoly}). We then compare different polygon mesh discretization methods (Section~\ref{subsec:mesh:poly}). We further study how does mesh quality affect the efficiency (Section~\ref{subsec:mesh:quality}) and how do different meshes perform under extreme qualities (Section~\ref{subsec:mesh:best-worst}).

\subsubsection{Polygonal Meshes and Simplicial Meshes}
\label{subsec:mesh:tripoly}
We compare the performance of polygonal and simplicial meshes generated from the corresponding polygonal meshes using the triangulation methods introduced in Section \ref{sec:meshing}. In every comparison, we choose the Voronoi polygonal mesh with 20 iterations (VP4) and the displaced polygonal mesh with uniform distribution (DP2), which has the best mesh quality, and run experiments on both the direct solver and iterative solver. We show the results of PS\#1-SC, PS\#3-UD, and PB-LS and use four different triangulation methods. 

\begin{figure}
     \centering\footnotesize
     \parbox{.02\linewidth}{~}\hfill\hfill
     \parbox{.32\linewidth}{\centering PS\#1-SC}\hfill
     \parbox{.32\linewidth}{\centering PS\#3-UD}\hfill
     \parbox{.32\linewidth}{\centering PB-LS}\par
     \parbox{.02\linewidth}{\rotatebox{90}{\centering Direct Solver}}\hfill\hfill
     \parbox{.32\linewidth}{\includegraphics[width=\linewidth]{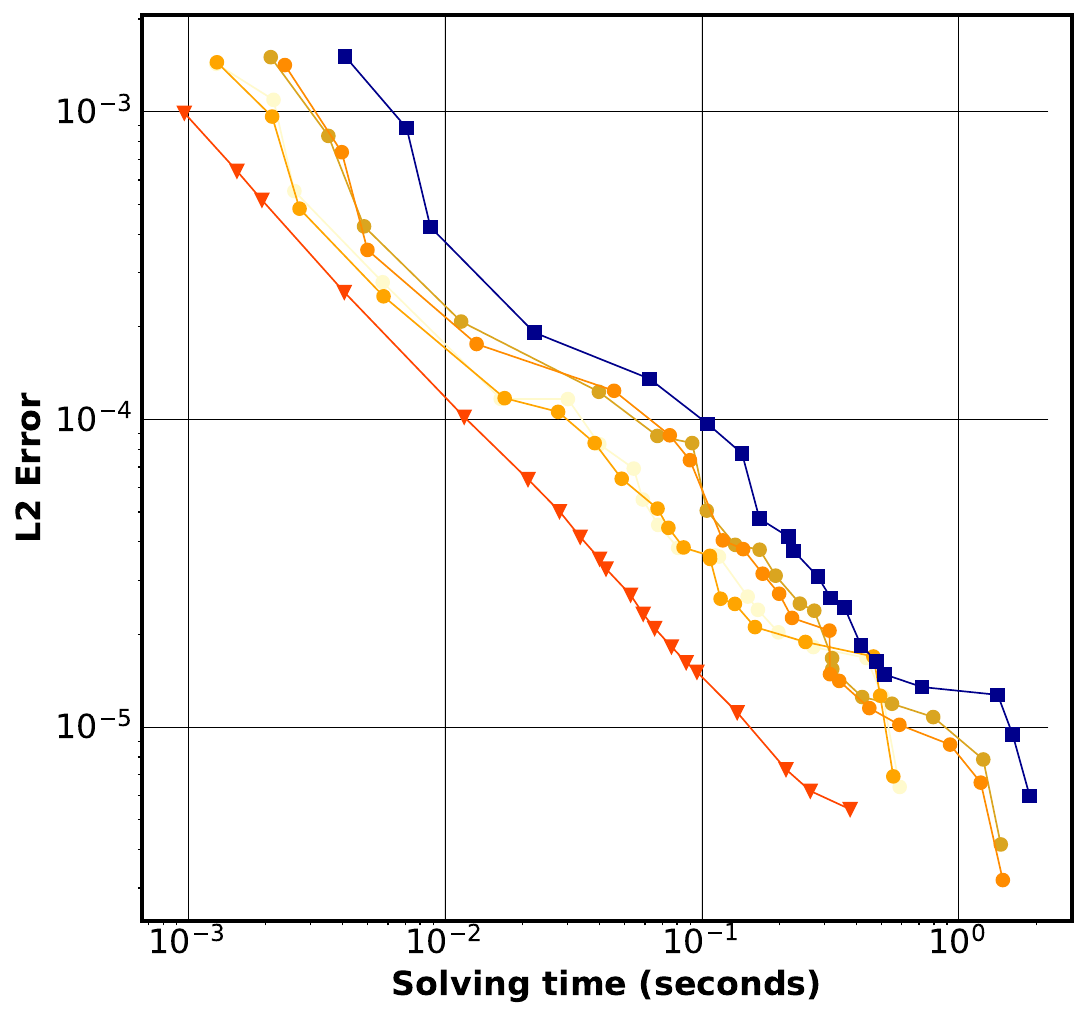}}\hfill
     \parbox{.32\linewidth}{\includegraphics[width=\linewidth]{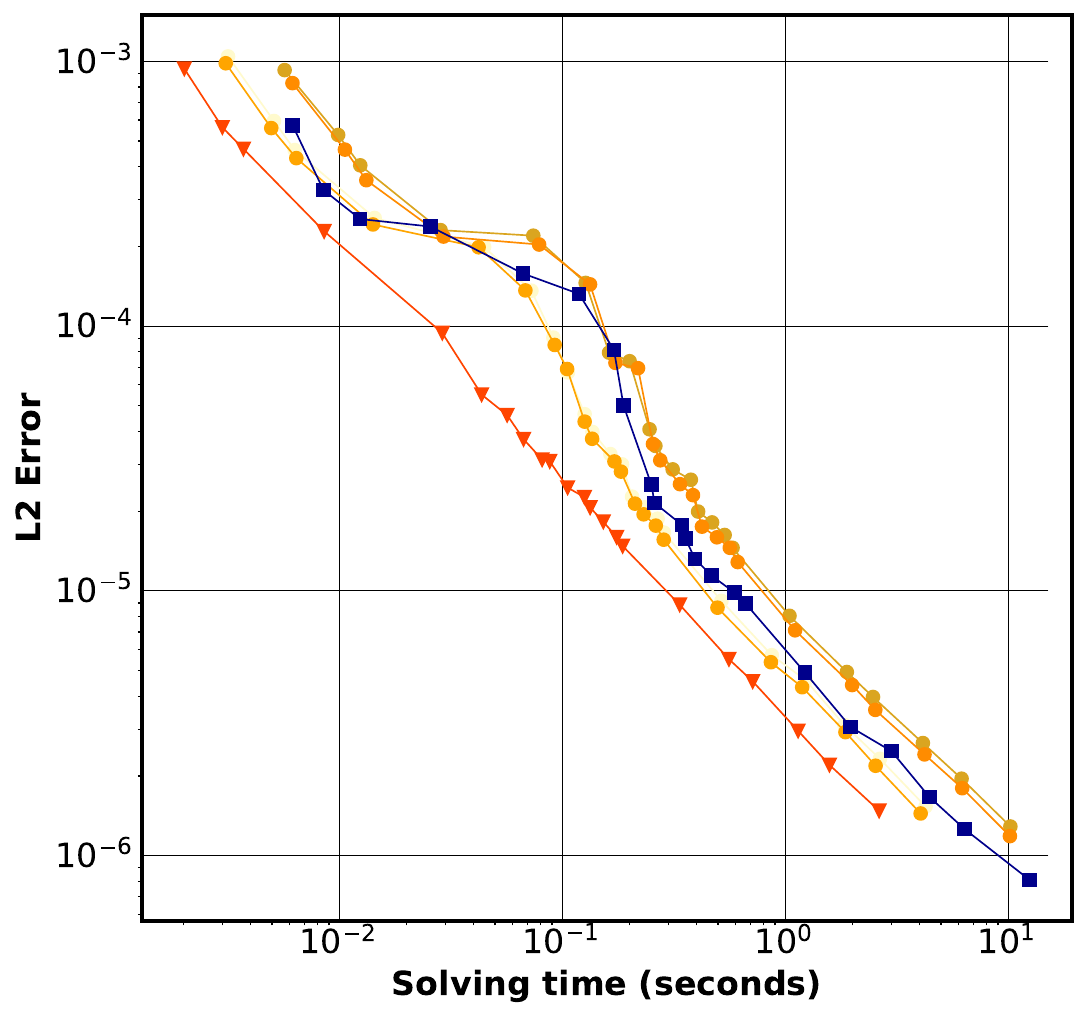}}\hfill
     \parbox{.32\linewidth}{\includegraphics[width=\linewidth]{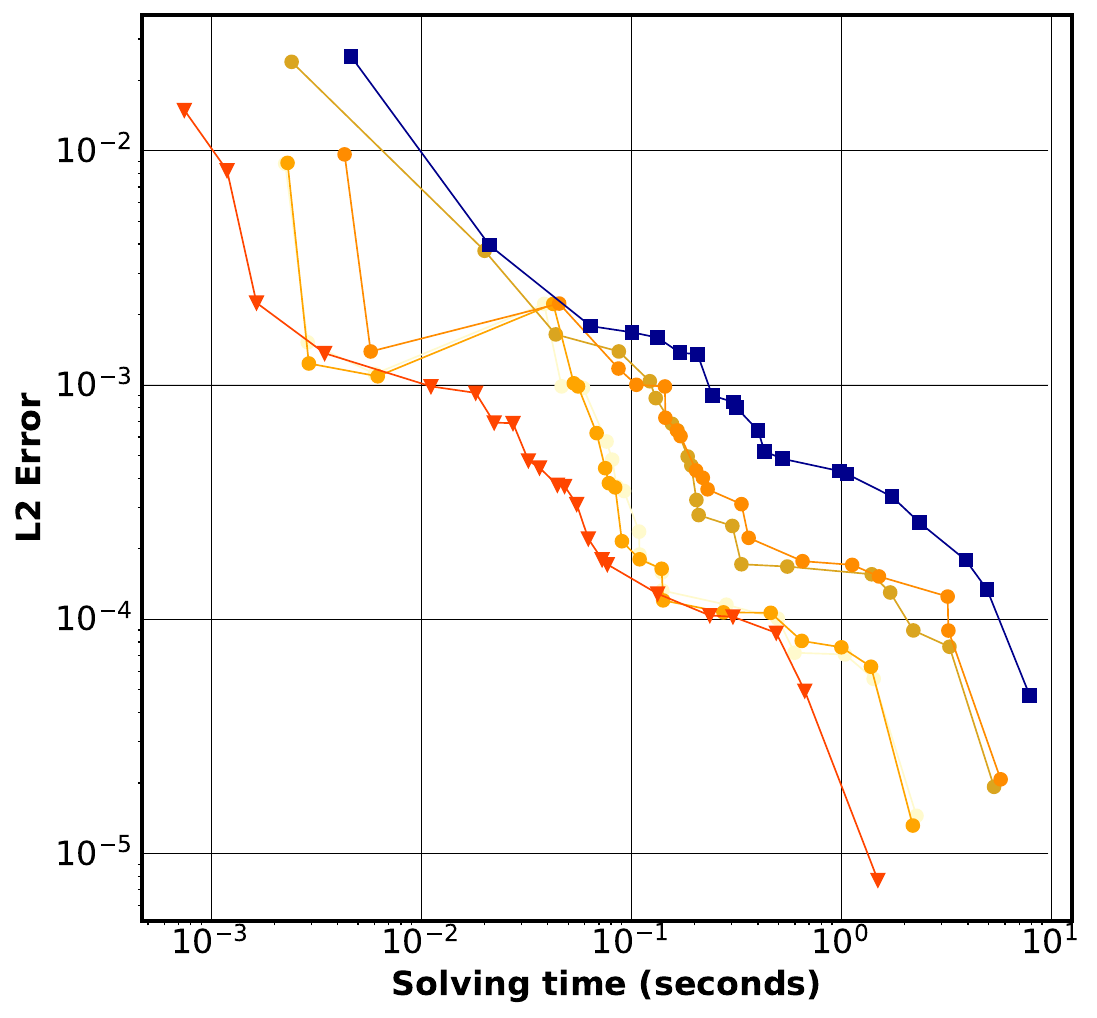}}\par
     \parbox{.02\linewidth}{\rotatebox{90}{\centering Iterative Solver}}\hfill\hfill
     \parbox{.32\linewidth}{\includegraphics[width=\linewidth]{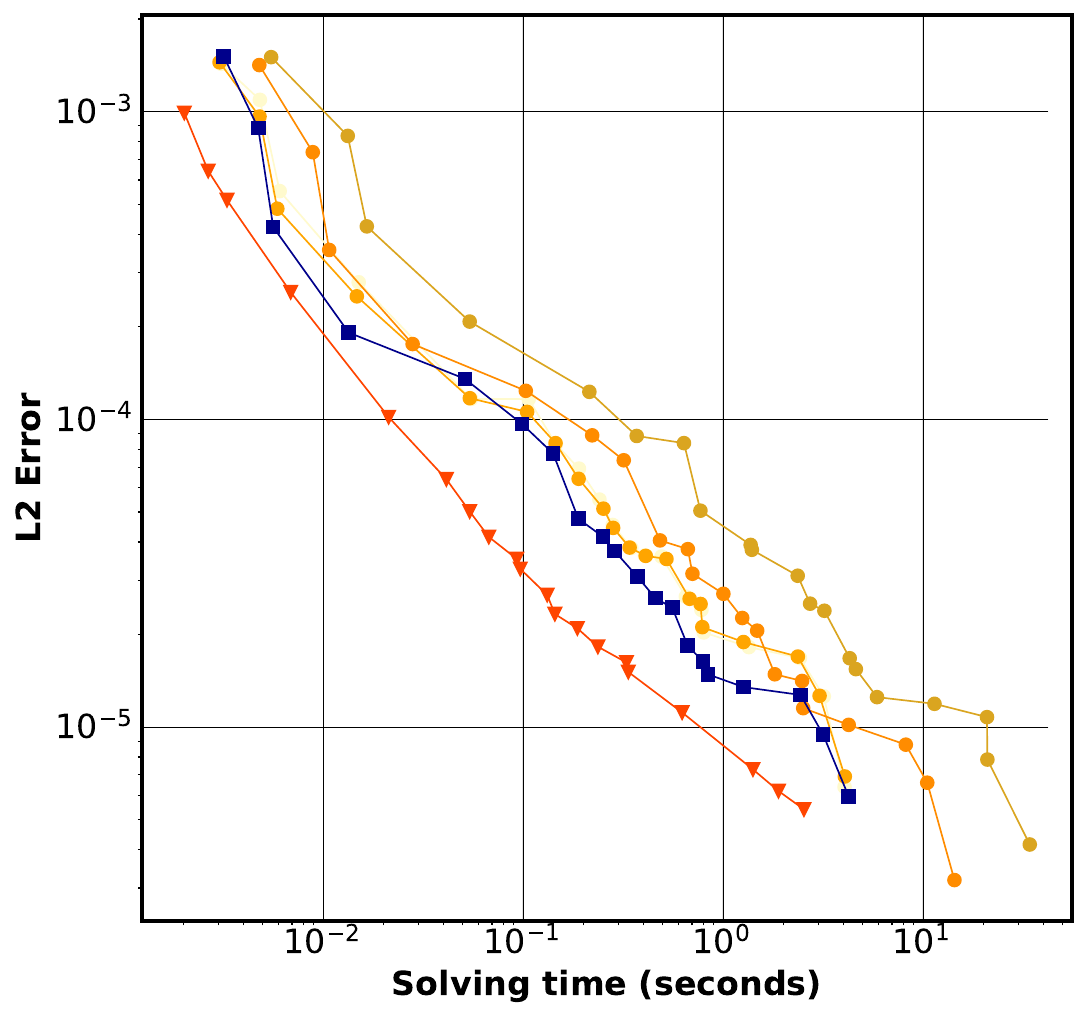}}\hfill
     \parbox{.32\linewidth}{\includegraphics[width=\linewidth]{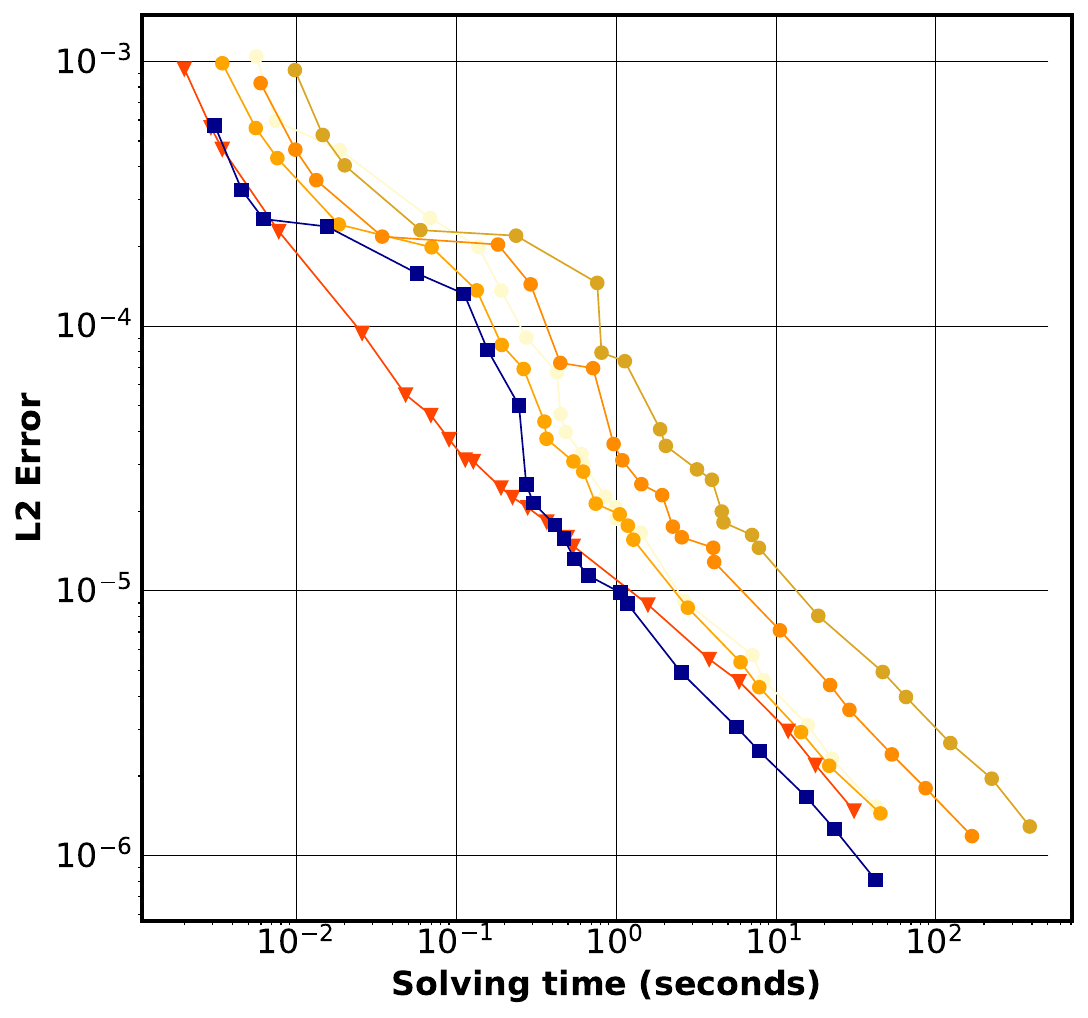}}\hfill
     \parbox{.32\linewidth}{\includegraphics[width=\linewidth]{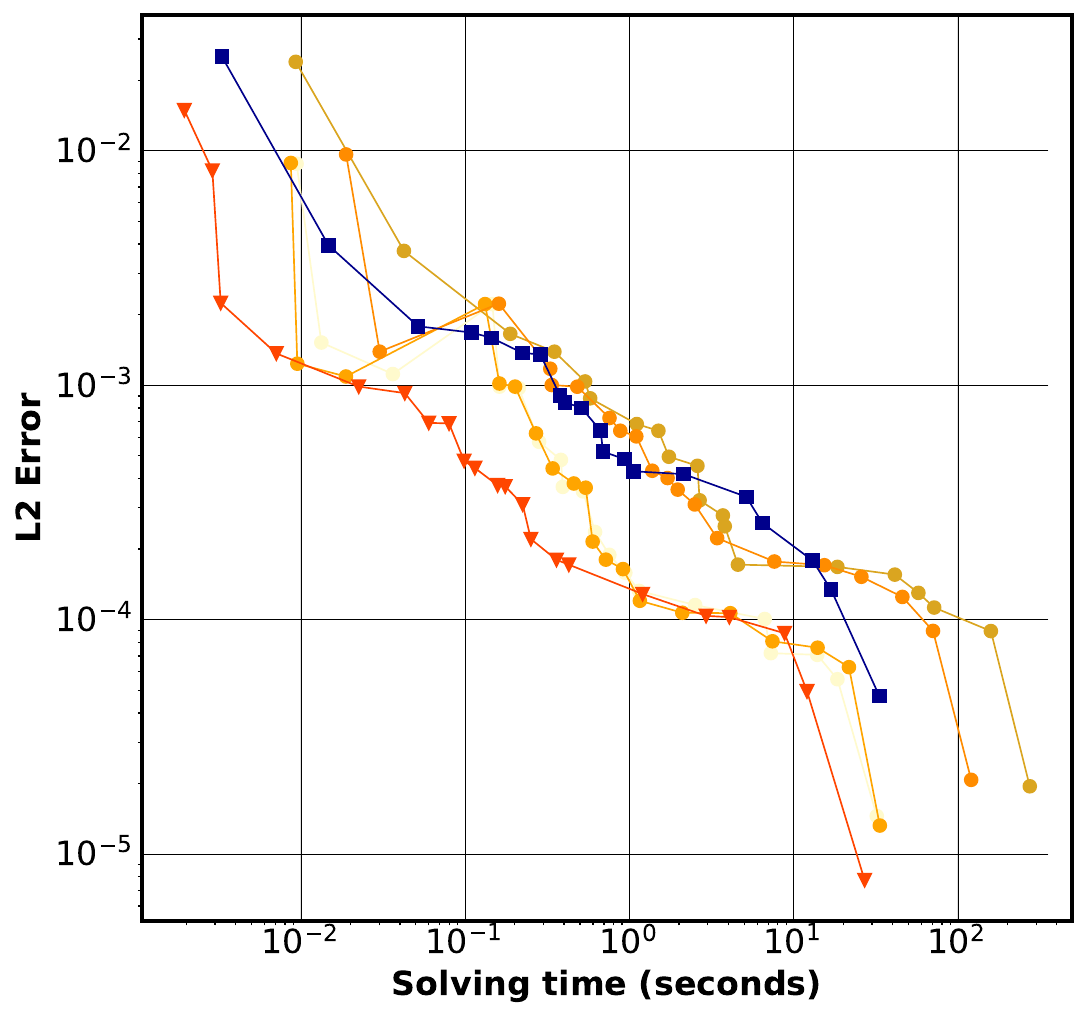}}\par
     \includegraphics[width=\linewidth]{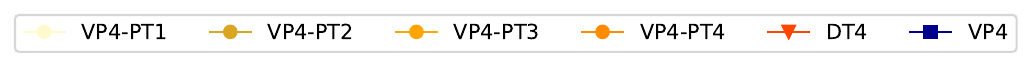}\par
     \caption{Solving time (s) versus $L_2$ error for VP4 to its corresponding dual and four triangulated meshes.}
     \label{fig:compare_poly_vs_simplicial_voro}
 \end{figure}

We first compare the Voronoi meshes with the dual triangular meshes and triangulated meshes. Figure \ref{fig:compare_poly_vs_simplicial_voro} shows the comparison of VP4 with the dual triangulation and four triangulations. For the results from the direct solver, we can observe that Voronoi meshes have poorer performance than almost all triangulation, which contain triangular meshes with relatively low quality, such as PT2. With the better performance of Voronoi polygonal meshes in iterative solvers, Voronoi polygonal meshes show competitive performance with most triangular meshes. However, the dual triangular meshes still perform better than the Voronoi polygonal meshes.

\begin{figure}
     \centering\footnotesize
     \parbox{.02\linewidth}{~}\hfill\hfill
     \parbox{.32\linewidth}{\centering PS\#1-SC}\hfill
     \parbox{.32\linewidth}{\centering PS\#3-UD}\hfill
     \parbox{.32\linewidth}{\centering PB-LS}\par
     \parbox{.02\linewidth}{\rotatebox{90}{\centering Direct Solver}}\hfill
     \parbox{.32\linewidth}{\includegraphics[width=\linewidth]{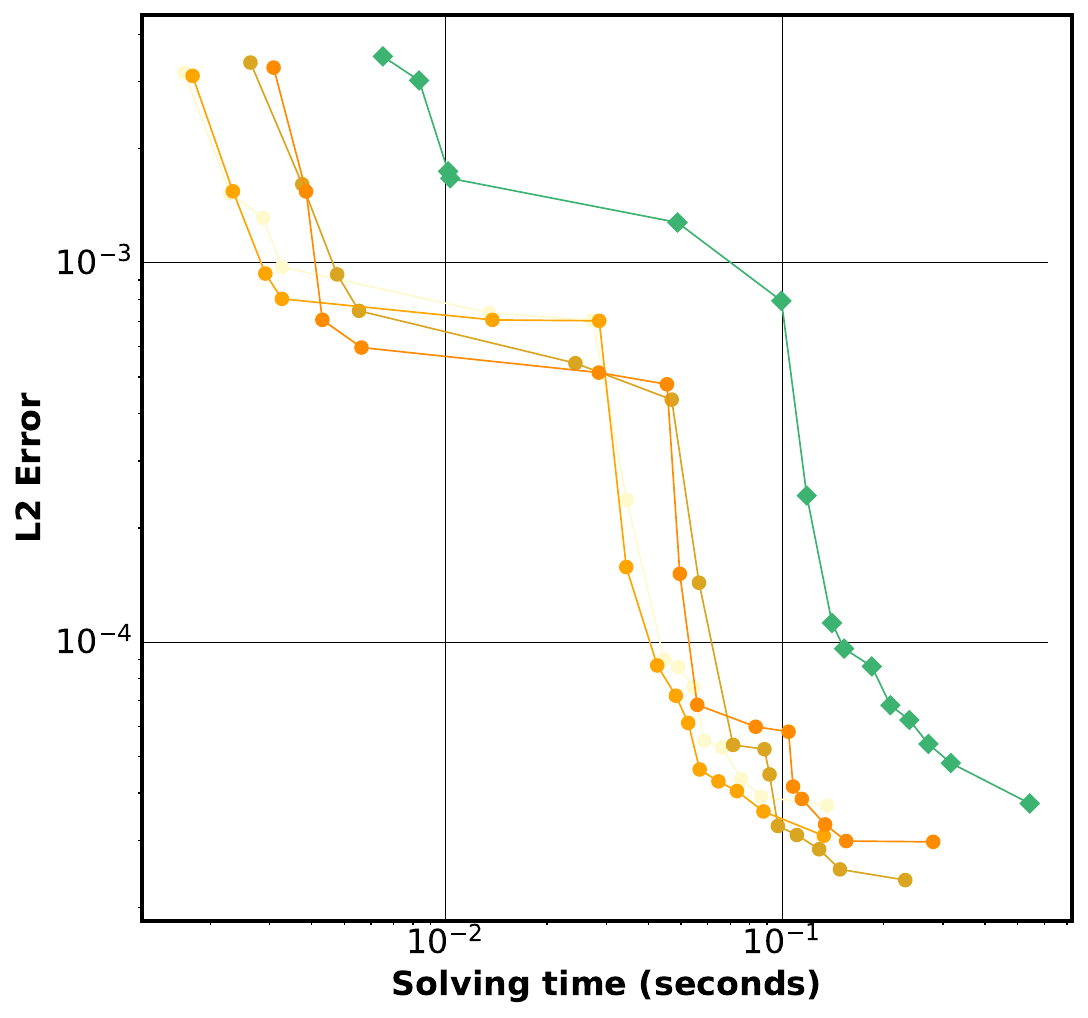}}\hfill
     \parbox{.32\linewidth}{\includegraphics[width=\linewidth]{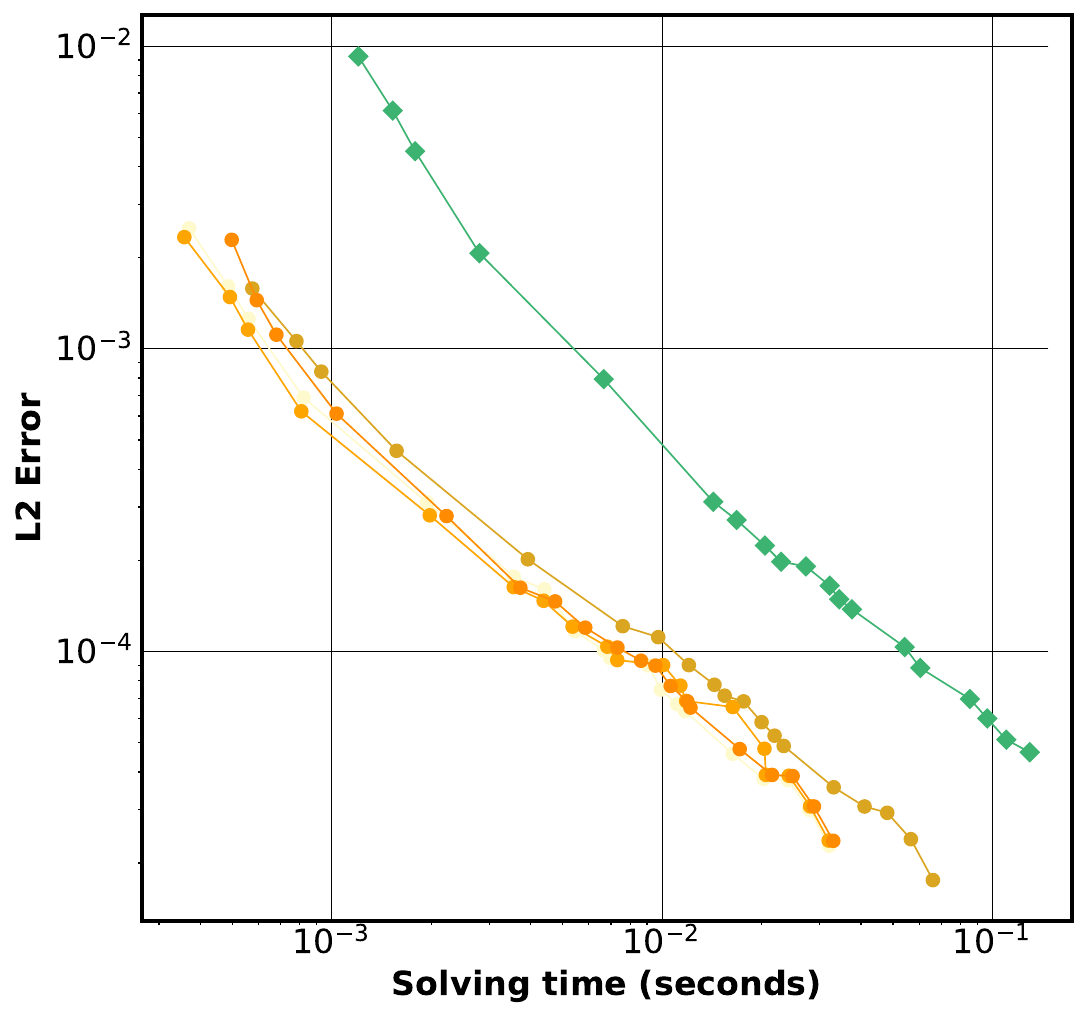}}\hfill
     \parbox{.32\linewidth}{\includegraphics[width=\linewidth]{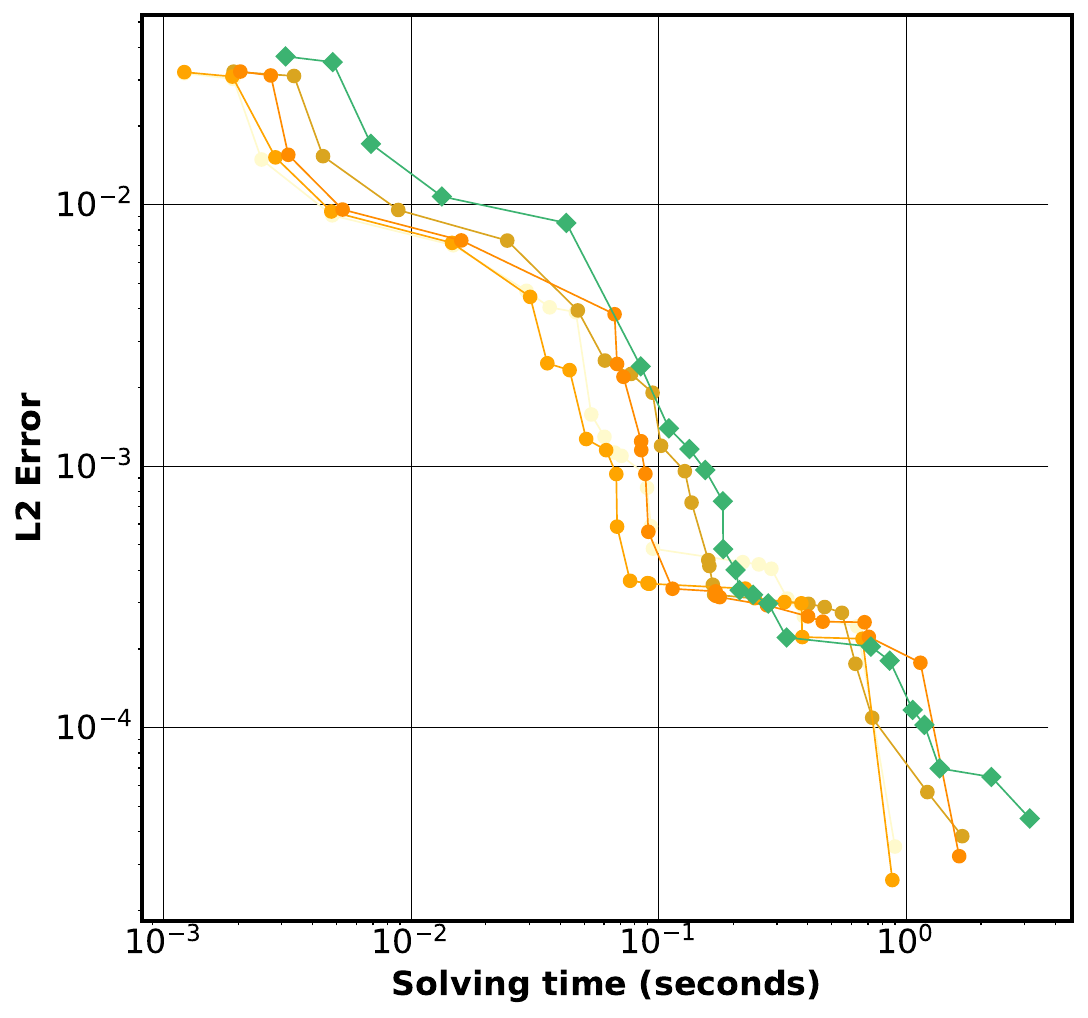}}\par
     \parbox{.02\linewidth}{\rotatebox{90}{\centering Iterative Solver}}\hfill
     \parbox{.32\linewidth}{\includegraphics[width=\linewidth]{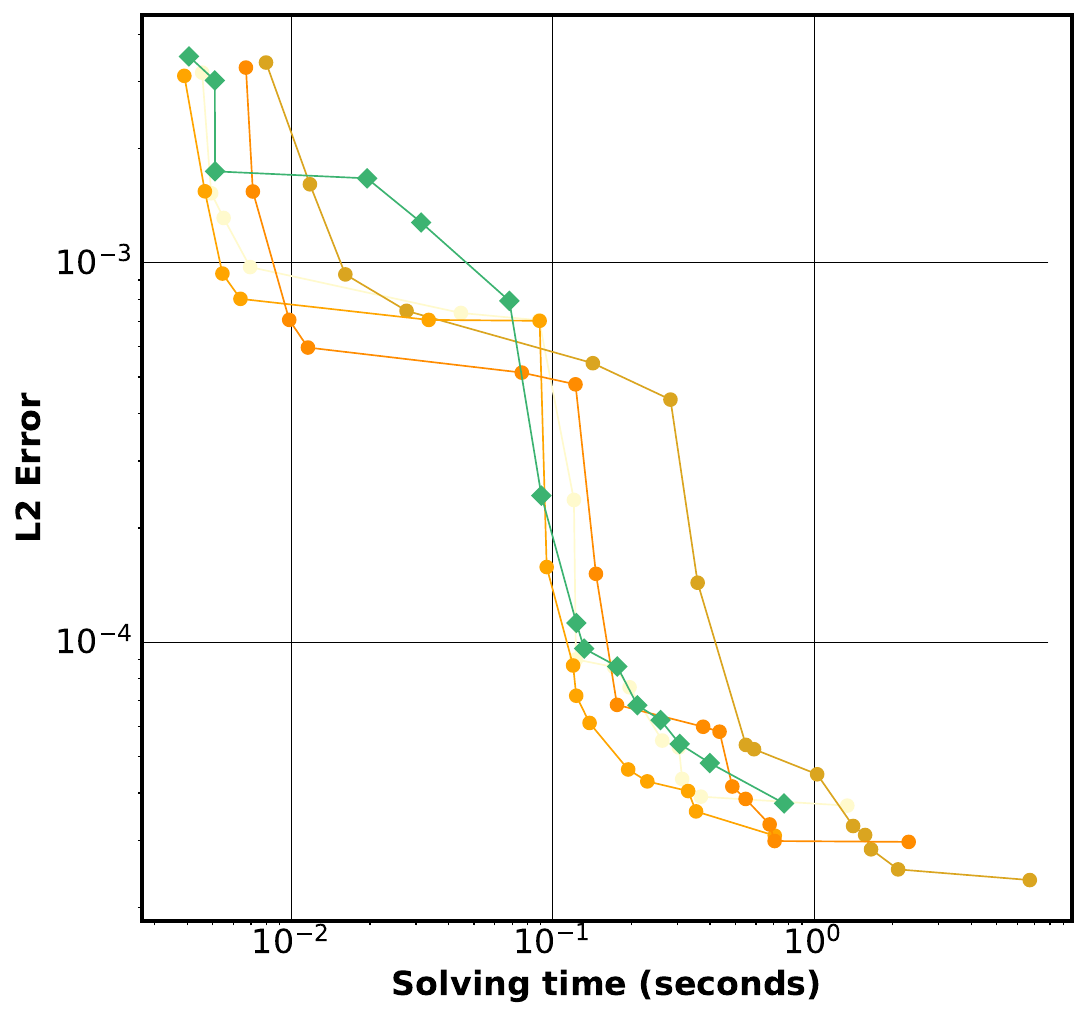}}\hfill
     \parbox{.32\linewidth}{\includegraphics[width=\linewidth]{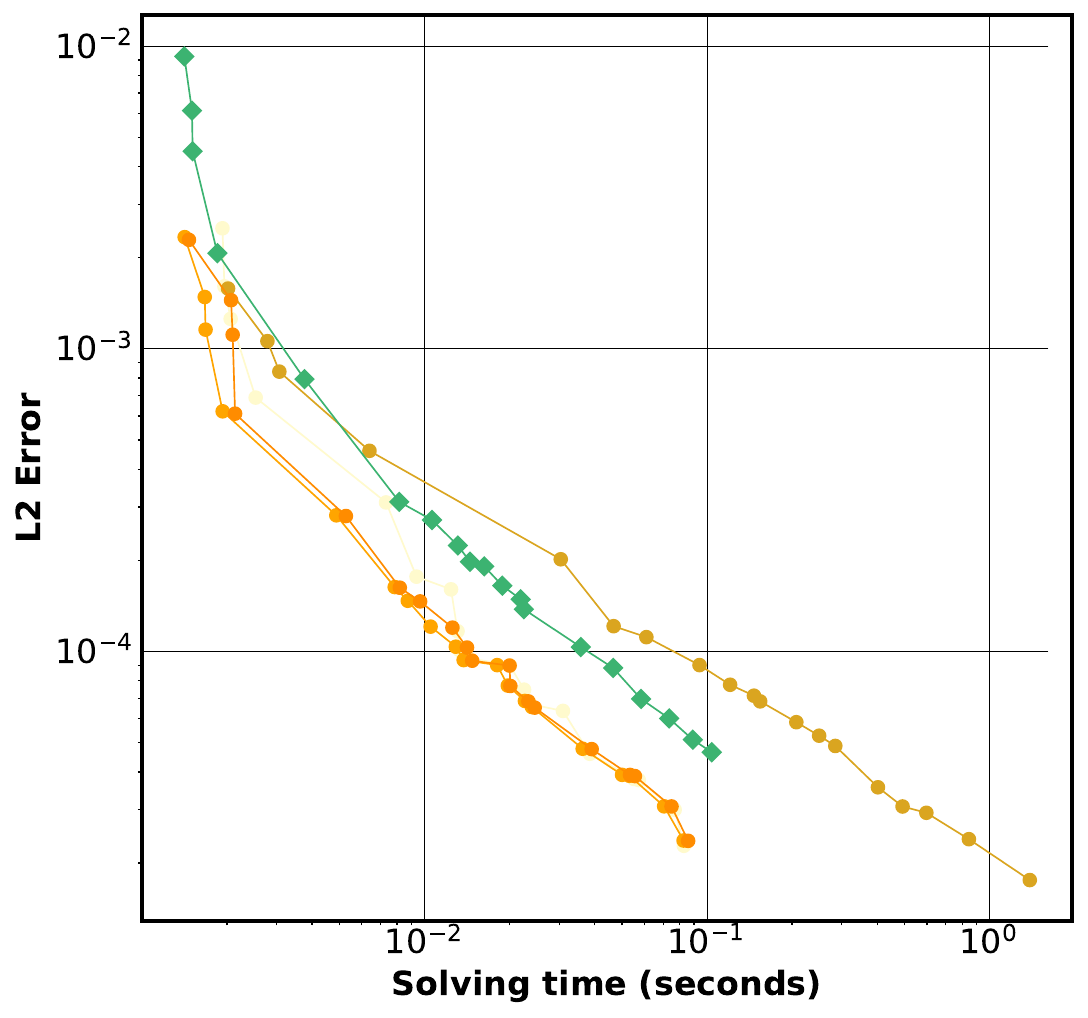}}\hfill
     \parbox{.32\linewidth}{\includegraphics[width=\linewidth]{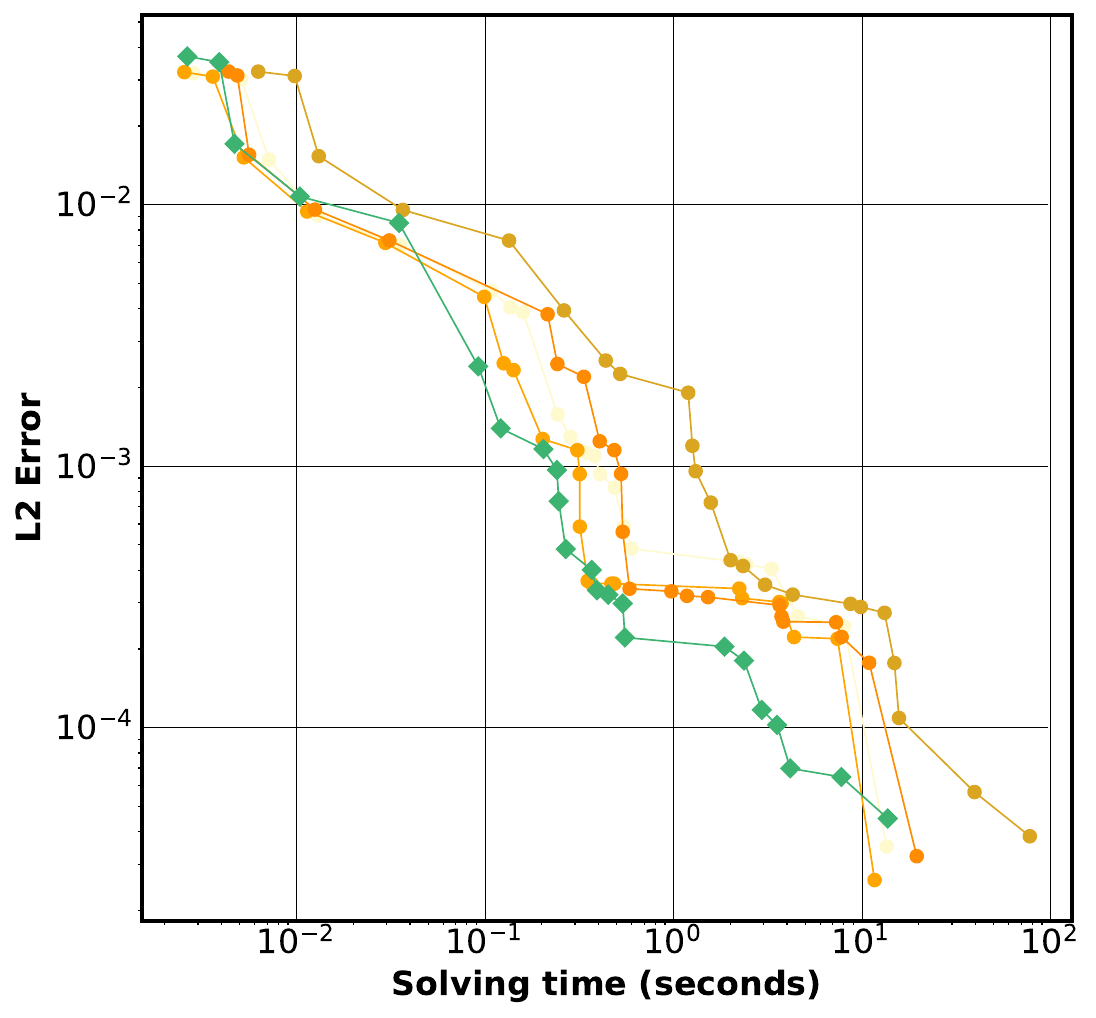}}\par
     \includegraphics[width=\linewidth]{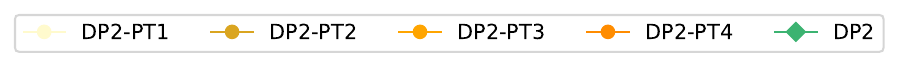}\par
     \caption{Solving time (s) versus $L_2$ error for DP3 to its four triangulated meshes.}
     \label{fig:compare_poly_vs_simplicial_displace}
 \end{figure}

Figure \ref{fig:compare_poly_vs_simplicial_displace} illustrates the comparison of DP2 meshes and the corresponding four triangulation meshes. For direct solvers, DP2 meshes show poorer performance than all triangular meshes for PS\#1-SC, but competitive performance than most triangular meshes for LEP-BE. In iterative solvers, DP2 performs better and generally shows competitive performance with triangular meshes. However, some triangulation (e.g., PT4) shows better performance). 

In conclusion, both VP4 and DP2 meshes show poorer performance than most triangulated meshes for direct solvers but can have relatively competitive performance for iterative solvers. Nevertheless, higher-quality triangular meshes can still perform better than polygonal meshes for both solvers. Thus, polygon meshes do not have much efficiency advantage compared to triangular meshes and could be more inefficient than high-quality triangular meshes. 

\subsubsection{Polygonal Meshes}
\label{subsec:mesh:poly}
\begin{figure}
     \centering\footnotesize
     \parbox{.02\linewidth}{~}\hfill\hfill
     \parbox{.32\linewidth}{\centering PS\#1-US}\hfill
     \parbox{.32\linewidth}{\centering PS\#1-SC}\hfill
     \parbox{.32\linewidth}{\centering LEP-BE}\par
     \parbox{.02\linewidth}{\rotatebox{90}{\centering Direct Solver}}\hfill
     \parbox{.32\linewidth}{\includegraphics[width=\linewidth]{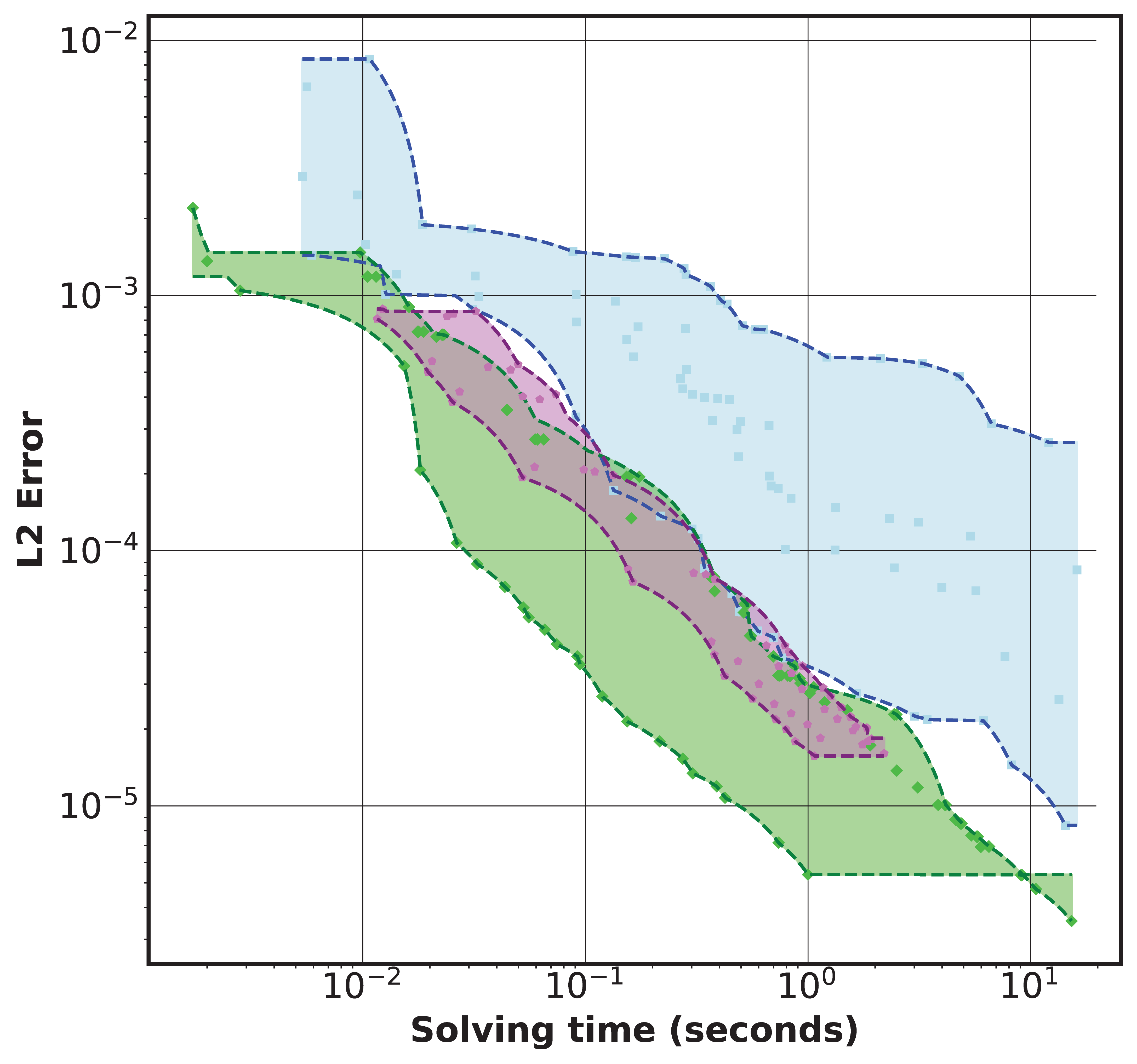}}\hfill
     \parbox{.32\linewidth}{\includegraphics[width=\linewidth]{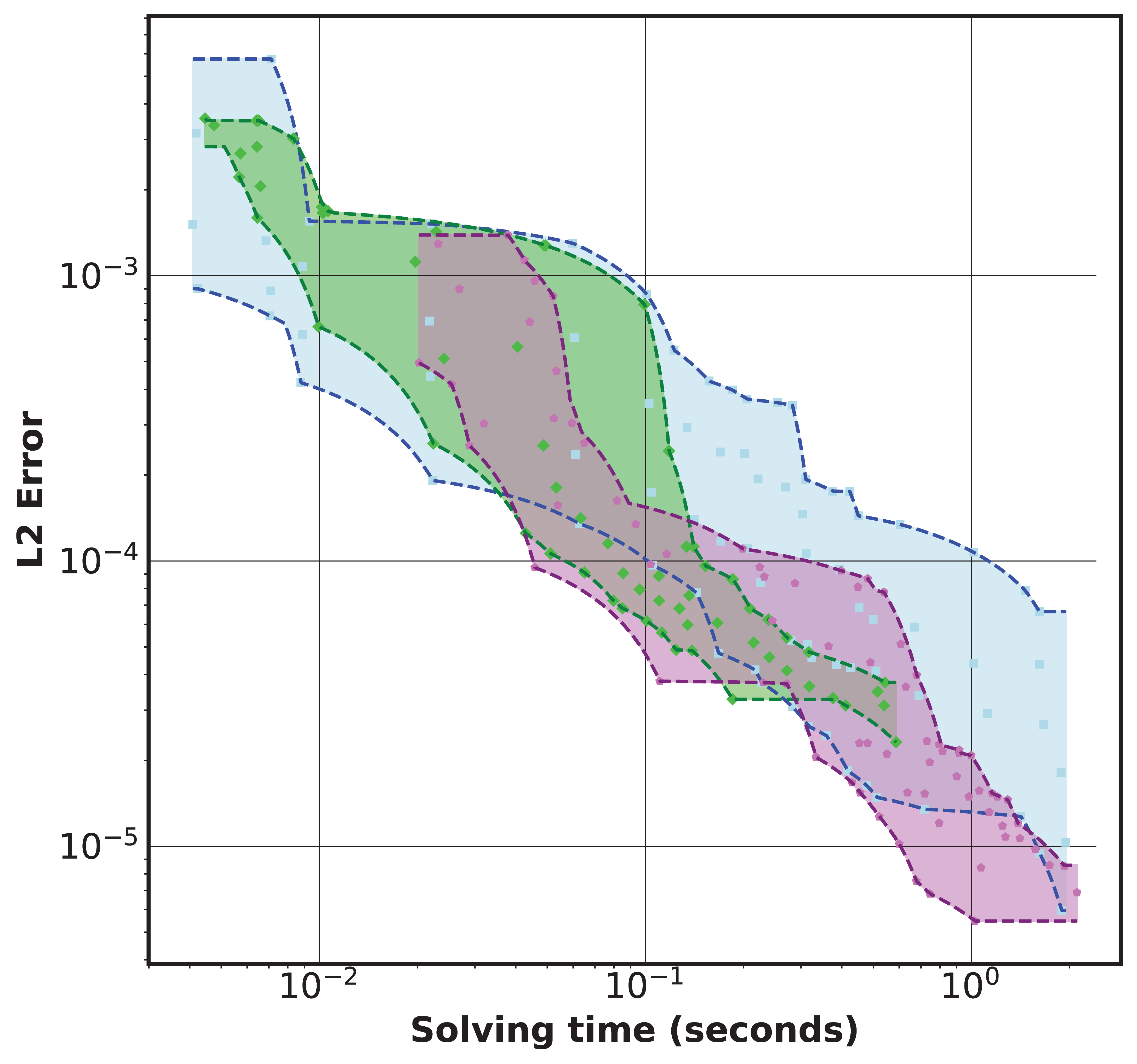}}\hfill
     \parbox{.32\linewidth}{\includegraphics[width=\linewidth]{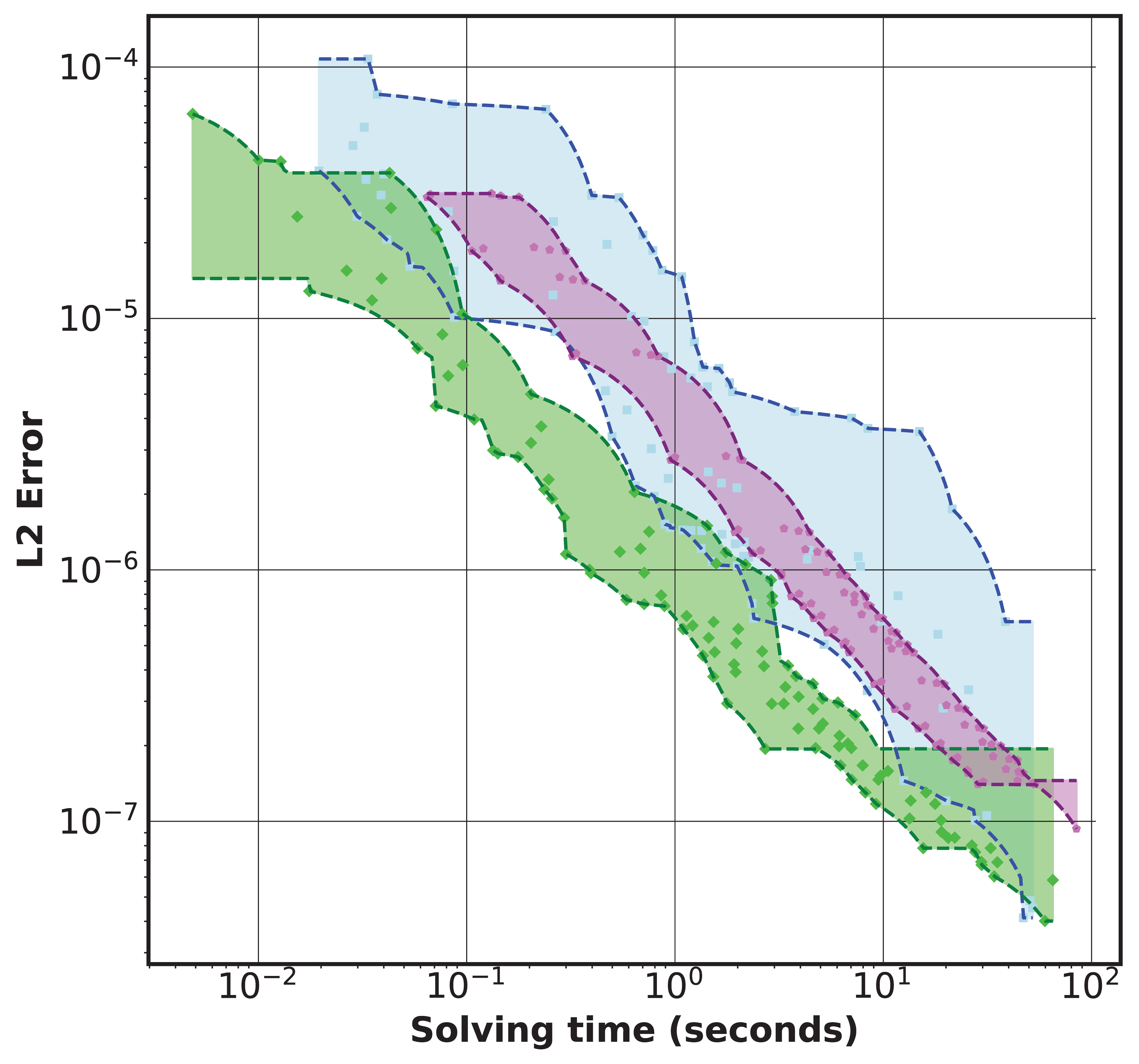}}\par
    \parbox{.02\linewidth}{\rotatebox{90}{\centering Iterative Solver}}\hfill
     \parbox{.32\linewidth}{\includegraphics[width=\linewidth]{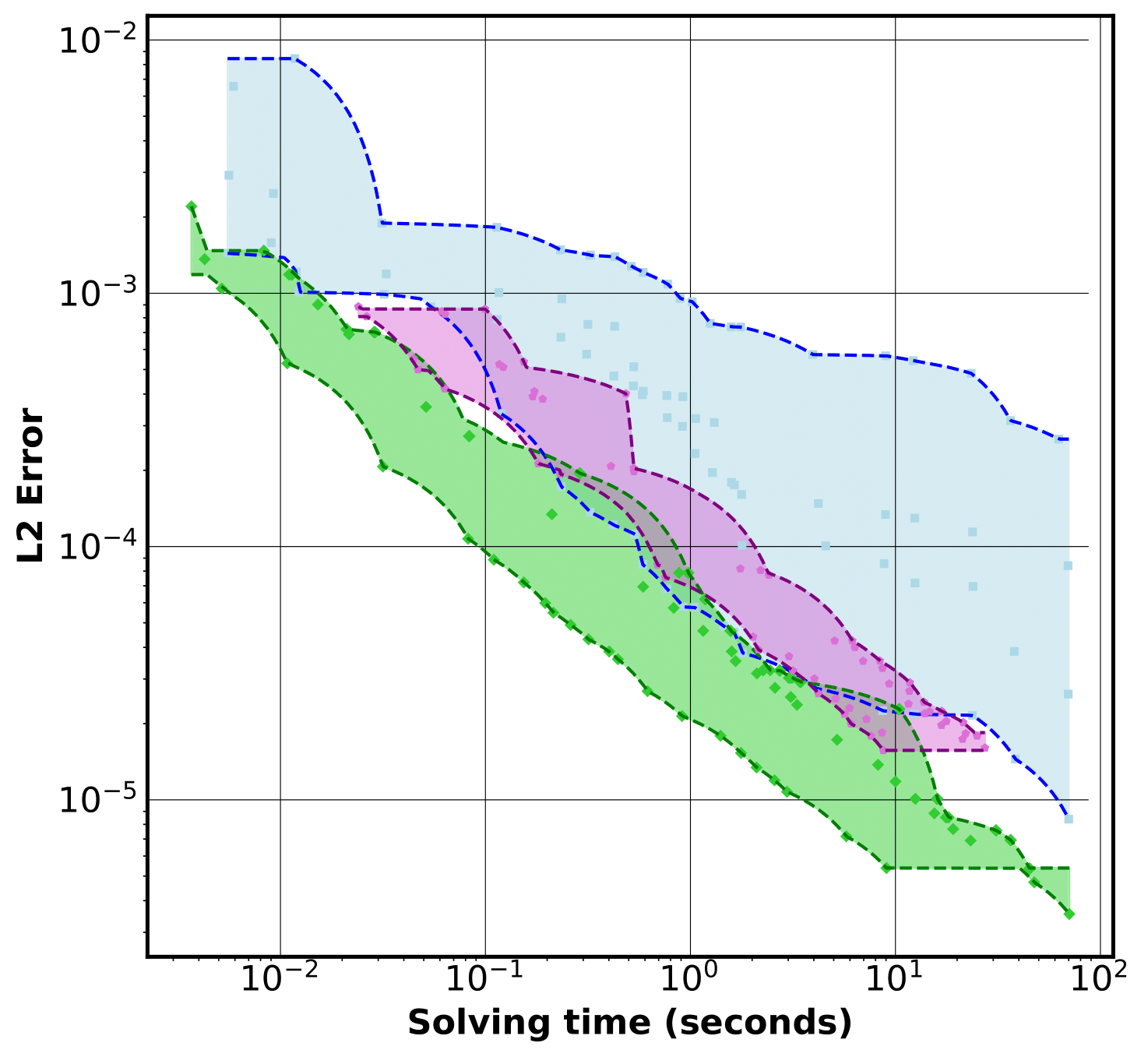}}\hfill
    \parbox{.32\linewidth}{\includegraphics[width=\linewidth]{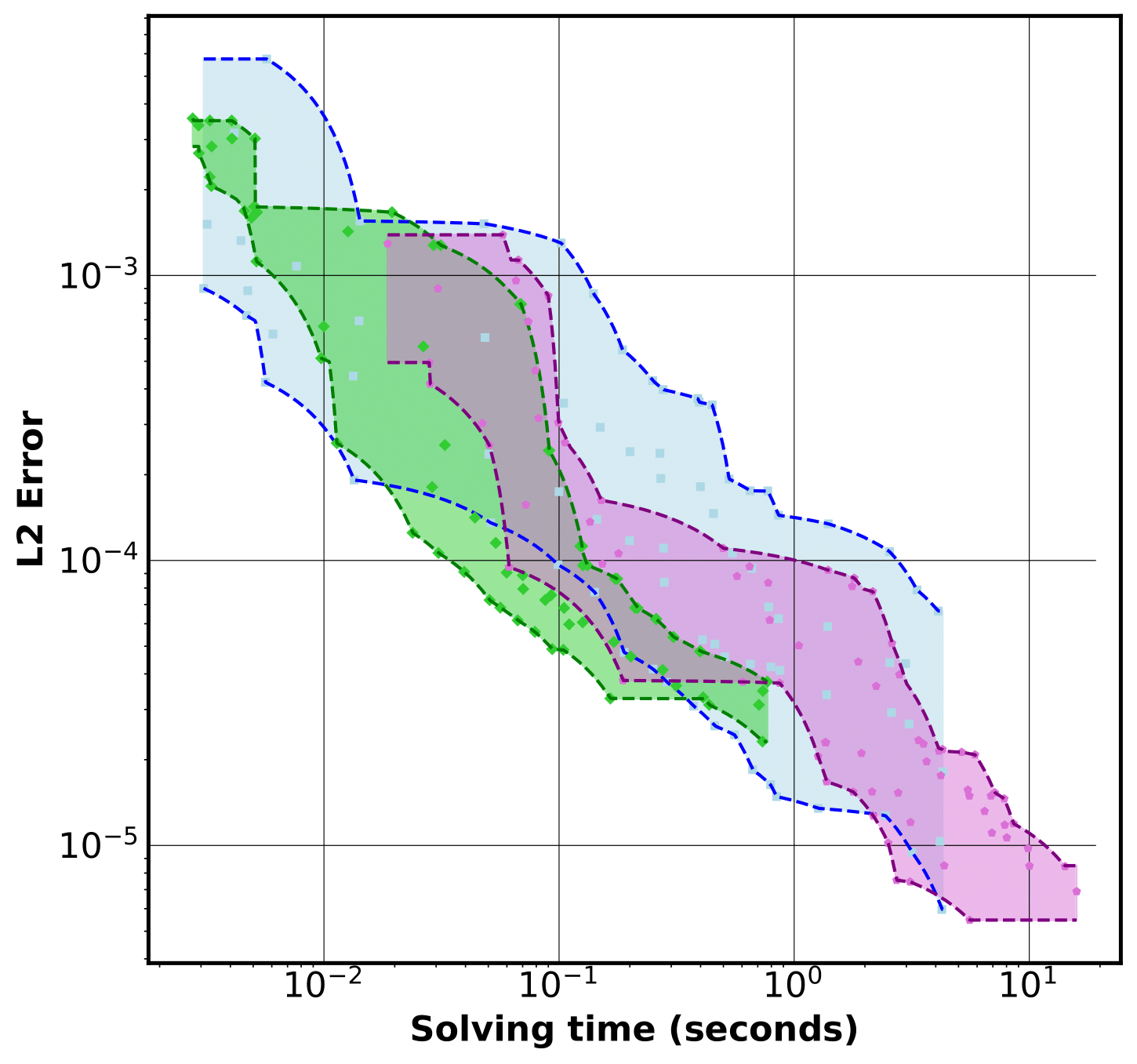}}\hfill
    \parbox{.32\linewidth}{\includegraphics[width=\linewidth]{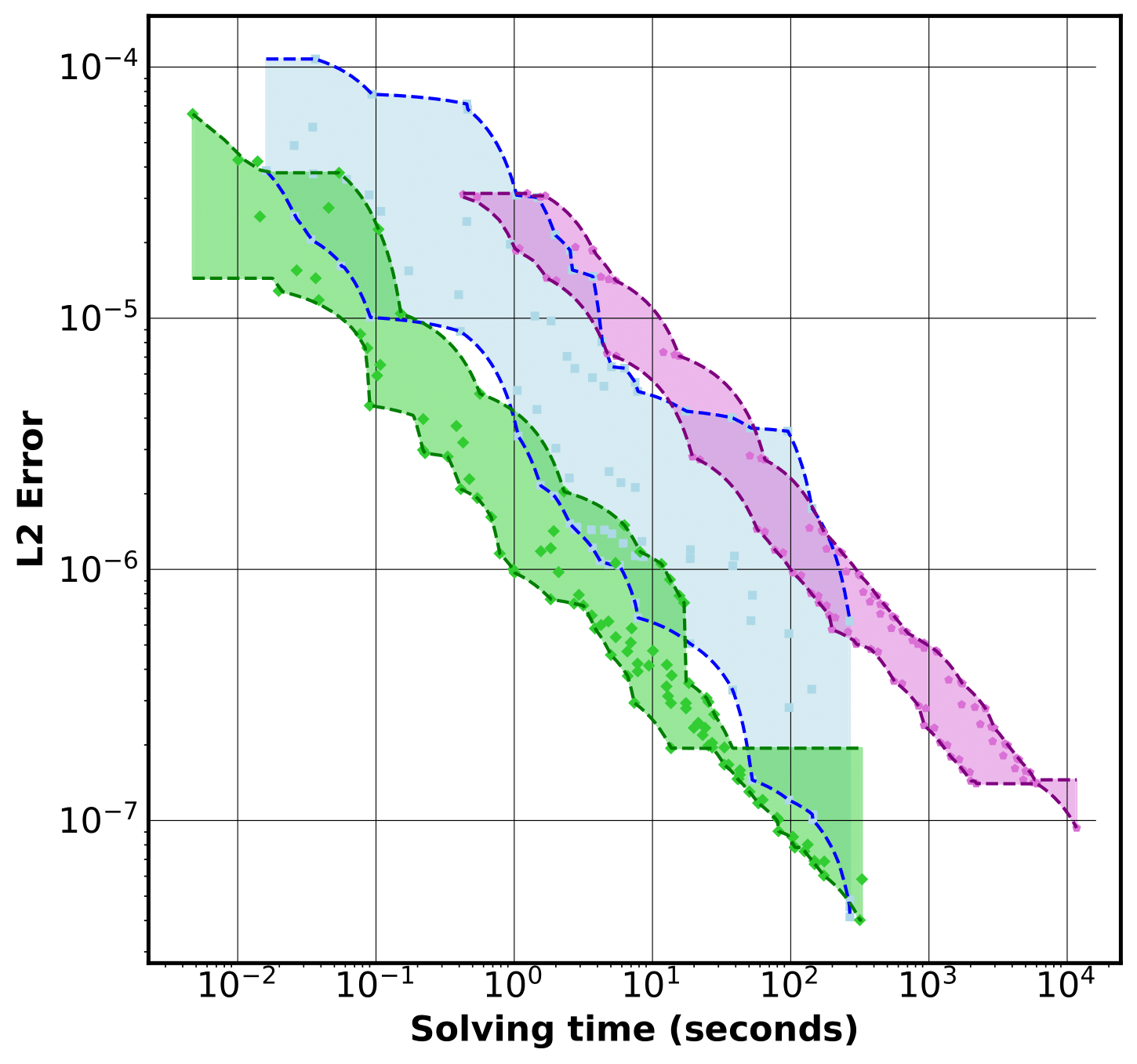}}\par
    \includegraphics[width=0.5\linewidth]{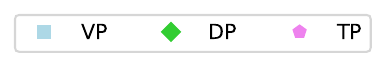}\par
     \caption{Comparison of time (s) versus error ($L_2$) for different polygonal meshes generated with different methods.}
     \label{fig:compare_poly}
 \end{figure}

We compare all polygonal meshes generated from different methods (Section \ref{sec:meshing}) to evaluate whether polygonal mesh generation methods make a difference in performance. We run experiments on both the iterative and direct solver and show the representative results for PS\#1-US, PS\#1-SC, and LEP-BE.
Figure \ref{fig:compare_poly} shows the results running on direct and iterative solvers. In direct solvers, displaced polygonal (DP) meshes (shown as green bands) generally have better or at least competitive performance than Voronoi polygonal (VP) meshes (shown as blue bands) and tiled polygonal meshes (TP) (shown as purple bands). For iterative solvers, both DP and VP meshes offer slightly better performance leaving the TP meshes with relatively poorer performance. In conclusion, displaced polygonal (DP) meshes generally have better efficiency than Voronoi polygonal (VP) and tiled polygonal meshes (TP), which might be explained by their simple element geometry (proximity to quad and hex meshes).

\subsubsection{Mesh Quality}\label{subsec:mesh:quality}
We compare the performance of different quality meshes to explore the impacts of quality on performance. Specifically, we compare different quality of the same category discretization method (polygon meshes and their triangulated meshes). We show the representative results of PS\#1-US,
and LEB-PH and divide the comparison into three groups for a clearer explanation.

\begin{figure}
     \centering\footnotesize
     \parbox{.8\linewidth}{
     \parbox{.49\linewidth}{\centering PS\#1-US}\hfill
     \parbox{.49\linewidth}{\centering LEP-BE}\par
     \includegraphics[width=.49\linewidth]{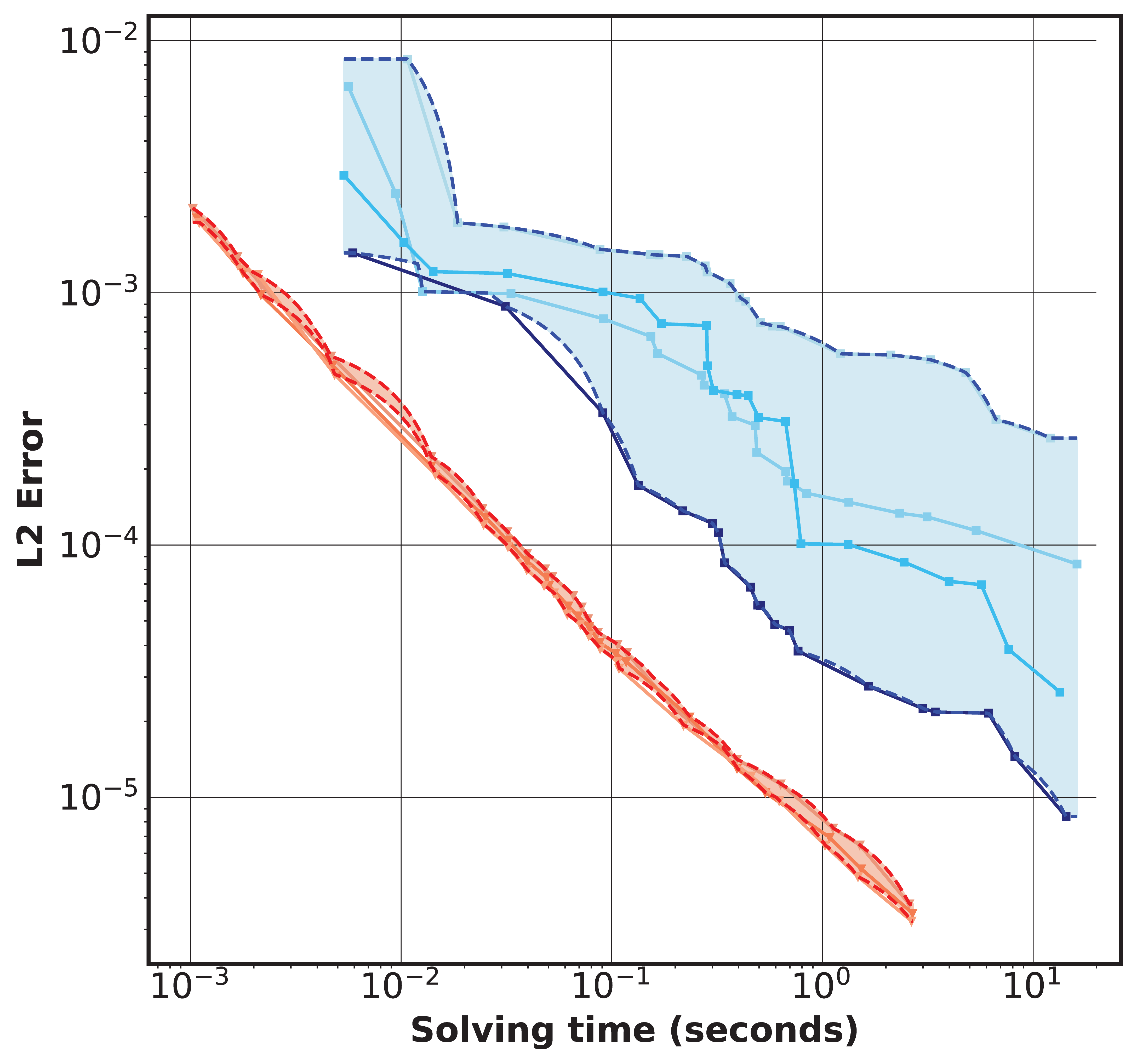}\hfill
     \includegraphics[width=.49\linewidth]{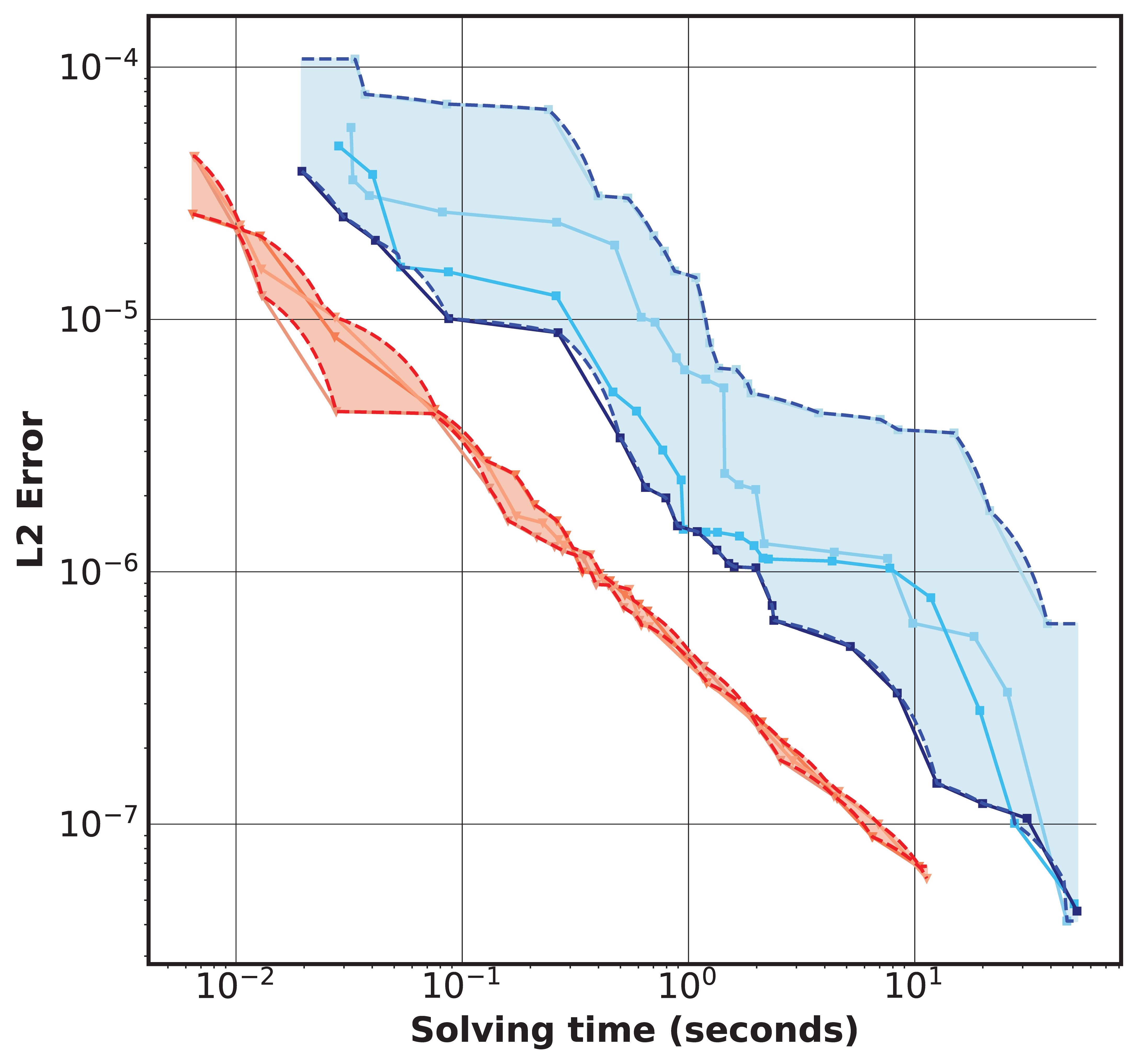}}\par
     \includegraphics[width=\linewidth]{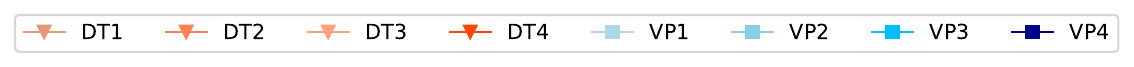}\par
     \caption{Solving time (s) versus $L_2$ error for different quality Voronoi polygonal meshes and the corresponding dual triangular meshes.}
     \label{fig:compare_quality_voro}
 \end{figure}

First, we compare the Voronoi polygonal meshes (VP) with different centroid iteration numbers and the corresponding dual triangular meshes (DT). In this case, we notice that the quality makes a big difference in Voronoi polygonal meshes (Figure \ref{fig:compare_quality_voro}). However, the corresponding dual triangular meshes do not show a significant difference in performance.

\begin{figure}
     \centering\footnotesize
     \parbox{.8\linewidth}{
     \parbox{.49\linewidth}{\centering PS\#1-US}\hfill
     \parbox{.49\linewidth}{\centering LEP-BE}\par
     \includegraphics[width=.49\linewidth]{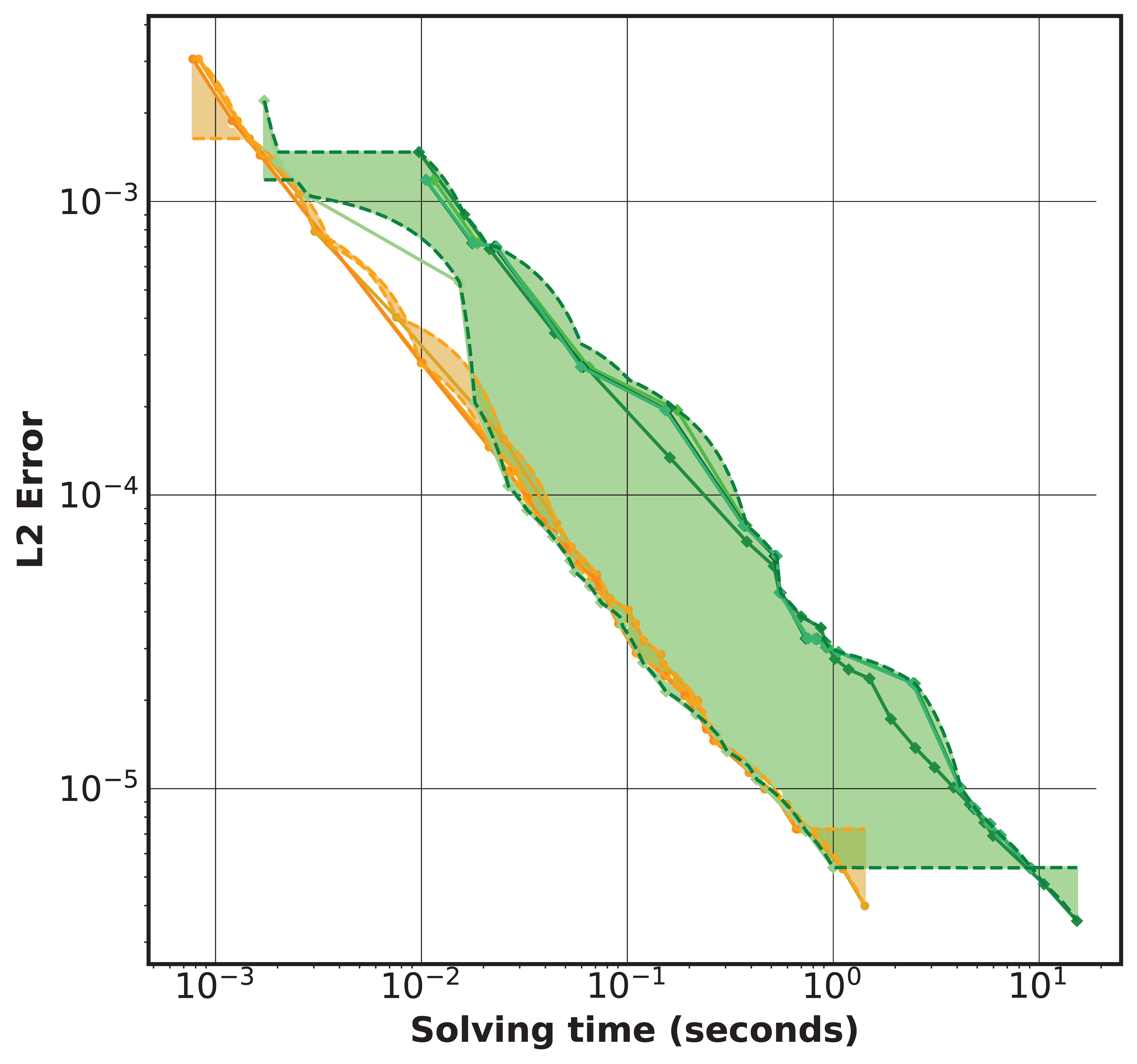}\hfill
     \includegraphics[width=.49\linewidth]{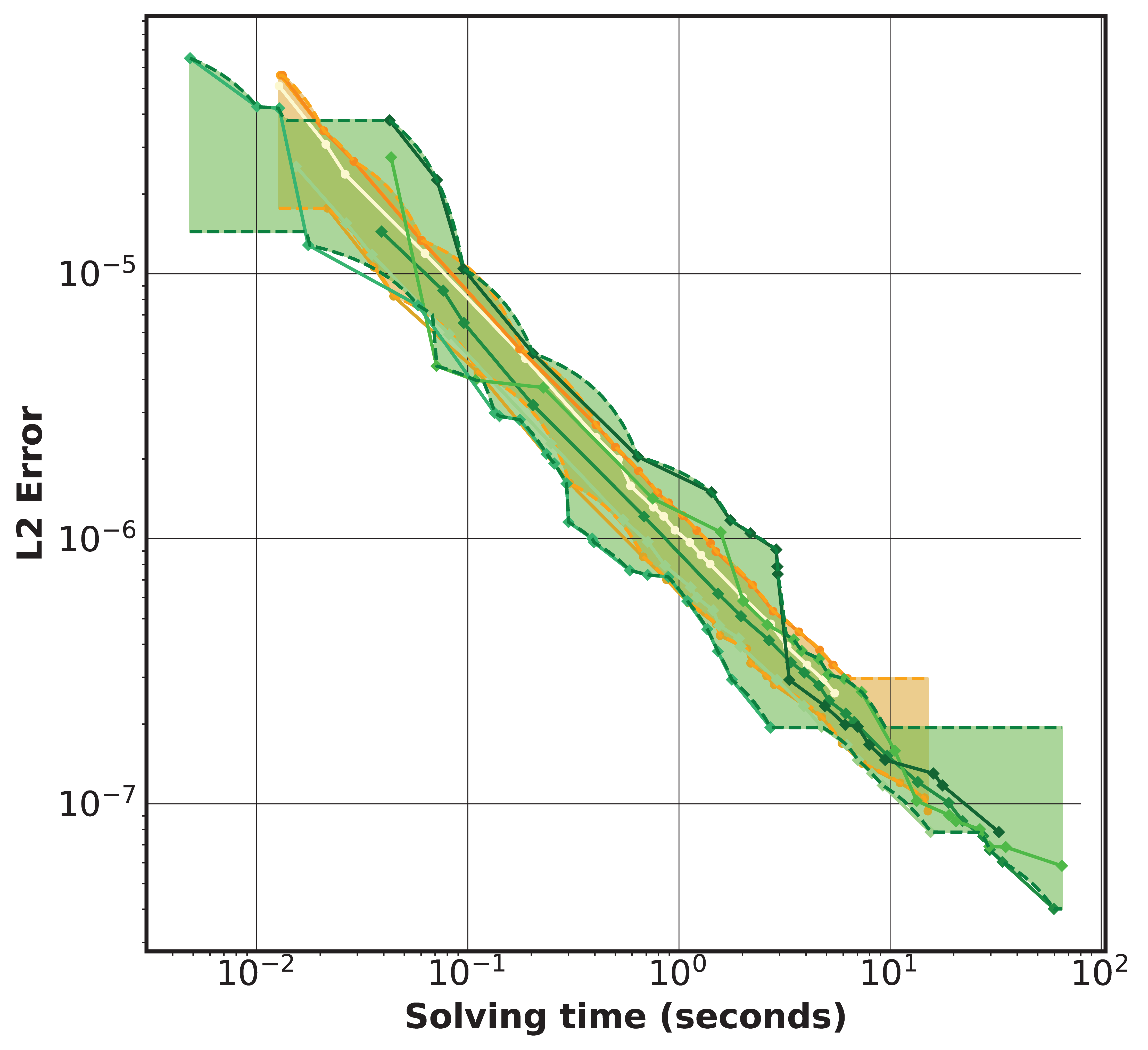}}\par
     \includegraphics[width=\linewidth]{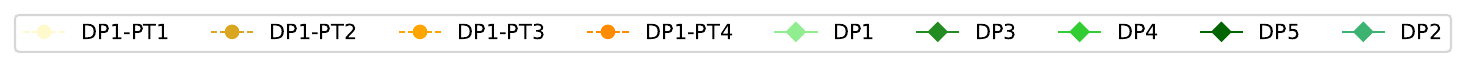}\par
     \caption{Solving time (s) vs $L_2$-error for different quality of displaced polygonal meshes and the corresponding triangulated meshes.}
     \label{fig:compare_quality_displace}
 \end{figure}

Second, we compare the different Displaced polygonal meshes (DP) and corresponding intra-triangulated meshes (PT).
Here we observe that the quality does not make much of a difference in some displaced polygonal meshes while having more significant differences in others (Figure \ref{fig:compare_quality_displace}). As for triangulation, the meshes do not show much difference even with very low-quality mesh elements (e.g., triangular meshes generated by randomly inserting points inside elements, PT2), and the performance of triangulated meshes depends on the original displaced polygonal meshes.

\begin{figure}
     \centering\footnotesize
     \parbox{.8\linewidth}{
     \parbox{.49\linewidth}{\centering PS\#1-US}\hfill
     \parbox{.49\linewidth}{\centering LEP-BE}\par
     \includegraphics[width=.49\linewidth]{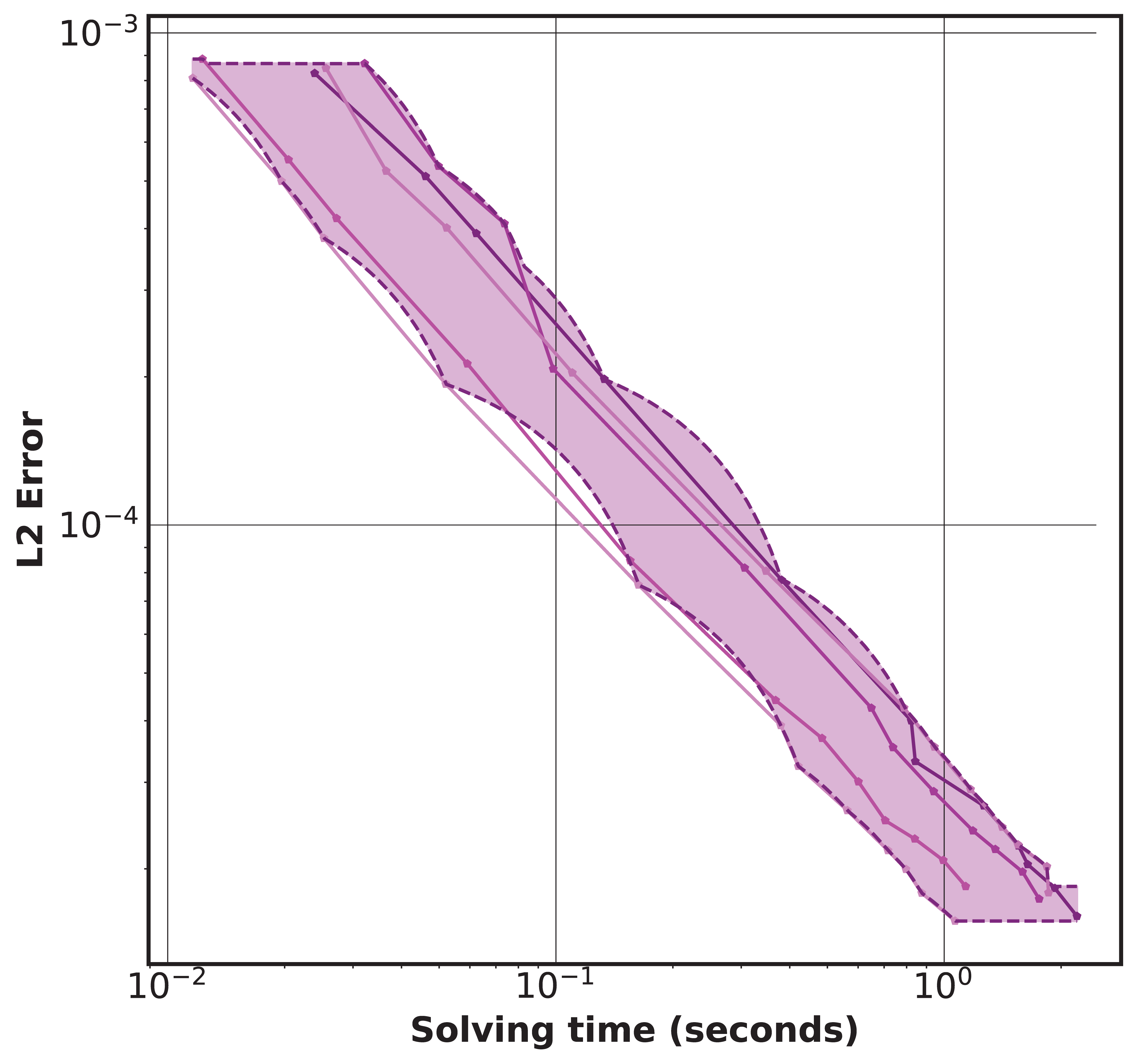}\hfill
     \includegraphics[width=.49\linewidth]{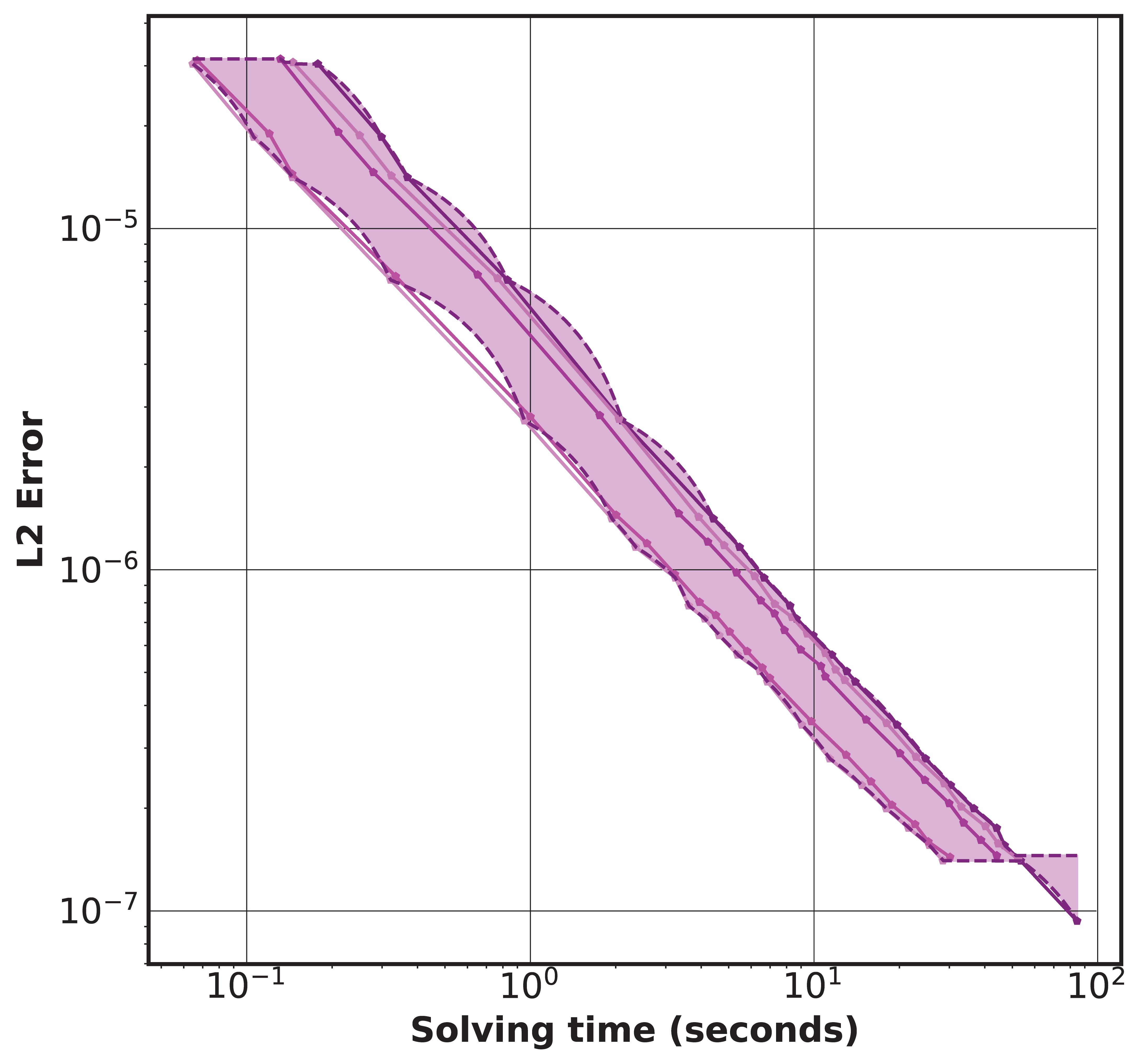}}\par
     \includegraphics[width=.8\linewidth]{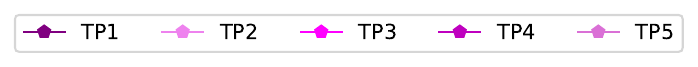}\par
     \caption{Solving time (s) versus $L_2$ error for different quality tiled polygonal meshes.}
     \label{fig:compare_quality_tile}
 \end{figure}

Third, we compare different tiled polygonal meshes (Figure \ref{fig:compare_quality_tile}), where we conclude that tiled polygonal meshes generated with 'Z' and 'U' shapes perform slightly better than the other three meshes, and the differences are generally small and consistent throughout different PDEs and domains.

 In conclusion, polygon meshes are generally more sensitive to mesh quality than their triangulated meshes when considering efficiency. A possible explanation could be that polygon meshes (especially Voronoi meshes) generally have very small edges which leads to such sensitivity. In addition, the assembled matrices for polygon meshes are generally denser than their triangulated meshes, which leads to less efficiency.

\subsubsection{Best/Worst Quality Meshes}
\label{subsec:mesh:best-worst}
We compare the polygonal meshes with the best (worst) quality and the triangular meshes with the best (worst) quality among all our meshes, to further evaluate how sensible are the different discretizations to meshing. We select VP4 and DP2 as the best-quality polygonal meshes and VP1 as the worst-quality polygonal meshes. For triangular meshes, we choose the dual of the Voronoi polygonal mesh with 20 iterations (DT4) as the best quality meshes and the triangulation meshes by inserting random points as the worst quality meshes PT2. We show the representative results on both solvers for PS\#1-US, LEP-BE, and LEB-PH.

\begin{figure}
     \centering\footnotesize
     \parbox{.02\linewidth}{~}\hfill\hfill
     \parbox{.32\linewidth}{\centering PS\#1-US}\hfill
     \parbox{.32\linewidth}{\centering LEP-BE}\hfill
     \parbox{.32\linewidth}{\centering LEB-PH}\par
     \parbox{.02\linewidth}{\rotatebox{90}{\centering Direct Solver}}\hfill\hfill
     \parbox{.32\linewidth}{\includegraphics[width=\linewidth]{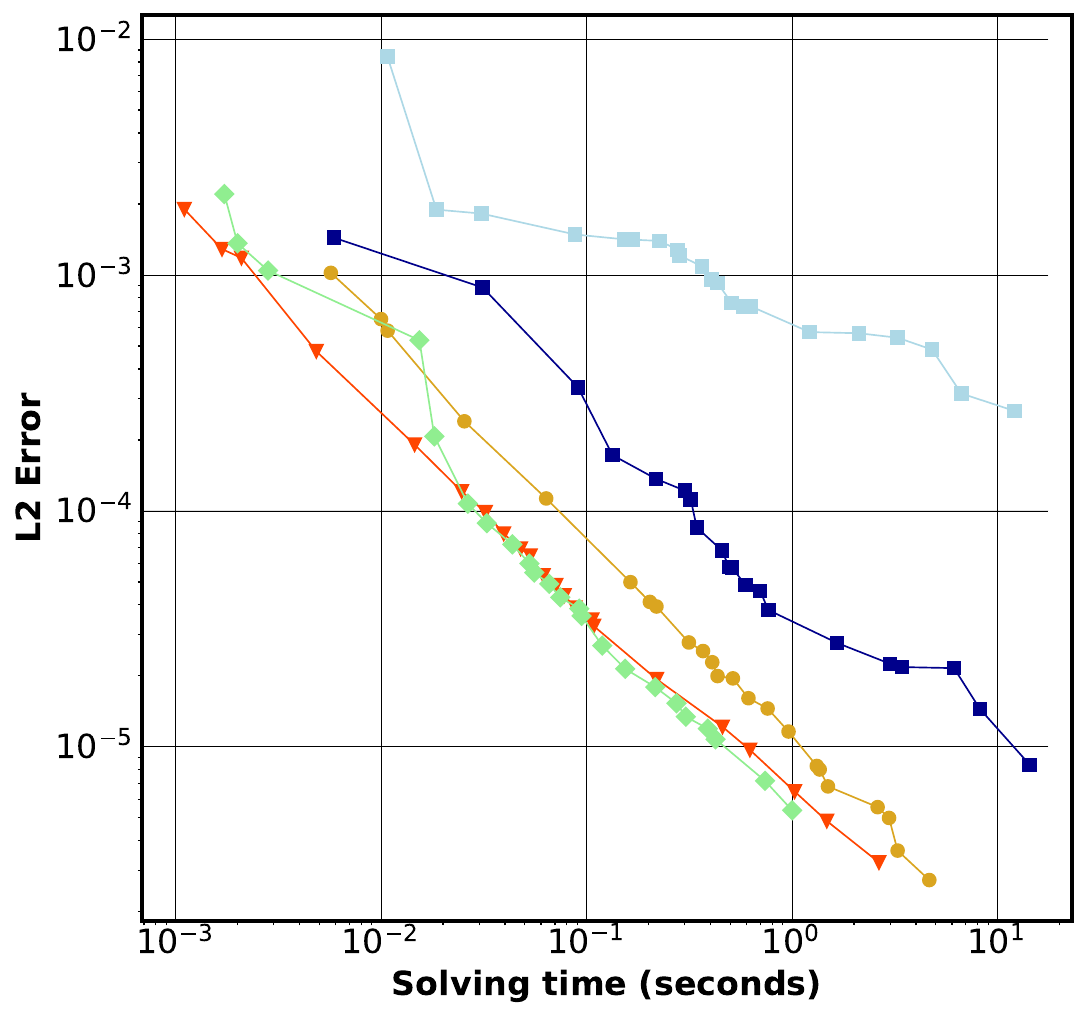}}\hfill
     \parbox{.32\linewidth}{\includegraphics[width=\linewidth]{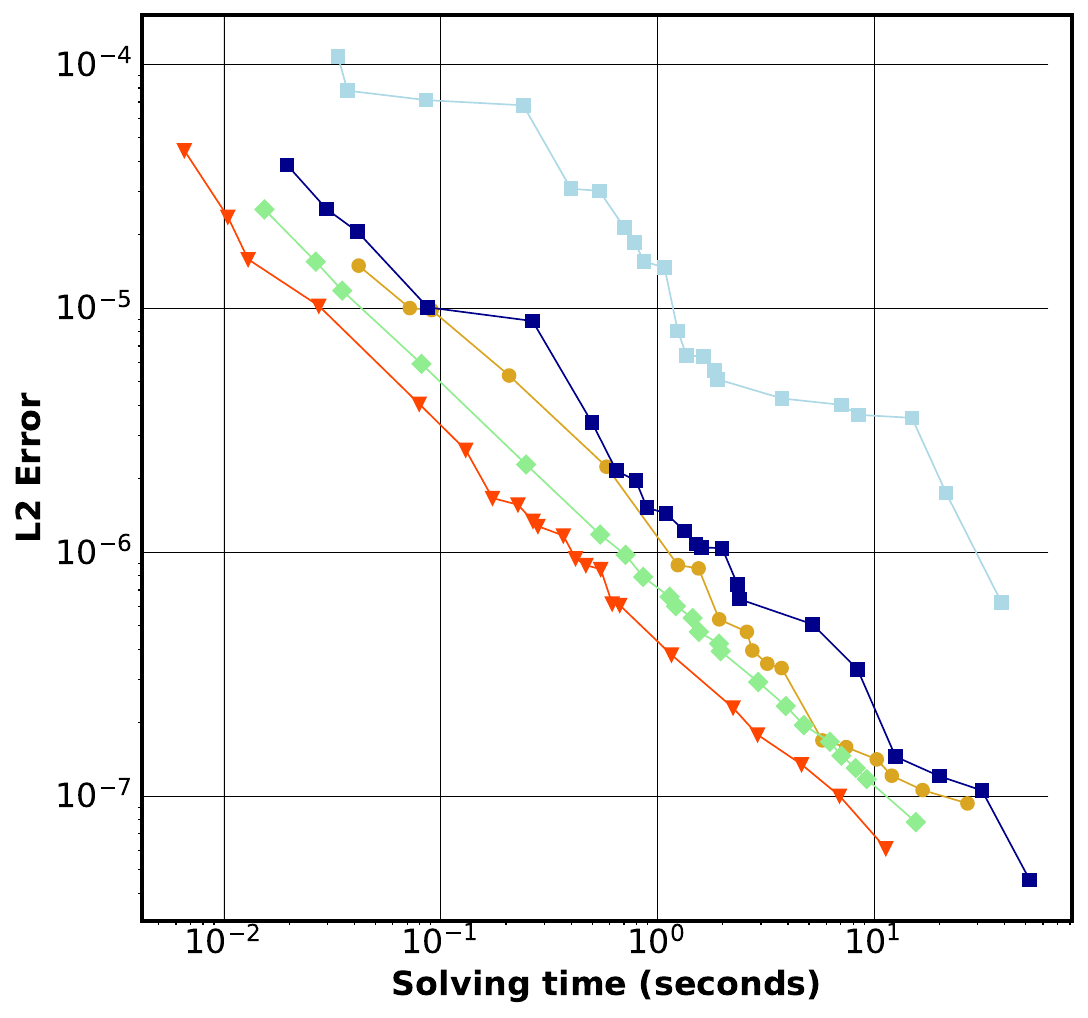}}\hfill
     \parbox{.32\linewidth}{\includegraphics[width=\linewidth]{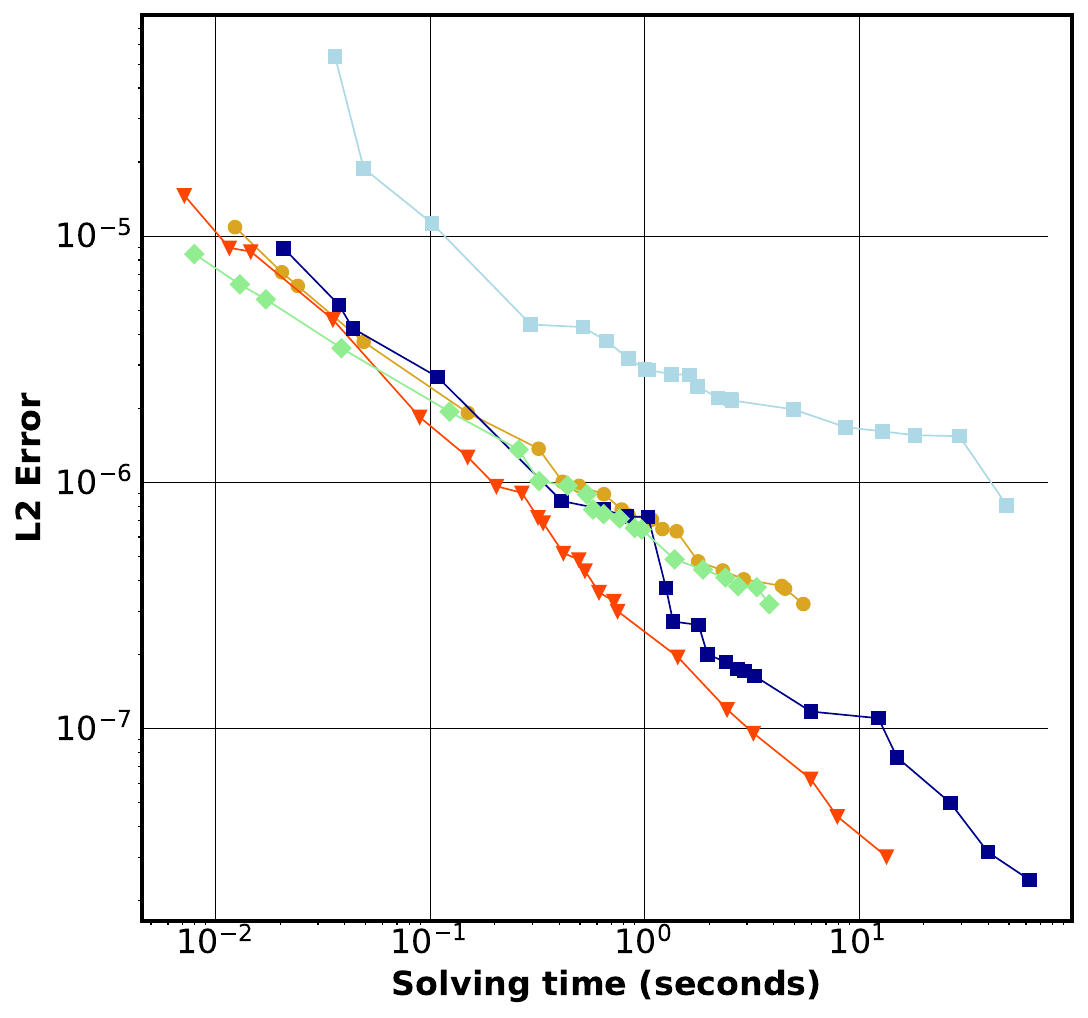}}\par
    \parbox{.02\linewidth}{\rotatebox{90}{\centering Iterative Solver}}\hfill\hfill
     \parbox{.32\linewidth}{\includegraphics[width=\linewidth]{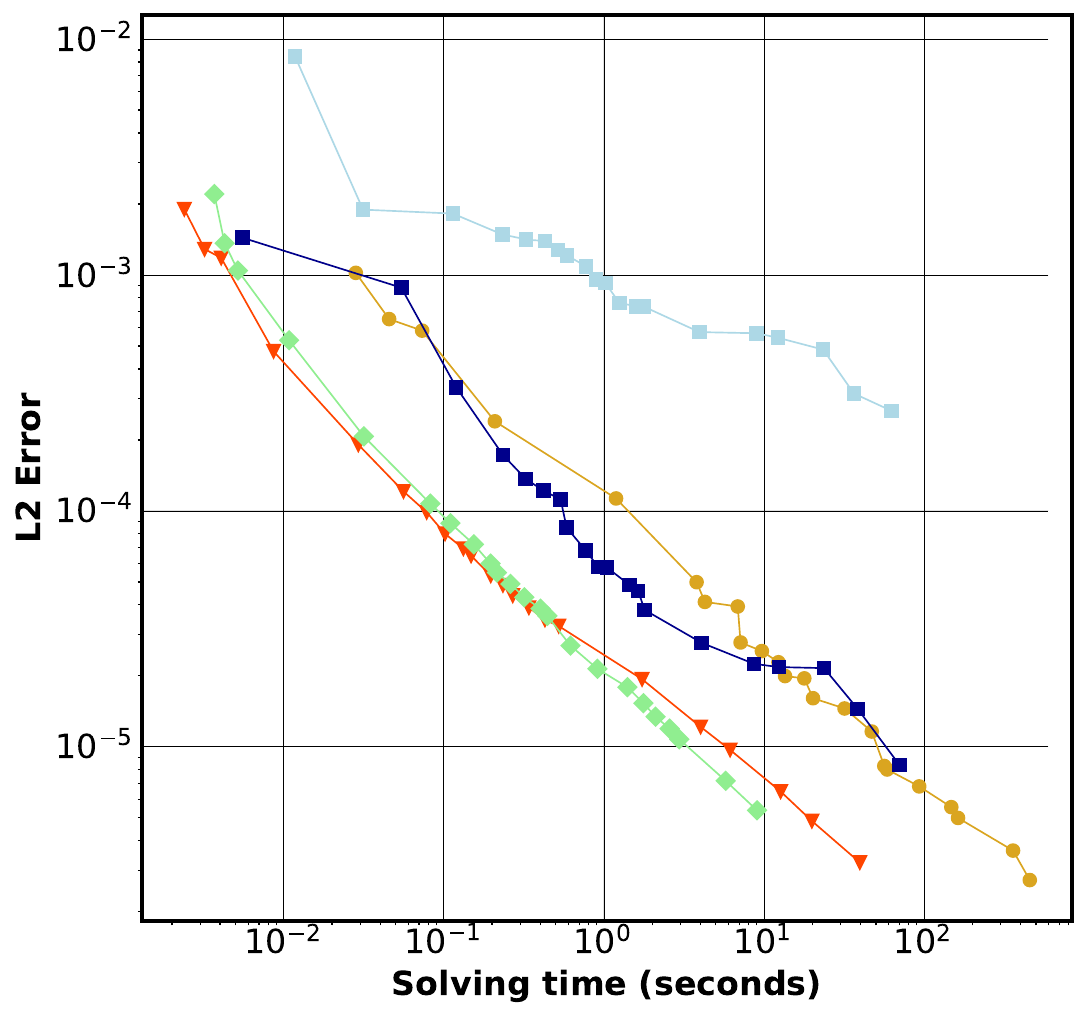}}\hfill
     \parbox{.32\linewidth}{\includegraphics[width=\linewidth]{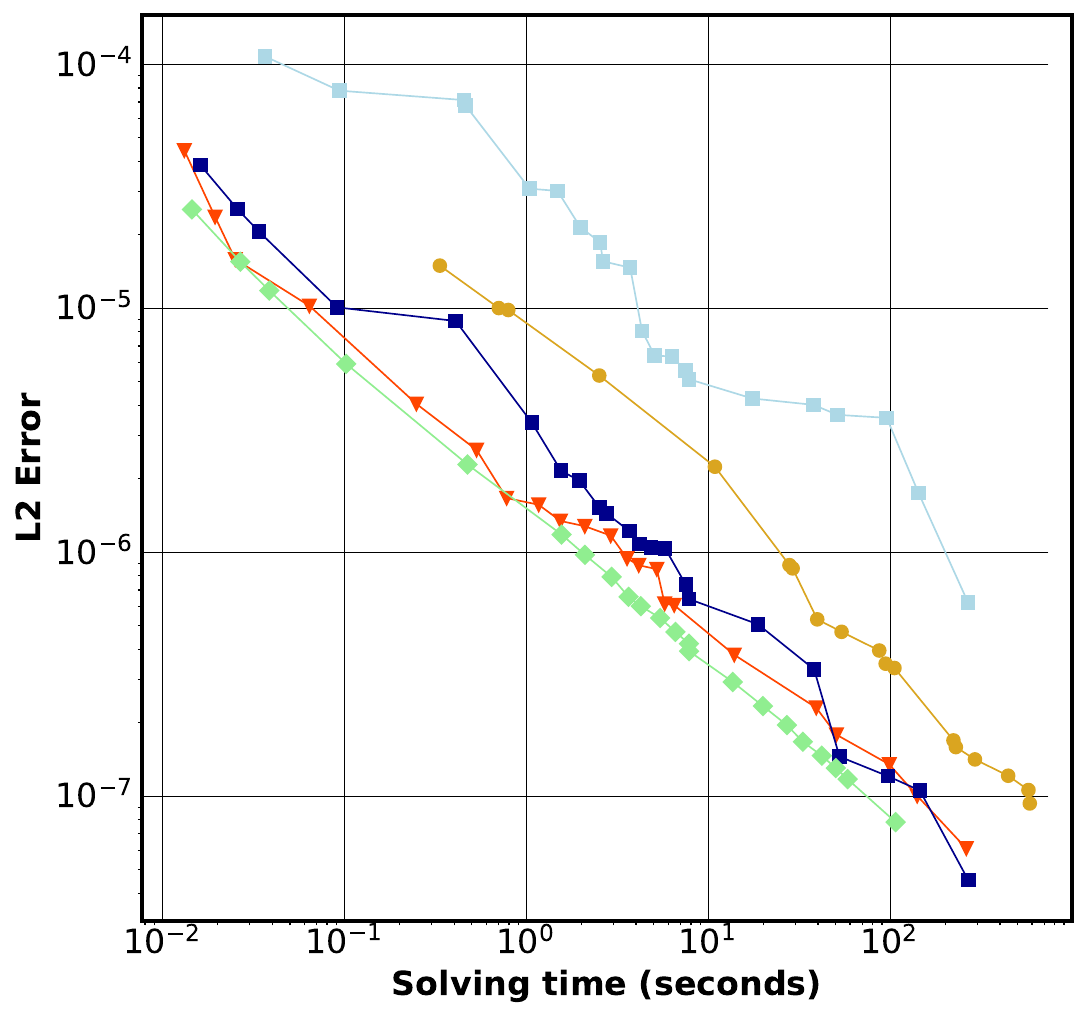}}\hfill
     \parbox{.32\linewidth}{\includegraphics[width=\linewidth]{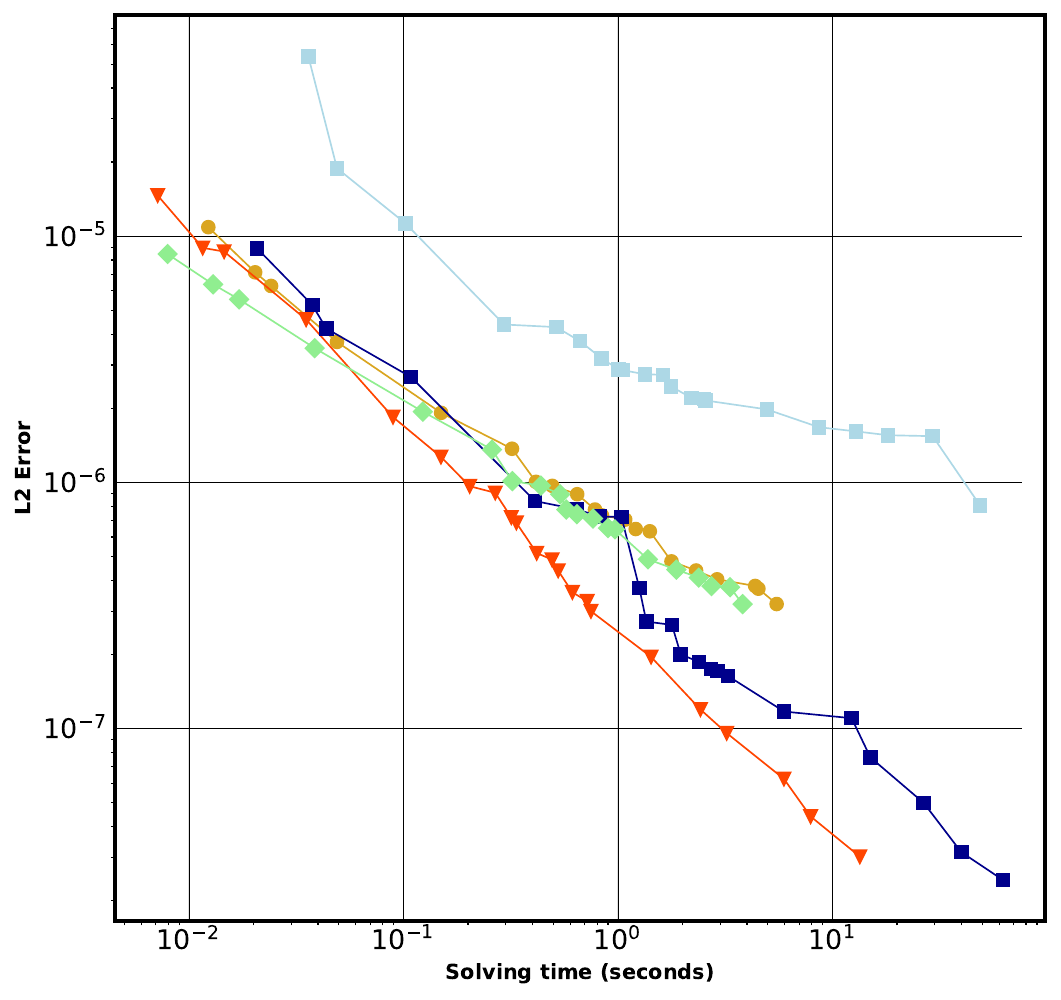}}\par
     \includegraphics[width=.8\linewidth]{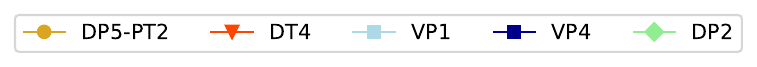}\par
     \caption{Comparison of the best/worst quality polygonal meshes with best/worst quality triangular meshes for time versus error.}
     \label{fig:compare_bb_ww}
 \end{figure}

For direct solvers (Figure \ref{fig:compare_bb_ww}), DP2 has a similar and competitive performance with DT4, while VP4 has a poorer performance. VP1 performs significantly worse than PT2. For the iterative solvers, DP2 also has a similar and competitive performance with DT4, while the gap between VP4 and DT4 decreases. VP1 still performs poorer than PT2, though the gap is reduced.

 In conclusion, the best quality triangular mesh performs better than the best quality polygon meshes, and the worst quality triangular meshes perform no worse than the worst quality polygon meshes. This finding further suggests that triangular meshes are generally more efficient regardless of quality.

\subsection{Simulation Methods}
\label{subsec:simulation}
We report our comparisons of different simulation methods. We would like to clarify that by comparing the simplicial meshes with polygonal meshes in Section~\ref{subsec:mesh:tripoly}, we already compare the performance of linear FEM and linear VEM (the meshes and the most basic simulation methods are correlated). Thus, we conduct additional comparisons of other simulation techniques to further explore their impacts on time efficiency. We discuss the impact of basis order in Section~\ref{subsec:simulation:high-order} by comparing linear FEM/VEM to higher-order FEM. We compare the Barycentric FEM with the VEM on solving with polygonal meshes in Section~\ref{subsec:simulation:bc-vs-vem}, followed by comparing the Barycentric FEM with traditional FEM on solving simplicial meshes in Section~\ref{subsec:simulation:bc-vs-fem}. At last, we discuss the impact of different stabilization techniques of VEM when solving linear elasticity problems in Section~\ref{subsec:simulation:stabilization}. 

\subsubsection{Linear FEM/VEM and Higher-Order FEM}
\label{subsec:simulation:high-order}
We compare the linear FEM/VEM with higher-order FEM on triangular meshes to evaluate the impact of basis orders. Our experiments focus on quadratic bases for the higher-order FEM basis. We show the representative results for PS\#1-US and PS\#1-SC.

\begin{figure}
    \centering\footnotesize
     \parbox{.02\linewidth}{~}\hfill
     \parbox{.48\linewidth}{\centering PS\#1-US}\hfill
     \parbox{.48\linewidth}{\centering PS\#1-SC}\par
    \parbox{.02\linewidth}{\rotatebox{90}{\centering DT}}\hfill
    \parbox{.48\linewidth}{\includegraphics[width=\linewidth]{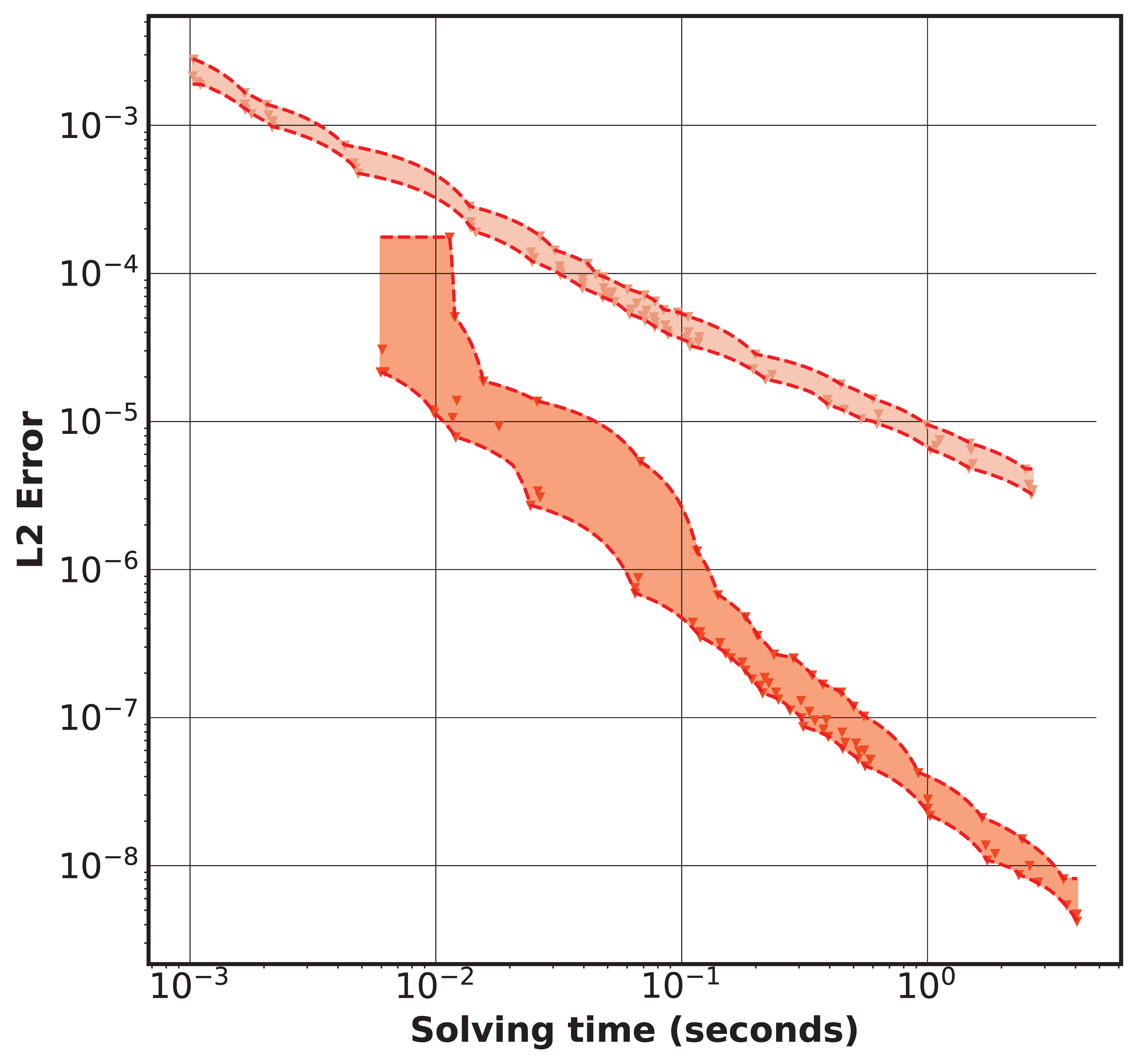}}\hfill
    \parbox{.48\linewidth}{\includegraphics[width=\linewidth]{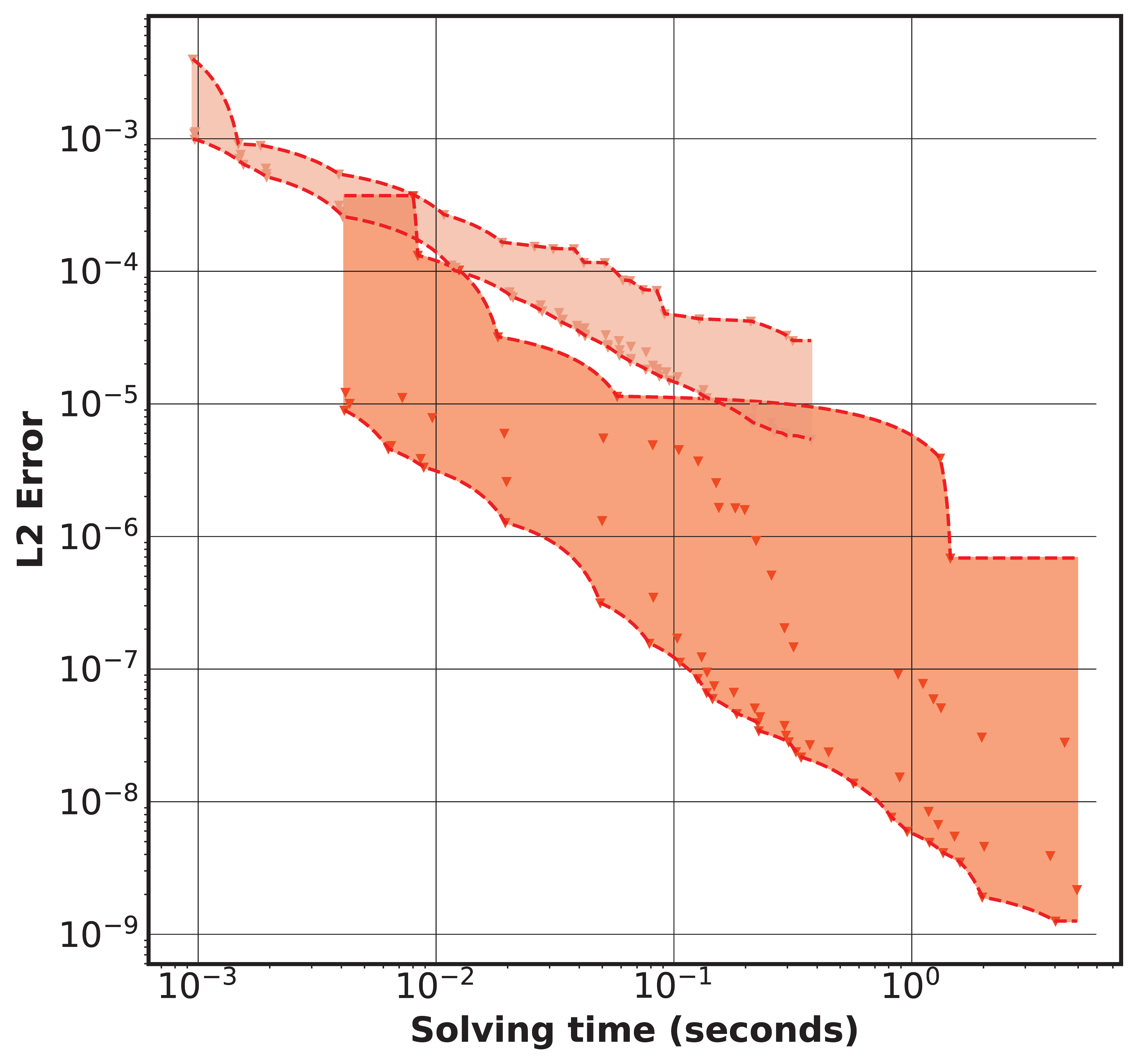}}\par
    \includegraphics[width=0.5\linewidth]{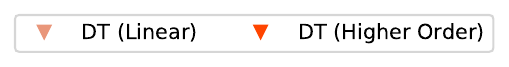}\par
    \parbox{.02\linewidth}{\rotatebox{90}{\centering VP4-PT}}\hfill
    \parbox{.48\linewidth}{\includegraphics[width=\linewidth]{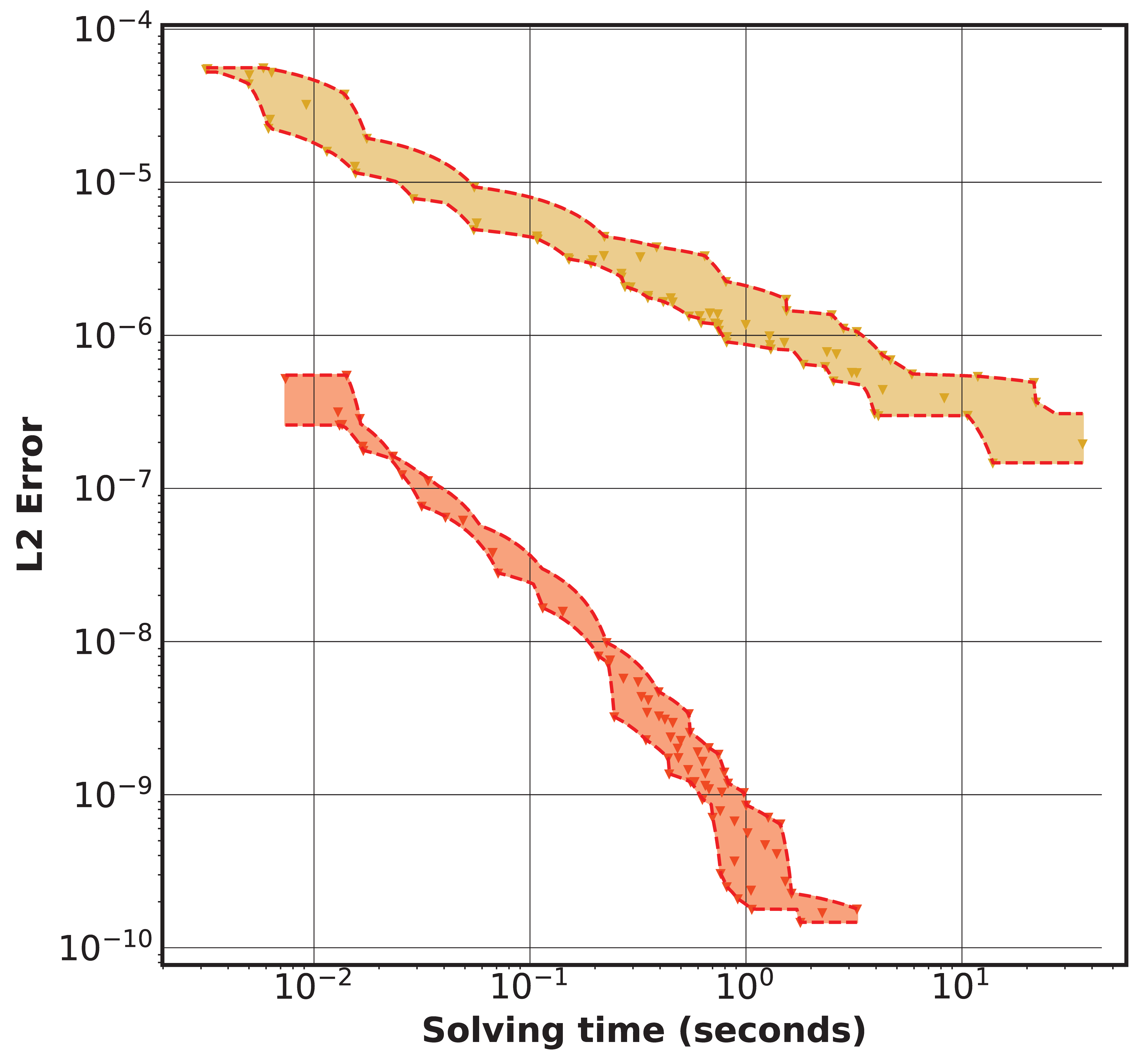}}\hfill
    \parbox{.48\linewidth}{\includegraphics[width=\linewidth]{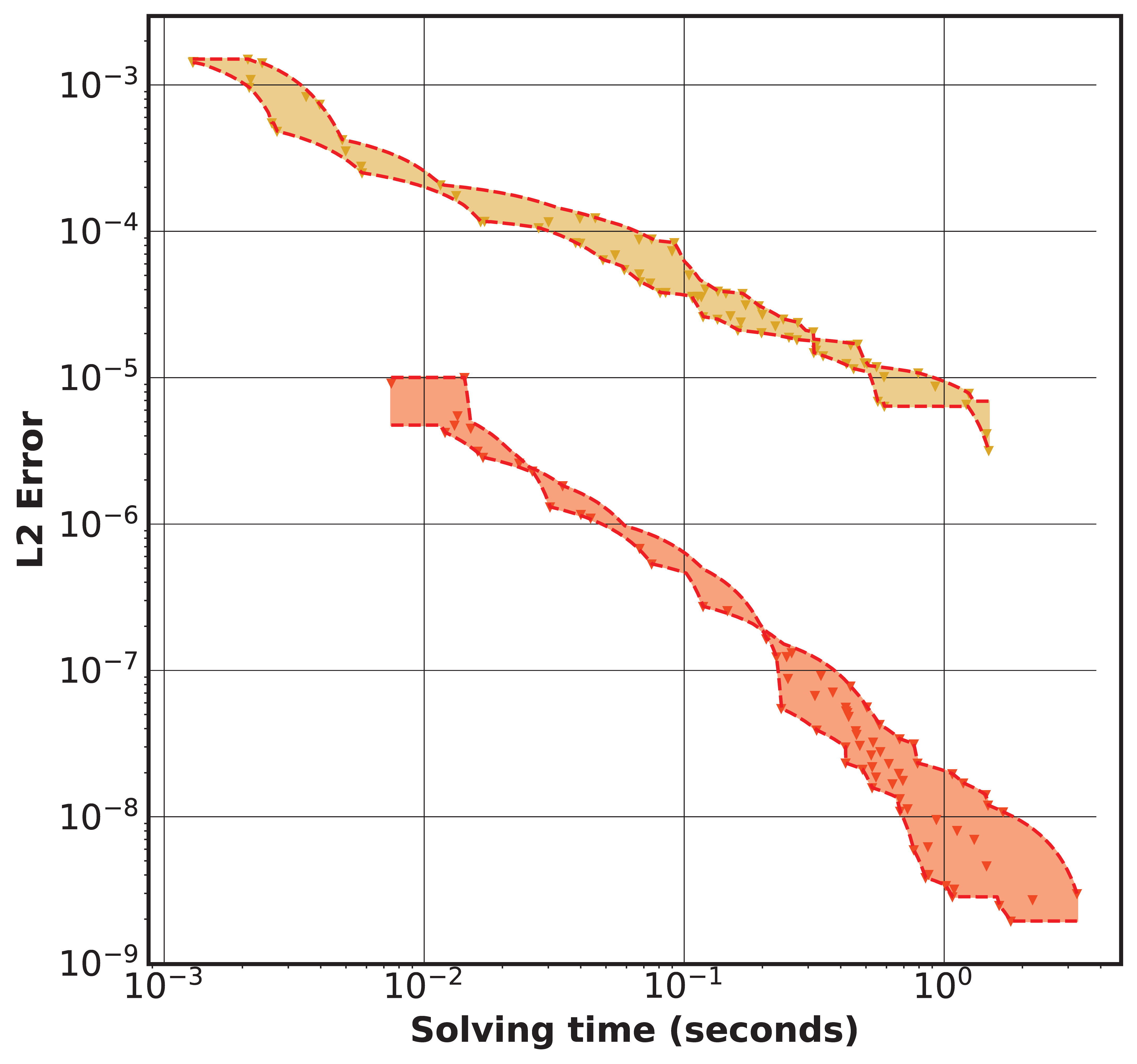}}\par
    \includegraphics[width=0.5\linewidth]{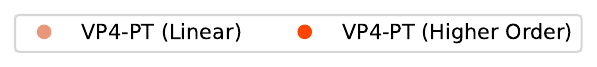}\par
    \parbox{.02\linewidth}{\rotatebox{90}{\centering DP-PT}}\hfill
    \parbox{.48\linewidth}{\includegraphics[width=\linewidth]{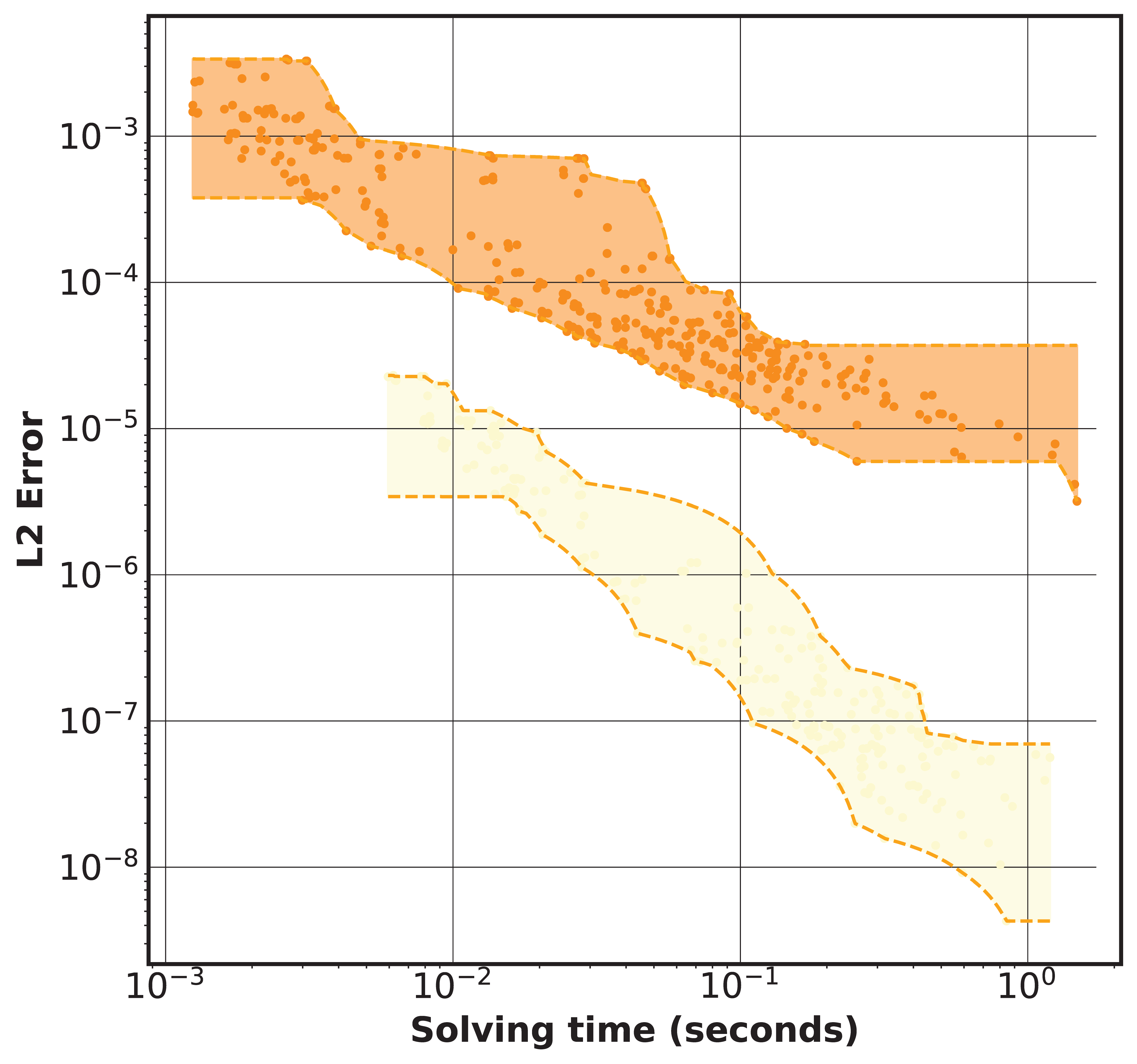}}\hfill
    \parbox{.48\linewidth}{\includegraphics[width=\linewidth]{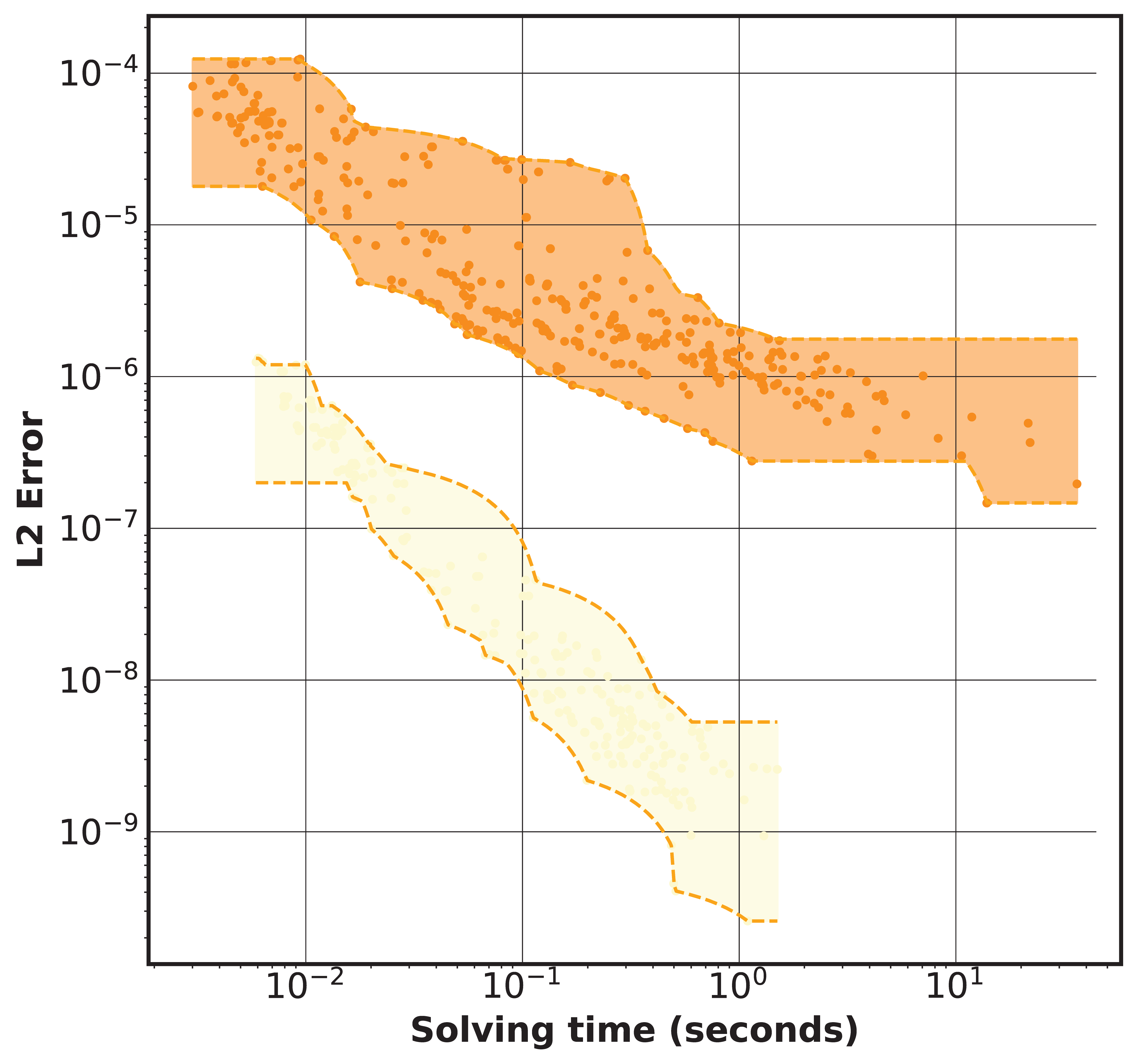}}\par
    \includegraphics[width=0.5\linewidth]{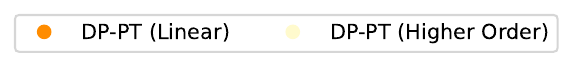}\par
    \caption{Comparison (time versus error) of linear FEM and higher-order FEM on triangular meshes.}
    \label{fig:high-order}
\end{figure}

For all simplicial meshes, both dual and triangulated (Figure~\ref{fig:high-order}), higher-order FEM shows a significant performance advantage over linear FEM/VEM and has a faster convergence rate (cubic versus linear). This result is consistent with the findings introduced in \cite{tet-vs-hex}.

\subsubsection{Barycentric Coordinate and VEM}
\label{subsec:simulation:bc-vs-vem}
We compare using barycentric coordinates (Mean Value \cite{Barycentric-MeanValue-2D,Barycentric-MeanValue-3D} and Wachspress \cite{Barycentric-Wachspress1,Barycentric-Wachspress2}) as bases with VEM on polygonal meshes to evaluate whether VEM has any performance advantage against traditionally established polytopal methods. Wachspress coordinates are well-defined only for concave polygons, so we only run it on concave polygonal meshes. We run experiments for PS\#1-US, PS\#2-SC, and PS\#3-UD.

\begin{figure}
    \centering\footnotesize
     \parbox{.02\linewidth}{~}\hfill
     \parbox{.48\linewidth}{\centering PS\#1-US}\hfill
     \parbox{.48\linewidth}{\centering PS\#1-SC}\par
    \parbox{.02\linewidth}{\rotatebox{90}{\centering VP}}\hfill
    \parbox{.48\linewidth}{\includegraphics[width=\linewidth]{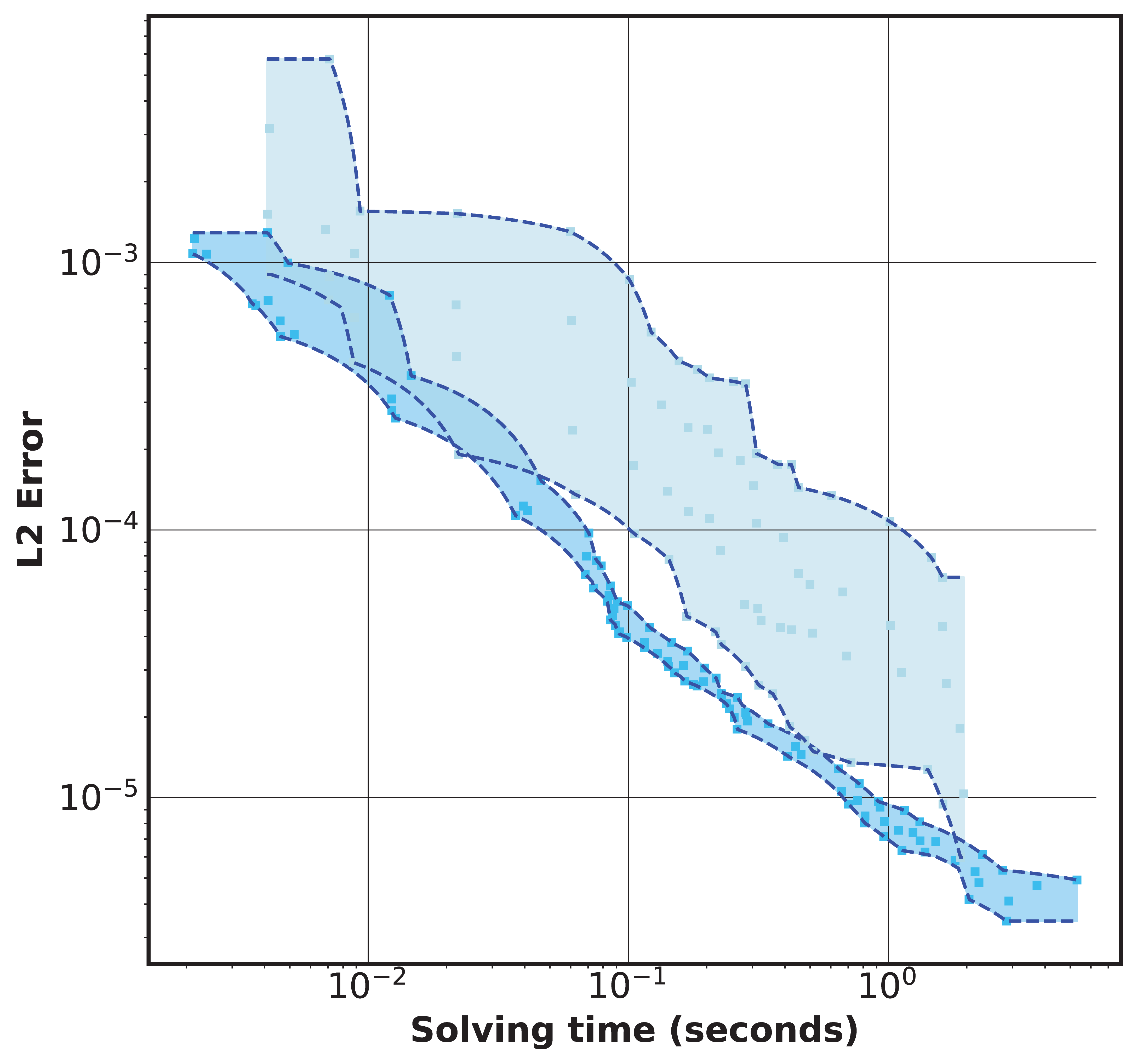}}\hfill
    \parbox{.48\linewidth}{\includegraphics[width=\linewidth]{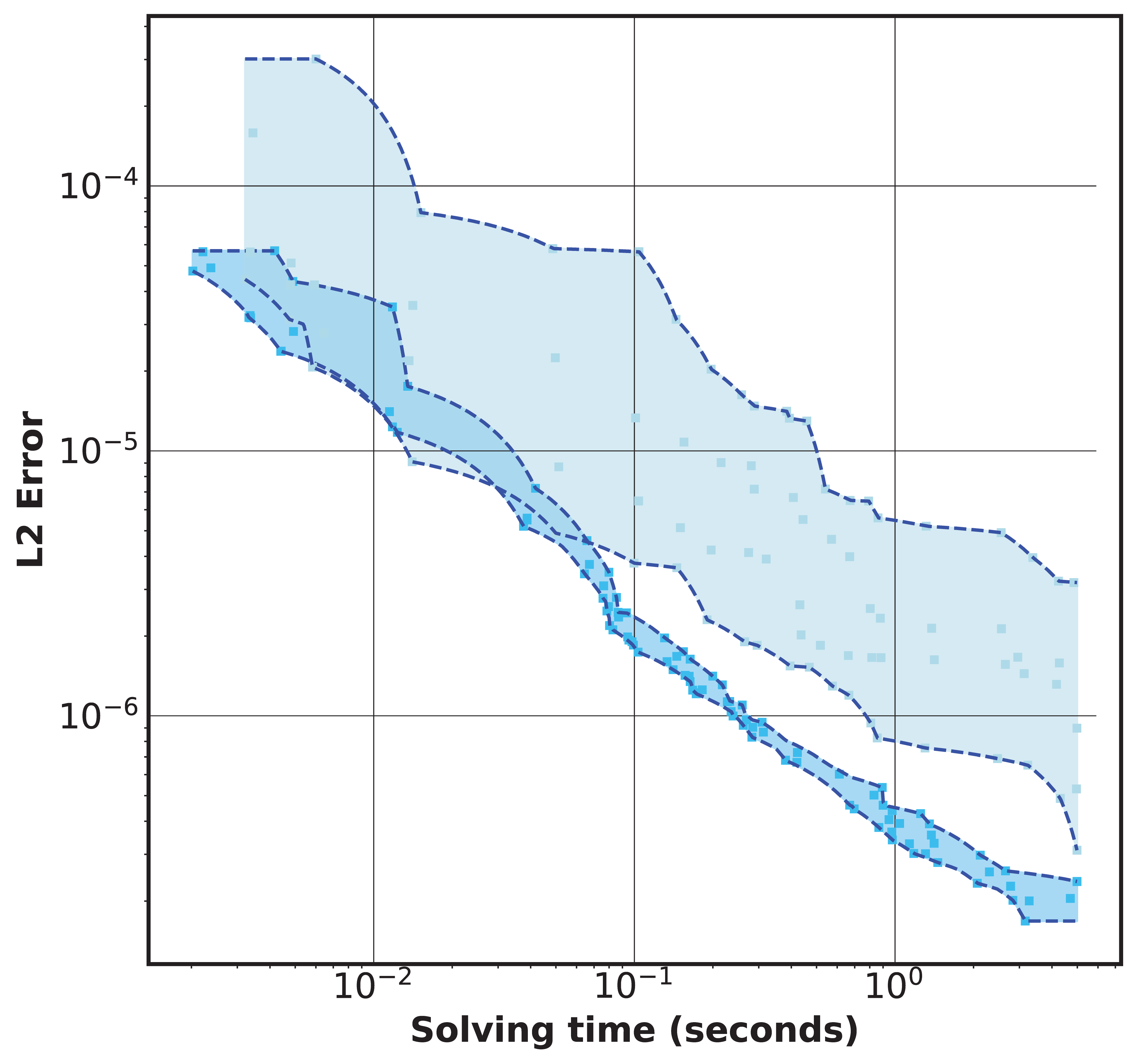}}\par
    \includegraphics[width=0.5\linewidth]{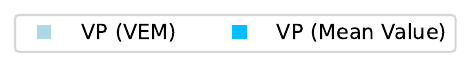}\par
    \parbox{.02\linewidth}{\rotatebox{90}{\centering DP}}\hfill
    \parbox{.48\linewidth}{\includegraphics[width=\linewidth]{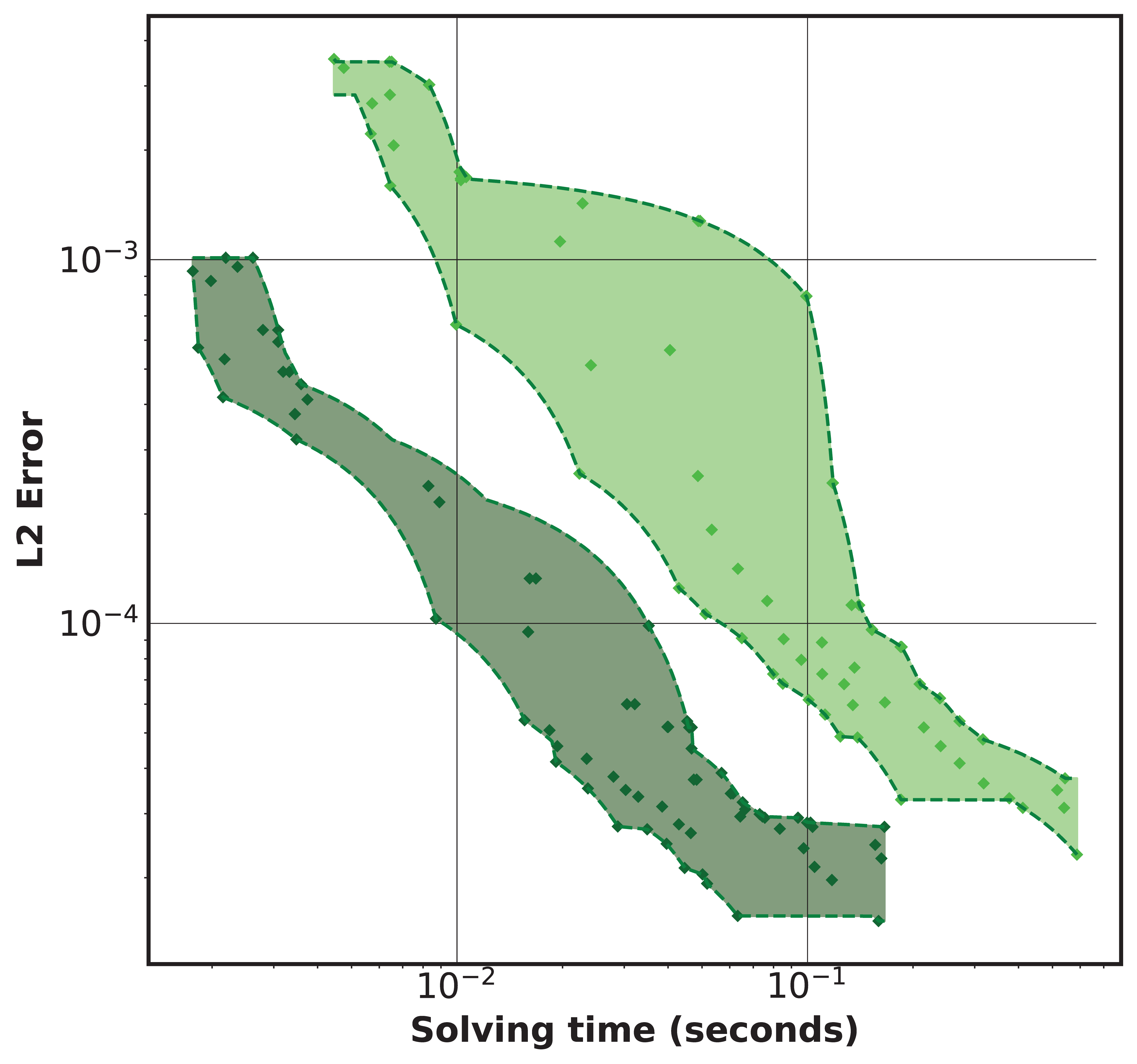}}\hfill
    \parbox{.48\linewidth}{\includegraphics[width=\linewidth]{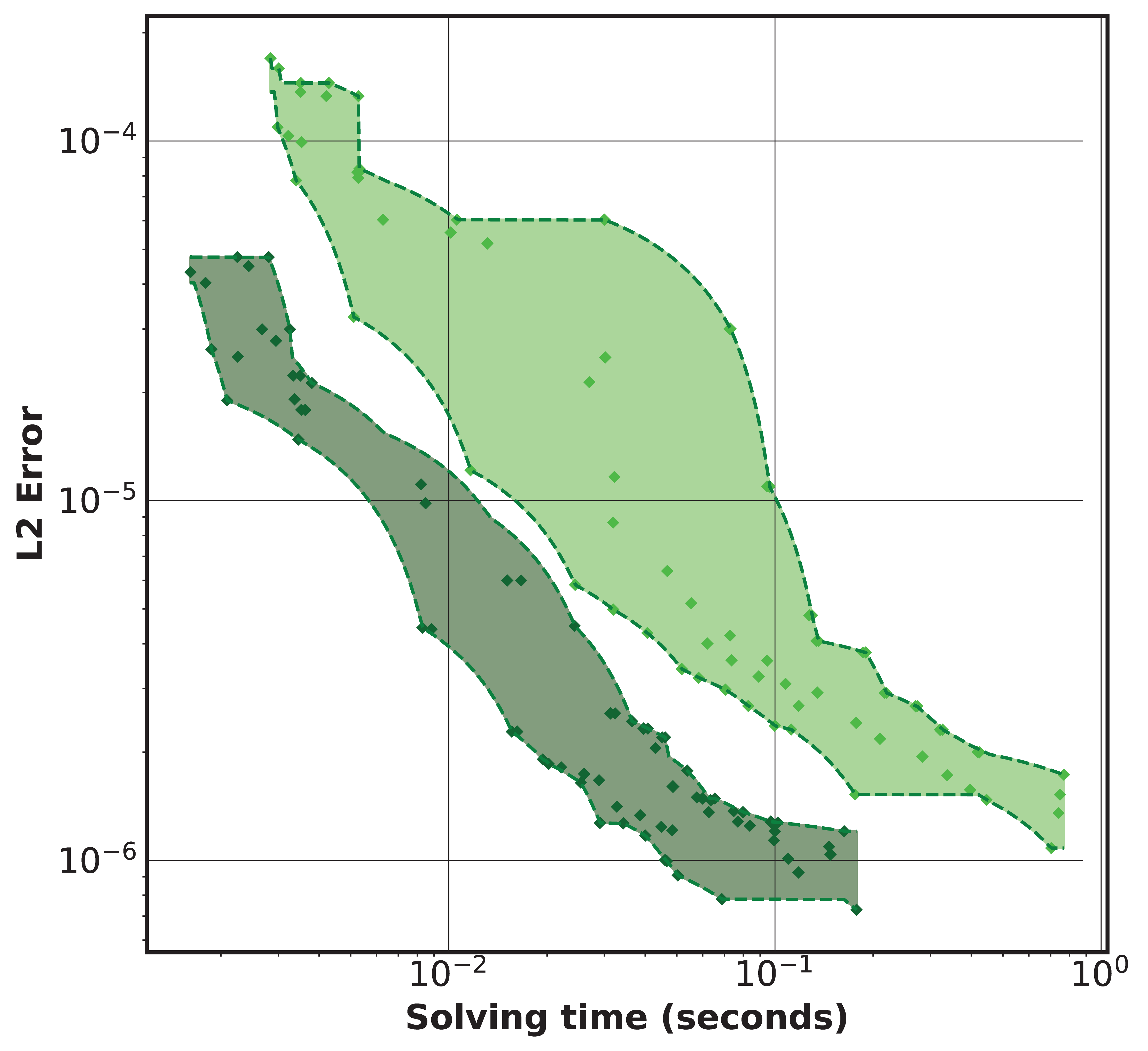}}\par
    \includegraphics[width=0.5\linewidth]{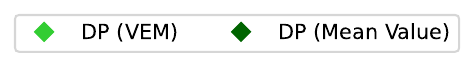}\par
    \parbox{.02\linewidth}{\rotatebox{90}{\centering TP}}\hfill
    \parbox{.48\linewidth}{\includegraphics[width=\linewidth]{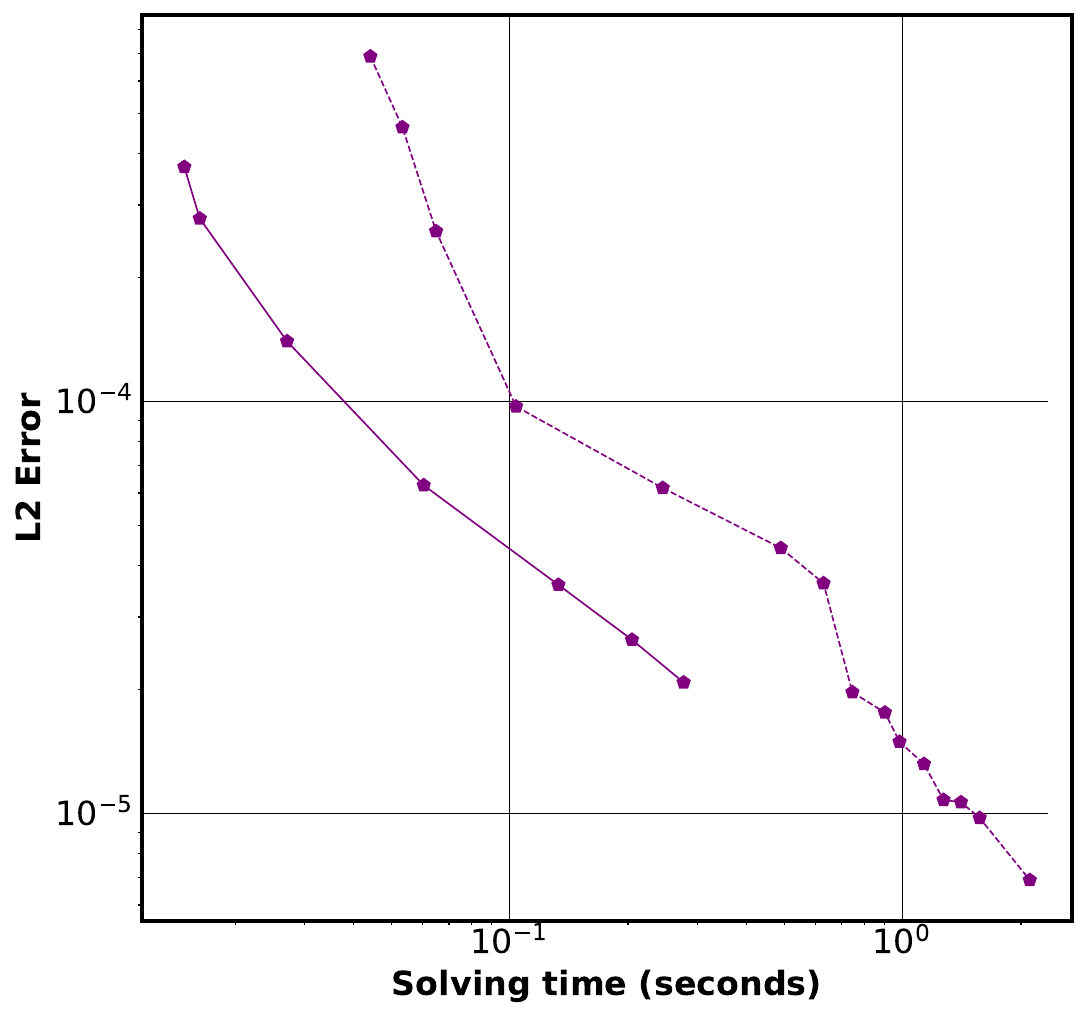}}\hfill
    \parbox{.48\linewidth}{\includegraphics[width=\linewidth]{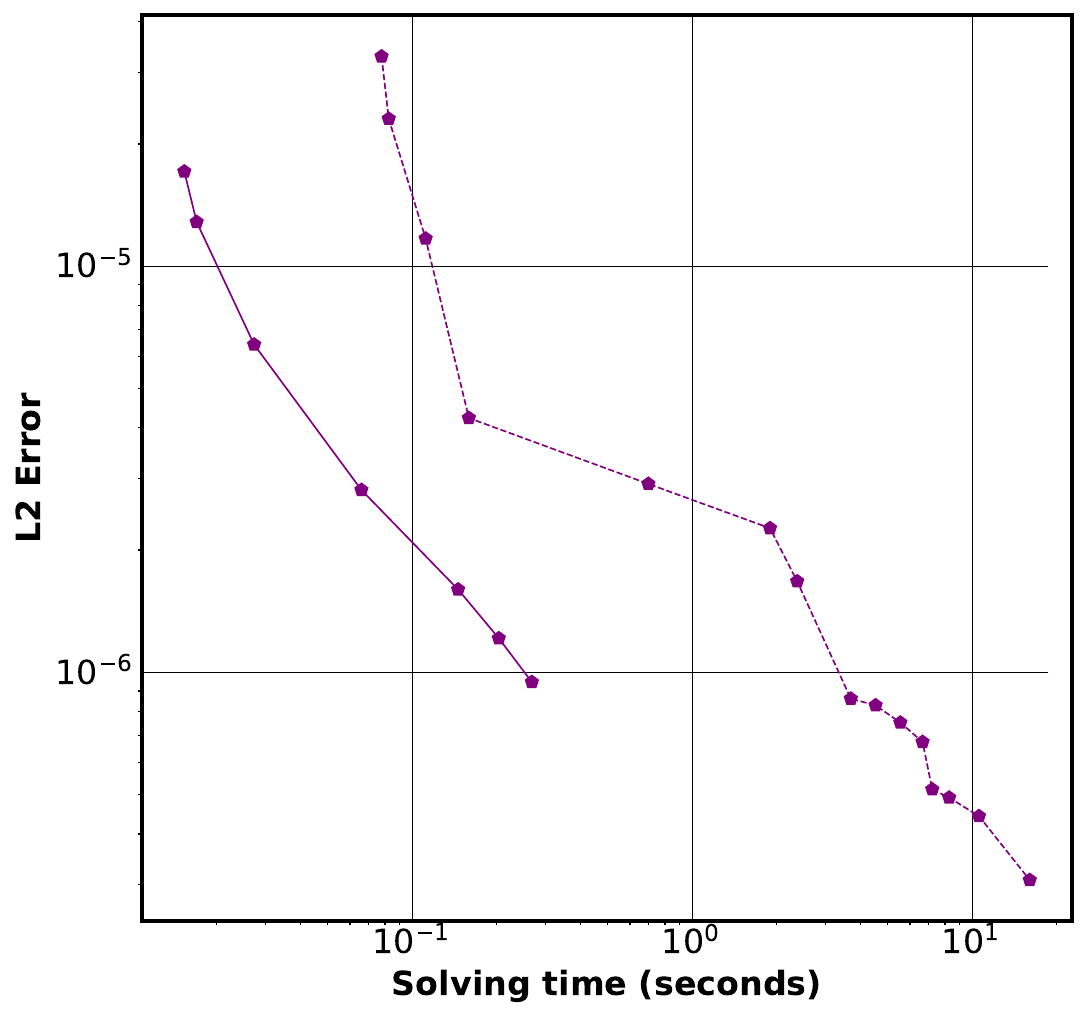}}\par
    \includegraphics[width=0.5\linewidth]{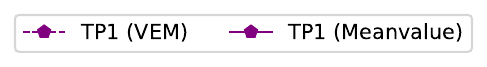}\par
    \caption{Comparison of barycentric coordinate and VEM on polygonal meshes.}
    \label{fig:bary-vem}
\end{figure}

For Voronoi polygonal meshes (VP in Figure~\ref{fig:bary-vem}), barycentric coordinates show at least competitive (PS\#1-SC) or slightly better (PS\#1-US) performance than VEM. However, we observe that when Voronoi mesh resolutions are high, the convergence rate decreases, which suggests that barycentric coordinates may not be optimal for very dense Voronoi polygonal meshes. For Displaced and Tiled polygonal meshes (DP and VP in Figure~\ref{fig:bary-vem}), barycentric coordinates generally show better performance than VEM.

In conclusion, FEM with barycentric bases could generally be more efficient than VEM on polygon meshes. However, during the experiment, we notice that barycentric coordinates could fail on non-conforming meshes (which occasionally happens due to the unrobustness of different meshing software) in our large mesh dataset, and we omitted those failure cases in the plotting. VEM could properly handle these non-conforming meshes without catastrophic failure.

\subsubsection{Barycentric FEM and Traditional FEM}
\label{subsec:simulation:bc-vs-fem}
\begin{figure}
    \centering\footnotesize
     \parbox{.02\linewidth}{~}\hfill
     \parbox{.48\linewidth}{\centering PS\#1-US}\hfill
     \parbox{.48\linewidth}{\centering PS\#2-SC}\par
    \parbox{.02\linewidth}{\rotatebox{90}{\centering VP and DT}}\hfill
    \parbox{.48\linewidth}{\includegraphics[width=\linewidth]{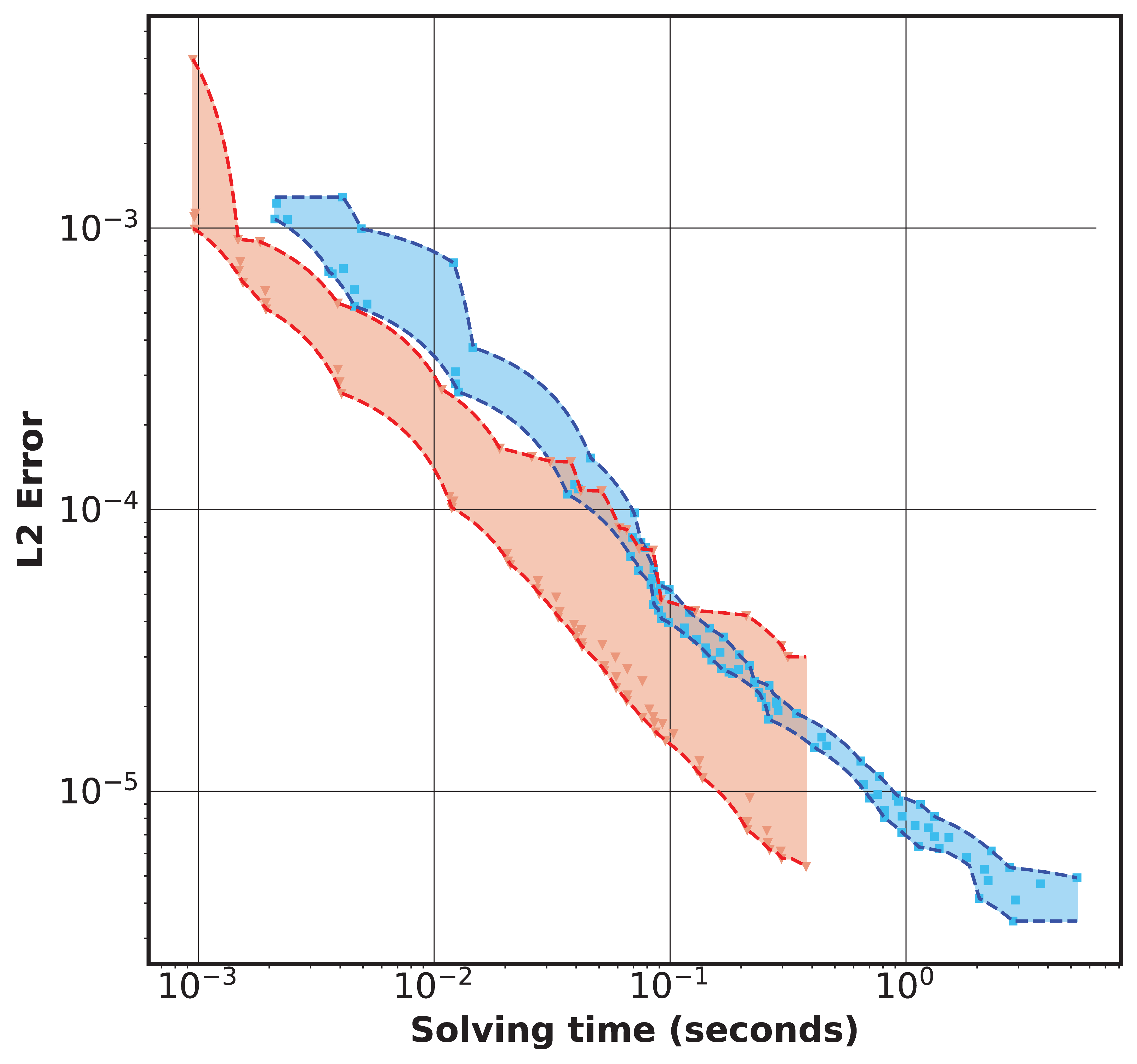}}\hfill
    \parbox{.48\linewidth}{\includegraphics[width=\linewidth]{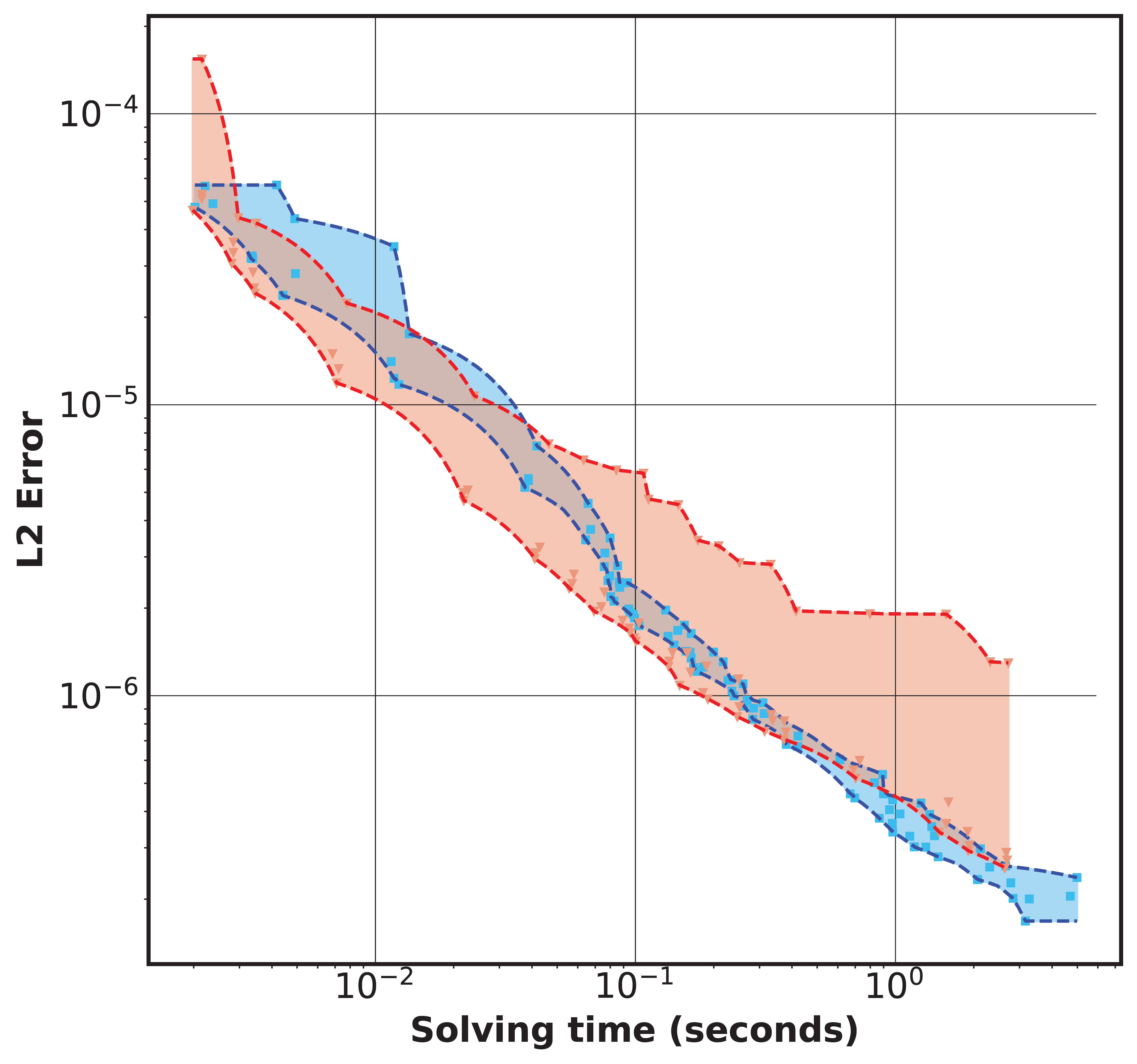}}\par
    \includegraphics[width=0.5\linewidth]{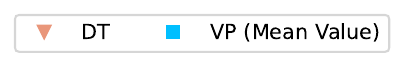}\par
    \parbox{.02\linewidth}{\rotatebox{90}{\centering DP and PT}}\hfill
    \parbox{.48\linewidth}{\includegraphics[width=\linewidth]{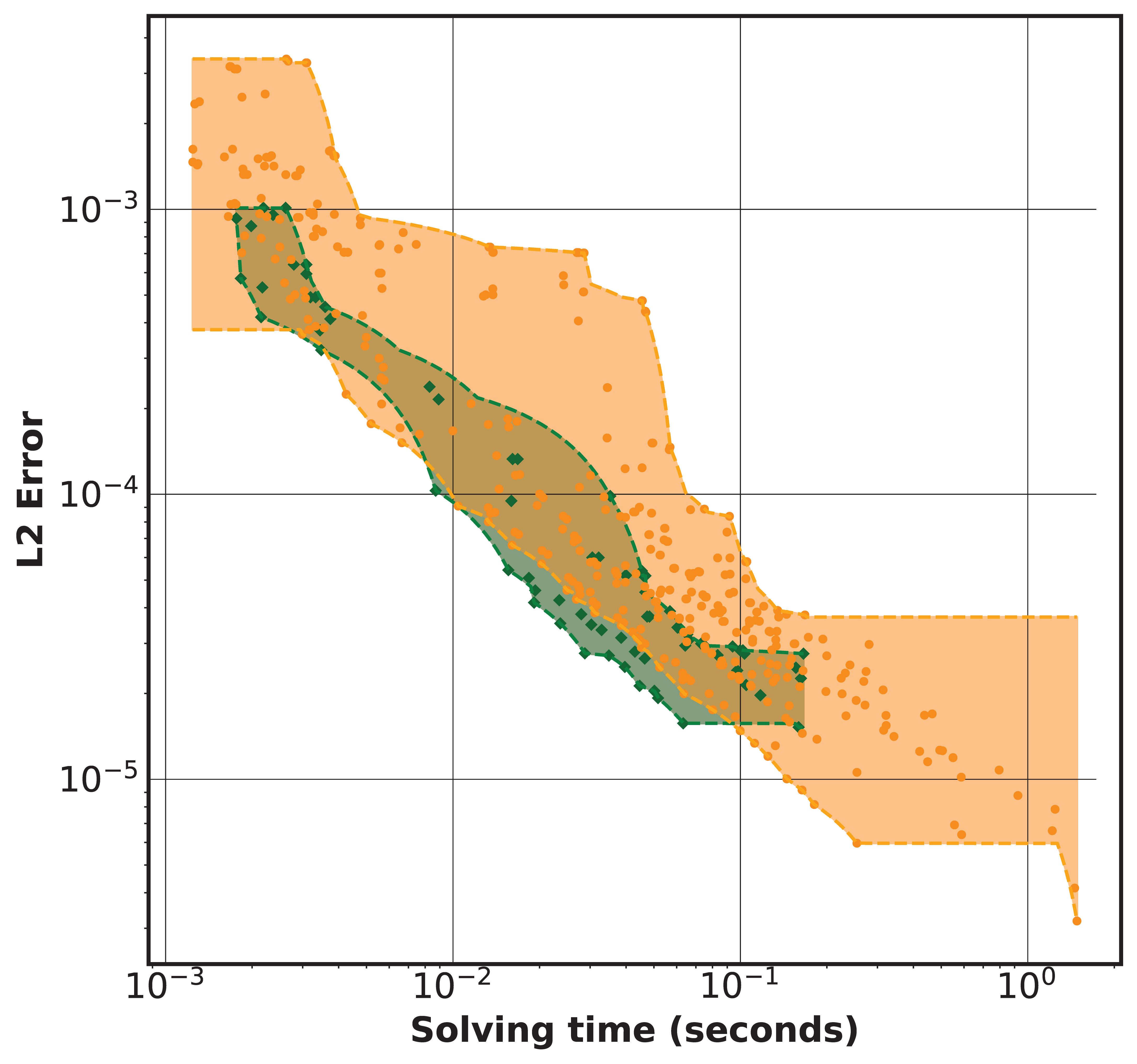}}\hfill
    \parbox{.48\linewidth}{\includegraphics[width=\linewidth]{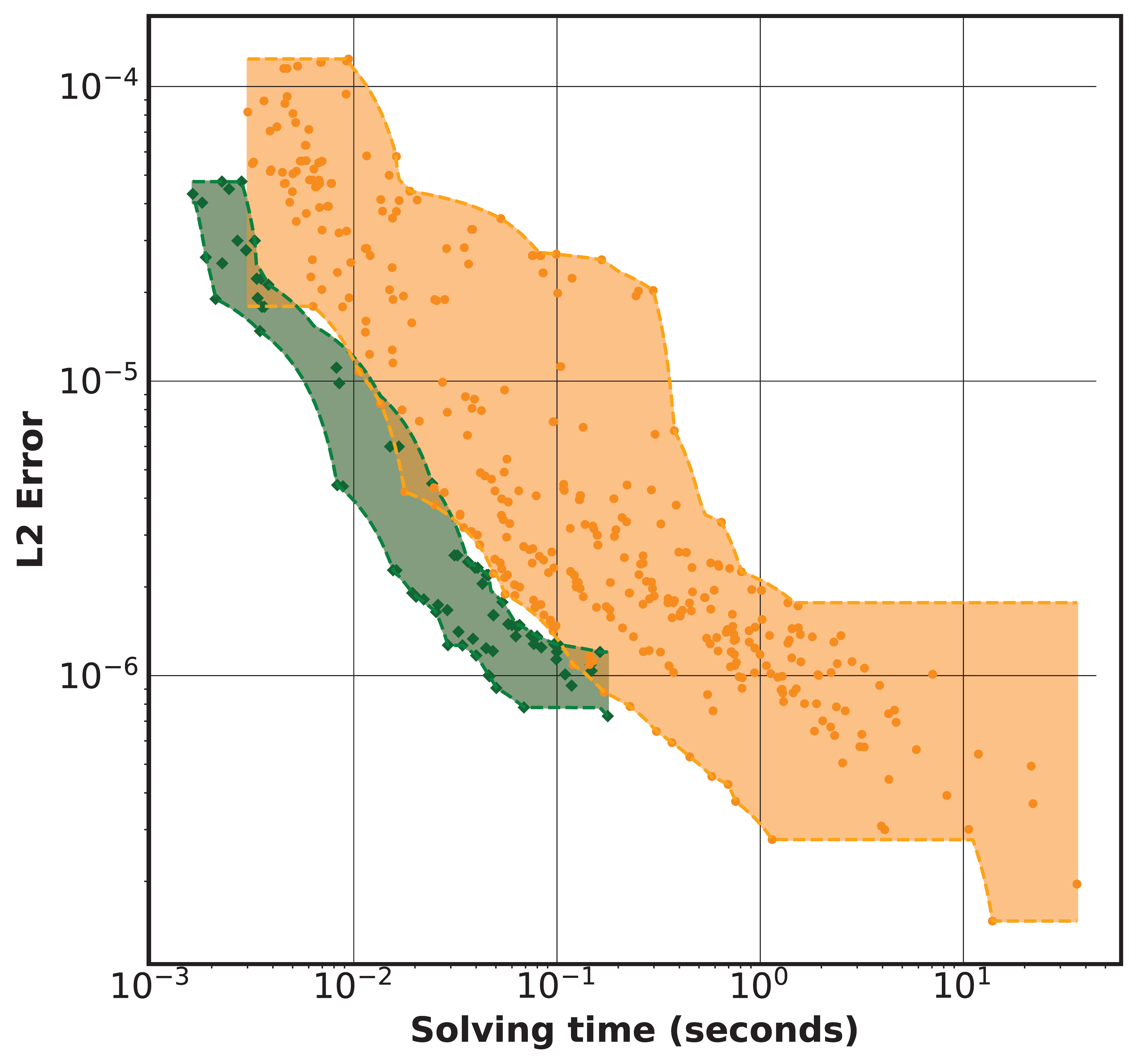}}\par
    \includegraphics[width=0.5\linewidth]{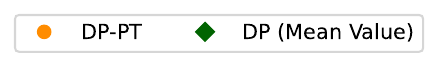}\par
    \caption{Comparison of barycentric FEM on polygon meshes and FEM on triangular meshes.}
    \label{fig:bary-vem-tri}
\end{figure}

We compare using barycentric coordinates (Mean Value \cite{Barycentric-MeanValue-2D,Barycentric-MeanValue-3D} and Wachspress \cite{Barycentric-Wachspress1,Barycentric-Wachspress2}) as bases on polygonal meshes with FEM on triangular meshes to evaluate whether the better-performed barycentric coordinates on polygon meshes can compete with the triangular meshes. We run experiments for PS\#1-US and PS\#2-SC (Figure~\ref{fig:bary-vem-tri}).
For Voronoi polygonal meshes and their dual triangular meshes, most polygonal meshes generally show competitive performance with the triangular meshes. For Displaced polygonal meshes and their triangulated meshes, the polygonal meshes show competitive and even better performance than triangle meshes. These results suggest that using barycentric coordinates basis on polygonal meshes could achieve similar and even better computation efficiency than linear FEM triangular meshes.

\subsubsection{VEM Stabilization Mechanisms in Linear Elasticity}\label{subsec:simulation:stabilization}
We compare different stabilization mechanisms used in VEM on linear elasticity problems, including the default implementation using the modified Diagonal Recipe~\cite{VEM-stabilization-modified-DRecipe}, stabilization term introduced by \citet{VEM-stabilization-Gain}, and Diagonal Recipe \cite{VEM-stabilization-DRecipe}. The choice of stabilization mechanisms might affect the convergence rate of VEM on different shaped polygonal/polyhedral meshes \cite{VEM-stabilization-DRecipe}. We focus on LEP-BH and LEB-PH and run experiments for the four stabilization mechanisms.

On both LEP-BH and LEB-PH, the three different stabilization mechanisms show similar results. We leave as future work any other stabilization mechanisms that are not included in our experiments. We note that other mechanisms might produce different results since they play crucial roles in the VEM. We do not include any results for VEM without stabilization since it is widely acknowledged that they perform poorly \cite{VEM,VEM-stability}.

 \begin{figure}
     \centering\footnotesize
     \parbox{.02\linewidth}{~}\hfill
     \parbox{.32\linewidth}{\centering Gain et al.}\hfill
     \parbox{.32\linewidth}{\centering Diagonal Recipe}\hfill
     \parbox{.32\linewidth}{\centering modified Diagonal Recipe}\par
     \parbox{.02\linewidth}{\rotatebox{90}{\centering LEP-BE}}\hfill
     \parbox{.32\linewidth}{\includegraphics[width=\linewidth]{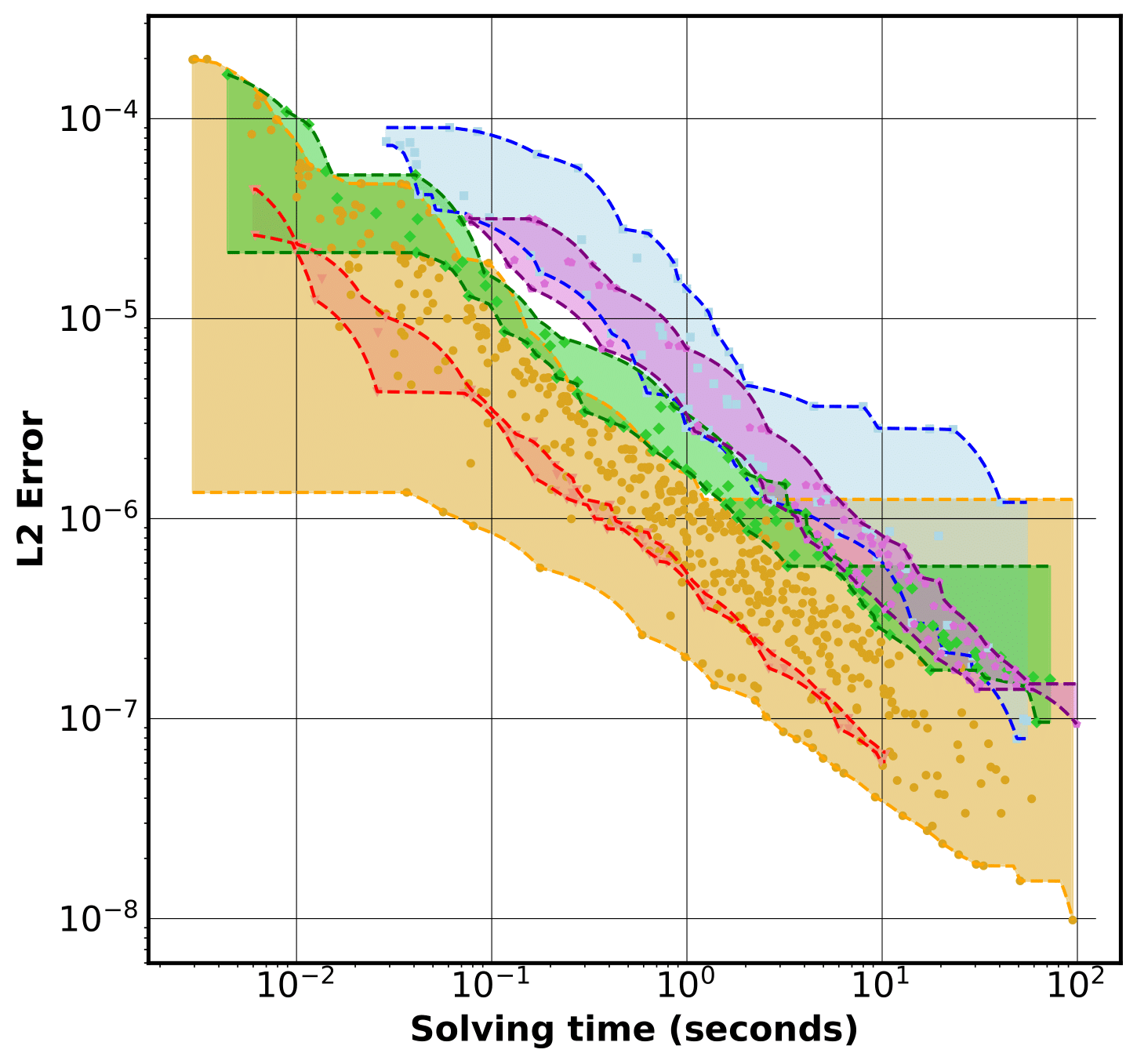}}\hfill
     \parbox{.32\linewidth}{\includegraphics[width=\linewidth]{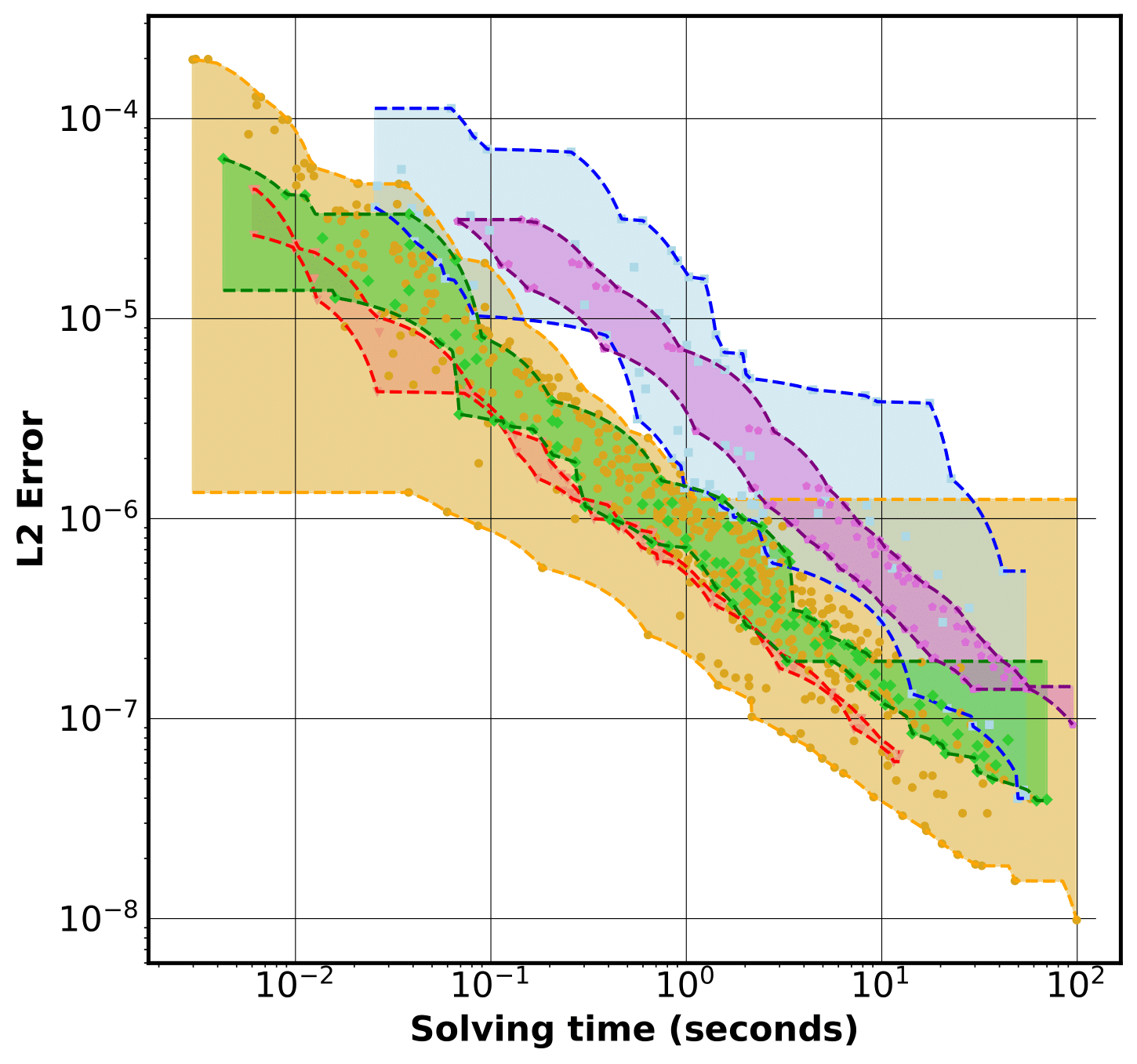}}\hfill
     \parbox{.32\linewidth}{\includegraphics[width=\linewidth]{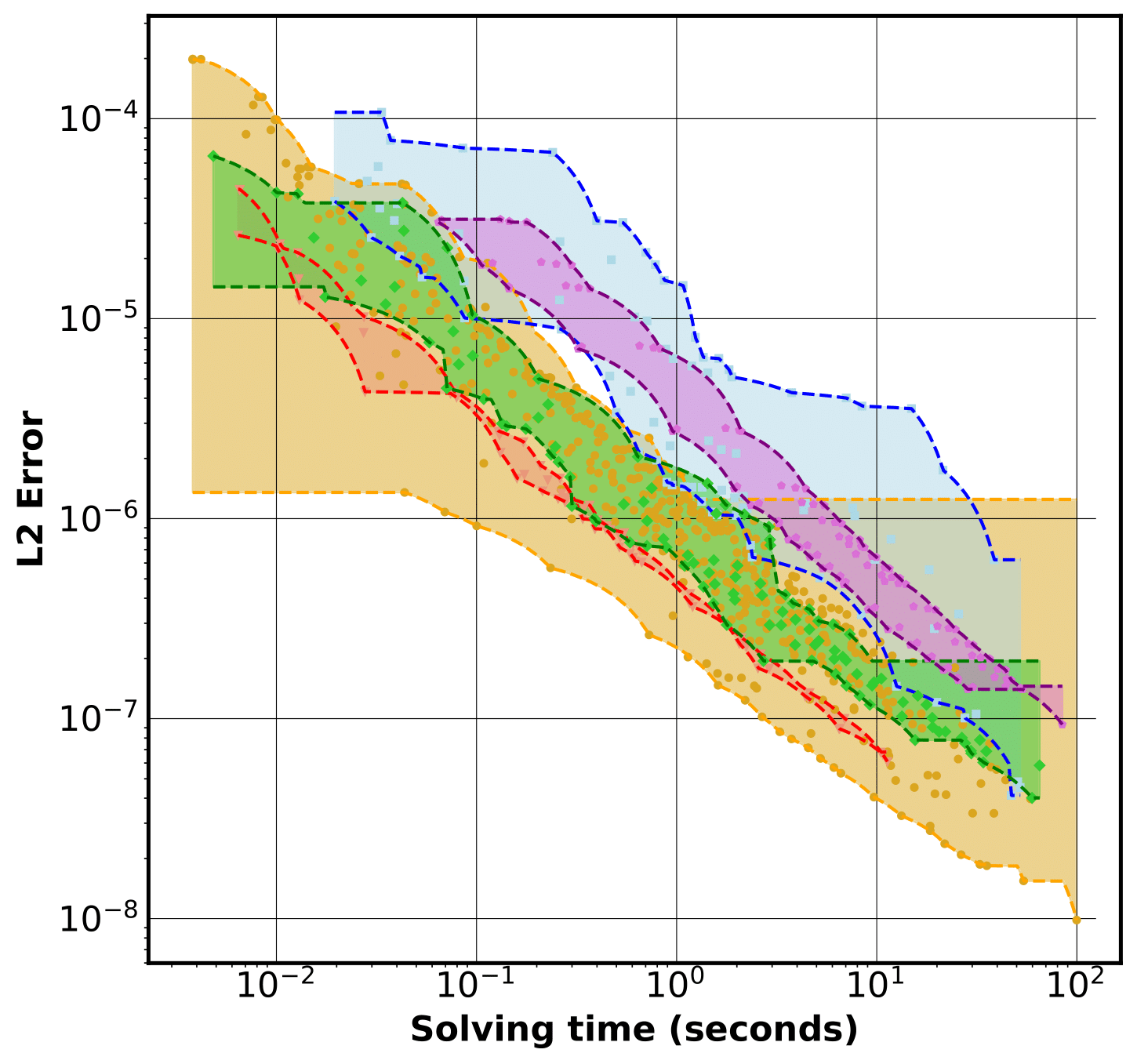}}\par
     \parbox{.02\linewidth}{\rotatebox{90}{\centering LEB-PH}}\hfill
     \parbox{.32\linewidth}{\includegraphics[width=\linewidth]{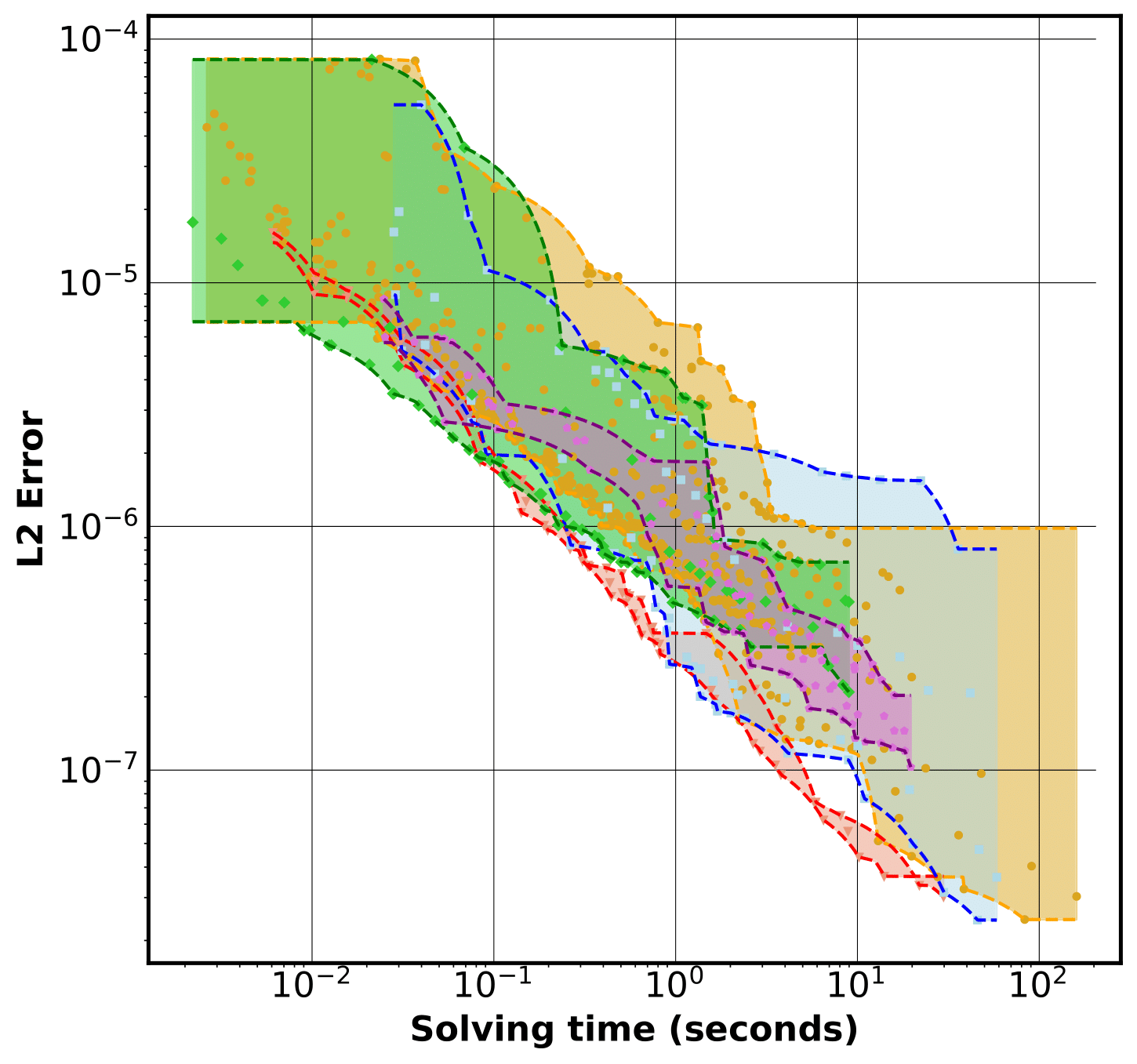}}\hfill
     \parbox{.32\linewidth}{\includegraphics[width=\linewidth]{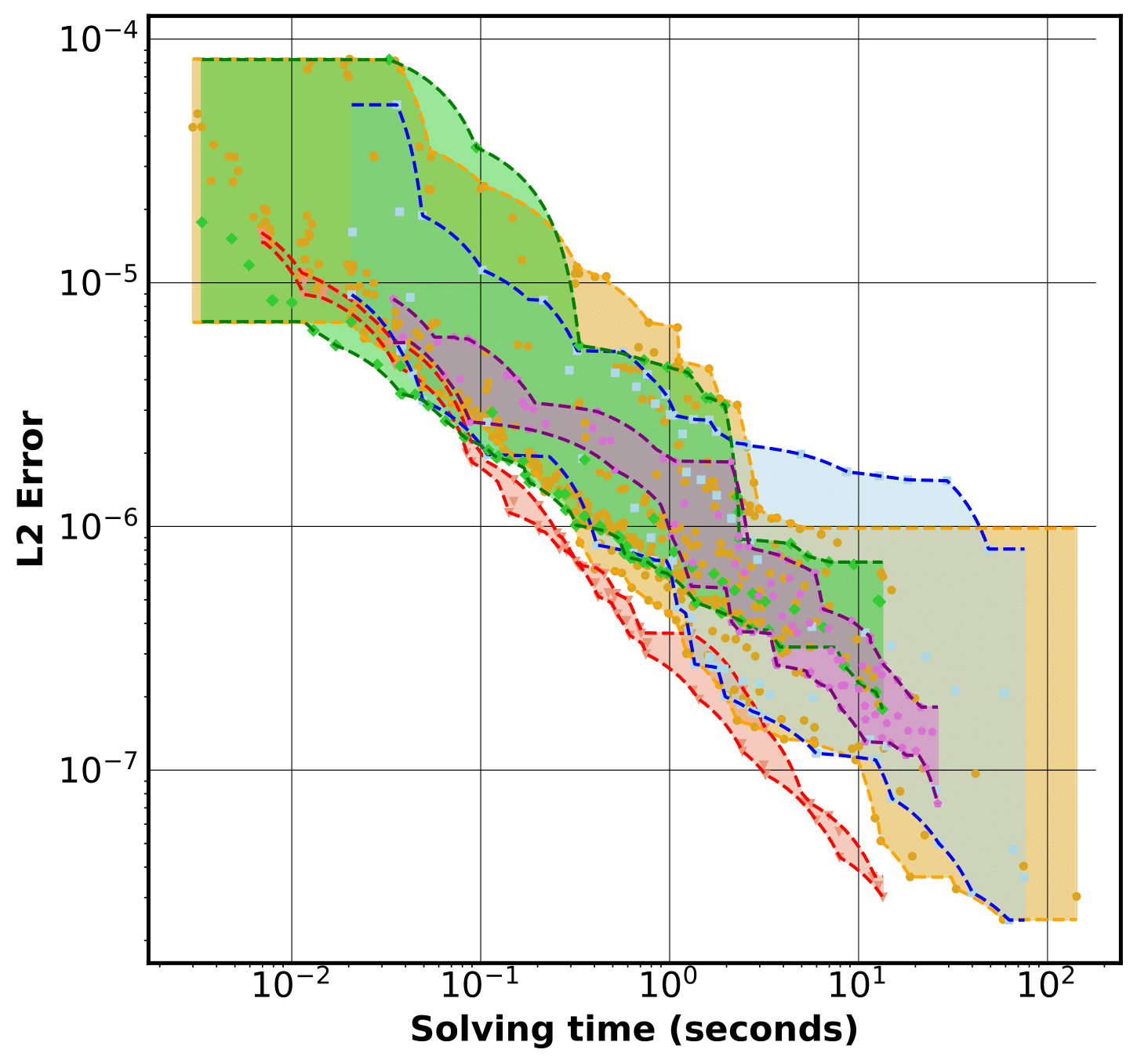}}\hfill
     \parbox{.32\linewidth}{\includegraphics[width=\linewidth]{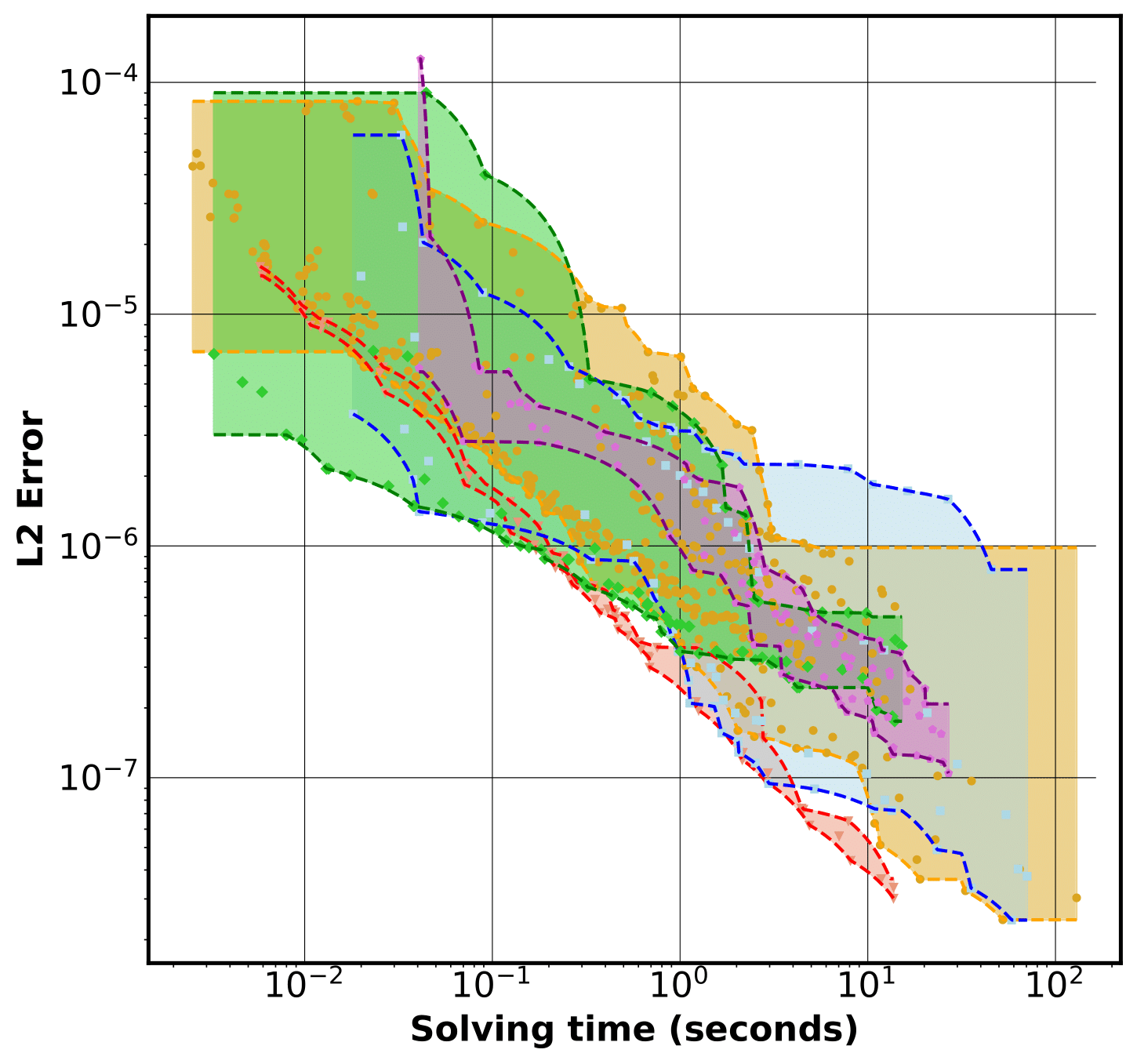}}\par
     \includegraphics[width=0.5\linewidth]{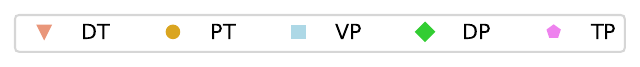}\par
     \caption{Comparison of different stabilization mechanisms of VEM on Linear Elasticity Problems.}
     \label{fig:compare_VEM_stabilization}
 \end{figure}

\subsection{Extensive Comparisons}
\label{subsec:extensive}
We conduct extensive comparisons to showcase the generalization ability of our findings in other simulation decisions and into 3D domains. We first evaluate the impact of using direct and iterative solvers in Section~\ref{subsec:extensive:direct-vs-iterative}. We then showcase how do different PDEs affect our findings in Section~\ref{subsec:extensive:pde}. We show our comparisons of Poisson problems on 3D domains in Section~\ref{subsec:extensive:3d}.

\subsubsection{Linear Solver}
\label{sec:res:solver}
\label{subsec:extensive:direct-vs-iterative}
In our study, we use both a direct linear solver (through the \textit{linsolve} function in Matlab) and an iterative solver (\textit{pcg}) in Matlab. For the iterative solver, the linear system is preconditioned to the system to improve convergence and stability. Since the stiffness matrix is symmetric, we use an incomplete LU factorization with partial pivoting as a preconditioner and use preconditioned conjugate gradient (PCG) \cite{Iterative-solver-2}. The conclusions of our study show mostly similar results on both types of solvers.

From our experiments, we notice that the relative difference between polygonal meshes (VP and DP) and triangulated ones, decreases when using PCG instead of direct solvers. In contrast, most triangular meshes and tiled polygonal meshes (which are triangle dominant) have similar performance when using both solvers.

Figure \ref{fig:compare_solver} shows the solving time of PS\#1-US, PS\#1-SC, and LEP-BE. We select four $L_2$ error levels ($10^{-2}, 10^{-3}, 10^{-4}, \text{and } 10^{-5}$) and compare the solving time of each mesh. In each subplot, the x-axis shows different mesh types and the y-axis represents solving time. The only obvious change when comparing the use of direct and iterative solvers is that the blue and green candles (i.e., Voronoi meshes and displaced polygonal meshes) shift downwards (i.e., less solving time) compared to other dots when using an iterative solver.

\input{img/tex/solving-time.tex}

\subsubsection{Different PDEs}
\label{subsec:extensive:pde}
We compare the performance of different PDE problems on different domains to discuss whether the different PDEs can bring different performance results. By comparing PS\#1-US, PS\#1-SC, LEP-BE, and LEB-PH (Figure \ref{fig:compare_PDEs}), we find that most problems show similar results, except for LEP-PH where the high-quality Voronoi meshes have better performance.

\input{img/tex/diff-pde.tex}

\subsubsection{3D Domain on Poisson Problems}\label{subsec:extensive:3d}
\begin{figure}\centering\footnotesize
    \parbox{.02\linewidth}{~}\hfill\hfill
    \parbox{.48\linewidth}{\centering PS\#2-US}\hfill
    \parbox{.48\linewidth}{\centering 3D}\par
    \parbox{.02\linewidth}{\rotatebox{90}{\centering Direct Solver}}\hfill\hfill
    \parbox{.48\linewidth}{\includegraphics[width=\linewidth]{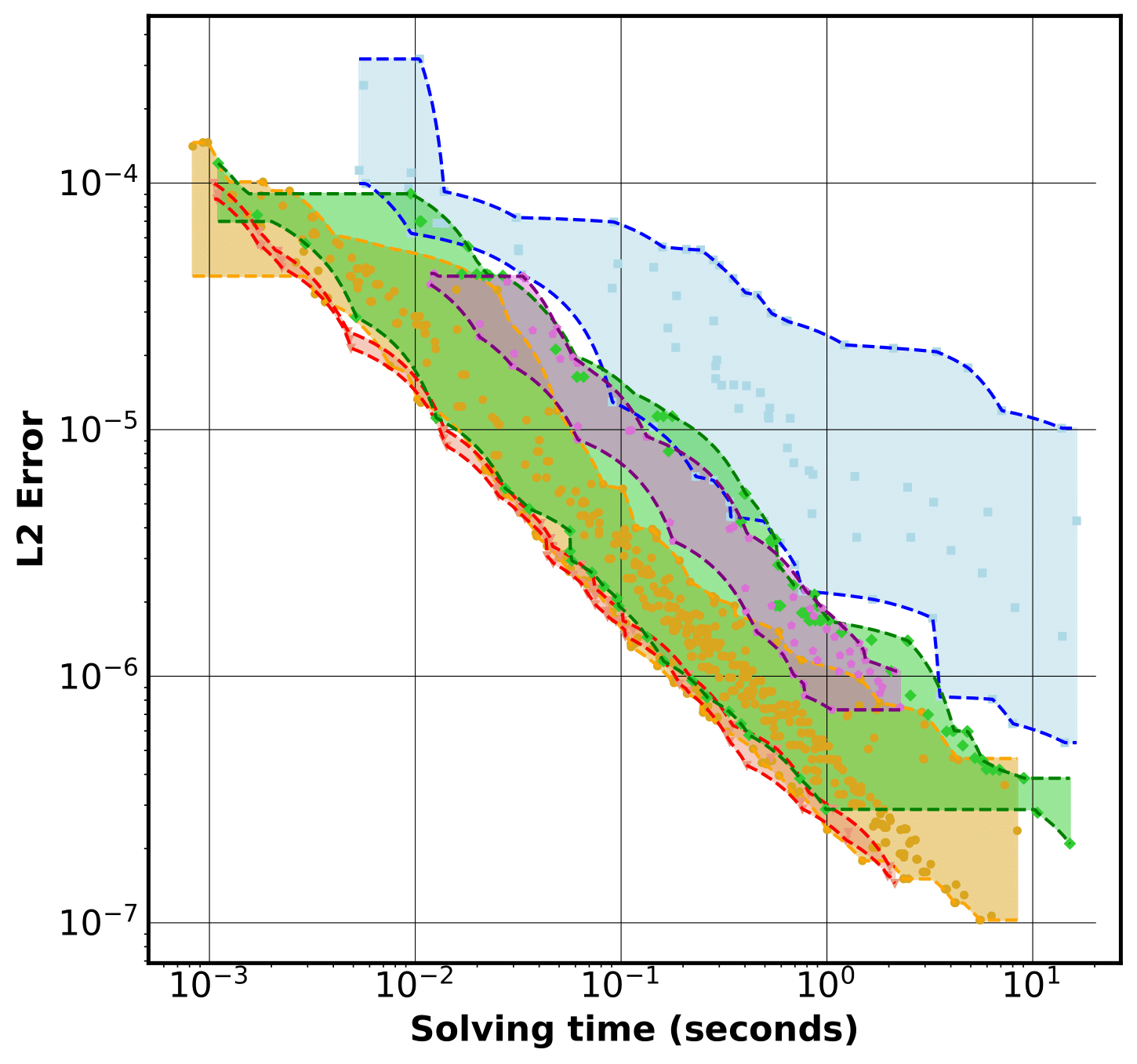}}\hfill
    \parbox{.48\linewidth}{\includegraphics[width=\linewidth]{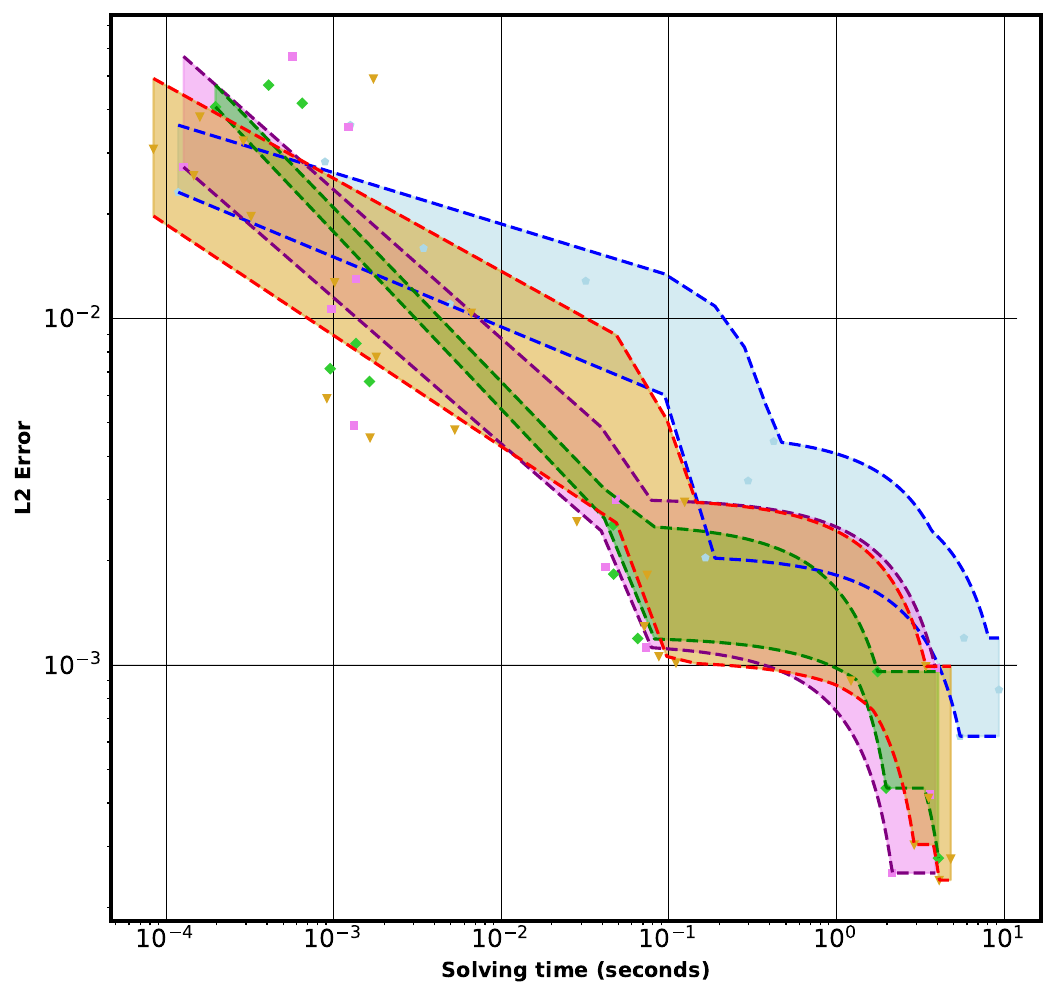}}\par
    \parbox{.02\linewidth}{\rotatebox{90}{\centering Iterative Solver}}\hfill\hfill
    \parbox{.48\linewidth}{\includegraphics[width=\linewidth]{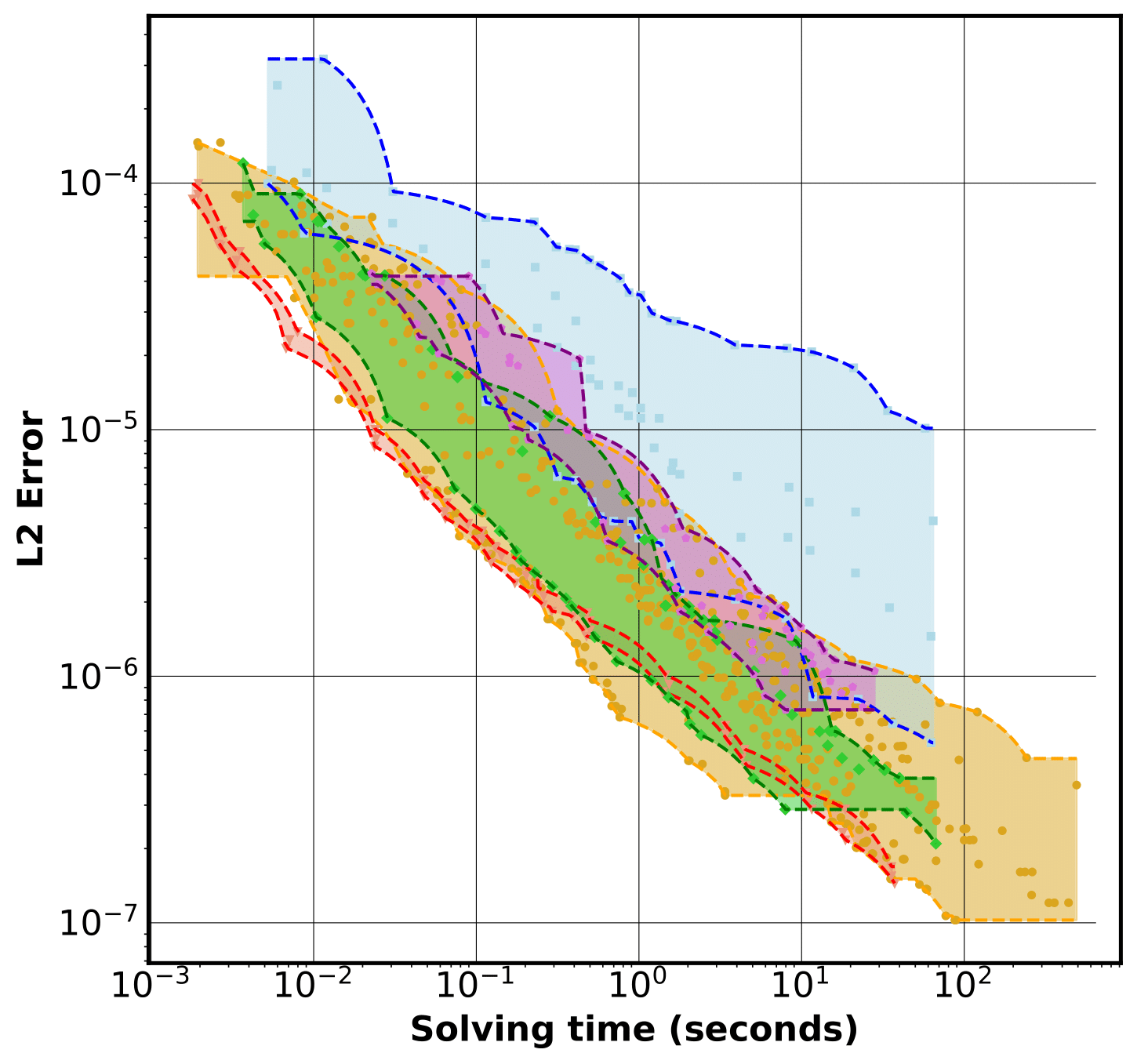}}\hfill
    \parbox{.48\linewidth}{\includegraphics[width=\linewidth]{img/results_new/3D/direct/pde1_3D_L2}}\par
    \parbox{.02\linewidth}{~}\hfill\hfill
    \parbox{.48\linewidth}{\includegraphics[width=\linewidth]{img/results_new/legend/plot_all}}\hfill
    \parbox{.48\linewidth}{\includegraphics[width=\linewidth]{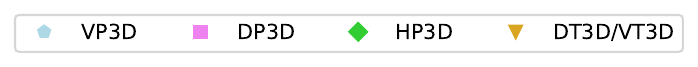}}\par
    \caption{Comparison of 2D and 3D domain on Poisson problems.}
    \label{fig:compare_2d_vs_3d}
\end{figure}

\begin{figure}
    \centering\footnotesize
    \parbox{.02\linewidth}{~}\hfill
    \parbox{.32\linewidth}{\centering PS3D\#1}\hfill
    \parbox{.32\linewidth}{\centering PS3D\#2}\hfill
    \parbox{.32\linewidth}{\centering PS3D\#3}\par
    \parbox{.02\linewidth}{\rotatebox{90}{\centering Direct Solver}}\hfill
    \parbox{.32\linewidth}{\includegraphics[width=\linewidth]{img/results_new/3D/direct/pde1_3D_L2}}\hfill
    \parbox{.32\linewidth}{\includegraphics[width=\linewidth]{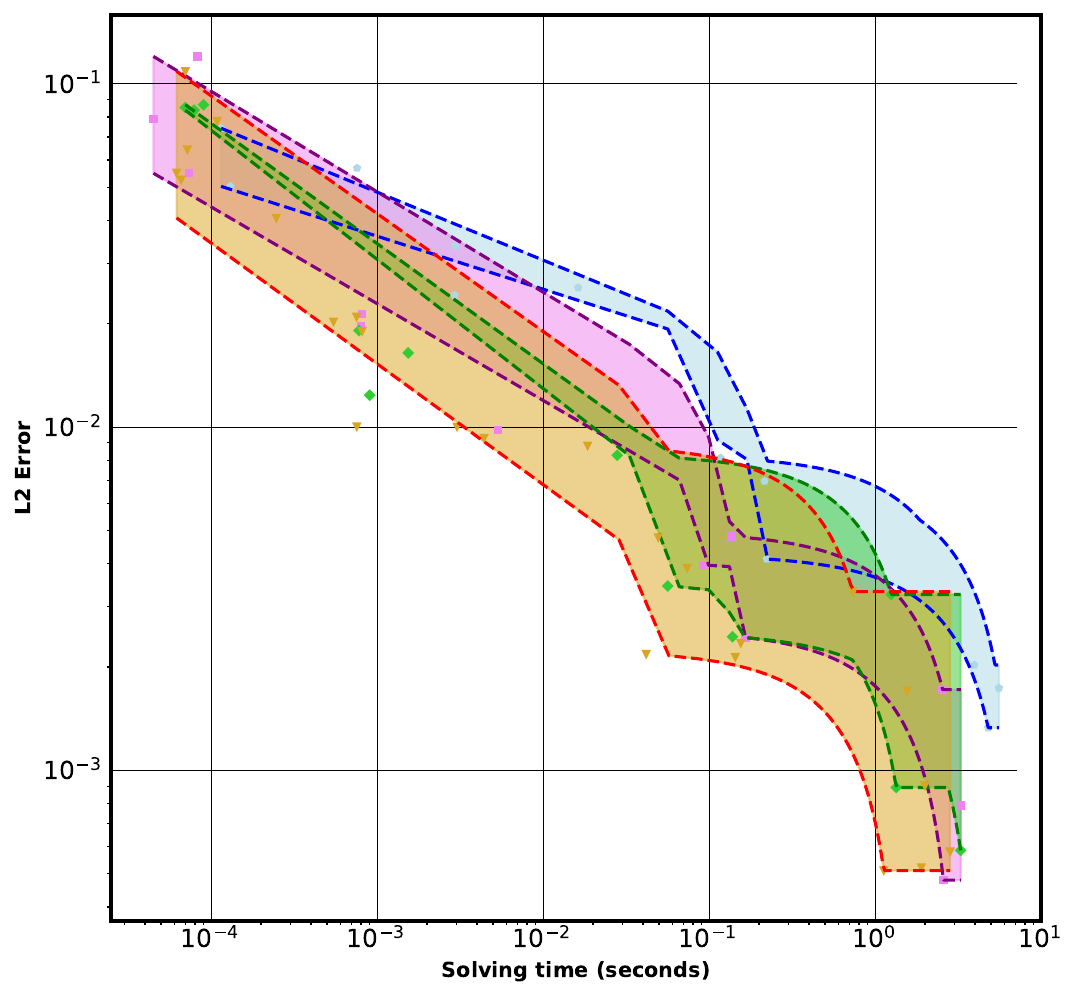}}\hfill
    \parbox{.32\linewidth}{\includegraphics[width=\linewidth]{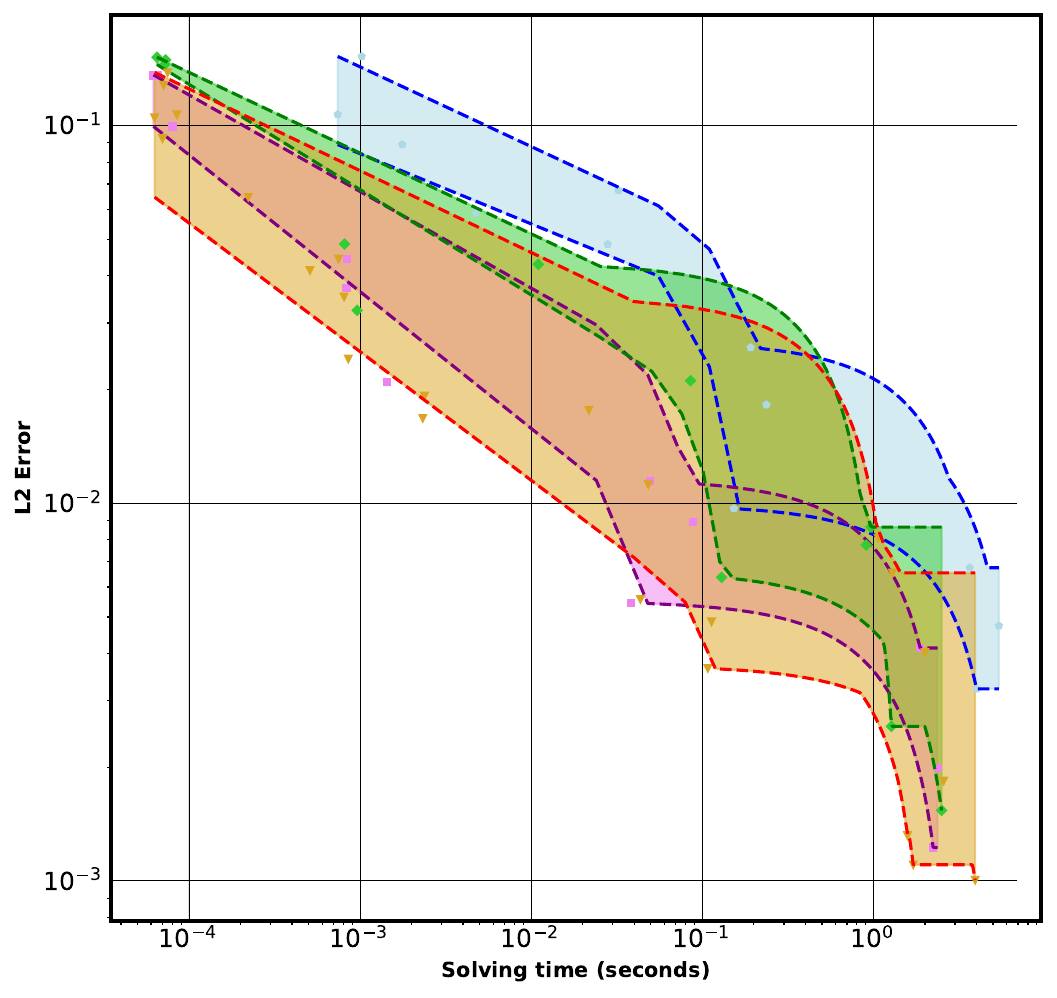}}\par
    \parbox{.02\linewidth}{\rotatebox{90}{\centering Iterative Solver}}\hfill
    \parbox{.32\linewidth}{\includegraphics[width=\linewidth]{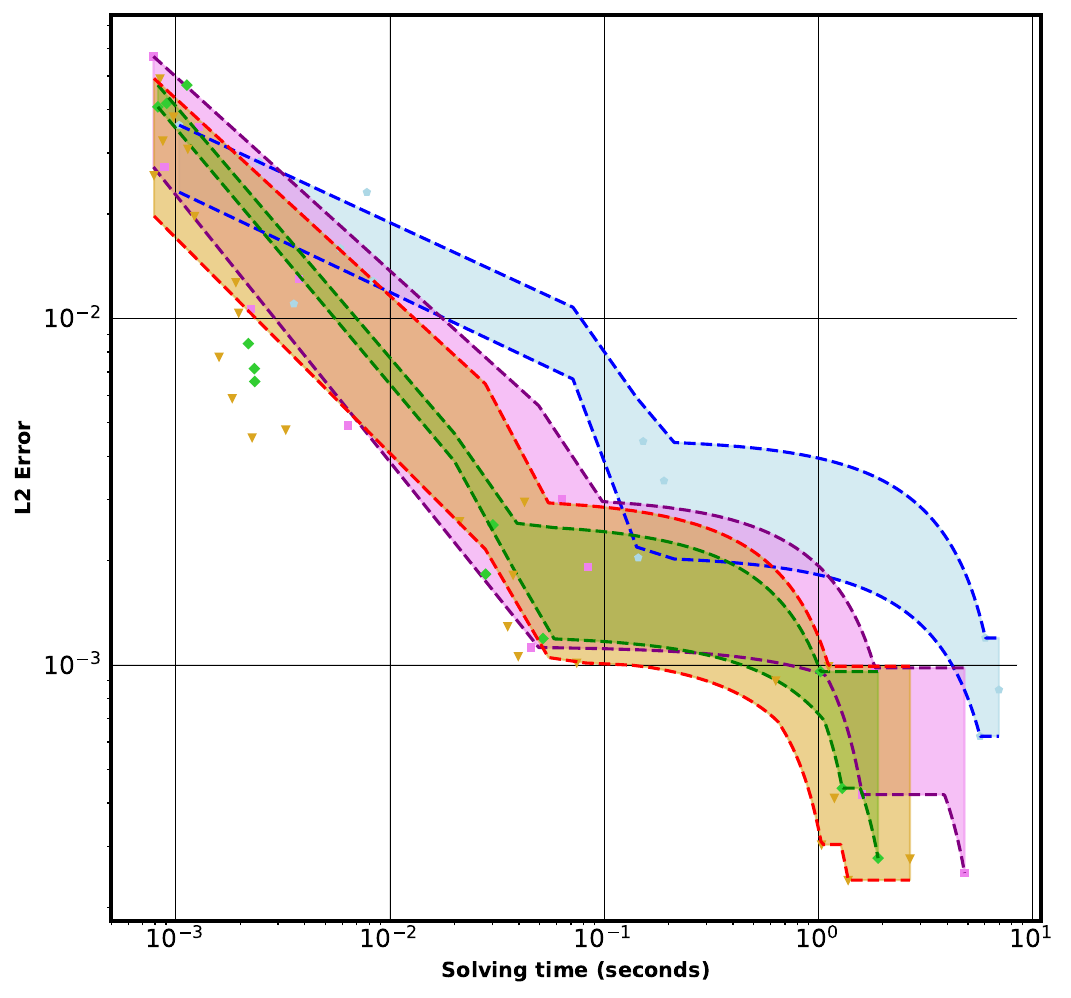}}\hfill
    \parbox{.32\linewidth}{\includegraphics[width=\linewidth]{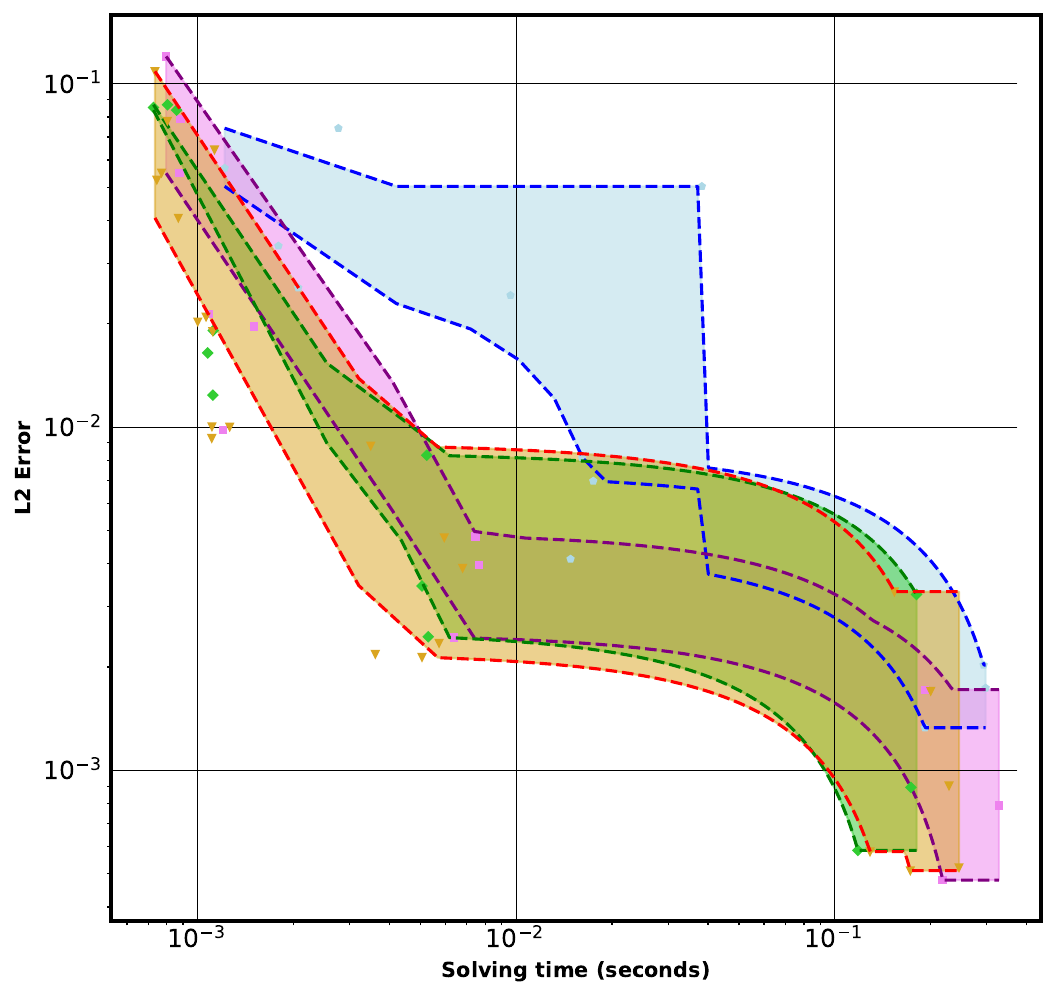}}\hfill
    \parbox{.32\linewidth}{\includegraphics[width=\linewidth]{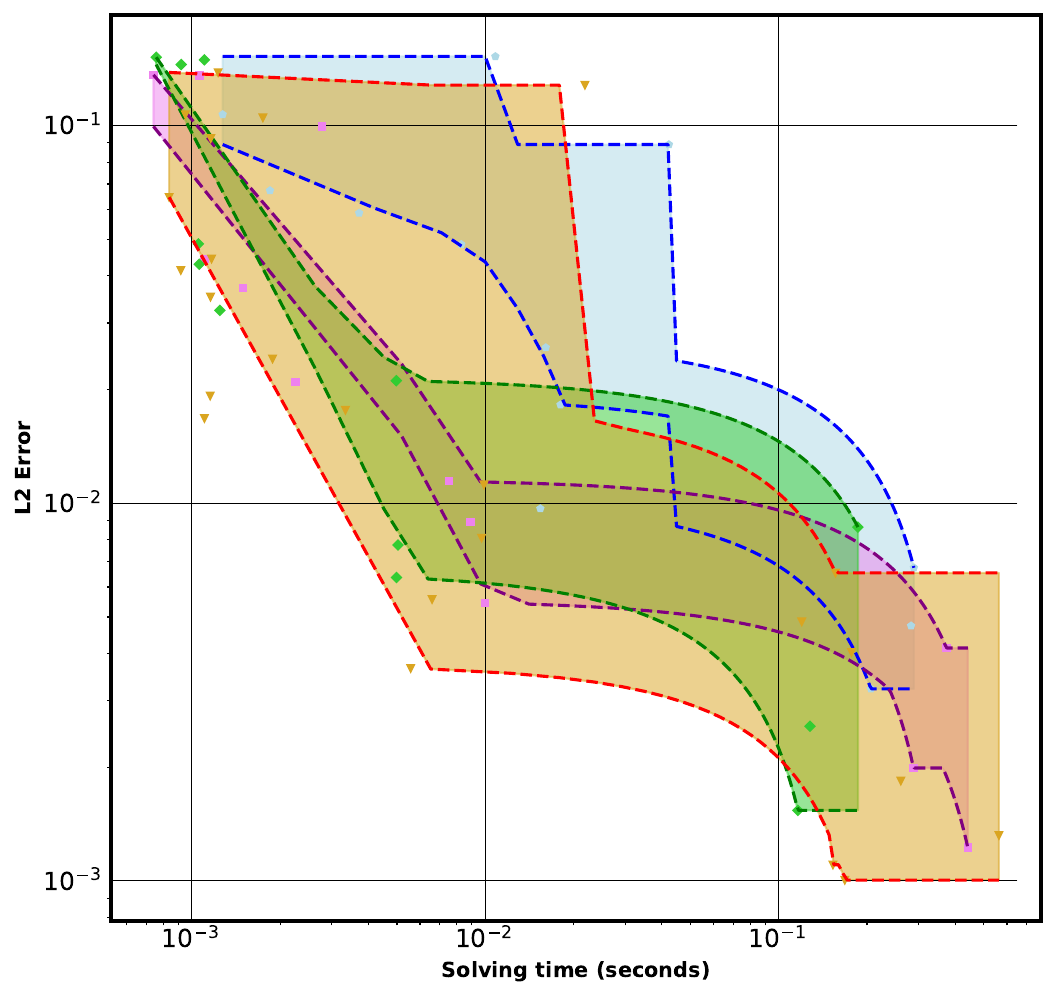}}\par
    \includegraphics[width=0.5\linewidth]{img/results_new/legend/3D}\par
    \caption{Comparison of 3 different 3D Poisson problem on cubes.}
    \label{fig:3D_poisson}
\end{figure}

We run a study on 3D domains to evaluate whether our above observations extend to 3D. The mesh dataset we use is introduced in \cite{Polyheral-quality-indicator}, and its detailed discretization choices are presented in \ref{sec:meshing}. We use a 3D VEM MATLAB codebase mVEM \cite{mVEM} and use PS3D\#1, while in 2D, we use PS\#2. We run the experiments on both the direct solver and the iterative solver. The solver settings in MATLAB are exactly the same as all the experiments on 2D to enable a fair comparison of solving time.
We have very similar findings for the 3D cases as 2D. For both the direct solvers and iterative solvers, the VP shows the worst performance compared to other meshes, especially when the mesh has low quality (Figure \ref{fig:compare_2d_vs_3d}).
Figure~\ref{fig:3D_poisson} shows that, as in 2D, simplicial meshes are superior to polyhedral meshes, and Voronoi meshes exhibit the largest error.
Other observations summarized above also show similar trends in both 2D and 3D. We thus expect that our 2D observations similarly hold for 3D problems.

%% file: img/tex/solving-time.tex

   \begin{figure}
     \centering\footnotesize
     \parbox{.02\linewidth}{~}\hfill\hfill
     \parbox{.32\linewidth}{\centering $10^{-2}$}\hfill
     \parbox{.32\linewidth}{\centering $10^{-3}$}\hfill
     \parbox{.32\linewidth}{\centering $10^{-4}$}\par
     \parbox{.02\linewidth}{\rotatebox{90}{\centering Direct Solver}}\hfill\hfill
     \parbox{.32\linewidth}{\includegraphics[width=\linewidth]{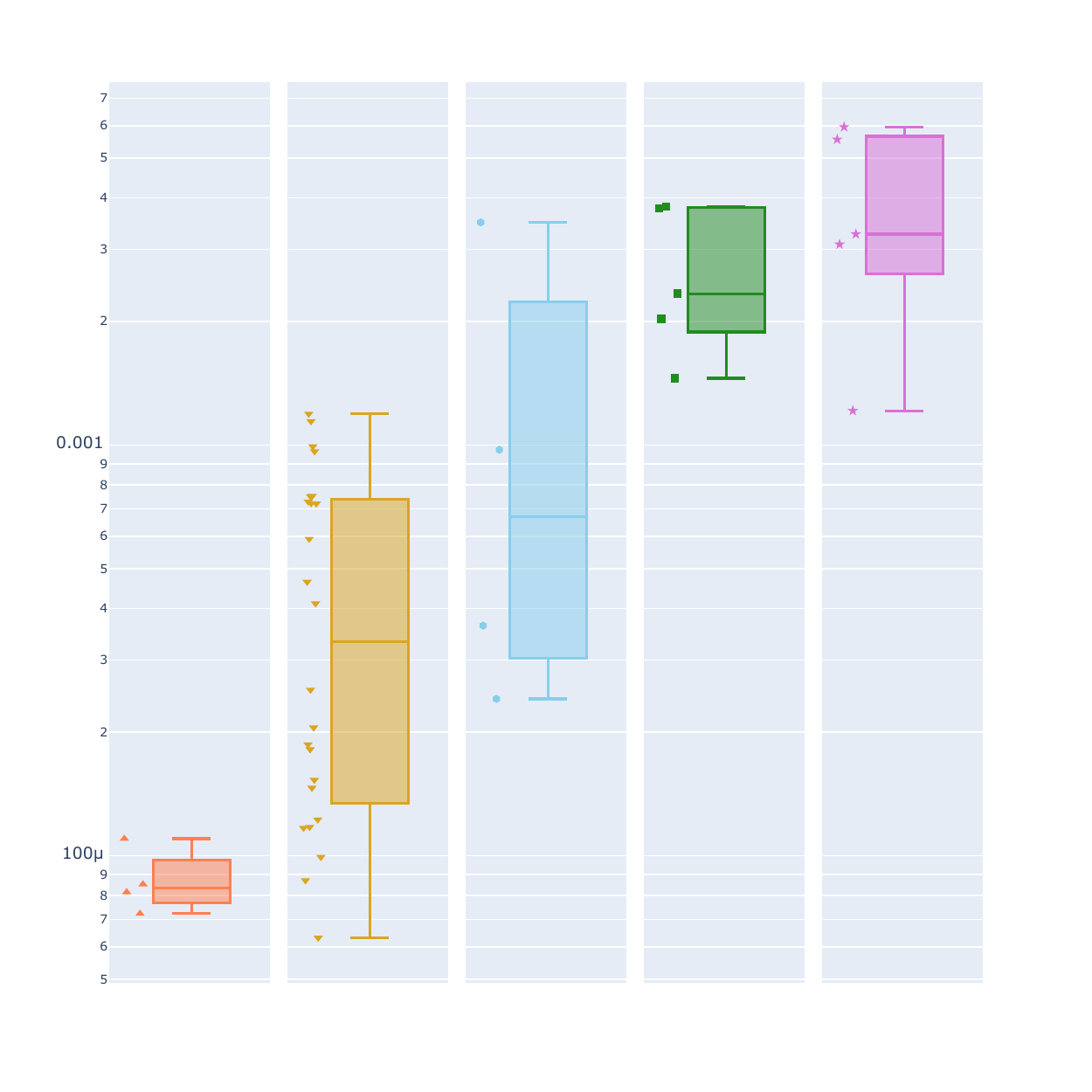}}\hfill
     \parbox{.32\linewidth}{\includegraphics[width=\linewidth]{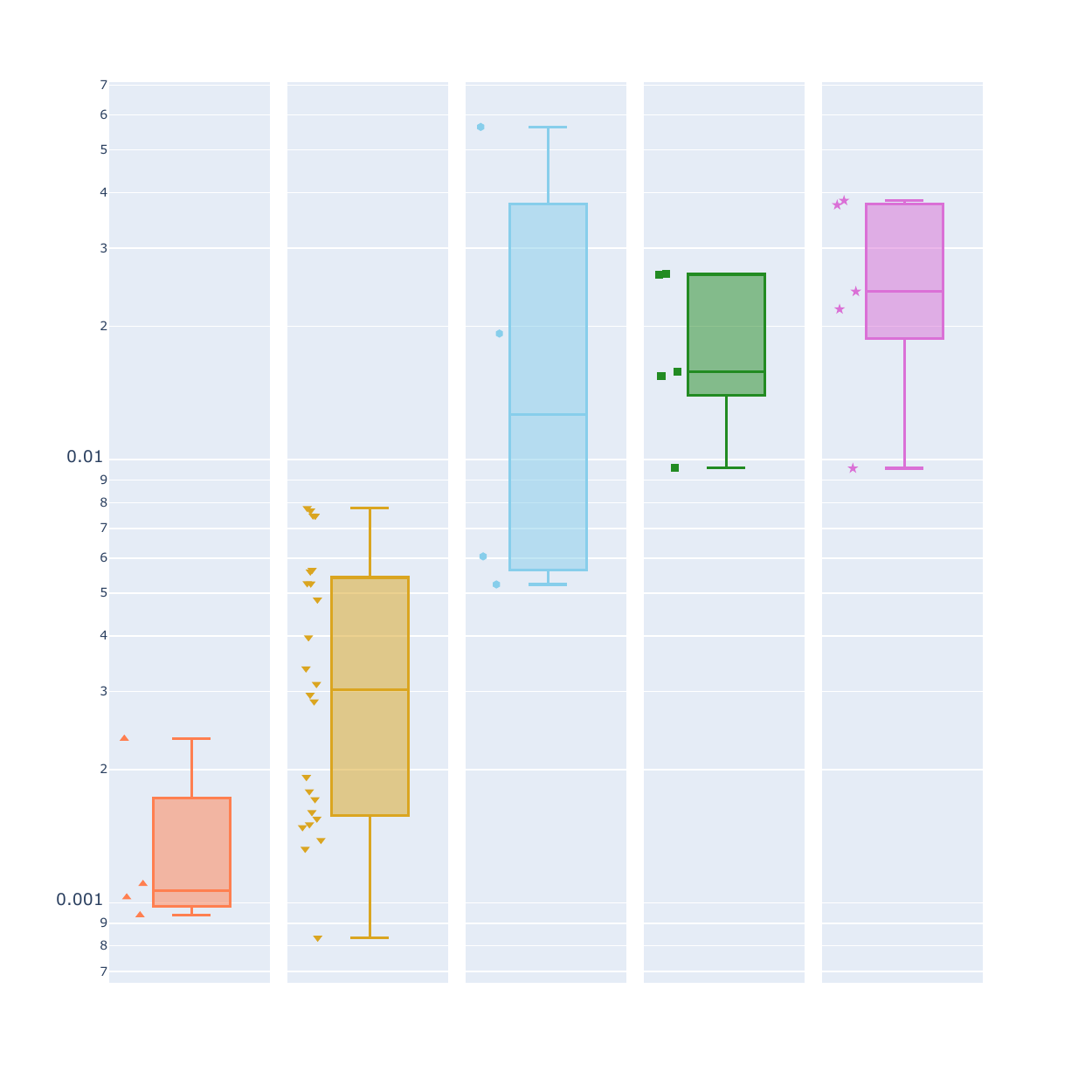}}\hfill
     \parbox{.32\linewidth}{\includegraphics[width=\linewidth]{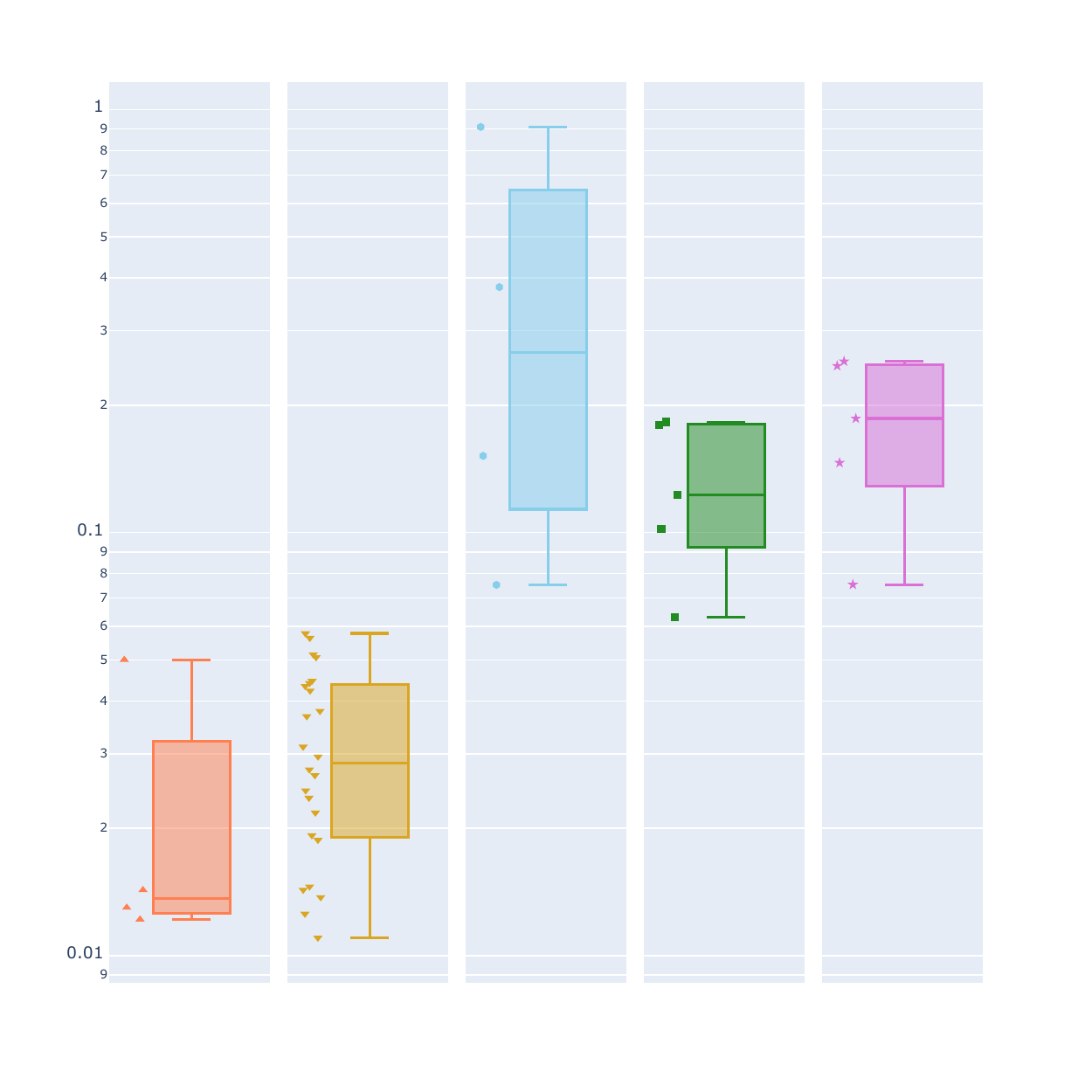}}\par
     \parbox{.02\linewidth}{\rotatebox{90}{\centering Iterative Solver}}\hfill\hfill
     \parbox{.32\linewidth}{\includegraphics[width=\linewidth]{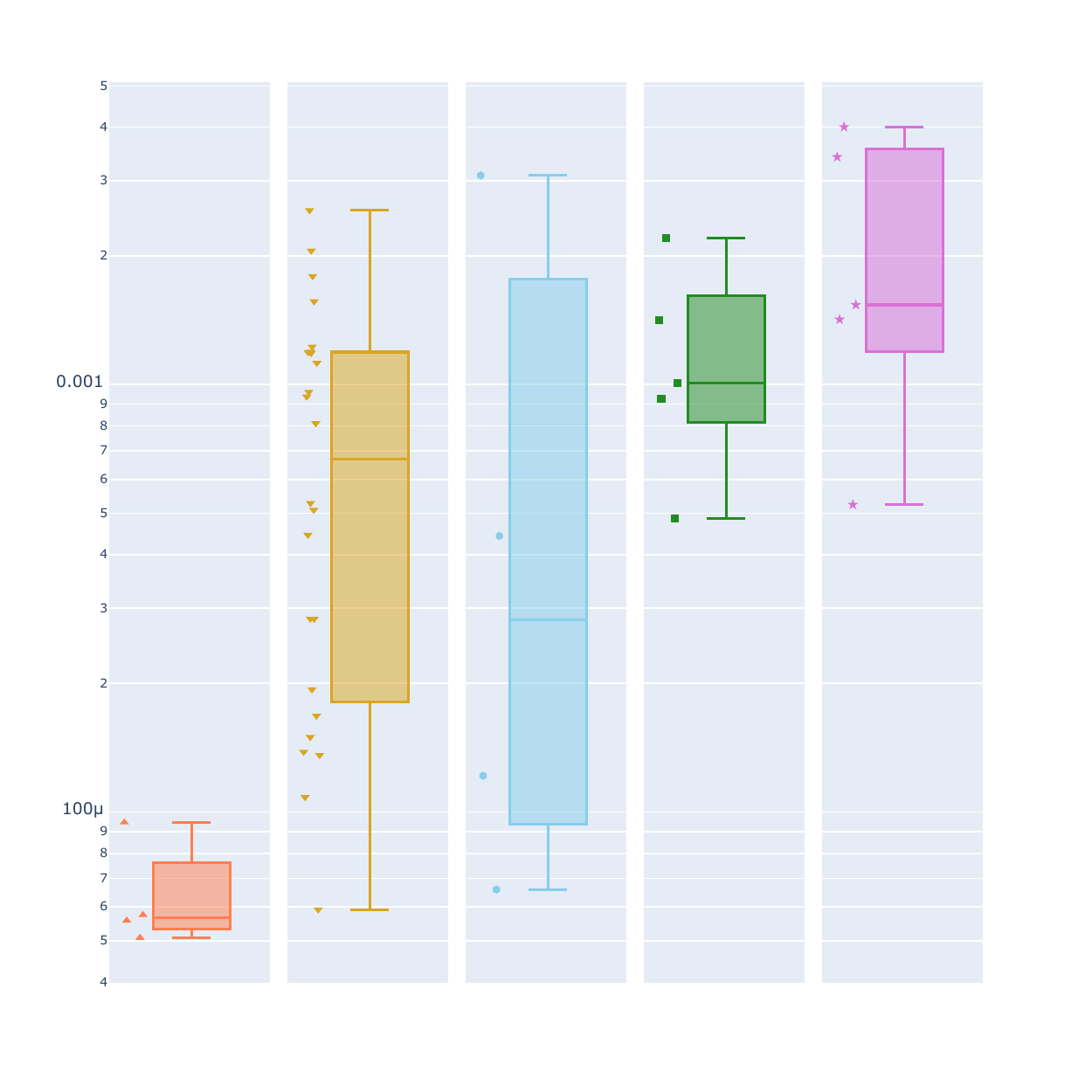}}\hfill
     \parbox{.32\linewidth}{\includegraphics[width=\linewidth]{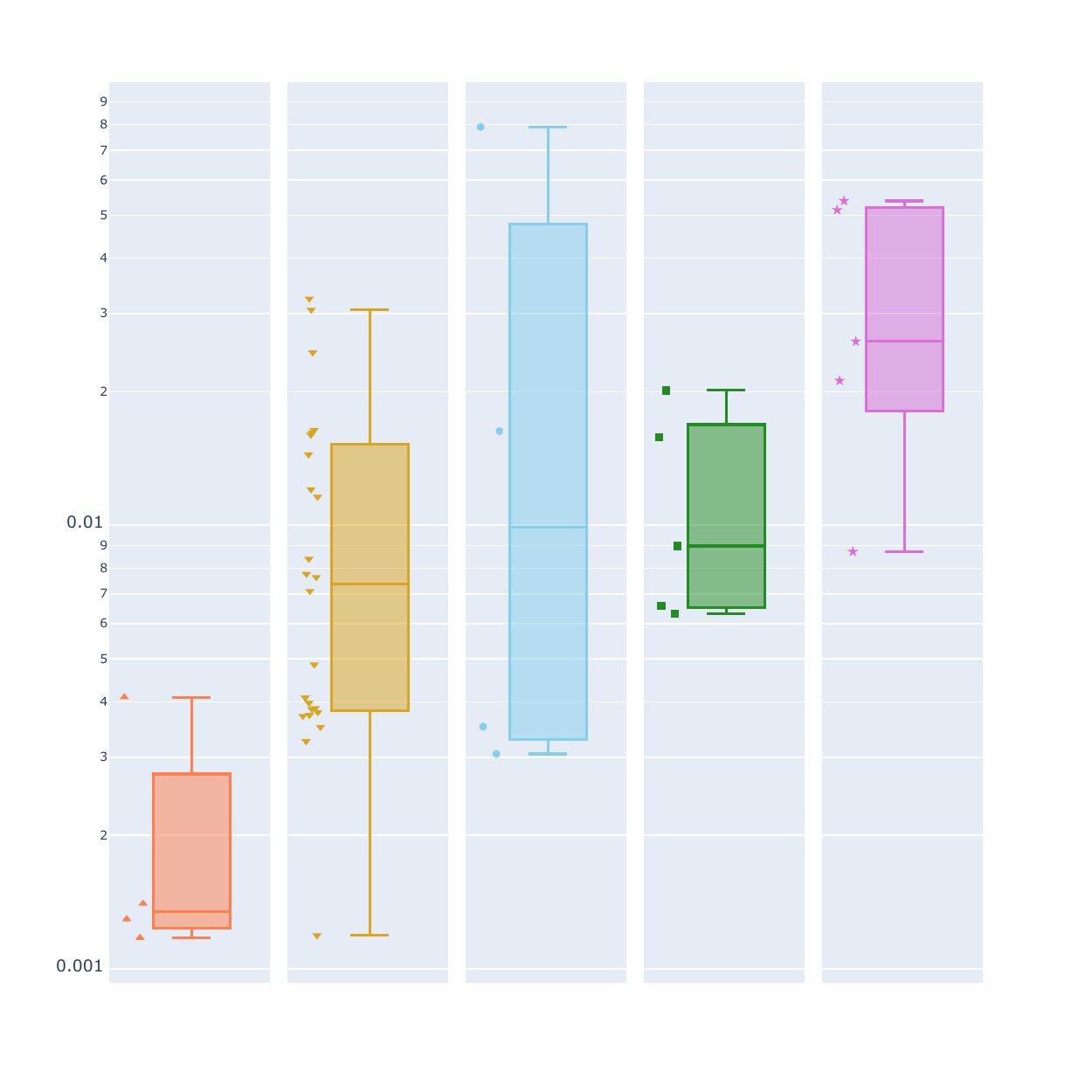}}\hfill
     \parbox{.32\linewidth}{\includegraphics[width=\linewidth]{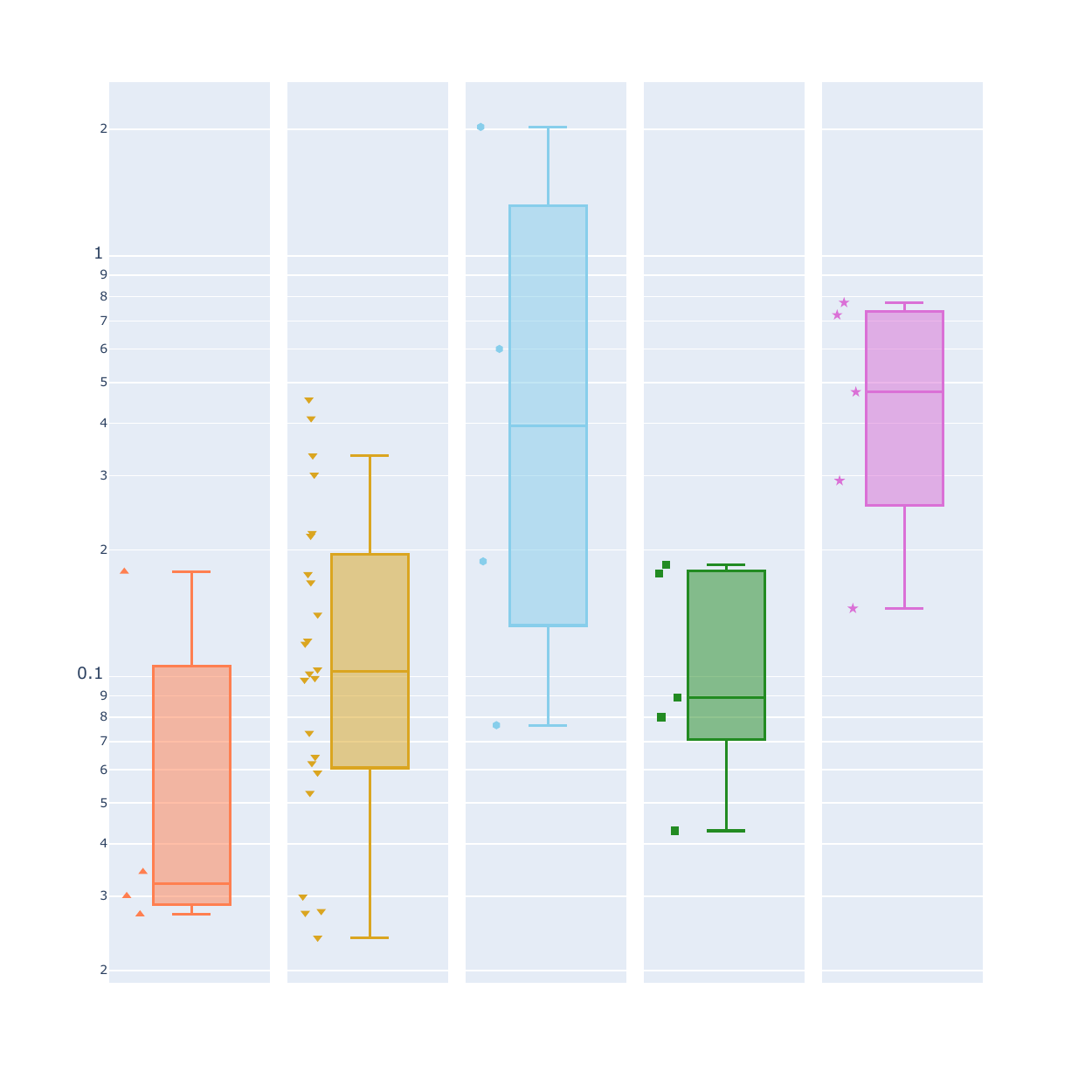}}\par
     \includegraphics[width=0.7\linewidth]{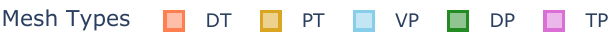}\par
     \caption{Comparison of PS1-SC solving time at 3 error levels with direct solvers and iterative solvers.}
     \label{fig:compare_solver}
 \end{figure}

%% file: img/tex/diff-pde.tex
\begin{figure}
   \centering\footnotesize
   \centering Poisson\par
   \parbox{.49\linewidth}{
      \centering
      Direct Solver\par
      \includegraphics[width=.49\linewidth]{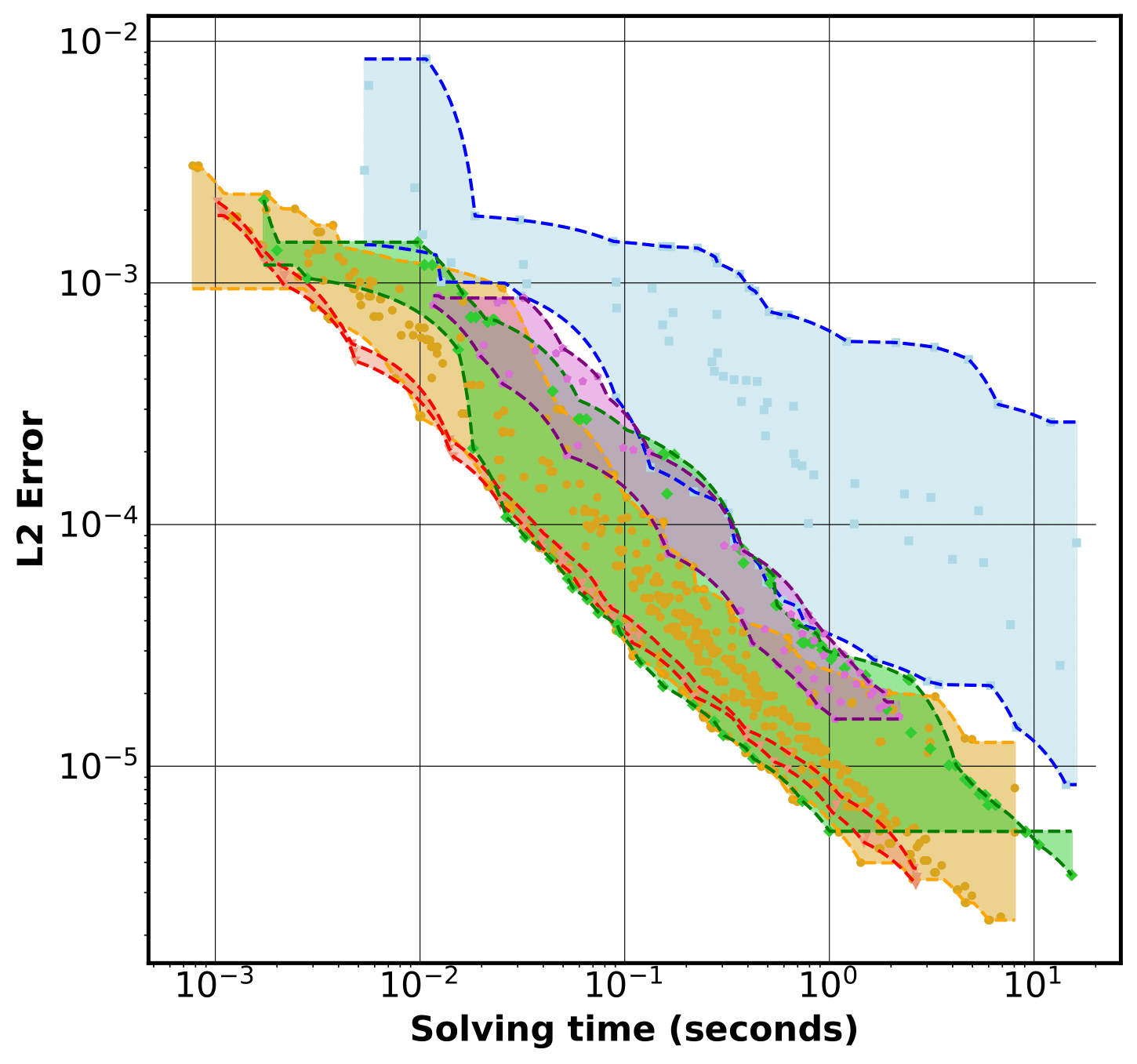}\hfill
      \includegraphics[width=.49\linewidth]{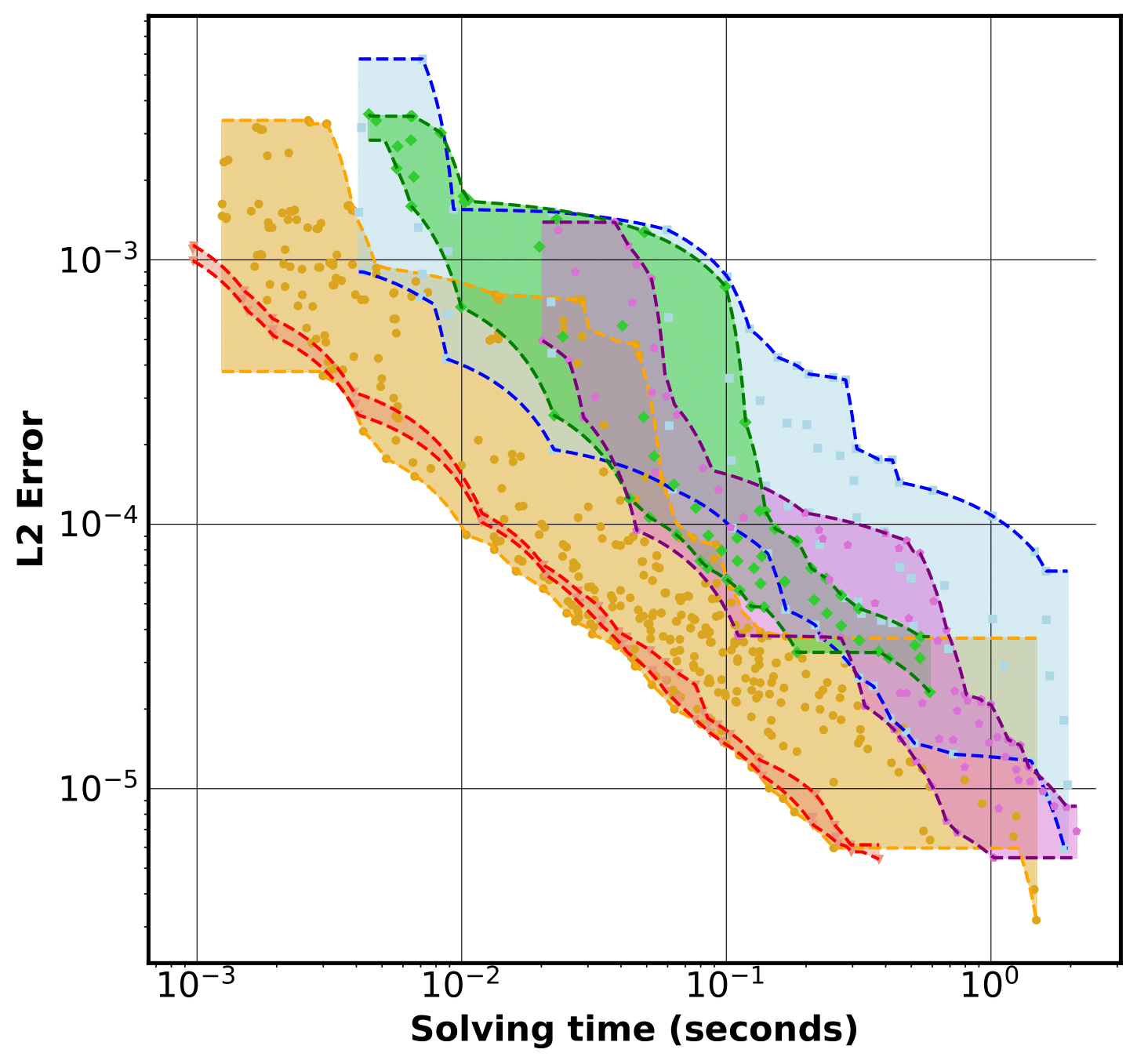}\par
      \parbox{.49\linewidth}{\centering PS\#1-US}\hfill
      \parbox{.49\linewidth}{\centering PS\#1-SC}
   }\hfill
  \parbox{.49\linewidth}{\centering
  Iterative Solver\par
   \includegraphics[width=.49\linewidth]{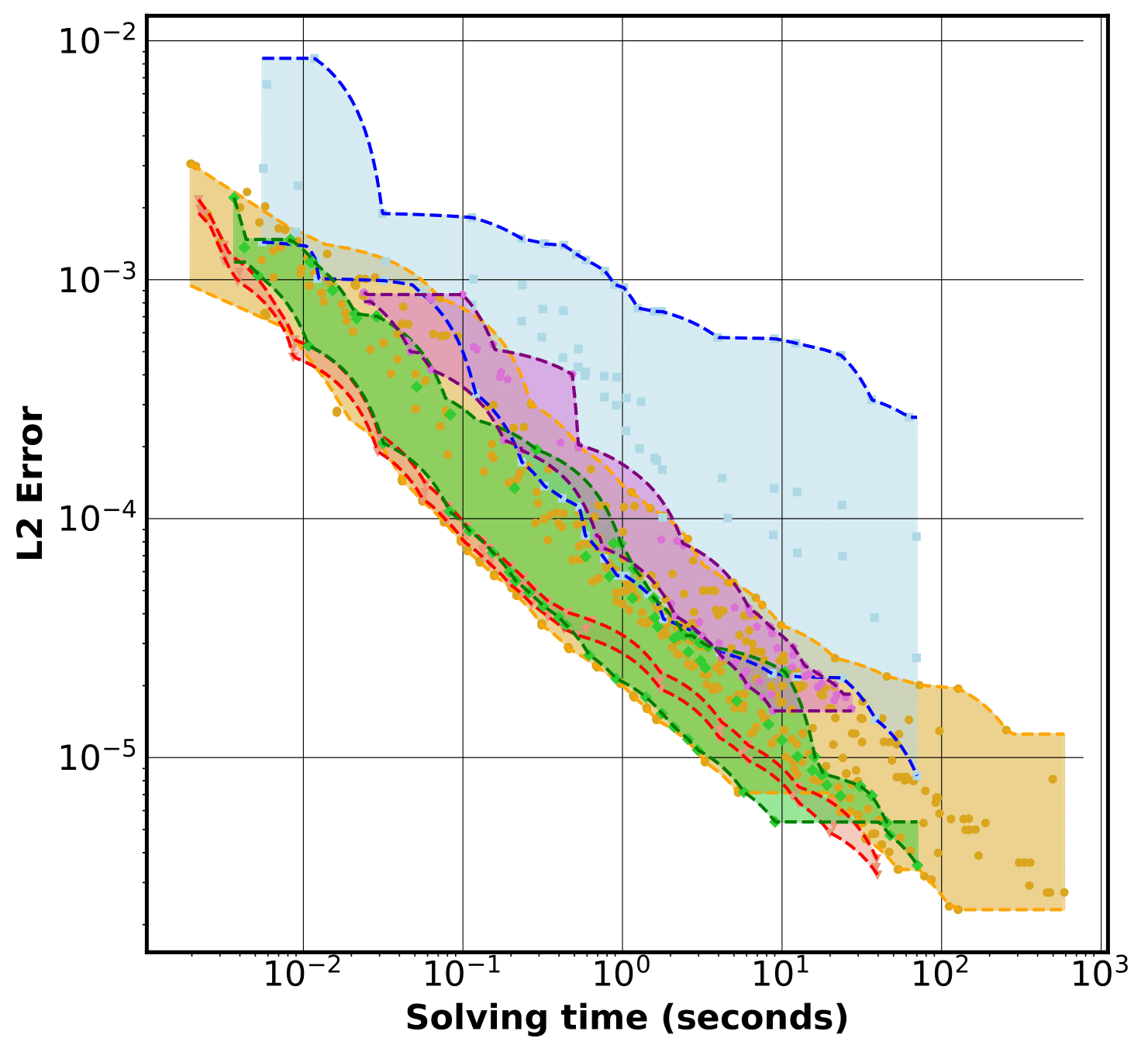}\hfill
   \includegraphics[width=.49\linewidth]{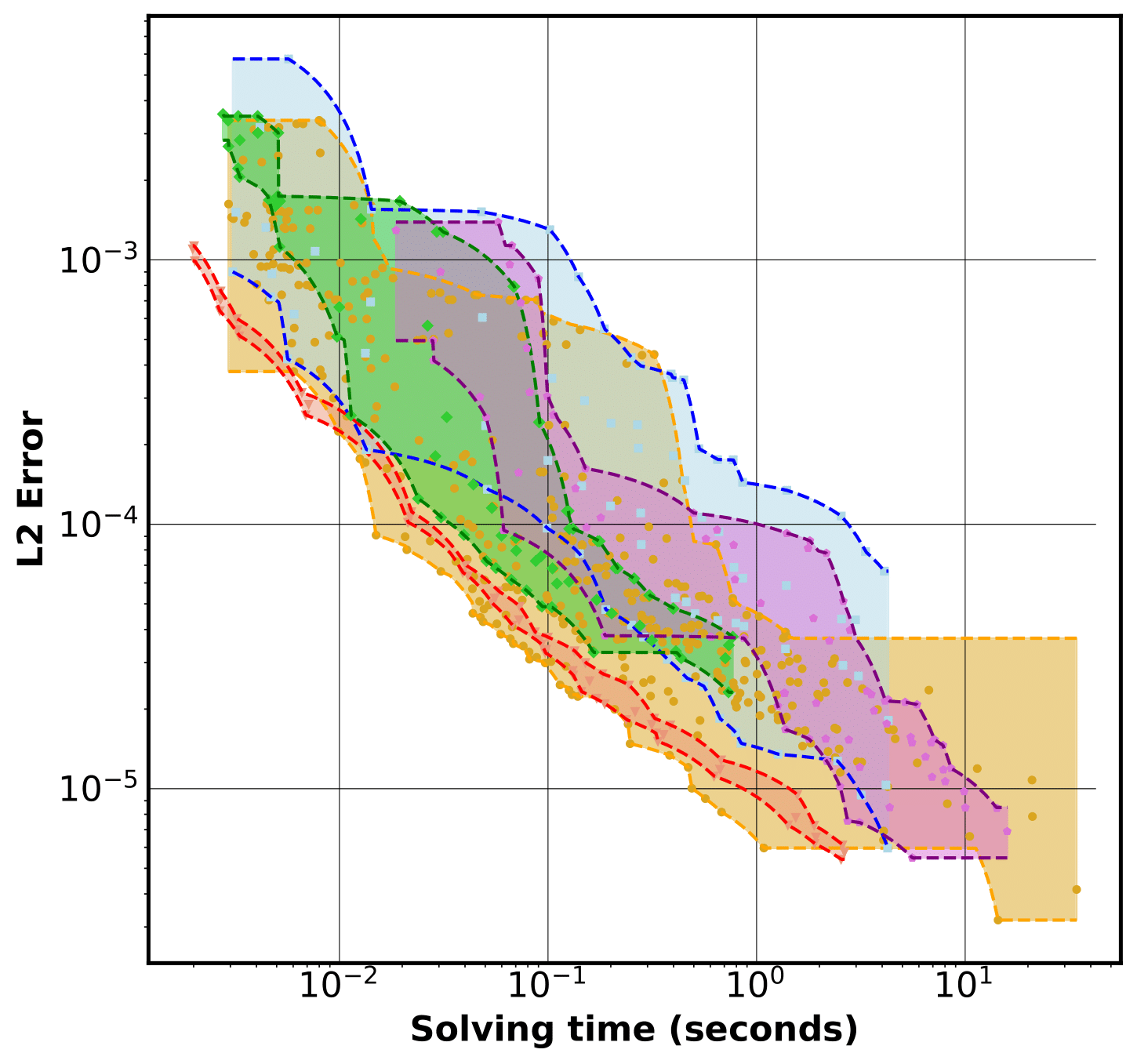}\par
   \parbox{.49\linewidth}{\centering PS\#1-US}\hfill
   \parbox{.49\linewidth}{\centering PS\#1-SC}
   }\\[1em]
  \centering Linear Elasticity\par
  \parbox{.49\linewidth}{\centering
  Direct Solver\par
      \includegraphics[width=.49\linewidth]{img/results_new/VEMLab/direct/plot_all/LEP-BE_L2_norm}\hfill
      \includegraphics[width=.49\linewidth]{img/results_new/VEMLab/direct/plot_all/LEB-PH_L2_norm}
      \parbox{.49\linewidth}{\centering LEP-BE}\hfill
      \parbox{.49\linewidth}{\centering LEB-PH}
  }\hfill
  \parbox{.49\linewidth}{\centering
  Iterative Solver\par
      \includegraphics[width=.49\linewidth]{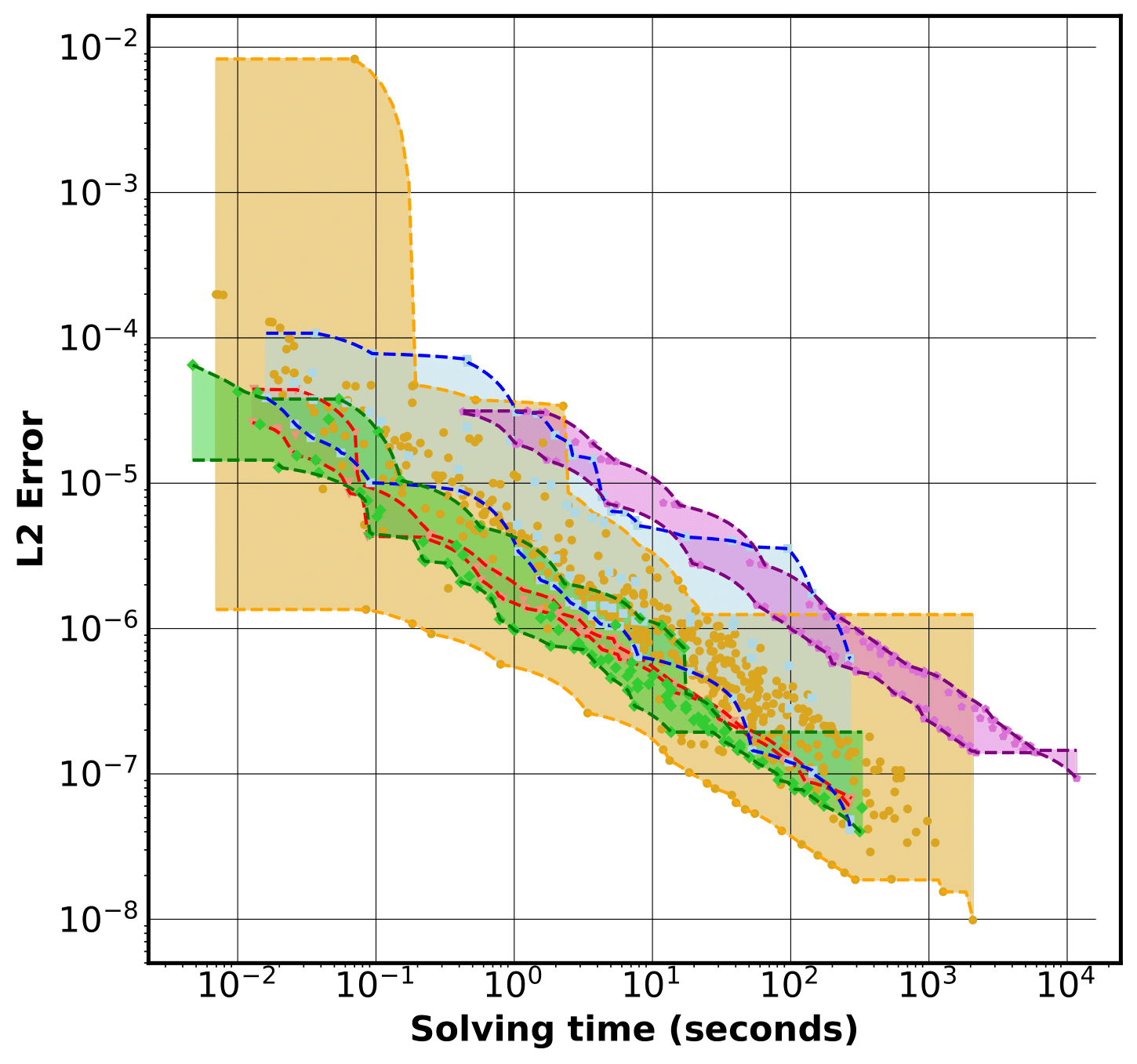}\hfill
      \includegraphics[width=.49\linewidth]{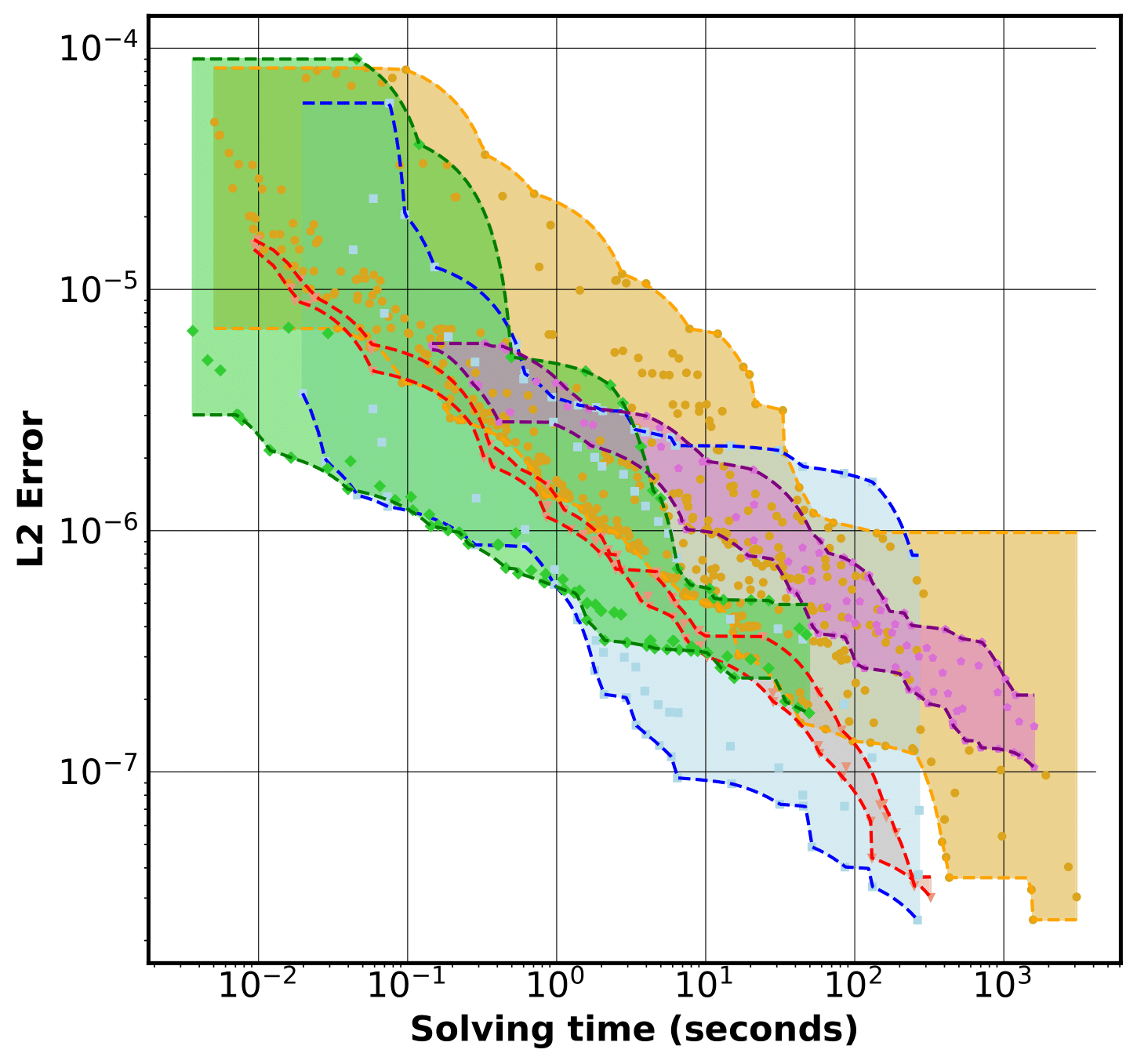}
      \parbox{.49\linewidth}{\centering LEP-BE}\hfill
      \parbox{.49\linewidth}{\centering LEB-PH}
  }\par
  \includegraphics[width=0.7\linewidth]{img/results_new/legend/plot_all}\par
   \caption{Comparison of different PDEs; the $x$-axis shows the time in seconds, while the $y$-axis shows the $L_2$ error.}
   \label{fig:compare_PDEs}
\end{figure}

%% file: 051-complex.tex
\section{Complex Examples}\label{sec:complex}

To showcase the performance and usability of polygonal meshes for real applications, we integrate them with IPC \cite{IPC} to perform elastodynamic simulation of non-linear Neo-Hookean materials with contact and friction. We run all experiments in this section using mean value as bases on an AMD Ryzen Threadripper pro 64 cores 2.1GHz CPU with 512Gb of memory, limiting the execution to 8 threads, and using Pardiso \cite{pardiso-7.2a,pardiso-7.2b,pardiso-7.2c} LDLT direct solver.

Figure~\ref{fig:teaser} left shows that a complex dynamic deformation for a high-quality Voronoi mesh or a tiled mesh is similar to a triangle mesh (see Appendix~\ref{app:complex} for the complete setup). The Voronoi meshes require around 2 times more Netwon iterations, while the tiled mesh requires only half (Figure~\ref{fig:teaser} top right), while the linear solver is about the same speed (Figure~\ref{fig:teaser} bottom right).

An advantage of polygonal meshes is that they can describe the whole domain as a single element. In Figure~\ref{fig:ball}, we simulate a softball hitting an obstacle (see Appendix~\ref{app:complex} for the entire setup). The Delaunay mesh clearly shows catastrophic artifacts coming from the coarseness of the mesh. In this case, both simulations exhibit a similar number of non-linear iterations (Figure~\ref{fig:ball} top right), while Pardiso is about 3.7 times slower as the polygonal mesh leads to a dense matrix (Figure~\ref{fig:ball} bottom right).

\begin{figure}
    \centering\footnotesize
    \parbox{.81\linewidth}{\centering
    \includegraphics[width=.33\linewidth]{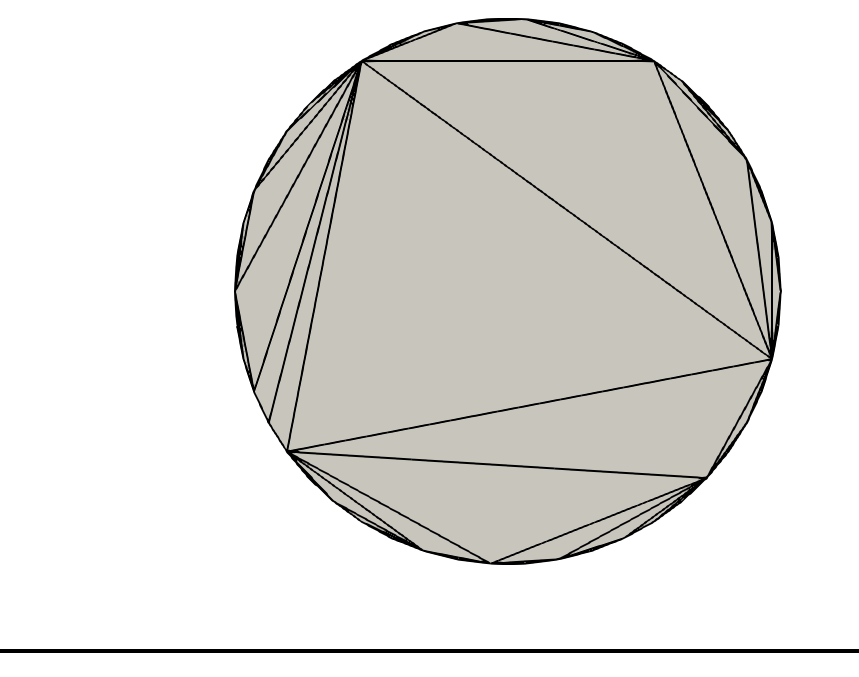}\hfill
    \includegraphics[width=.33\linewidth]{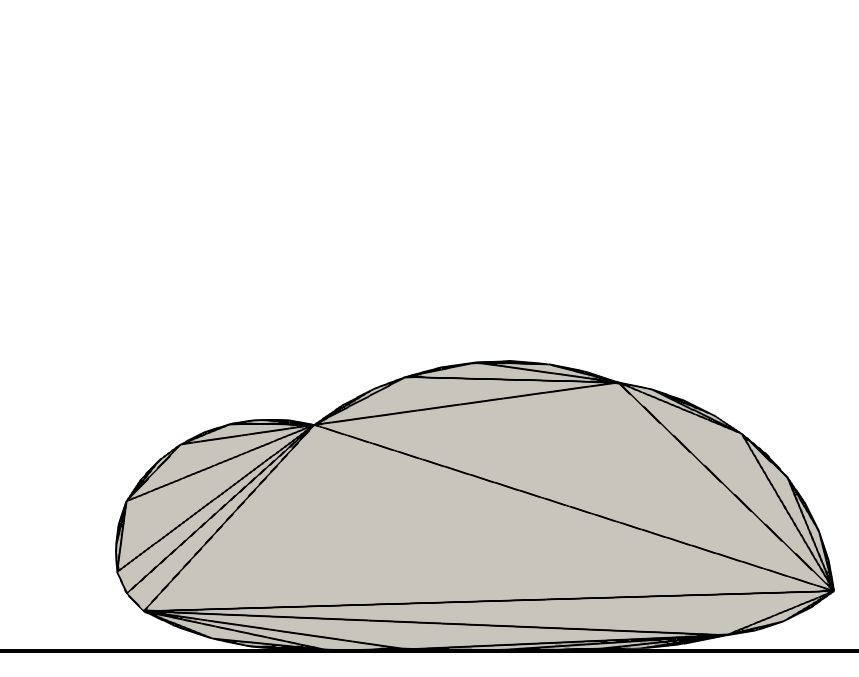}\hfill
    \includegraphics[width=.33\linewidth]{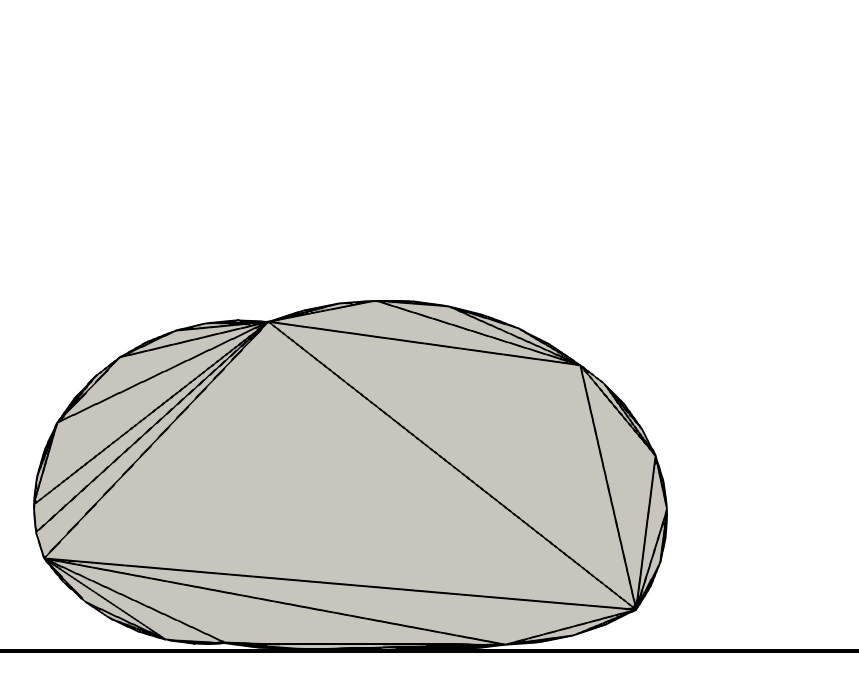}\par
    \includegraphics[width=.33\linewidth]{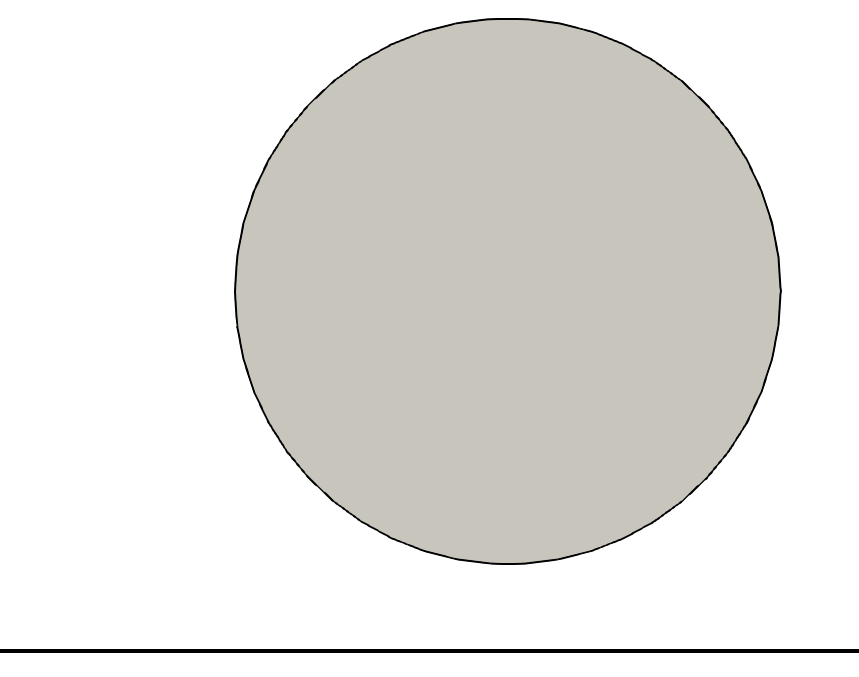}\hfill
    \includegraphics[width=.33\linewidth]{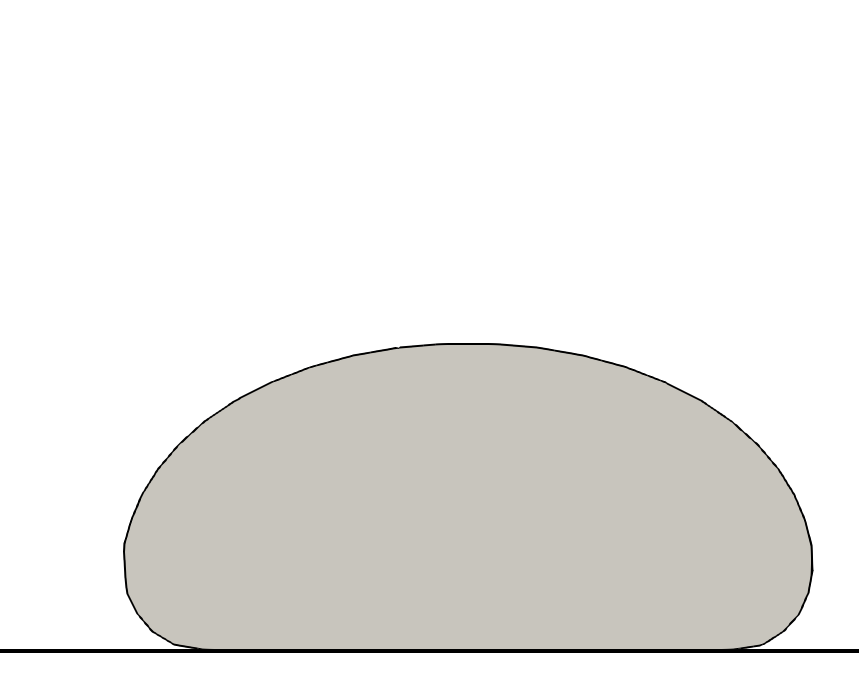}\hfill
    \includegraphics[width=.33\linewidth]{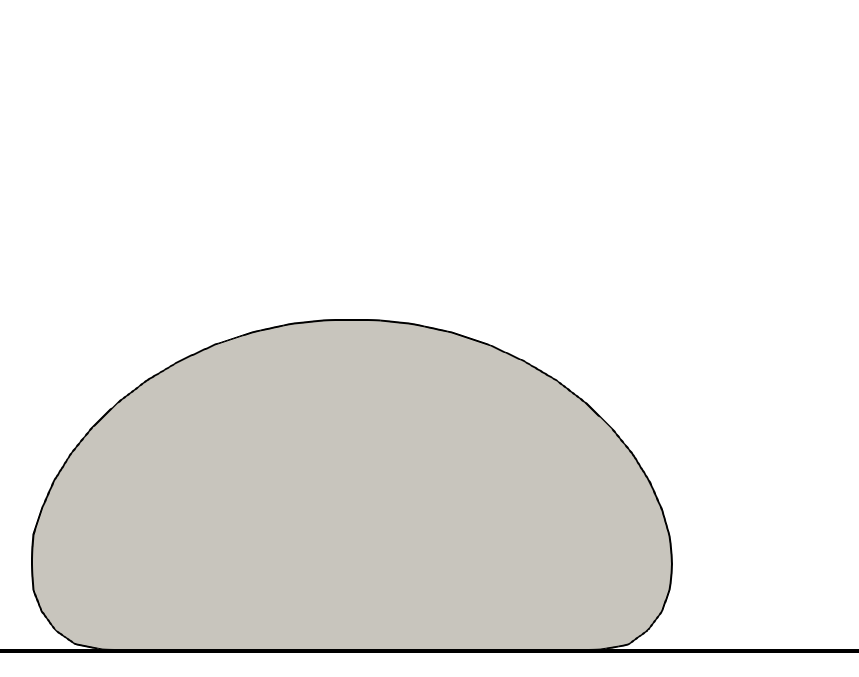}\par
    \parbox{.33\linewidth}{\centering $t=0$\si{\second}}\hfill
    \parbox{.33\linewidth}{\centering $t=0.75$\si{\second}}\hfill
    \parbox{.33\linewidth}{\centering $t=1.5$\si{\second}}
    }\hfill
    \parbox{.18\linewidth}{\centering
    Iterations\\
    \includegraphics[width=\linewidth]{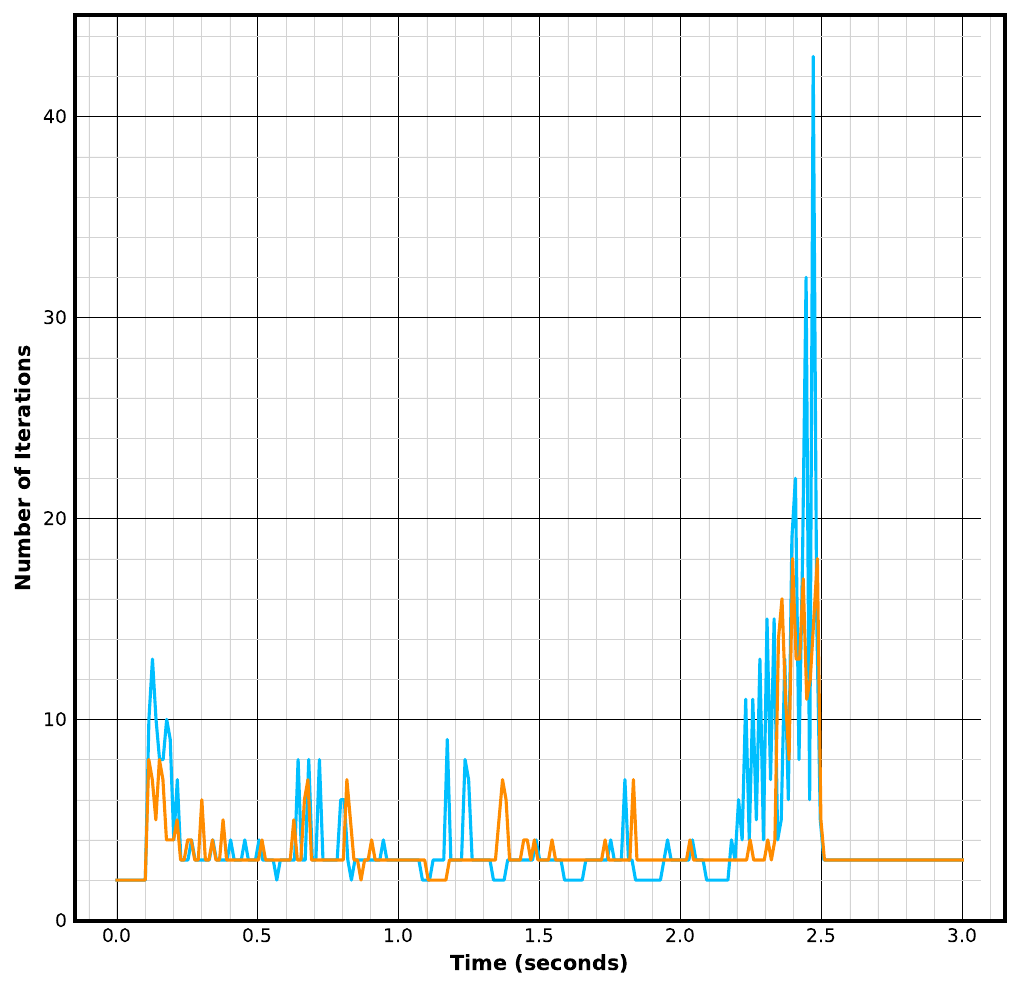}\\[0.5em]
    Time\\
    \includegraphics[width=\linewidth]{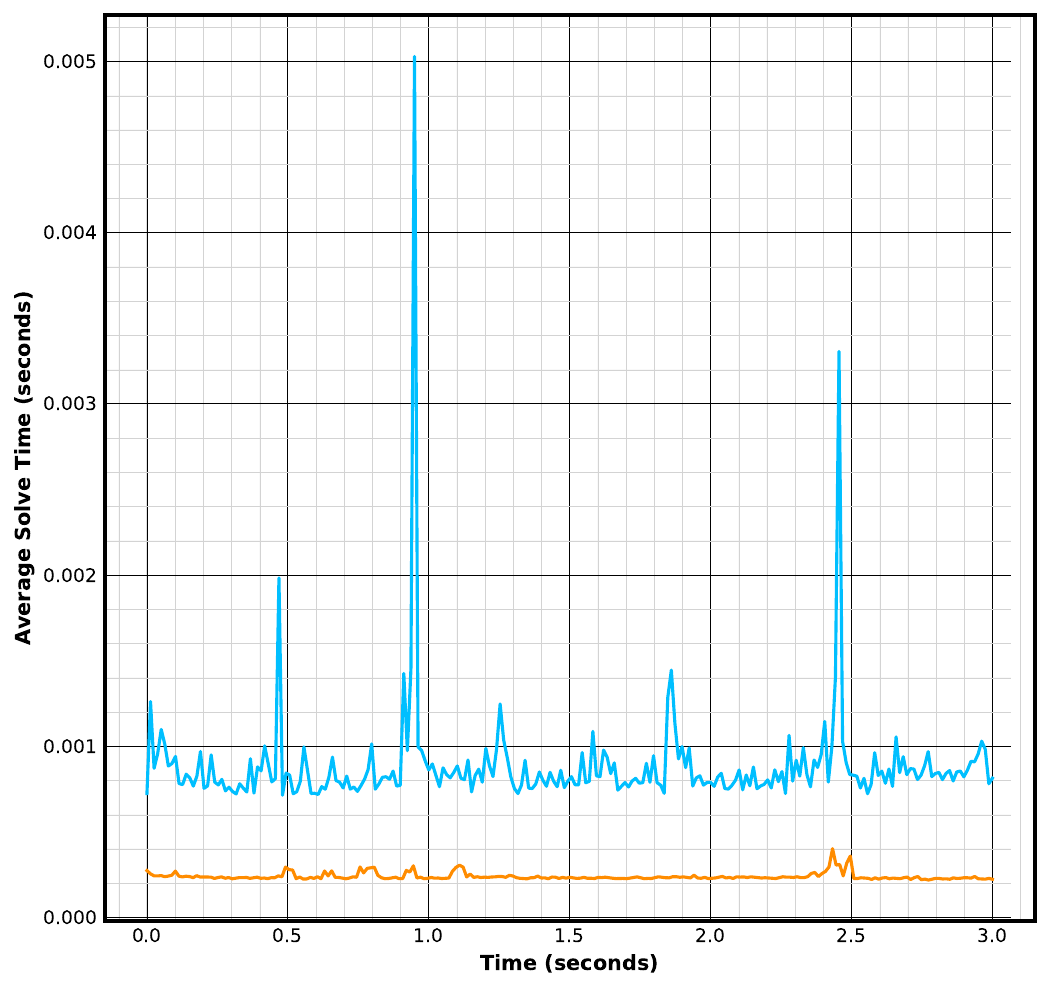}
    }
    \caption{Several frames of a ball piece falling under gravity using a triangular mesh (top left) and a Voronoi mesh (bottom right). Number of iterations (top right) and average solve time (bottom right) for every time-step.}
    \label{fig:ball}
\end{figure}

\begin{figure}
    \centering\footnotesize
    \includegraphics[width=.24\linewidth]{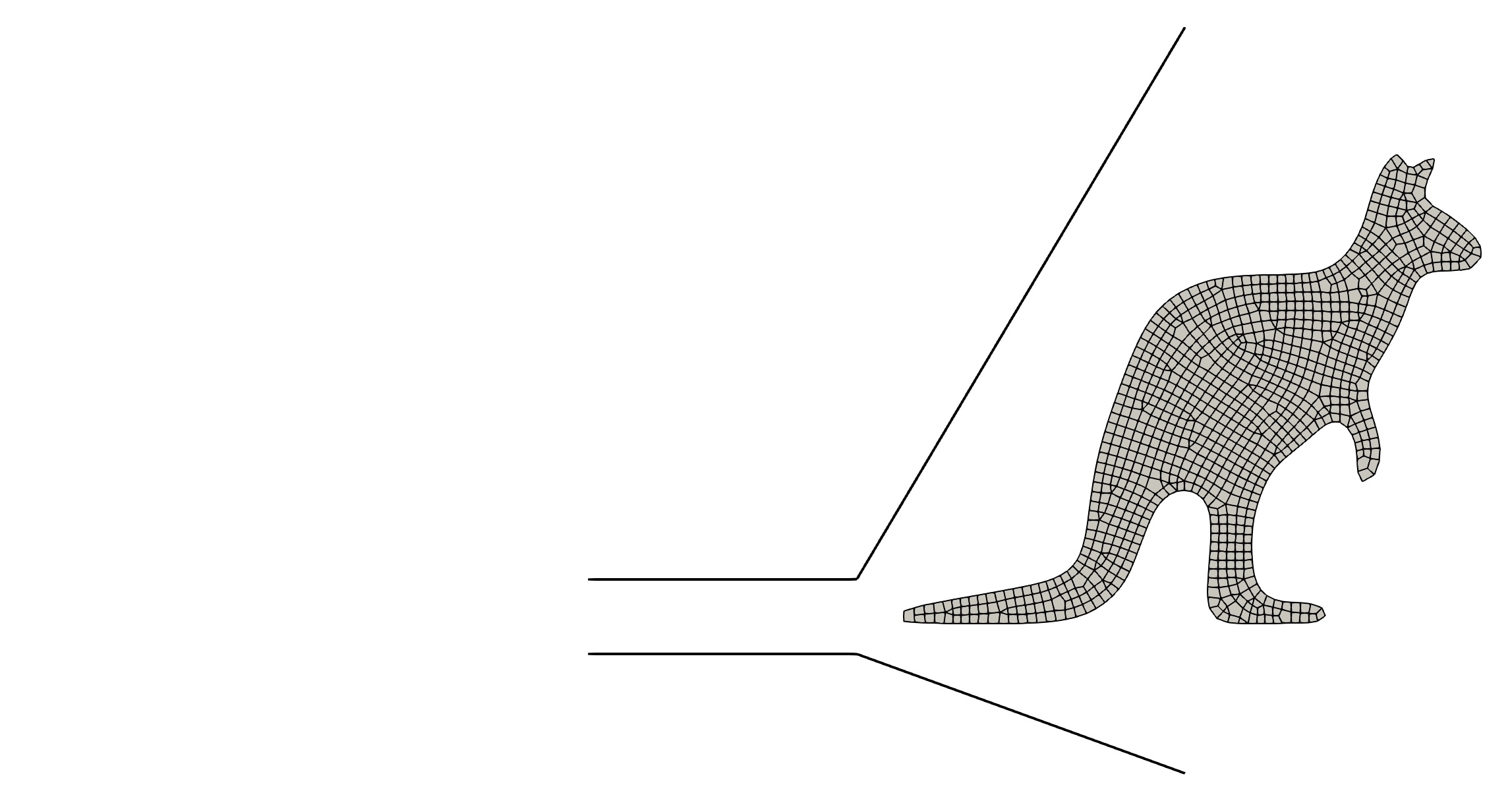}\hfill
    \includegraphics[width=.24\linewidth]{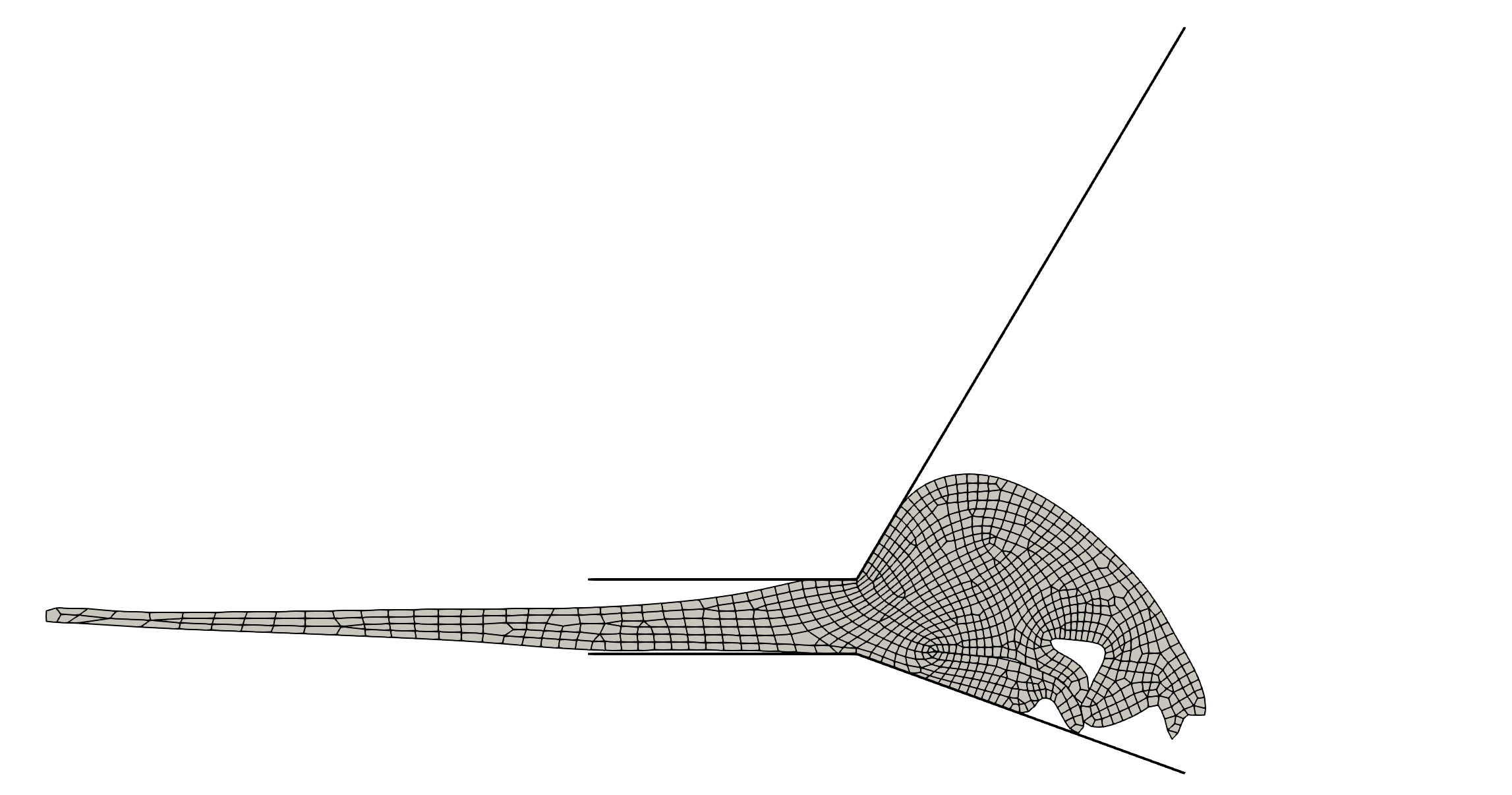}\hfill
    \includegraphics[width=.24\linewidth]{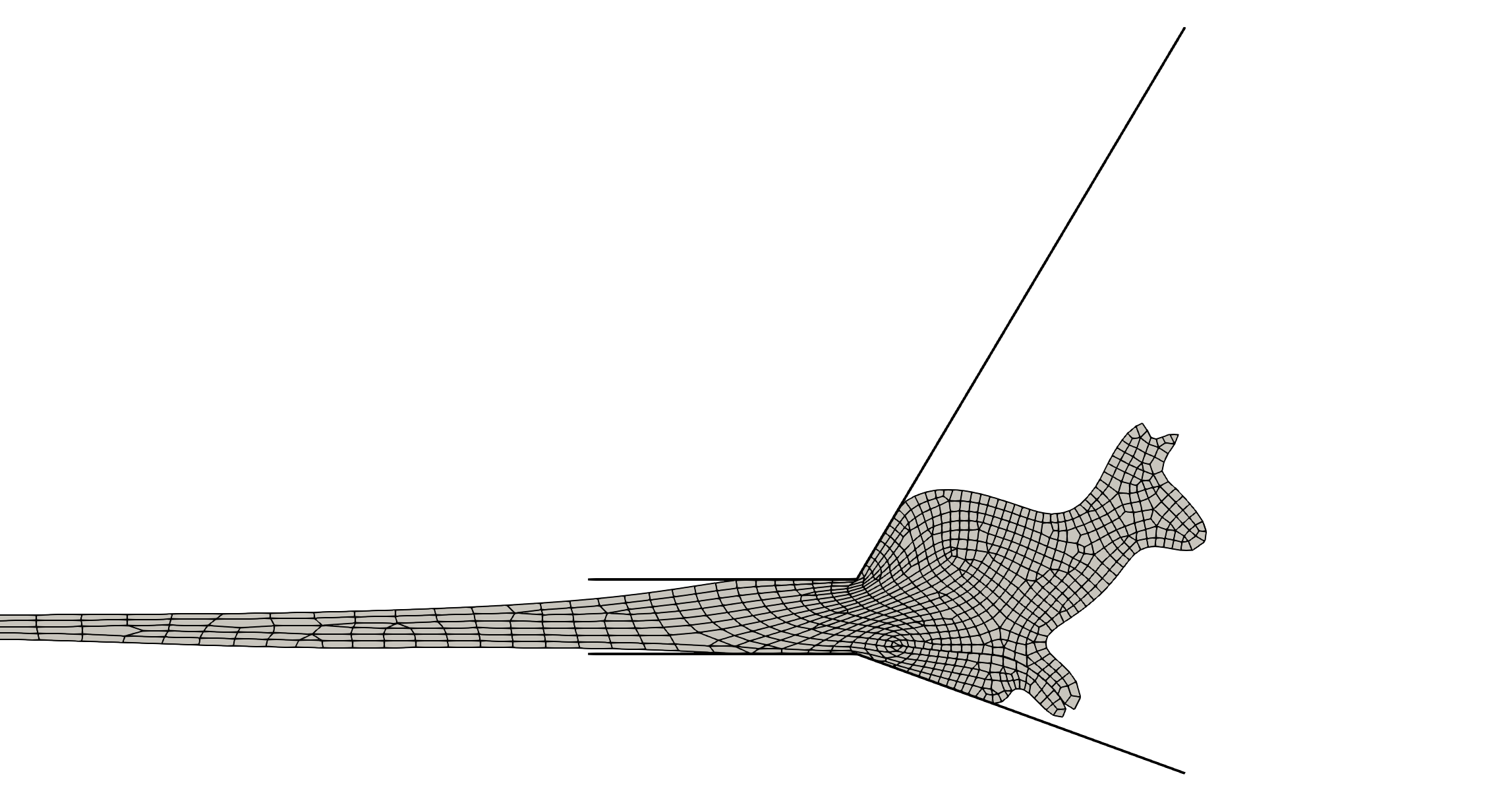}\hfill
    \includegraphics[width=.24\linewidth]{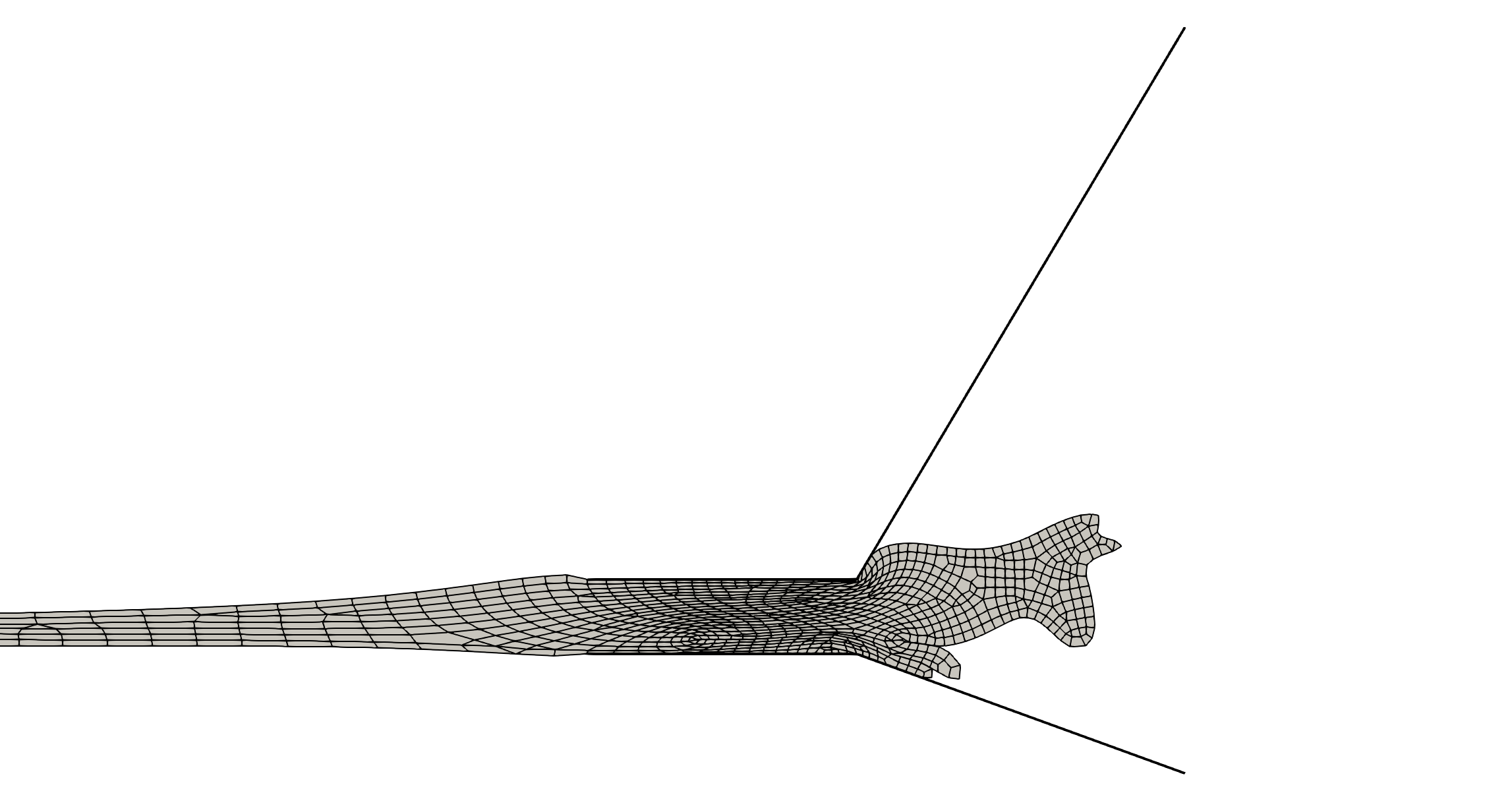}\par
    \parbox{.24\linewidth}{\centering $t=0$\si{\second}}\hfill
    \parbox{.24\linewidth}{\centering $t=0.575$\si{\second}}\hfill
    \parbox{.24\linewidth}{\centering $t=1.15$\si{\second}}\hfill
    \parbox{.24\linewidth}{\centering $t=1.725$\si{\second}}\par
    \caption{Simulation of a kangaroo passing through a funnel.}
    \label{fig:kangaroo}
\end{figure}

Generating pure quadrilateral (or hexahedral) meshes is challenging \cite{Quadratic-Basis}, and allowing the mesh to have a few polygons simplifies the problem. Figure~\ref{fig:kangaroo} shows how a quad-dominant mesh generated with Instant Meshes \cite{Instant-Meshes} can be simulated using MeanValue for the few polygons it contains
(see Appendix~\ref{app:complex} for the full setup).

%% file: 06-concluding.tex
\section{Conclusions}\label{sec:conclusions}

We introduced a benchmark to compare both meshing and analysis methods for polygonal meshes. 

Our conclusion is that, with the current state of the art, the benefit of using polygonal meshes for elliptic PDE is unclear; for most graphics applications, the use of simplicial meshes is favorable, as the meshing technology is more robust and mature, and the performance of both BFEM and VEM is comparable to linear Lagrangian finite elements on a triangular/tetrahedral mesh, which is the most common discretization used in graphics.

There are many exciting avenues to explore for extending our study: (1) consider higher-order barycentric coordinate constructions, such as \cite{Quadratic-Basis}, (2) add additional 3D examples and especially consider very large problems requiring distributed computation, where the different matrix structure might make a performance difference in a bandwidth-limited setting, and (3) consider more complex systems, for example, a surgery simulator, where the connectivity changes in between solves.

%% file: 21-related-addiitonal.tex
\section{Two major approaches in PEMs}
\label{app:two-approaches-PEM}
Generalized Barycentric coordinates are introduced by \citet{Generalized-Barycentric} to define bases on polygons since they reproduce linear polynomials.
Several definitions exist, for instance:
Triangulation \cite{Barycentric-Triangulation}, Wachspress \cite{Barycentric-Wachspress1,Barycentric-Wachspress2}, Sibson and Laplace \cite{Barycentric-Sibson-Laplace1,Barycentric-Sibson-Laplace2}, Mean Value \cite{Barycentric-MeanValue-2D,Barycentric-MeanValue-3D}, Harmonic \cite{Barycentric-Harmonic1,Barycentric-Harmonic2}, Maximum Entropy \cite{Barycentric-Maximum-Entropy}, Moving Least Squares \cite{Barycentric-Moving-Least-Squares}, Surface Barycentric \cite{Barycentric-Surface}, etc. 
These coordinates have different properties; for instance, Mean Value and Harmonic can handle both convex and concave polygons, while Wachspress, Sibson, and Laplace are defined only for convex polygons. The geometric criteria required to estimate the linear error are also different, as a maximum interior angle is necessary for both the Triangulation and Wachspress but optional for Sibson \cite{Barycentric-Error-Estimate1,Barycentric-Error-Estimate2}. 

Mimetic Finite Differences \cite{Mimetic-Finite-Difference,MFEM-elliptic} introduces an algorithm sidestepping the explicit specification of basis functions. Such an idea is generalized into the Virtual Element Method (VEM) \cite{VEM}, which is combined with the Galerkin framework, becoming an ultimate generalization of standard FEM enabling the use of arbitrary polygonal or polyhedral meshes \cite{FEM-vs-VEM-engineering}. 
The essential difference of VEM when compared to the FEM-based polygonal/polyhedral methods is that the local shape function space in each element is defined implicitly and does not need to be determined or evaluated in practice \cite{VEM-50lines-code}. Instead, these virtual functions are processed and used solely by specific defining properties of the element space and their corresponding degrees of freedom. Together with the discrete bilinear form, they are selected with care to ensure the direct and exact computation of the stiffness matrix \cite{FEM-vs-VEM-elasticity}. 
VEM has been proven to be robust to mesh distortion \cite{VEM-distortion-robust}, which saves the time for constant re-meshing and refinement process, and, as many polygonal methods, it can directly handle non-conforming discretizations \cite{FEM-vs-VEM-engineering}. 
However, VEM inevitably requires more complex calculations when assembling the stiffness matrix \cite{FEM-vs-VEM-elasticity} and often needs stabilization techniques to ensure steady performance \cite{VEM-stability}.

\section{Comparisons Between Different Elements of FEM in Specific Physical Problems}
\label{app:compare-fem-physical}
\citet{tet-vs-hex-simple-structure} conclude that quadratic tetrahedral meshes show similar accuracy and time as linear hexahedral meshes in simple structural problems. \citet{tet-vs-hex-elastic} reports the superiority of linear hexahedral meshes to linear tetrahedral meshes on elastoplastic experiments. More recent works like \citet{tet-vs-hex-footwear} focus on footwear modeling with nonlinear incompressible materials under shear loads to show that trilinear hexahedral meshes have superior performance than linear tetrahedral meshes, and conclude that quadratic tetrahedral meshes have more expensive computation than trilinear hexahedral elements but with higher accuracy. \citet{tet-vs-hex-linearstatic} conduct experiments on linear static problems with the conclusion that quadratic hexahedral meshes require more expensive computation to achieve similar accuracy with quadratic tetrahedral meshes.

\section{The Impact of Polygonal/polyhedral Meshes on the Performance of VEM solutions}
\label{app:poly-impact-vem}
\citet{VEM-quality-indicator} investigates the convergence rates of VEM on 2D polygonal meshes with different levels of regularity and proposes a quality indicator that correlates mesh regularity with the performance of a VEM solution. The experiments show that VEM can have an almost optimal convergence rate even with a significant breaking of regularity assumptions (e.g., adding randomly shaped concave elements into meshes).  The following work, \cite{Polyheral-quality-indicator}, expands a similar investigation into 3D polyhedral meshes, showing that VEM can also have a decent convergence rate on irregular polyhedral meshes.
\citet{VEM-quality-benchmark} proposes a benchmark to address the correlation between general 2D polygonal meshes and VEM solvers on Poisson equations. This study provides an in-depth exploration of shape regularity by using 17 geometric metrics on meshes that are composed of parametric polygons and random polygons. The results show that although no single metric can drive the correlation, a combination of several metrics in this benchmark can play relevant roles in indicating the VEM solution accuracy. 

\section{Finite Difference Methods in Stokes Problems}

Orthogonal to our work on finite element methods on polyhedral elements, we also briefly survey the extensive literature on Finite Difference Methods \cite{FDM1,FDM2,FDM3} (FDMs) on general polygonal meshes, mainly in stokes problems and fluid simulation. FDMs only use surface representations of the discretization to construct the stiffness matrices of the linear systems, instead of extending inside the elements and dealing with the challenges of choosing basis functions in FEMs \cite{MFD-stoke}. Such characteristics enable FDMs to smoothly handle general polygonal and polyhedral meshes, which are commonly used in stokes-related problems, including diffusion \cite{FDM-diffusion}, convection-diffusion \cite{FDM-convection-diffusion}, and fluid flows \cite{FDM-flow-1,FDM-flow-2,FDM-flow-3}. These stokes problems require meshing elements that have more DOFs (general polygonal/polyhedral meshes) than general simplicial (triangular) meshes and quadrilateral meshes since simplicial meshes could often lead to degenerate elements (i.e., elements that have 180 degrees between faces) in thinning stokes layers \cite{MFD-stoke}. 

%% file: 42-problems.tex
\section{PDE Definition}
\label{sec:pde-definition}

Given a domain $\Omega\subset\mathbb{R}^d, d\subset\{2,3\}$ with boundary $\partial\Omega$, our aim is to solve:
\begin{equation}
    \begin{aligned}
        F(x,u,\nabla u,D^2u)=b,\text{ subject to}\\
        u=d\text{ on }\partial\Omega_D \text{  and  } \nabla u \cdot n = f\text{ on }\partial\Omega_N
    \end{aligned}
\end{equation}
where $D^2$ is the second derivative matrix, $b$ is the right-hand side, $\partial\Omega_D\in\partial\Omega$ is the Dirichlet boundary conditions, and $\partial\Omega_N\in\partial\Omega$ is the Neumann boundary conditions (with $\partial\Omega_D \cap \partial\Omega_N = \emptyset$).

In our case study, we use the Poisson and linear elasticity PDE to test the performance of different meshing; thus
\[
F(x,u,\nabla u,D^2u) = -\Delta u,
\]
for the Poisson equation,  while for elasticity
\[
F(x,u,\nabla u,D^2u) = \Div \sigma[u],
\]
with
\begin{equation*}
\begin{aligned}
\sigma[u] &= 2\mu\epsilon[u] + \lambda \tr \epsilon [u] I,\\
\epsilon[u] &= 1/2(\nabla u^T + \nabla u),
\end{aligned}
\end{equation*}
where $\epsilon[u]$ is the strain tensor, $\mu$ the shear modulus, and $\lambda$ the first Lame parameter.

\paragraph{Bases}\label{sec:based}
For simplicial meshes, we use standard (linear and quadratic) Lagrange bases, while for polygons, we use VEM and barycentric coordinates. We use VEMLab~\cite{vemlab} for simplicial and VEM bases, while we run PolyFEM~\cite{polyfem} for barycentric and higher order Lagrange bases. As the two codebases are different (and use different languages), we export the matrices from PolyFEM and use the \emph{same} Matlab solver as for VEM. Since error computations require knowledge about the bases, each individual library is responsible to compute them (and we confirmed using simplicial meshes, supported by both libraries, that the errors match).

\section{Complex examples setup}\label{app:complex}

For all simulations, we use $\partial t = 0.025$\si{\second}; for the puzzle and ball, we simulate to $6$\si{\second}, while the kangaroo stops at $2.2$\si{\second}.
All materials have a density $1000$ \si{kg/m^3} and use the NeoHookean model. The puzzle and ball are soft materials with $E=2\times 10^4$\si{\pascal}; the kangaroo is a bit stiffer and has $E=2\times 10^5$\si{\pascal}. All materials have the same Poisson ratio of $0.2$. The puzzle and ball only have a body force corresponding to gravity, while the kangaroo is pulled to the left with a Dirichlet condition of $-5t$.

All simulations share the same solver settings: $10^{-5}$ gradient norm for the Newton tolerance and $\hat d = 0.001$\si{\meter} for IPC.

%% file: 41-tables.tex
\section{Benchmark Study}\label{append:sec:dataset-tables}

We use tables to demonstrate the notations for our large-scale benchmark study. The detailed description of the dataset is in Section~\ref{sec:dataset} and our case study results that utilize these notations are in Section~\ref{sec:results}

\begin{table}[H]
    \centering 
    \begin{tabular}{|c|c|c|c|c|c|c|c|}
        \hline
             & PS\#1 & PS\#2 & PS\#3 & PB \\ \hline
        US   & PS\#1-US & PS\#2-US & PS\#3-US & PB-US \\ \hline
        UD   & PS\#1-UD & PS\#2-UD & PS\#3-UD & PB-UD \\ \hline
        BE   & PS\#1-BE & PS\#2-BE & PS\#3-BE & PB-BE \\ \hline
        PH   & PS\#1-PH & PS\#2-PH & PS\#3-PH & PB-PH \\ \hline
        SC   & PS\#1-SC & PS\#2-SC & PS\#3-SC & PB-SC \\ \hline
        LS   & PS\#1-LS & PS\#2-LS & PS\#3-LS & PB-LS \\ \hline
    \end{tabular}
    \caption{Combinations of Poisson problems and domains} 
    \label{tab:poisson-problem-domain}
\end{table}

\begin{table}[H]
    \centering 
    \begin{tabular}{|c|c|c|c|c|c|c|c|}
        \hline
             & LEP & LEB \\ \hline
        US   & LEP-US & LEB-US \\ \hline
        UD   & LEP-UD & LEB-UD \\ \hline
        BE   & LEP-BE & LEB-BE \\ \hline
        PH   & LEP-PH & LEB-PH \\ \hline
        SC   & LEP-SC & LEB-SC \\ \hline
        LS   & LEP-LS & LEB-LS \\ \hline
    \end{tabular}
    \caption{Combinations of Elasticity problems and domains} 
    \label{tab:elasticity-problem-domain}
\end{table}

\begin{table}[H]
    \centering 
    \begin{tabular}{|c|c|c|c|c|}
        \hline
            & VP1     & VP2     & VP3     & VP4 \\ \hline
        DT1 & VP1-DT1 & VP2-DT1 & VP3-DT1 & VP4-DT1 \\ \hline
        DT2 & VP1-DT2 & VP2-DT2 & VP3-DT2 & VP4-DT2 \\ \hline
        DT3 & VP1-DT3 & VP2-DT3 & VP3-DT3 & VP4-DT3 \\ \hline
        DT4 & VP1-dT4 & VP2-DT4 & VP3-DT4 & VP4-DT4 \\ \hline
\end{tabular}
    \caption{Triangulations via the dual of Voronoi tessellations} 
    \label{tab:mesh:voro-dual}
\end{table}

\begin{table}[H]
    \centering 
    \begin{tabular}{|c|c|c|c|c|}
        \hline
            & VP1     & VP2     & VP3     & VP4 \\ \hline
        PT1 & VP1-PT1 & VP2-PT1 & VP3-PT1 & VP4-PT1 \\ \hline
        PT2 & VP1-PT2 & VP2-PT2 & VP3-PT2 & VP4-PT2 \\ \hline
        PT3 & VP1-PT3 & VP2-PT3 & VP3-PT3 & VP4-PT3 \\ \hline
        PT4 & VP1-PT4 & VP2-PT4 & VP3-PT4 & VP4-PT4 \\ \hline
    \end{tabular}
    \caption{Triangulations within elements of Voronoi tessellations} 
    \label{tab:mesh:voro-tri}
\end{table}

\begin{table}[H]
    \centering 
    \begin{tabular}{|c|c|c|c|c|c|}
        \hline
        & DP1 & DP2 & DP3 & DP4 & DP5 \\ \hline
        PT1 & DP1-PT1 & DP2-PT1 & DP3-PT1 & DP4-PT1 & DP5-PT1 \\ \hline
        PT2 & DP1-PT2 & DP2-PT2 & DP3-PT2 & DP4-PT2 & DP5-PT2 \\ \hline
        PT3 & DP1-PT3 & DP2-PT3 & DP3-PT3 & DP4-PT3 & DP5-PT3 \\ \hline
        PT4 & DP1-PT4 & DP2-PT4 & DP3-PT4 & DP4-PT4 & DP5-PT4 \\ \hline
    \end{tabular}
    \caption{Triangulations within elements of Displacement polygon meshing} 
    \label{tab:mesh:dp-tri}
\end{table}